%% file: thesis.tex
\documentclass[a4paper,10pt]{report}

\usepackage{natbib}
\usepackage{graphicx}
\usepackage{fullpage}
\usepackage{multirow}
\usepackage{subfigure}
\usepackage{rotating}
\usepackage{supertabular}
\usepackage[version=3]{mhchem}
\usepackage{url}
\usepackage{booktabs}
\usepackage{subfigure}
\usepackage{lscape}
\usepackage[table]{xcolor}
\usepackage{color}
\usepackage{amsmath,amssymb,amsfonts}
\usepackage{longtable} 
\bibpunct{(}{)}{;}{a}{,}{,}

\newcommand{\luna}{{\tt LUNA}}

\begin{document}

\begin{titlepage}
\pagenumbering{arabic}
\begin{center}
\vspace*{1.5in}
{\huge The Transits of Extrasolar  \\ \vspace{2mm}
        Planets with Moons \\ \vspace{5mm}}
\par
\vspace{0.5in}
{\Large David Mathew Kipping}
\par
\vfill
{\Large Submitted for the degree of Doctor of Philosophy}
\par
\vspace{0.1in}
{\Large Department of Physics and Astronomy}
\par
\vspace{0.1in}
{\Large University College London}
\par
\vspace{0.1in}
{\Large March 2011}
\end{center}
\end{titlepage}

\clearpage

\setlength\topmargin{4in}
\begin{verse}
\begin{center}
I, \textit{David Mathew Kipping}, confirm that the work presented in this thesis
 is my own.  \\
Where information has been derived from other sources, I confirm that this has 
been 
indicated in the thesis.
\end{center}
\end{verse}

\clearpage

\setlength\topmargin{0in}

\begin{abstract}

The search for extrasolar planets is strongly motivated by the goal of 
characterizing how frequent habitable worlds and life may be within the Galaxy. 
Whilst much effort has been spent on searching for Earth-like planets, large 
moons may also be common, temperate abodes for life as well. The methods to 
detect extrasolar moons, or ``exomoons'' are more subtle than their planetary 
counterparts and in this thesis I aim to provide a method to find such bodies 
in transiting systems, which offer the greatest potential for detection.

Before one can search for the tiny perturbations to the planetary signal, an 
understanding of the planetary transit must be established. Therefore, in 
Chapters~\ref{ch:Chapt3} to \ref{ch:Chapt5} I discuss the transit model and 
provide several new insights. Chapter~\ref{ch:Chapt4} presents new analytic 
expressions for the times of transit minima and the transit duration, which will 
be critical in the later search for exomoons. Chapter~\ref{ch:Chapt5} discusses 
two sources of distortion to the transit signal, namely blending (with a focus 
on the previously unconsidered self-blending scenario) and light curve smearing 
due to long integration times. I provide methods to compensate for both of these 
effects, thus permitting for the accurate modelling of the planetary transit 
light curve.

In Chapter~\ref{ch:Chapt6}, I discuss methods to detect exomoons through their 
gravitational influence on the host planet, giving rise to transit timing and 
duration variations (TTV and TDV). The previously known TTV effect is updated 
with a new model and the associated critical problems are outlined. I then 
predict a new effect, TDV, which solves these problems, making exomoon detection 
viable. Chapter~\ref{ch:Chapt7} presents a feasibility study for detecting 
habitable-zone exomoons with \emph{Kepler}, where it is found that moons down to 
0.2\,$M_{\oplus}$ are detectable. Finally, conclusions and future work are 
discussed in Chapter~\ref{ch:Chapt8}.

\end{abstract}

\chapter*{Acknowledgements}

I am immensely grateful to those who have supported me during both my PhD 
studies and those who guided and nurtured my insatiable curiosity about 
the natural world since my earliest memories.
 
Firstly, I thank my official supervisors, Giovanna Tinetti and Alan
Aylward, for their unerring support and guidance throughout the last three
and a half years. I feel fortunate to have had supervisors who afforded
me the freedom to pursue my own ideas and encouraged such thinking
throughout. Special thanks to Jonathan Tennyson and Hugh Jones, my thesis
examiners, for taking the time to so carefully read this work.

I owe an enormous debt of gratitude to Giovanna Tinetti who always
found time for me and provided a level of support which has been the envy
of other graduate students. I feel privileged to have been the student of
such a talented scientist and I extend my warmest thanks to her for 
everything she has done for me.

I offer a special thanks to G\'asp\'ar Bakos, who invited me to the CfA
as a predoc. I never met anyone who has such a formidable knowledge 
of so many different aspects of the art of detecting exoplanets and it was
a privilege to have worked with G\'asp\'ar. I would also like to acknowledge
the continual support and friendship of Jean-Philippe Beaulieu who has made me
smile many times during those darker days of a graduate student's life.

I am grateful to the teachers of Twycross House School who taught me how
to apply and focus my mind in a way which changed the path of my life. In 
particular, I thank Michael Fox who nurtured my passion for physics and 
the Kirkpatricks for their patience and kindness. During my time at Cambridge,
I would like to thank Bill Nolan for his guidance and support. Additionally,
I am grateful to Roland G\"{a}hler for inviting me to work at L'Institut
Laue Langevin and his immense hospitality during that time.

I would like to thank my mother and father for their love and immeasurable
support throughout my life. Without them, none of my achievements would
have been possible. I also thank my sister, Louisa, and grandparents,
Rene\'e and Alec, for all they have done for me. Finally, I thank Emily,
the sunshine of my life, who makes me feel like I can achieve anything. I 
cannot find words to express my love for you.

\setcounter{page}{4}

\clearpage

\setlength\topmargin{4in}
\begin{verse}
\begin{center}
{\large\it  I dedicate this thesis to Emily,\\
my morning light and brightest star \\}
\end{center}
\end{verse}

\clearpage

\setlength\topmargin{0in}

\tableofcontents
\listoffigures
\listoftables

\input{Chapt1}

\input{Chapt2}

\input{Chapt3}

\input{Chapt4}
\input{Chapt5}
\input{Chapt6}
\input{Chapt7}
\input{Chapt8}

\appendix
\input{AppenA}


\bibliographystyle{plainnat}
\bibliography{sources}

\end{document}

%% file: Chapt1.tex
\chapter{Introduction}
\label{ch:Chapt1}

\vspace{1mm}
\leftskip=4cm

{\it ``
To consider the Earth as the only populated world in infinite space is as 
absurd as to assert that in an entire field sown with millet, only one 
grain will grow.''} 

\vspace{1mm}

\hfill {\bf --- Metrodorus of Chios, 4th century BC} 

\leftskip=0cm


\section{Are We Alone?}
\label{sec:arewealone}


It is no exaggeration to say that a great deal of the motivation behind 
searching for extrasolar planets is embodied by the above question. 
Since time immemorial, humans have turned their gaze towards the vast, 
illimitable, velvet darkness peppered with oases of light and wondered 
whether other beings, such as ourselves, also populate the ocean of 
space and time.

With the Earth representing the only environment where life is known
to reside, a natural place to begin a search for extraterrestrial life are
environments similar to this sole example. Consequently, planets, and in
particular ``Earth-like'' planets, are believed to offer the best 
locations for life to begin. Just like the ancient fascination with
extraterrestrial life, planets have long captured the imagination of human
minds.

Since antiquity, the seven \emph{asteres planetai}, or the ``wandering 
stars'', were known to many cultures across the world. These bodies, all
visible to the naked eye, were noted as abnormal heavenly objects in that
they did not appear in the same part of the sky each night. The seven are
the Sun, the Moon, Mercury, Venus, Mars, Jupiter and Saturn, which give us
the seven days of the week. It was the irregular \emph{motions} of these objects 
which distinguished them as planets and this provides a natural place for
us to start our exploration too.

\section{Motions of the Planets}
\label{sec:planetmotions}

\subsection{Celestial Mechanics}
\label{sec:CM}

Whilst the positions of the planets were rigorously recorded and tracked
by the Babylonians as early as the second millennium BC, it was the Greeks
who introduced a geometric model of their motions. A geocentric view of
the Universe was employed, presumably since it was natural to assume
the Earth was at rest, leading to elaborate models to describe the
planetary motions. The pinnacle of these theories was presented in 
\emph{Almagest} by \citet{ptolemy} in the $2^{\mathrm{nd}}$ century AD utilizing
a complex arrangement of planetary spheres and epicycles. Ptolemy's theory
was so successful at predicting these motions it remained the definitive
text for thirteen centuries.

It was not until the European renaissance that the heliocentric model of 
Copernicus, Galileo and Kepler gained favour in the $16^{\mathrm{th}}$ 
century and the definition of a planet changed from something which
orbited the Earth to something which orbited the Sun, demoting the
number of planets from seven to six\footnote{Although the Moon and the Sun were
no longer classed as planets, the Earth was now seen as a planet, to give six}.
Galileo Galilei's telescope design in 1609 paved the way for the future 
discovery of Uranus \citep{herschel1781}, Neptune (Le Verrier (1845), 
\citet{galle1846}) and Pluto (Tombaugh 1930) (as well as the moons of these 
planets), ushering in the modern view of the Solar System.

The other revolution came with Sir Isaac Newton's \emph{Philosophiae 
Naturalis Principia Mathematica}, first published on the $5^{\mathrm{th}}$ July 
1687 \citep{newton}. Whilst Johannes Kepler had earlier provided his three laws 
of planetary motion which so accurately predicted the motions of the planets,
the laws were simply empirical observations without a fundamental 
understanding as to why the three laws worked. Newton's theory showed that
planets move in ellipses due to a single force, the force of gravity, and
that a simple inverse-square force potential could completely explain all 
of the observed motions of the planetary bodies.

\subsection{Einstein's General Theory of Relativity}
\label{sec:GR}

Despite the many great successes of the Newtonian theory, 
\citet{verrier1859} reported that the perihelion precession of Mercury's
orbit around the Sun could not be completely explained with Newton's theory.
Mercury's precession is measured to be 5600 arc seconds per century whereas 
Newtonian mechanics predicts a precession of 5557 arcseconds per century
\citep{clemence1947}. At the time, Le Verrier proposed an additional planet 
named ``Vulcan'', which he suggested resided between Mercury and the Sun.

However, the discrepancy was later resolved without the need to invoke
another planet when Albert Einstein published his General Theory of
Relativity in 1915 \citep{einstein1915}. The observed motions of the planetary 
bodies was finally completely understood by the early twentieth century.

\subsection{The Discovery of Neptune}
\label{sec:neptune}

The discovery of Neptune in many ways serves as a transitional stage 
between the methods used to detect Solar System planets and the methods 
later used to detect extrasolar planets. \citet{bouvard1821} first struck upon
the hypothesis of Neptune's existence when he tabulated the observations
of the orbit of Uranus and noticed substantial deviations from current
predictions. Bouvard hypothesized that an unknown outer planet was
perturbing the orbit of Uranus. Le Verrier (1845) and \citet{adams1846} went 
further by calculating what the orbit of this body must be. These 
predictions were sufficiently accurate that the detection of Neptune was
made soon after by \citet{galle1846}.

Neptune therefore represented a transitional stage between the classic
technique of directly observing the reflected starlight from planets and
moons, to detecting bodies through indirect techniques based upon the
predictions of celestial mechanics. Celestial mechanics had matured to
a sufficient point that it could be a tool for planetary
detection, which would later play a vital role in exoplanetary searches.

\section{Historical Overview of Efforts to Find Extrasolar Planets}
\label{sec:exohistory}

\subsection{From Planets to Exoplanets}
\label{sec:earlyexointro}

After the detection of Pluto in 1930 and the picture of the Solar System 
apparently complete, many astronomers set their sights on much more 
distant planets - those orbiting other stars. This ambitious goal would
not be achievable through the conventional techniques used to find Solar
System planets. 

These planets, including the seven \emph{asteres planetai}, were found
by detecting the sunlight reflected off their surface as they moved 
against the backdrop of a black canvas. Therefore, the oldest technique
used to find planets is what we would now call ``direct imaging''.

In contrast, the light reflected from an extrasolar planet, or
``exoplanet'', is very challenging to spatially resolve from the host 
star and is typically around one million to one billion times fainter (see 
\S\ref{sec:imaging}). In recognition of this, astronomers had to devise new 
indirect methods in their bid to detect an extrasolar planet.

\subsection{Orbital Perturbations}
\label{sec:orbitalpert}

As discussed earlier in the case of Neptune, the second technique to 
yield success in detecting Solar System planets was by searching for 
deviations in the motions of a known orbiting body, 
betraying the presence of a unseen perturber. The use of this technique
reflects the extreme confidence astronomers now had in their understanding
of the motions of heavenly bodies, in large part due to the pioneering
work of Kepler, Newton, Lagrange, Hill, Einstein and many others.

These orbital deviations may be classed into two categories depending
upon the timescale over which they are observable. The first category
is orbit-to-orbit variations (e.g. Le Verrier's prediction of Neptune)
and the second is long-term secular (or apparently so) changes in the 
orbital elements of the system (e.g. the prediction of Vulcan, also by
Le Verrier).

In both cases, one requires at least two bodies in the system to detect 
the third. Naturally, this limits the use of the technique to either a
multiple-star system or a star for which exoplanets are already known. In 
the nineteenth century, the latter of these was of course not available.
Therefore, following the success of Le Verrier's work with Neptune and
the known existence of numerous bright multiple star systems, it should
perhaps come as no surprise that one of the first scientific claims of an 
exoplanet detection was for a binary star system.

\begin{figure}
\begin{center}
\includegraphics[width=15.0 cm]{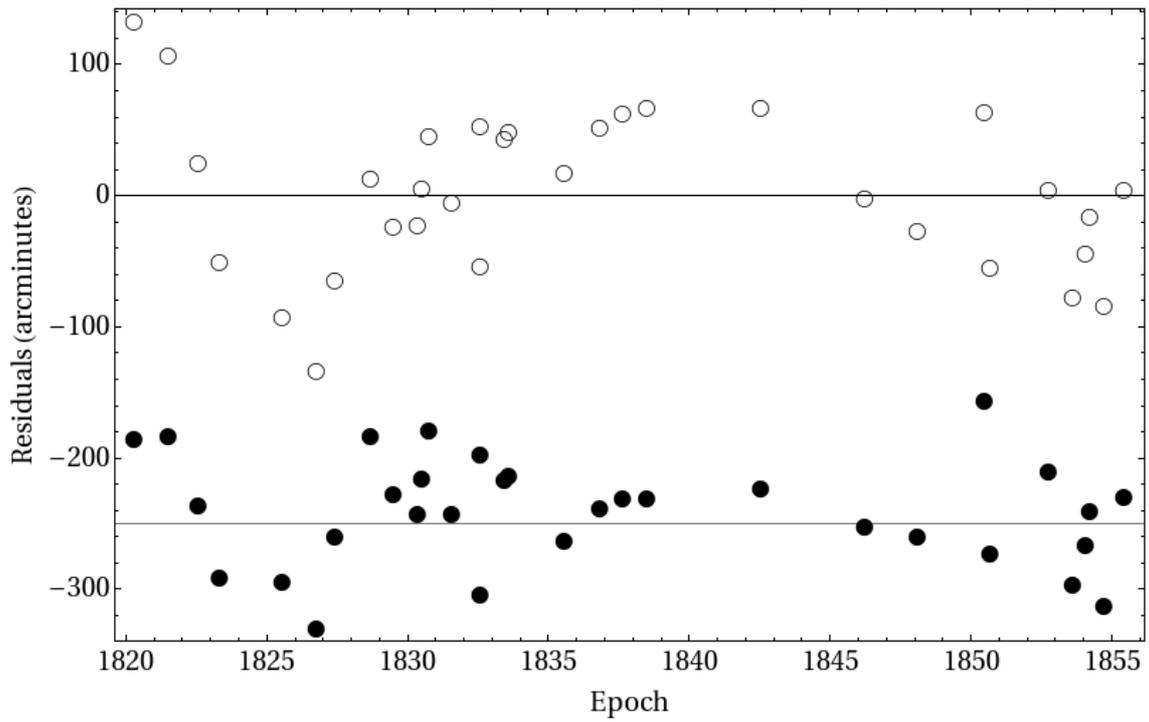}
\caption[Original data of Capt. W. S. Jacob (1855); the first claim for
an exoplanet detection]{\emph{Original 1855 data of Capt. W. S. 
Jacob, comparing the residuals from a two-body model (open circles) 
for 70 Ophiuchi and a three-body model (filled circles, offset by -250 
arcminutes for clarity). The decreased residuals of the three-body 
model was used to make one of the first claims for the existence of 
a exoplanet. A modern statistical analysis using an F-test finds the 
three-body model to be accepted with 1.9-$\sigma$ confidence, 
below the standard detection threshold.}}
\label{fig:jacob}
\end{center}
\end{figure}

In 1855, Capt. W. S. Jacob at the East India Company's Madras Observatory
reported that orbital anomalies in the binary star 70 Ophiuchi made it 
``highly probable'' that there was a ``planetary body'' in this system. 
\citet{jacob1855} compared observations of the positions of the binary 
components to a calculated model of a two-body and three-body system and noted 
that the average residuals decreased from 49 arcminutes to 35 arcminutes over 
the entire data set (see Figure~\ref{fig:jacob}). The statement 
``highly probable'' was not quantified in the way which would be mandatory
by modern standards, but Jacob can be forgiven, given that much of seminal 
work in observational statistics by Karl Pearson (Pearson's chi squared 
test in 1900, \cite{pearson1900}), Ronald Fischer (the F-test in 1920's) and 
others had not been devised yet. Using Jacob's
original data, a modern F-test would find the three-body model to be 
accepted over the two-body with a confidence of 1.9-$\sigma$, which is
below the oft-cited detection threshold of 3-$\sigma$.

Many interesting aspects of nineteenth century astronomy are visible
when one reads through Jacob's paper with a modern perspective. For
example, Jacob remarks that the deviations ``might arise from an erroneous
assumption of the universal application of the law of gravitation''. Jacob
essentially concludes that only two hypotheses can explain the 
observations; either Newton's laws of gravity requires modification or
there is a planet in orbit. Between these two hypotheses, Jacob quite
reasonably concludes the latter is much more likely. There are no other
hypotheses put forward; the idea that instrumental or human errors
could be responsible for the observations is notably absent. By
virtue of the fact this third hypothesis is not even mentioned, it 
therefore follows that Capt. Jacob considered a systematic error much less
likely than his second hypothesis - that the laws of gravitation 
are not universal - a concept which would be viewed as grossly aberrant to the 
modern exoplanetary scientist.

A few decades later, \citet{see1896} made a stronger claim for the 
existence of a dark companion in this system. However, Forest Ray Moulton,
who had been a graduate student of See at the University of Chicago, soon
published a paper proving that a three-body system with the specified
orbital parameters would be highly unstable \citep{moulton1899}. The claims by 
Jacob and See have both been shown to be erroneous \citep{heintz1988}.

\subsection{Astrometry}
\label{sec:astrometry}

One of the oldest and most famous scientific techniques used to look for 
an exoplanet is that of astrometry. Unlike the orbital perturbations
method, there is no example of a Solar System planet ever being discovered
through astrometry and so in many ways astrometry was quite a radical
technique. What is astrometry and why was it proposed so early on without
any successes in the Solar System?

\begin{figure}
\begin{center}
\includegraphics[width=15.0 cm]{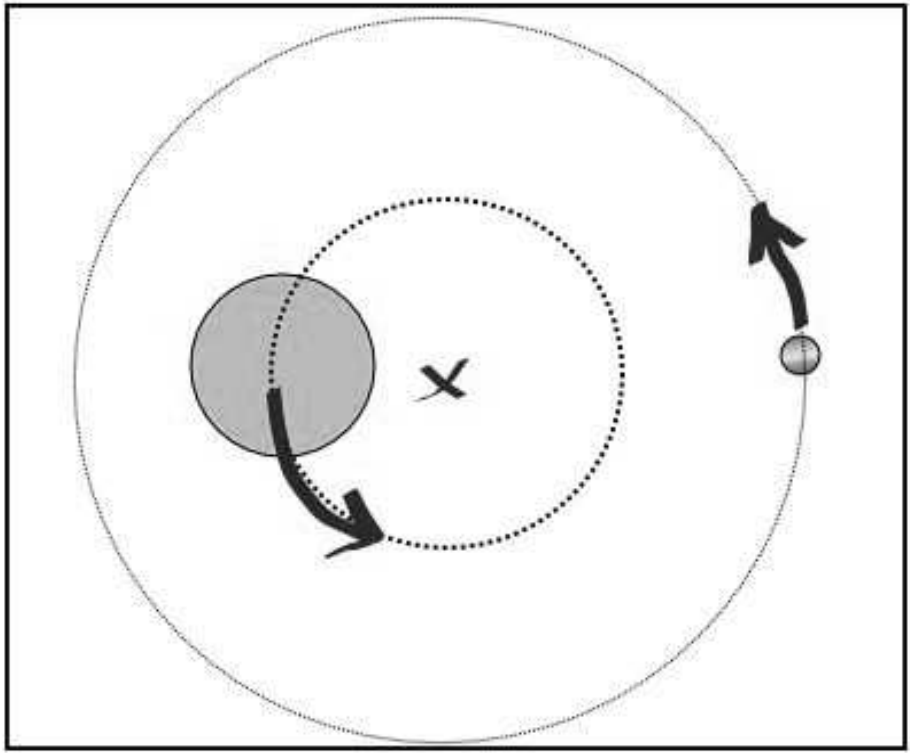}
\caption[Cartoon of the two-body problem]
{\emph{Cartoon illustrating the reflex motion of a star in the 
inertial frame of the barycentre. The planet's presence gives rise to 
variations in the position (used for astrometry and pulsar timing) and the
velocity (used for radial velocity) of the host star, which may used to 
detect an exoplanet.}}
\label{fig:centreofmass}
\end{center}
\end{figure}

From Newton's theory, two masses in an inertial frame orbit a common 
centre of mass. This means that a planet doesn't just have an orbit,
the star also has an orbit around the centre-of-mass, albeit of much 
smaller amplitude. Figure~\ref{fig:centreofmass} provides 
an illustration of this and \citet{murray1999} give a detailed discussion of 
the two-body problem. The star's ``reflex motion'' to the orbiting mass 
could therefore potentially be used to detect an exoplanet, if stellar 
positions could be measured to sufficient precision.

The variation in the sky position, $\alpha$, of a star of mass $M_*$ at a 
distance $d$ from the Earth, hosting a planet\footnote{I here introduce the
notation for several parameters related to the planet and star. In many other
texts, a subscript ``$P$'' is not placed on the planetary terms. However, in
this thesis I will be dealing with $>1$ planet and satellites and the advantage
of the ``$P$'' subscript will become apparent} of mass $M_P$ and semi-major axis 
$a_P$, can be estimated by Equation~(\ref{eqn:astrometrysim}). The equation 
reveals that $\sim 10^{-7}$\,arcseconds, or $\sim0.1$\,$\mu$as is
a reasonable estimate for the required precision in astrometry, which 
outlines just how challenging detecting an exoplanet truly is with this
method.

\begin{align}
M_P a_P &= M_* a_* \nonumber \\
\alpha &\simeq \frac{a_*}{d} \nonumber \\
\qquad &= (94.\mathrm{''}5 \times 10^{-9}) \Bigg(\frac{a_P}{\mathrm{AU}}\Bigg) \Bigg(\frac{10\,\mathrm{pc}}{d}\Bigg) \Bigg(\frac{M_P}{M_J}\Bigg) \Bigg(\frac{M_{\odot}}{M_*}\Bigg)
\label{eqn:astrometrysim}
\end{align}

Where ``AU'' denotes astronomical units, ``pc'' denotes parsec, $M_J$ is the
mass of Jupiter, $M_{\oplus}$ the Solar mass unit and $a_*$ represents the
semi-major axis of the star's reflex motion.

At the time when astrometry was first implemented, it was of course not the
only considered method. For example, transits of Venus and Mercury had 
long been known. The transit of Venus was first observed in 1032 by the 
Persian astronomer Avicenna, who used the observation to conclude that 
Venus is closer to the Earth than the Sun. In fact, one can go back to even
the work of Ptolemy, who wrote about the possibility of planetary transits
across the face of the Sun in his work ``Planetary Hypotheses''. He 
suggested that no transits had been observed either because planets such 
as Mercury were too small to see, or because the transits were too 
infrequent.

Despite the known existence of planetary eclipses, transits would not have
seemed an appealing method of detecting exoplanets in the nineteenth and
early twentieth century. For a Jupiter-sized planet transiting a Sun-like
star, the flux decrement would be $\sim1$\% (see \S\ref{sec:transits}). Using 
just the human eye or even later with photographic plates, this level of 
precision was simply not feasible. However, astronomers were very skilled in 
measuring positions of objects in the sky since the late eighteenth 
century and early nineteenth century, able to reach better than an 
arcsecond precision, through improved telescope designs (for example 
\citet{bessel1838} measured the parallax of the binary star 61 Cygni to be 
300\,mas\footnote{Note: ``mas'' denotes milli-arcseconds}).

Other reasons why transits were likely seen as unfeasible, even in the 
age of photomultipliers after 1934, was that transits are low probability
events, both geometrically and temporally, meaning that a very large 
dedicated survey would have been required at 
great expense by nineteenth and early twentieth century standards.
Finally, the expectation was to find planets on large distant orbits,
based upon the configuration of the Solar System, which maximizes the 
astrometric amplitude\footnote{And also minimizes the radial velocity
amplitude} and minimizes the geometric probability of a transit.

Therefore, astrometry can be seen to have been destined to emerge as one 
of the first methods to look for exoplanets. Observers were no doubt
spurred on by the early successes of Friedrich Bessel. \citet{bessel1844}
used astrometry to detect the presence of unseen dark companions to
Sirius and Procyon (astrometric amplitudes of 3.7\,mas and 1.8\,mas 
respectively), although these were unseen stars rather than 
exoplanets. It is worth pointing out that these detections did withstand
the scrutiny of later observations.

\begin{figure}
\begin{center}
\includegraphics[width=15.0 cm]{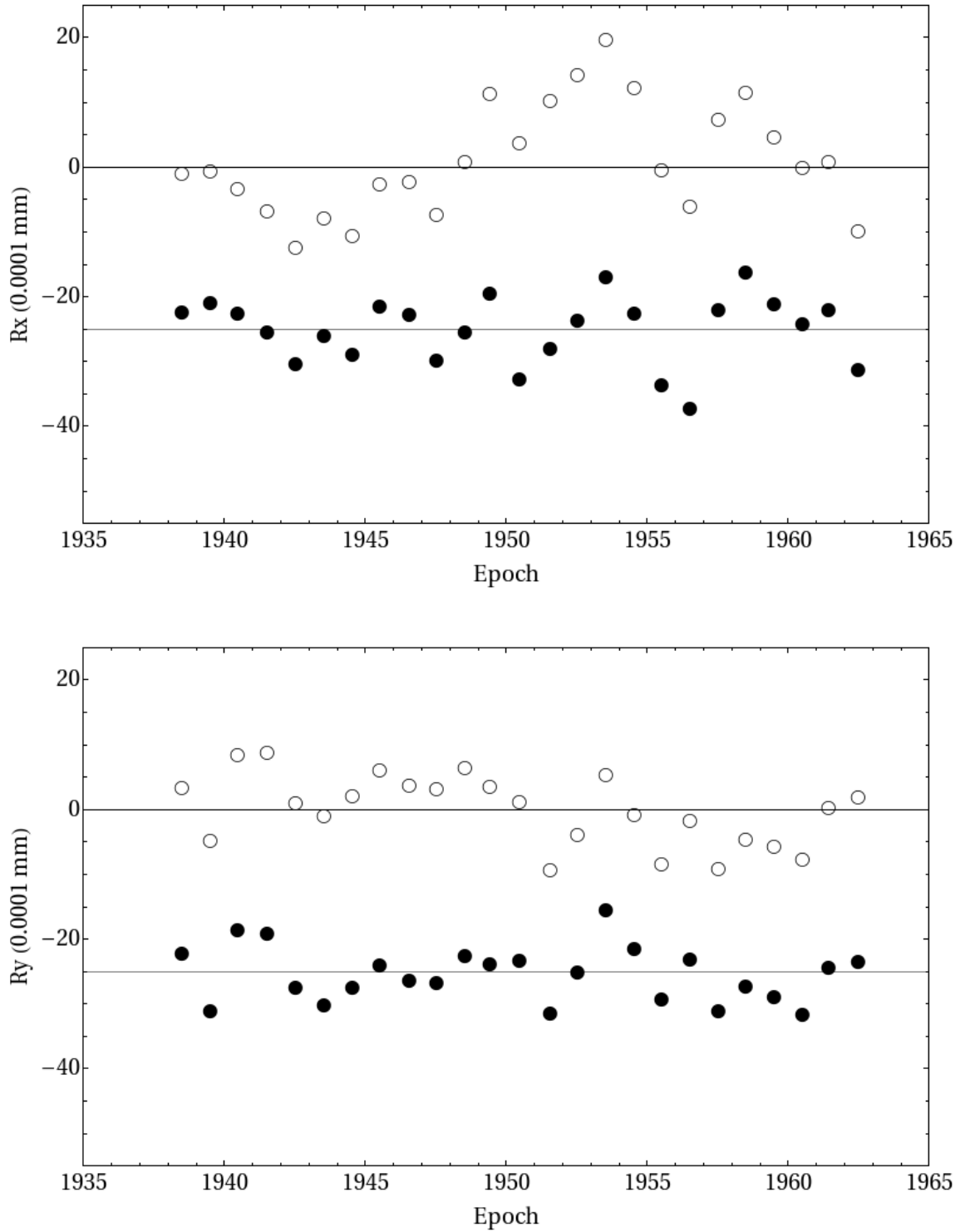}
\caption[Original data of Peter van de Kamp (1963), claiming the
detection of exoplanet around Barnard's star]
{\emph{Original 1963 data of Peter van de Kamp, comparing the residuals 
from a two-body model (open circles) for Barnard's star and a three-body 
model (filled circles, offset by -25$\mu$m for clarity). The 
decreased residuals of the three-body model was used to claim the presence
of a 1.6\,$M_J$ exoplanet in a 4.4\,AU orbit. A modern 
statistical analysis using an F-test finds the three-body model to be 
accepted with 1.7-$\sigma$ confidence, below the standard detection 
threshold.}}
\label{fig:barnardclaim}
\end{center}
\end{figure}

Over a hundred years after Bessel's groundbreaking work in astrometry,
the stage was set for one of the most famous stories in the tale of
exoplanets. In 1963, Peter van de Kamp claimed to have detected an 
exoplanet orbiting a nearby M-dwarf known as ``Barnard's star'' (named 
after the discoverer of its very high proper motion). \citet{kamp1963} argued 
for a 1.6 Jupiter mass ($M_J$) planet on a 4.4\,AU orbit electrifying the 
astronomical community (see Figure~\ref{fig:barnardclaim} for the original 
data).

In a familiar pattern though, the signal was soon identified as spurious. 
In subsequent observations, \citet{gatewood1973} failed to verify 
the planetary signal and around the same time \citet{herschey1973} showed that 
that changes in the astrometric field of various stars correlated to the timing 
of adjustments and modifications that had been carried out on the refractor 
telescope's objective lens. As a result, the planetary ``discovery'' was an 
artefact of maintenance and upgrade work on the telescope and is an
excellent example of an uncorrected systematic effect masquerading as 
a planet.

Barnard's star may have seemed an attractive target as it has a low mass, 
being an M-dwarf, and is one of the nearest stars to the Solar System 
meaning the claimed signal of Van de Kamp was $(24.5 \pm 2.0)$\,mas. 
However, Barnard's star also has the highest proper motion out of any star in 
the sky at a formidable 10.3 arcseconds per year. This means over the 
putative 24 year period of Van de Kamp's planet, the star exhibits a 
systematic motion over 10,000 times larger than the signal due to the 
proposed planet.

It is worth noting that despite the early problems with astrometry, it has
since been exploited on several targets known to already host planets for
the purpose of characterizing the orbit and demonstrating the feasibility
of astrometry in general. An excellent example comes from \citet{benedict2002},
who use HST astrometry to determine the mass of GJ 876b.

\subsection{Pulsar Timing}
\label{sec:pulsar}

After the debacle over Barnard's star, it is easy to see how confidence
may been lost in the hunt for exoplanets. By the beginning of
1990, over 130 years had passed of claimed detections of exoplanets
which were consistently debunked by later studies.

In the same decade that Van de Kamp wrote his infamous
paper on Barnard's star, Burnell and Hewish detected the first
pulsar (a highly magnetized rotating neutron star) in 1967 \citep{hewish1968}, 
which they designated ``Little Green Men 1'' or LGM01 (now less poetically known 
as PSR B1919+21). The original designation came
from the half tongue-in-cheek proposition that the measured regular
radio beats were artificial in nature, as there was
previously no concept of a rotating star beaming electromagnetic radiation
in this manner. The few dozen pulsars subsequently discovered were
subject to frequent timing measurements, in order to pin down their 
positions and spin-down parameters.

The 6.2\,ms pulsar PSR B1257+12 became an object of interest in the early
1990's due to irregularities in the timing of the pulses, as measured by
the team of Aleksander Wolszczan. The object was studied further by the
giant 305\,m Arecibo telescope by the same group and led to the inescapable
conclusion that deviations in the times-of-arrival (TOA) of the pulsar
must be due to the presence of two perturbing planets. \citet{wolszczan1992}
published their results soon after in what would become 
historically acknowledged as the first unambiguous detection of an extrasolar
planet.

On a brief aside, \citet{wolszczan1992} were under the impression that
their detection was not the first at the time of writing. A year earlier,
\citet{bailes1991} published a letter to Nature claiming a 10\,$M_{\oplus}$ 
planet orbiting the pulsar PSR B1829-10 with an orbital period of nearly exactly
6 months. The authors later retracted the planet upon realising they failed
to properly account for the Earth's elliptical orbit.

The method of looking for TOA deviations is very similar to astrometry.
Due to the changing position of the star, as a result of its reflex 
motion, the distance between the observer and the pulsar changes 
periodically. These changes in distance result in a shorter and longer
light travel times.

\begin{align}
\Delta(\mathrm{TOA}) &\simeq \frac{a_*}{c} \sin i_P \nonumber \\
\qquad &= \frac{a_P M_P}{M_* c} \sin i_P \nonumber \\
\qquad &= (4.74 \times 10^{-4}\,\mathrm{s}) \sin i_P \Bigg(\frac{a_P}{\mathrm{AU}}\Bigg) \Bigg(\frac{M_P}{M_J}\Bigg) \Bigg(\frac{M_{\odot}}{M_*}\Bigg)
\label{eqn:pulsarsim}
\end{align}

\begin{figure}
\begin{center}
\includegraphics[width=15.0 cm]{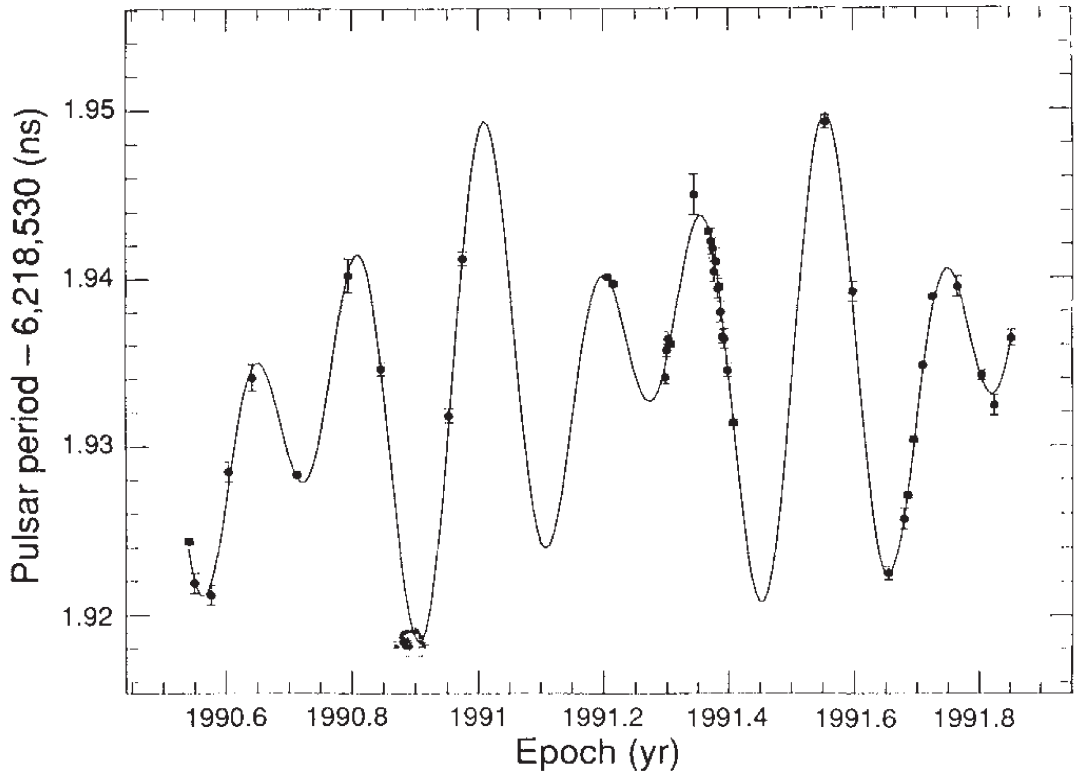}
\caption[Original data of \citet{wolszczan1992}, making the
first unambiguous detection of an exoplanet.]
{\emph{Original data of \citet{wolszczan1992}, showing a two-planet
fit to the measured pulsar times of PSR B1257+12. In contrast the 
earlier claims of \citet{jacob1855} (see Figure~\ref{fig:jacob}) and 
\citet{kamp1963} (see Figure~\ref{fig:barnardclaim}), the signals
of each planet are overwhelming obvious and have a significance 
of roughly 13-$\sigma$. Figure taken from \citet{wolszczan1992}.}}
\label{fig:pulsarplanet}
\end{center}
\end{figure}

Where $i_P$ is the orbital inclination of the planet.
Equation~(\ref{eqn:pulsarsim}) shows a typical signal is around half a
millisecond which was much larger than the typical uncertainties of 
a few hundred nanoseconds being achieved for millisecond pulsars by the time of 
1990. As a result, the signal seen by \citet{wolszczan1992}, which I
show in Figure~\ref{fig:pulsarplanet}, was extremely secure from a
statistical perspective.

Although the significance was not quoted in the original paper, I have
estimated this by digitizing the figure of the orbital fit and comparing
the root mean square (RMS) between the two-planet model and the
null hypothesis of a static system. The result is a false-alarm-probability
(FAP) of $4.3 \times 10^{-40}$ or 13.3-$\sigma$, which gives a sense
of just how secure the detection was relative to that of \citet{jacob1855}
and \citet{kamp1963}. Note that \citet{wolszczan1992} also considered whether 
their signal could be a systematic error by verifying that the observed 
periodicities were independent of radio frequency and that other millisecond 
pulsars routinely observed with the same equipment did not show such 
periodicities either.

The \citet{wolszczan1992} paper is recognized as the first unambiguous
detection of a planet orbiting another star. What made the paper even
more remarkable was that the planets detected were in the terrestrial mass
regime at 2.8\,$M_{\oplus}$ and 3.4\,$M_{\oplus}$ and what would be
considered ``long-period'' from a modern perspective, with respective 
orbital periods of 98.2\,d and 66.6\,d.

However, despite the low masses and long periods, pulsars would be
a hellish environment for life and there was still a strong motivation to
find a planet around a ``normal'' star, i.e. main-sequence. In many ways,
the pulsar detection spurred on the exoplanet hunt tremendously since
now astronomers knew stars hosted planets and yet the real prize of a 
planet around a Sun-like star remained undiscovered.

\subsection{Radial Velocity}
\label{sec:rv}

\subsubsection{Principles of RV}

In many ways, the age of exoplanet discovery did not truly begin until the 
detection of 51-Pegasi b, discovered using the radial velocity technique. Radial
velocity, or RV, is very similar to the astrometric method except that instead
of looking for the reflex motion of the star in terms of changes in position,
one looks for changes in velocity (along the line of sight). Since a
planet, and thus the star's corresponding reflex motion, orbit with a higher
speed when in tighter orbits, RV favours planets on very short orbital periods.
This is in contrast to astrometry which favours long-period planets yielding
a large planet-star separation and thus large amplitude spatial motion
of the host star. I summarise these very important points below.

\begin{itemize}
\item[{\tiny$\blacksquare$}] Astrometry: Measures the sky-projected 
\textbf{position} of a star: Prefers long-period companions
\item[{\tiny$\blacksquare$}] Radial Velocity: Measures the line-of-sight 
\textbf{velocity} of a star: Prefers short-period companions
\end{itemize}

Therefore, RV and astrometry are highly complementary to one another in 
parameter space. Additionally, the combination of the two allows for
a unique solution for the planetary mass \citep{tuomi2009}. This
complementarity between a position-based phenomenon and a velocity-based
phenomenon will have important ramifications later when I discuss detecting
exomoons and so I stress the point early on. The reflex velocity semi-amplitude
can be easily written down for a circular orbit to give a feeling as to the 
feasibility of such an enterprise. One may again exploit the fact that the 
planetary motion and the stellar motion must balance out in the inertial frame:

\begin{align}
M_P v_P &= M_* v_* \nonumber \\
K_* &\simeq v_* \sin i_P = \frac{M_P \sin i_P}{M_*} \frac{2 \pi a_P}{P_P} \nonumber \\
\qquad &= M_P (2 \pi G)^{1/3} M_*^{-2/3} P_P^{-1/3} \nonumber \\
\qquad &= (28.4\,\mathrm{m/s}) \sin i_P \Big(\frac{M_P}{M_J}\Big) \Big(\frac{M_*}{M_{\odot}}\Big)^{-2/3} \Big(\frac{P_P}{\mathrm{years}}\Big)^{-1/3} 
\label{eqn:rvsim}
\end{align}

On the second line, I have used $K_*$ to denote the RV semi-amplitude,
as is standard in the exoplanet literature. On the third line, I have used
Newton's version of Kepler's Third Law to remove the dependency on $a_P$.
Given that by 1995 typical RV errors could reach 20\,m/s, 
Equation~(\ref{eqn:rvsim}) reveals that for giant planets on an orbit of a year
or less, RV would be a viable detection technique.

The RV method requires measurements of a star's velocity along the 
line-of-sight, which can be achieved by measuring the Doppler shifts of
said star's spectral lines. This naturally requires a very stable spectrograph,
with highly sensitive calibration and a rich forest of lines to measure. To make
the method feasible for an exoplanet search, the process has to become
somehow automated to expedite what would otherwise be a very 
tedious task by hand (as well as prone to inaccuracies).

\subsubsection{A shot in the dark}

For RV to succeed, the key hinge was whether gas giants would exist in orbits
of a year or less and thus be detectable. Planetary formation models 
\citep{pollack1996} of the Solar System perpetrated that gas giants formed 
beyond the snow-line (where the equilibrium temperature is less than 150\,K 
which means 2-3\,AU typically). Whilst inward migration was conceived, 
inspection of the positions of the Solar System gas giants alluded that this 
mechanism was not very effective. Thus, even after the \citet{wolszczan1992} 
discovery, little resources were devoted to looking for exoplanets using the RV 
method. Nevertheless, some groups did take the gamble, which ultimately paid off
for \citet{mayor1995}.

Before discussing the famous paper of \citet{mayor1995}, I point out that
some tentative successes had already transpired for the RV method before 1995.
Notably, seven years prior, \citet{campbell1988} cautiously claimed to have 
detected a companion to Gamma Cephei which was challenged a few years later
by \citet{walker1992}. It was therefore not generally considered a confirmed
planet until validated in \citet{hatzes2003}. Another example comes from
\citet{latham1989} who detected a companion of minimum mass 11\,$M_J$ around
HD 114762, but concluded the object was more likely to be a brown dwarf than
a planet.

The story of \citet{mayor1995} centres around ``51-Peg''; a bright ($V=5.5$) G2- 
to G4-type star relatively close at 15.6\,pc in the Pegasus constellation. The 
star was part of a exploratory survey being conducted by a Swiss group based in 
Geneva, led by Michel Mayor looking at 142 bright K and G dwarfs for radial 
velocity variations with a sensitivity of $\sim$13\,m/s.

\begin{figure}
\begin{center}
\includegraphics[width=15.0 cm]{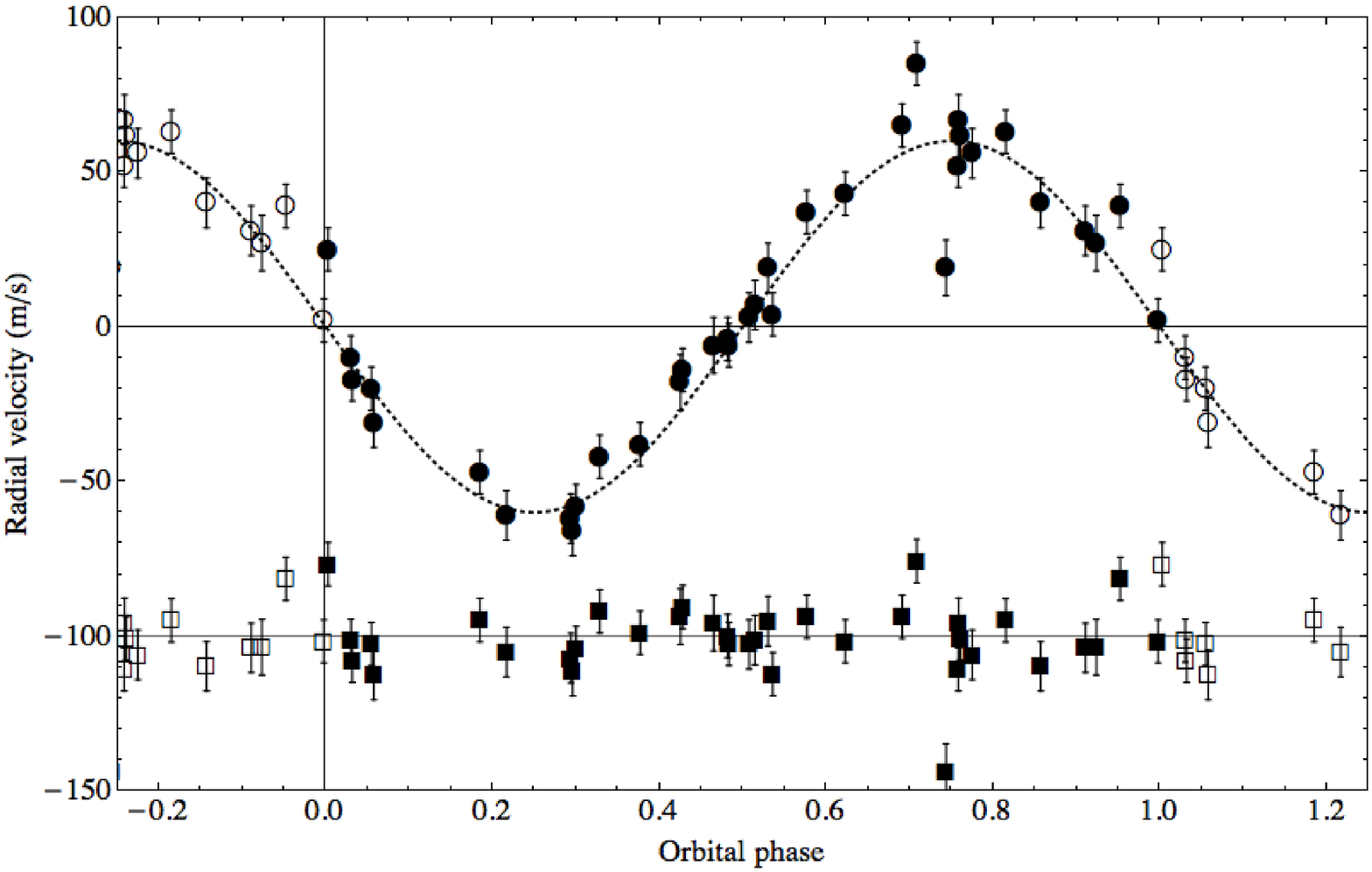}
\caption[Original data of \citet{mayor1995}, making the
first unambiguous detection of an exoplanet around a
normal star]
{\emph{Original data of \citet{mayor1995} (circles), showing a one-planet
fit (continuous dotted line) to the measured radial velocities of 51-Pegasi. 
In contrast to the earlier claims of \citet{jacob1855} (see 
Figure~\ref{fig:jacob}) and \citet{kamp1963} (see 
Figure~\ref{fig:barnardclaim}), the signal of the planet is
very clear and has a significance of roughly 9-$\sigma$. Squares show 
the offset (-100\,m/s) residuals of the one planet circular-orbit fit. Empty
points represent a repetition of the data to illustrate to signal more 
clearly.}}
\label{fig:rvplanet}
\end{center}
\end{figure}

Although not quoted in the original paper, it is simple to download the
original data, refit a circular orbit model (no linear drift) and compute
the significance of the fit. The result is a false alarm probability (FAP)
of $7.1 \times 10^{-19}$ or 8.9-$\sigma$, easily above the standard
detection threshold of 3-$\sigma$. It is also visible that \citet{mayor1995}
spend considerable effort excluding alternate hypotheses for the signal, for 
example from a blended eclipsing binary by virtue of a bisector analysis.

\subsubsection{An explosion of discoveries}

Just six days after the \citet{mayor1995} discovery, Geoff Marcy and Paul Butler
of San Francisco State University independently confirmed the signal and then 
charged on with two new discoveries for 70 Virginis \citep{marcy1996} and 
47 Ursae Majoris \citep{butler1996}. And so began the age of exoplanetary 
science with new discoveries coming in on a regular basis from here on. Today, 
over 500 exoplanets are known to exist with over 80\% coming from RV surveys.

Due to the bias of RV surveys to high mass, short-period planets, a large
fraction of these bodies are previously unanticipated ``hot-Jupiters''. The fact 
that hot-Jupiters exist was a major boon for RV surveys in the early years,
but modern high resolution spectroscopy can now measure stellar velocities
to better than 1\,m/s meaning planets down to a few Earth-masses (Super-Earths)
at long periods can be found (e.g. \citet{vogt2010}).

One interesting consequence of the RV discoveries was that many of the 
exoplanets being found had unanticipated short-periods (a few days) and thus had
a much higher chance ($\sim10$\%) of transiting the face of the star than 
previously expected. After the paper of \citet{mayor1995}, astronomers started 
to think about the technique which had long been considered implausible.

\subsection{Transiting Planets}
\label{sec:transits}

Despite RV surveys churning out plenty of exoplanet discoveries, RV provides 
frustratingly little information about these bodies. For example, the true mass 
of the planet is not actually known, just $M_P \sin i_P$. The only other 
information one can glean is the orbital period, $P_P$, the orbital 
eccentricity, $e_P$ and the argument of periastron, $\omega_P$. For the ultimate 
goal of detecting signs of life on another planet, this was simply not enough. 
If one of the RV detected planets happened to transit across the star though, 
vast amounts of detail could be inferred about the planet's size, composition, 
oblateness, atmosphere, temperature, albedo, surface coverage and ultimately 
habitability\footnote{Note that not all of these possibilities were known early 
on}. Transits are the window into the soul of an exoplanet.

The principles of the transit method are very simple at a first-order level. The
star has a nominal flux level which temporarily decreases due the planet 
blocking out a fraction of the projected stellar surface as it passes in front. 
The depth of the transit\footnote{I here describe the geometric transit depth. 
The observed transit depth is, in general, deeper than this due to limb 
darkening effects} is given by the ratio of the sky-projected area of the planet
and the sky-projected area of the star:

\begin{align}
\delta &= \frac{\pi R_P^2}{\pi R_*^2} \nonumber \\
\qquad &= 1.03\,\Big(\frac{R_P}{R_J}\Big) \Big(\frac{R_{\odot}}{R_*}\Big) \, \%
\label{eqn:transitsim}
\end{align}

By the end of the twentieth century, even small telescopes (a few inches 
aperture) had photometric precisions better than 1\%, in large part due to the 
development of CCDs a few decades earlier. The key question was how likely was 
it that a transit would occur, given the plainly fortuitous geometric 
alignment required?

Assuming a planet is on a nearly circular orbit and the inclination angle is
uniformly distributed in space, the probability is simply given by $\sim R_*/a$.
For the hot-Jupiters being found from RVs, a good rough estimate was 10\% 
probability. In reality, it is larger than this since RV surveys have
a bias to aligned systems, but this conservative estimate is already 
encouraging. Consequently, success was soon coming.

\begin{figure}
\begin{center}
\includegraphics[width=15.0 cm]{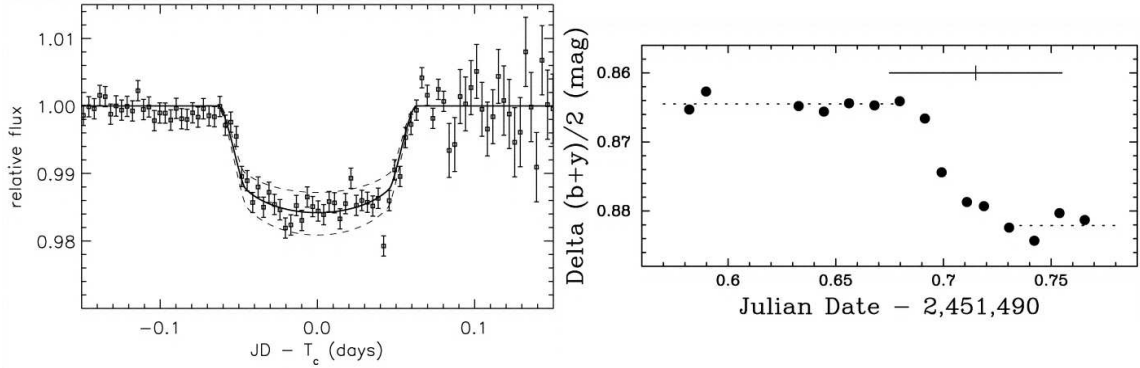}
\caption[First detected transit light curve of an exoplanet, made by
\citet{charbonneau2000} and \citet{henry2000} for HD 209458b]
{\emph{First detected transit light curve of an exoplanet, made by
\citet{charbonneau2000} (left) and \citet{henry2000} (right) for HD 209458b.
On the left, the solid line shows the best-fit model and the dotted line
shows the extrema of a planet which is 10\% smaller or larger in radius.
Left figure taken from \citet{charbonneau2000}. Right figure taken from
\citet{henry2000}.}}
\label{fig:1sttransit}
\end{center}
\end{figure}

\citet{charbonneau2000} and \citet{henry2000} detected the first
transit of an extrasolar planet for HD 209458b, a fairly typical hot-Jupiter
around a bright star (see Figure~\ref{fig:1sttransit}). 
\citet{charbonneau2000} used the 4-inch STARE telescope,
demonstrating just how inexpensive transit observations could potentially be 
(which opened the door for future ground-based transit surveys such as HATNet 
\citep{bakos2004} and WASP \citep{pollacco2006}). The resulting light curve 
showed a depth of around 1.5\% lasting for around 3 hours. The light curve 
allowed for a determination of the orbital inclination angle and thus the true 
planetary mass from the RV determined $M_P \sin i_P$. Also, since the transit 
depth is essentially $(R_P/R_*)^2$, the planetary radius could be calculated 
and thus the average density of HD 209458b was known. The power of transits had 
been demonstrated.

\begin{figure}
\begin{center}
\includegraphics[width=15.0 cm]{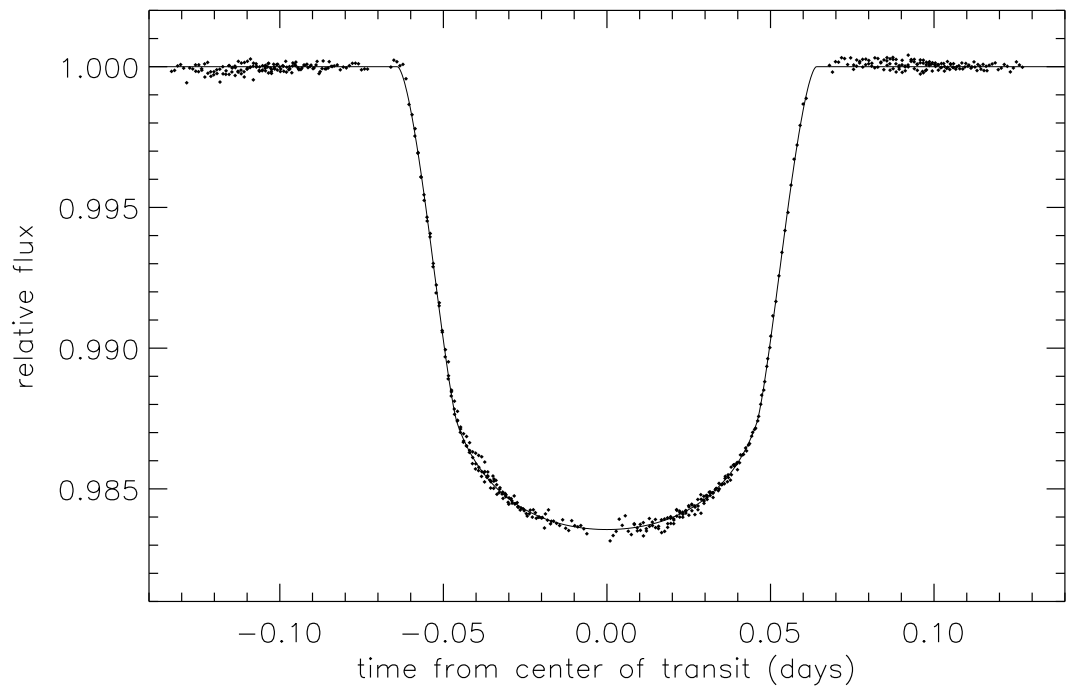}
\caption[First space-based transit light curve of an exoplanet, made by
\citet{brown2001} for HD 209458b]
{\emph{First space-based transit light curve of an exoplanet, made by
\citet{brown2001} for HD 209458b using the Hubble Space Telescope (HST). Data 
has been phase-folded from five partial transits. It was not possible to observe
a complete and uninterrupted transit, due to the 90 minute orbit of HST. Figure
taken from \citet{brown2001}.}}
\label{fig:HSTtransit}
\end{center}
\end{figure}

A spectacular follow-up paper by \citet{brown2001} used the STIS instrument
on HST to provide the first space-based transit light curve. The stunning
photometry, shown in Figure~\ref{fig:HSTtransit}, provided an exquisite
determination of the system parameters and toyed with the idea of constraining
the presence of rings and moons.

Having finally reached the discovery of transiting exoplanets, which constitutes 
the major topic of interest of this thesis, the relevant history of exoplanets 
has been covered, for the purposes of this thesis. However, for completion, I 
will briefly overview two other known successful methods to detect exoplanets. 
Methods based upon transit timing will be discussed later in the thesis, in 
Chapter~\ref{ch:Chapt6}.

\subsection{Gravitational Microlensing}
\label{sec:label}

\begin{figure}
\begin{center}
\includegraphics[width=15.0 cm]{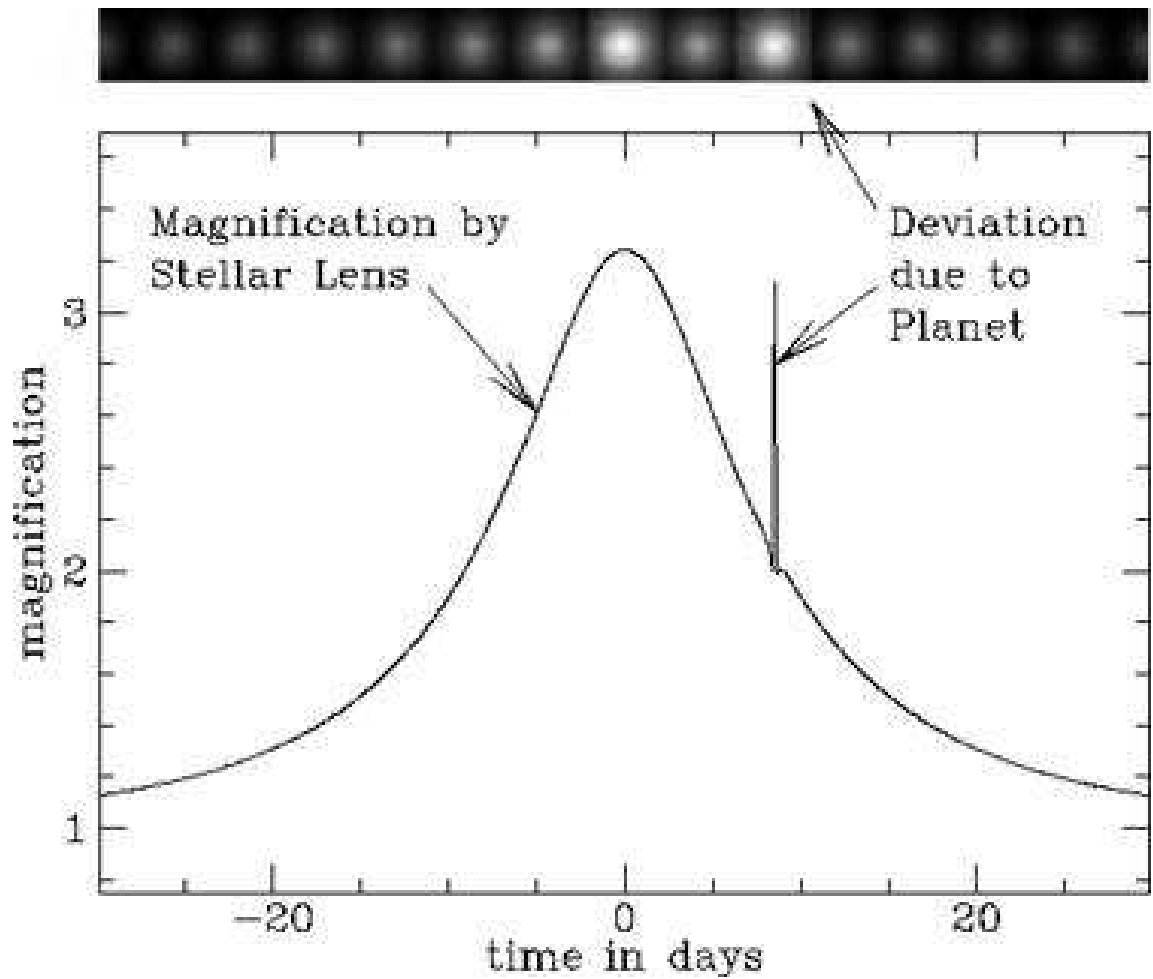}
\caption[Example microlensing light curve due to a star with a planet.]
{\emph{Example microlensing light curve of a star with a planet. One can see
the brightening due to the lensing of the star and later the effect of the 
planet is visible. Image taken from the Microlensing Planet Search Project 
homepage\footnote{http://bustard.phys.nd.edu/MPS/}.}}
\label{fig:microlensing}
\end{center}
\end{figure}

Gravitational microlensing is based upon the fact that massive bodies bend the
apparent path of light, in essence acting as a lens. If a star passes in front
of another more distant luminous body, it causes the luminous body to 
dramatically increase in brightness for a few days or weeks, depending on the
configuration. The same is true for a star with a planet except that one sees 
two (or more) increases in brightness; one large increase due to the more
massive star and one smaller increase due to the planet (see 
Figure~\ref{fig:microlensing} for an illustration).

Microlensing favours massive planets with a wide separation leading to the 
detection of cold or frozen planets. However, with a space-based mission in
the form of EUCLID or WFIRST, microlensing could detect many Earth-mass
habitable-zone planets and provide comprehensive statistics on the frequency of
habitable bodies (often dubbed $\eta_{\oplus}$). On the down side, microlensing
events are inherently transient and once the lensing event is over there are no
further opportunities for follow-up.

Microlensing hit upon its first success in 2004 with the discovery of 
OGLE 2003-BLG-235/MOA 2003-BLG-53, named jointly as it was discovered by both
the OGLE and MOA surveys. \citet{bond2004} reported the discovery of 1.5\,$M_J$
at 3\,AU and since this time several more have been found.

\subsection{Direct Imaging}
\label{sec:imaging}

The latest technique to broach success could also be considered the oldest
technique used to detect a planet - direct imaging. Direct imaging seeks to 
either spatially, or spectrally, resolve the light from the planet from that of 
the star. The light from the planet is either reflected or thermal
emission and so hot planets at a wide separation, meaning young planetary 
systems, are favourable. For reflected light only, the contrast ratio between
the disk-averaged intensities of the planet and star is given by:

\begin{align}
\frac{I_P}{I_*} \,\mathrm{d}\lambda &= \mathfrak{A}_G C(f_P) \frac{\pi R_P^2}{4 \pi a_P^2}\,\mathrm{d}\lambda \nonumber \\
\qquad&= 5.71\times10^{-8} \, \Big(\frac{R_P}{R_J}\Big) \Big(\frac{\mathrm{AU}}{a_P}\Big) \mathfrak{A}_G C(f_P) \,\mathrm{d}\lambda
\label{eqn:reflectcontrast}
\end{align}

Where $\mathfrak{A}_G$ is the geometric albedo, defined as 
the flux reflected by the planet when viewed at opposition (full phase), and 
$C(f_P)$ can be between 0 and 1 depending upon the true anomaly of the planet. 
Equation~(\ref{eqn:reflectcontrast}) therefore reveals that the contrast ratios
are likely to be quite extreme.

Matters are further exacerbated by the challenge of resolving this light.
The star and planet have an angular separation of $a_P/d$, where $d$ is again
the distance to the star from the observer. Typically, telescopes have their
angular resolution constrained by the diffraction limit, given by
$1.22 \lambda/D$, where $\lambda$ is the wavelength of the light one is using
and $D$ is the diameter of the telescope. Therefore, the smallest separation
one can detect is:

\begin{align}
a_{P,min} &= \frac{1.22 d \lambda}{D} \nonumber \\
a_{P,min} &= (1.38\,\mathrm{AU}) \Big(\frac{d}{10\,\mathrm{pc}}\Big) \Big(\frac{\lambda}{550\,\mathrm{nm}}\Big) \Big( \frac{1\,m}{D}\Big)
\label{eqn:imagingapprox}
\end{align}

So it can be seen that direct imaging naturally prefers to use large telescopes
with short wavelengths to look around nearby stars with wide companions. 
Coronographic techniques can null the stellar flux leaving the planets alone to
identify and track over the course of their orbit (e.g. \citet{heap2000}).

\begin{figure}
\begin{center}
\includegraphics[width=15.0 cm]{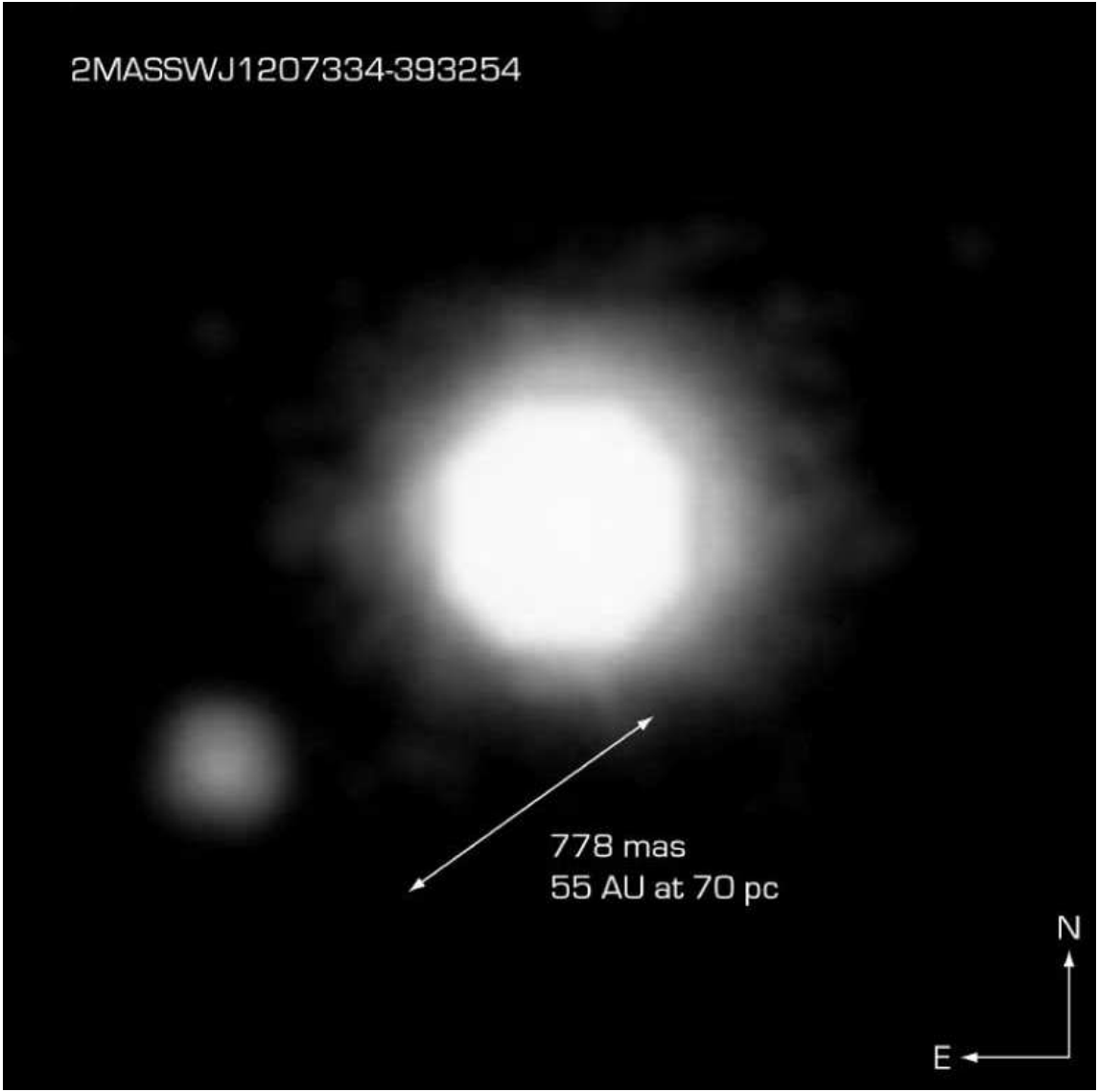}
\caption[First image of an extrasolar planet by \citet{chauvin2004}.]
{\emph{First image of an extrasolar planet by \citet{chauvin2004}. 2M1207b
orbits its primary, a brown dwarf named 2M1207a, at around 50\,AU.}}
\label{fig:directimage}
\end{center}
\end{figure}

The first image of an extrasolar planet is 2M1207b by \citet{chauvin2004}. The
primary is a brown dwarf meaning the contrast ratio between the planet and the
``star'' is quite favourable. Also, the planet orbits at $\sim$50\,AU around the 
primary and is hot, being in the range 1000-2000\,K. The pioneering first image
is shown in Figure~\ref{fig:directimage}.

\section{Lessons from History}
\label{sec:lessons}

\subsection{Context}
\label{sec:lessonscontext}

This tour of the history of exoplanet discovery does not only serve to 
introduce this thesis but clearly exposes several key lessons which one can, and
should, take away. In this thesis, I seek to develop ways to detect the moon of
an extrasolar planet, which, it must be admitted, is a challenging task. There 
is little question that the signals being dealt with will exhibit very 
low signal-to-noise, just like the early years of exoplanet searches. In such
a circumstance, how can one avoid the pitfalls of the early claims made for
planets?

In this section, I ask, what are the differences between the discovery papers,
methods and attitudes of those discoveries which were genuine and those which
were not?

\subsection{Detection Criteria of the Historically Genuine Discoveries}
\label{sec:differences}

\subsubsection{1 - Statistical significance}

Looking back to the earliest claim of \citet{jacob1855}, there are many 
differences in style, presentation and evident resources applied to that of
modern exoplanet detection papers. The first thing one can note is that the
statistical significance of Jacob's claim (1.9-$\sigma$) is marginal, at best, 
even assuming absolutely no systematic errors (which is of course never true).
A very similar situation is true for \citet{kamp1963} whose model fit I have
estimated to be accepted with 1.7-$\sigma$ confidence.

In contrast, the successful first detections of \citet{wolszczan1992}, 
\citet{mayor1995} and \citet{charbonneau2000} exhibit significances of 
13.3-, 8.9- and 31.8-$\sigma$ respectively. Although detecting a signal to
$\sim10$-$\sigma$ confidence does not guarantee a genuine detection (e.g.
the false positive of \citet{bailes1991}), it must certainly be considered
as a requisite along the chain of steps necessary to claim a detection. I 
identify this as my first detection criterion and use 3-$\sigma$ as an absolute
minimum detection threshold.

\subsubsection{2 - Systematic errors}

Another clear difference is that \citet{jacob1855} and \citet{kamp1963}
give little, if any, consideration to the hypothesis that the observations
are induced by some kind of systematic error(s)\footnote{Note that systematic 
errors can be either instrumental or astrophysical in origin}. In modern 
detection papers, this is a very regular practice. For example, RV detected 
planets are invariably accompanied with a bisector analysis to eliminate the 
hypothesis of a blended eclipsing binary (e.g. \citet{mayor1995}). Frequently, 
they also come with a photometric campaign too (e.g. \citet{vogt2010a}) to 
confirm the signal is not due to star spots or plages (e.g. \citet{queloz2001}).
Similarly, planets detected through the transit method are usually given a
thorough consideration of various blending scenarios and almost always come
with RV confirmation (e.g. \citet{bakos2010}).

I therefore identify a second criterion to be that an exploration of systematic
effects, both instrumental and astrophysical, should be performed.

\subsubsection{3 - Physicality of the claim}

For a two-body system, almost all configurations can be considered plausible
but once one moves into the realm of three-body systems then many configurations 
would quickly become dynamically unstable after a time span much shorter than
the age of the systems in question. As an example of this, the claim of 
\citet{see1896} was challenged by \citet{moulton1899} on the basis
that the system would rapidly become unstable. Not surprisingly, observations
later confirmed Moulton's warning \citep{heintz1988}. Modern exoplanet papers
regularly test the dynamical stability of the proposed system for multi-planet 
systems (e.g. \citet{vogt2010a}) as a criterion for the system being real. I
take this as my third criterion.

\subsubsection{4 - Suspicious periods}

\citet{bailes1991} provide an interesting example of a signal cleanly detected
to high significance but which ultimately proved to be false. The signal had a 
period of nearly exactly 6 months and turned out to be due to a calibration 
error for compensating for the Earth's orbit. Similarly, it can be seen to be 
prudent to check for signals which can be due to aliasing of the sampling 
frequency \citep{dawson2010}. This gives the fourth criterion.

\subsubsection{5 - Consistent instrumentation}

The claim of \citet{kamp1963} was eventually shown to be false as a result of
upgrades to the telescope over the course of the observations. These upgrades
meant essentially a different instrument was being used between the start and
end of the observations. If an instrument has an unknown systematic error which 
causes a constant offset then using it repeatedly actually has no bearing on the
results, so long as the offset is indeed constant. However, if one uses different
instruments, each one with its own unique (and unknown) systematic offset, then 
artificial signals will inevitably appear.

A lesson to take home is that it is strongly preferable to use the same
instrument as much as possible for the observational data in a detection paper.
This constitutes the fifth criterion.

\subsubsection{6 - Avoid observing targets with large systematic offsets in
the observed parameter}

The false claim of \citet{kamp1963} resonates with a later claim made for a 
different star, VB 10, in 2009. \citet{pravdo2009} claimed to have detected a 
planet around VB 10 using astrometry, in what would have been the first 
confirmed detection using this method. However, it was later shown by 
\citet{escude2010} and \citet{bean2010} that the predicted RV signal
did not exist and thus the astrometric signal was spurious.

What VB 10 and Barnard's star have in common is that both are objects with very
high proper motion (1'' and 10.3'' per year respectively). To perform 
astrometry, one therefore has to first remove this large systematic effect of
the proper motion. One can easily see that if there is a, say 0.01\%, error in 
this process then a residual signal of 0.1\,mas will persist in the case of
Barnard's star, which is the same size as the claimed planetary signal for that
system. 

Therefore, targeting objects with extremely high systematics (i.e. proper 
motion in the case of astrometry) can be instantly appreciated to be an unwise
strategy since it requires exceptionally accurate corrections for these effects
before a meaningful analysis can be performed. This constitutes the sixth
criterion.

%

\subsection{Summary}

I have identified six criteria which seem to discriminate between the
genuine and spurious exoplanet claims over the past decades. I caution that 
this is by no means comprehensive - counter examples of genuine systems will 
surely exist for any given criterion - but they do offer a general strategy 
to optimize the chances of finding authentic discoveries by exploiting the 
benefit of hindsight.

\begin{itemize}
\item[{\textbf{C1}}] Statistical significance
\item[{\textbf{C2}}] Systematic errors
\item[{\textbf{C3}}] Physicality of the claim
\item[{\textbf{C4}}] Suspicious periods
\item[{\textbf{C5}}] Consistent instrumentation
\item[{\textbf{C6}}] Avoid large systematics
\end{itemize}

\section{From Exoplanets to Exomoons}
\label{sec:planetstomoons}

To date, over 500 extrasolar planets have been detected\footnote{See 
www.exoplanet.eu}. The lower mass limits continue to drop with Super-Earth 
detections becoming routine and Earth-mass objects appearing imminent. As we 
approach the Earth-mass and radius limit, it is natural to consider the 
possibility of detecting the moons of exoplanets too.

Satellite systems represent mini Solar Systems, with a richness and 
diversity all of their own. The exploration of the moons of Jupiter and Saturn
has reaped immense understanding and shown us worlds beyond our imagination.
The prospect of studying thousands of new examples in utterly alien environments
would provide unforeseeable understanding of our place in the Universe.

The problem I therefore identify in this thesis, is how can we make this
ambitious vision a reality? How can we detect and characterize the moons of 
extrasolar planets?

\section{Thesis Outline}
\label{sec:thesisoutline}

In this thesis, I aim to provide a method for detecting the moons of transiting
extrasolar planets. To achieve this goal it will be necessary to first
understand the transit light curve of an exoplanet without a moon and so this
thesis will be broadly split into two objectives i) an exploration of methods to
model the transit light curve ii) description of a novel method to detect the 
moons of transiting planets.

In Chapter~\ref{ch:Chapt2}, I will discuss the motivations for such a search,
the expected properties of an exomoon and possible methods which could be used
to detect the moons of non-transiting planets. Having established what is being 
searched for and why, I move on to describe the basics of transit theory in
Chapter~\ref{ch:Chapt3}, which must be first understood. The method to detect
moons will be later be shown to rely on precise timing of transits and so in
Chapter~\ref{ch:Chapt4} I derive the times of transit minima and the transit
durations. Chapter~\ref{ch:Chapt4} marks the start of my own research into this
field with the other material covered thus far being predominantly introductory.
In Chapter~\ref{ch:Chapt5}, this continues with an exploration of two 
subtleties with modelling the transit light curve. I discuss how hot planets
can produce nightside emission which acts as a self-blend, which acts to dilute
the transit depth. Then, I explore how long-integration times, such as that used
by \emph{Kepler}'s long-cadence mode cause a smearing of the transit signal and
provide methods to compensate for the distortion.

With the theory of transiting planets well established by the end of 
Chapter~\ref{ch:Chapt5}, the stage is set to introduce the exomoon into the 
model. In Chapter~\ref{ch:Chapt6}, I present the transit timing variations (TTV)
method of detecting exomoons as first proposed by \citet{sartoretti1999}. I
then go on to discuss two critical problems with TTV, namely TTV only provides
the mass of the moon multiplied by the orbital distance, $M_S a_S$, and also
how many other phenomena can also cause TTVs leading to a signal ambiguity.
I then predict the existence of a new timing effect, velocity induced transit
duration variations (TDV-V), which solves both of these problems. TDV-V scales
as $M_S a_S^{-1/2}$ and thus the ratio of TDV-V to TTV solves for $M_S$ and 
$a_S$ separately. Further, TDV-V exhibits a $\pi/2$ phase shift from TTV
and thus provides a unique exomoon signature. The chapter closes by 
exploring a second order TDV effect, dubbed transit impact parameter induced
transit duration variations (TDV-TIP), which could be exploited to determine
an exomoon's sense of orbital motion with sufficient signal-to-noise.

Whilst the theory of transit timing effects is presented in 
Chapter~\ref{ch:Chapt6}, the issue of how feasible an exomoon detection would
be with these techniques remains absent. To address this, 
Chapter~\ref{ch:Chapt7} provides a feasibility study of detecting habitable-zone
exomoons with \emph{Kepler}-class photometry. In this chapter, I conclude that
exomoons down to 0.2\,$M_{\oplus}$ could be detected in the most favourable
scenarios and up to 25,000 stars within \emph{Kepler's} field-of-view are bright
enough to be surveyed for 1\,$M_{\oplus}$ habitable-zone exomoons. These
encouraging figures mean that TTV and TDV have been shown to be viable detection
techniques for finding exomoons.

I conclude in Chapter~\ref{ch:Chapt8} with a summary of the thesis
and a look at future work required in exolunar theory. Notably, a work in
preparation, which can model the transit dips of exomoons, is discussed and
the potential for it to characterize exomoons.

%% file: Chapt2.tex
\chapter{Extrasolar Moons}
\label{ch:Chapt2}

\vspace{1mm}
\leftskip=4cm

{\it ``
I think we're going to the Moon because it's in the nature of the human being to
face challenges.''} 

\vspace{1mm}

\hfill {\bf --- Neil Armstrong, Apollo mission press conference, 1969} 

\leftskip=0cm


\section{Motivations}
\label{sec:motivations}

Extrasolar moons, or ``exomoons'', offer a significant challenge to astronomers
given their anticipated\footnote{From a Copernican perspective} low masses and 
radii. However, there are many motivations for looking for such objects, which
I will overview here.

\subsection{Intrinsic Habitability}
\label{sec:intrinhab}

Based upon the only known examples, moons are likely to have 
either a solid or liquid surface (see \S\ref{sec:moonproperties} for a more 
detailed discussion of expected compositions). For that reason, exomoons have 
the same capacity as Earths and Super-Earths (planets of mass 
$\lesssim 10$\,$M_{\oplus}$) for harbouring a habitable environment. 

Habitability of exomoons is discussed in more detail in \citet{williams1997}, 
who find that such bodies can offer suitable conditions provided they are 
massive enough to maintain an atmosphere for Gyr. The authors argue that this 
cut-off occurs approximately at $\gtrsim 0.3$\,$M_{\oplus}$. 

\citet{scharf2006} explored the interesting possibility of moons outside of the
conventional habitable-zone being tidally heated to an extent where they could
maintain oceans of liquid water, which may be suitable for life.

\subsection{Extrinsic Habitability}
\label{sec:extrinhab}

Even if a moon is not habitable itself, a moon could enhance the probability of
a host planet maintaining a habitable environment. As an example, 
\citet{laskar1993} have argued that the presence of a large satellite around the 
Earth (i.e. the Moon) stabilizes the axial tilt of our planet and thus leads to 
a more favourable long-term climate for life. This may not significantly affect 
the probability of the emergence of life, but is thought to affect the 
probability of land-based life emerging \citep{benn2001}.

\citet{benn2001} provide a review of the various proposed beneficial influences
the Moon may have had on the emergence of life on the Earth:

\begin{itemize}
\item[{\tiny$\blacksquare$}] Stabilization of the Earth's axial tilt 
\citep{laskar1993}
\item[{\tiny$\blacksquare$}] Elimination of the primordial atmosphere 
(\citet{cameron1991}, \citet{taylor1992})
\item[{\tiny$\blacksquare$}] Generation of the Earth's magnetic field 
\citep{pearson1988}
\item[{\tiny$\blacksquare$}] Generation of large tides \citep{chyba1990}
\item[{\tiny$\blacksquare$}] Generation of longer-period tides \citep{rood1981}
\end{itemize}

\subsection{Planet/Moon Formation}
\label{sec:moonformA}

The formation theory of the planets and their moons is a topic of active debate.
Currently, two hypotheses of planet formation are considered the most
viable i) core accretion \citep{pollack1996} ii) gravitational 
instability \citep{boss2004}. 

Two mechanisms (although quite distinct from the planet formation models) have 
also been proposed for the formation of satellites. Moons are predicted to form 
either from the disk material surrounding a planet (``regular satellites''; 
always prograde) \citep{canup2002} or by gravitational capture/impacts/exchange 
interactions (``irregular satellites''; either prograde or retrograde) 
\citep{jewitt2007}. 

\begin{figure}
\begin{center}
\includegraphics[width=15.0 cm]{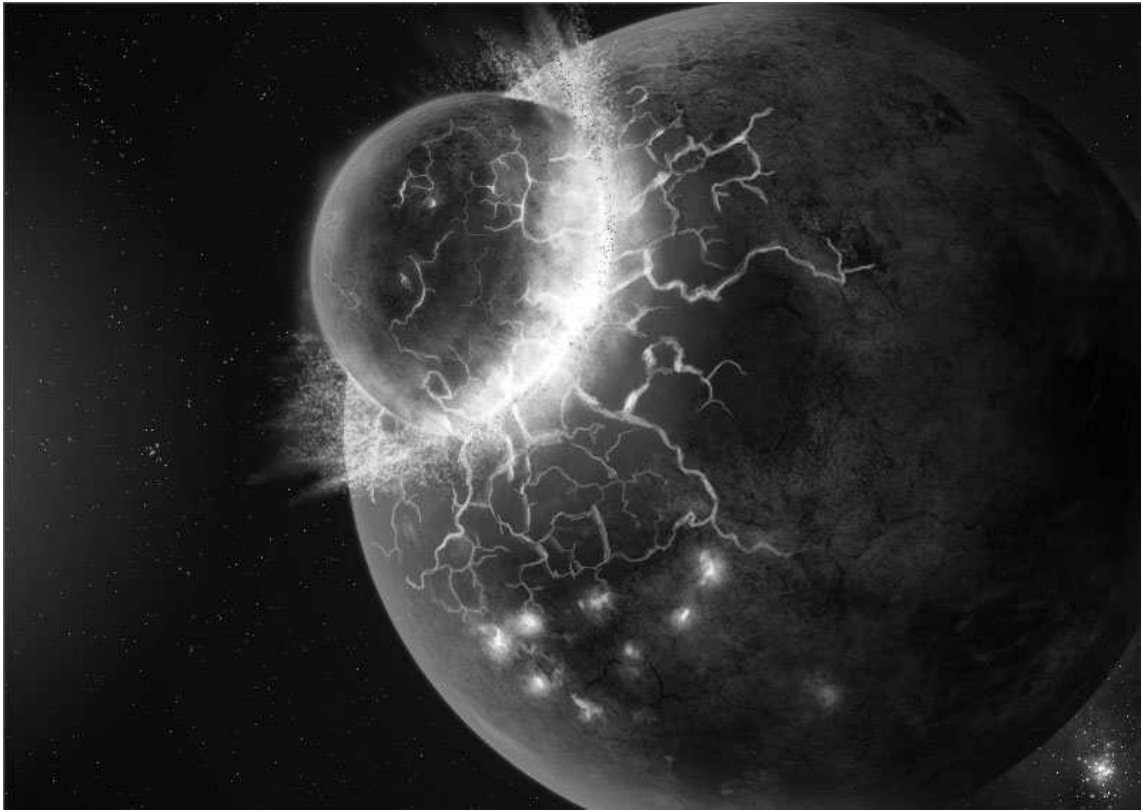}
\caption[Artist's impression of the Theia impact into the primordial Earth]
{\emph{Artist's impression of the impact of a Mars
sized planet (dubbed ``Theia'') into the primordial Earth. This ``Big Splash''
scenario is currently believed to be the most plausible mechanism for the 
formation of the Moon. Image credit Fahad Sulehria (www.novacelestia.com).}}
\label{fig:theia}
\end{center}
\end{figure}

The detection of many extrasolar examples of satellite systems would undoubtedly
enhance our understanding of the formation of planets and moons by providing a
greater population sample. This is particularly salient for the large irregular 
satellites, as currently only two examples are known to exist, the Moon and 
Triton. Whilst Triton is likely to have been gravitationally captured by 
Neptune \citep{agnor2006}, the most plausible mechanism for the formation of the 
Moon is quite distinct. In this case, a Mars-size planet, dubbed ``Theia'', is 
thought to have impacted the primordial Earth \citep{taylor1992} (see 
Figure~\ref{fig:theia}) and the prevalence of such encounters would impinge 
powerfully on models of young planetary systems. 

\section{Predicted Properties of Exomoons}
\label{sec:moonproperties}

Having established a clear motivation for searching for exomoons 
(\S\ref{sec:motivations}), one may begin to consider how they might be found. 
However, it is useful to first overview the expected properties of exomoons, 
which will inform the choice of detection method. 

\subsection{Formation}
\label{sec:moonformB}

As discussed in \S\ref{sec:moonformA}, current models of the formation
of satellites is naturally based upon observations of the Solar System. There
appear to be two broad classes of moons; regular and irregular satellites.
\citet{canup2006} have proposed that the maximum total mass of a satellite
system, formed as regular satellites, is $\sim2\times10^{-4}$ times the mass 
of the host planet. This is based upon the expected quantity of material in the 
disc surrounding a planet early in its history, which may subsequently be 
gravitationally bound to the planet and coalesce into solid moons. The region
in which a massive body, in orbit of a star, dominates the attraction of 
surrounding test particles is known as the Hill sphere (named after George 
William Hill who provided this definition). The radius of this
sphere is found by solving the restricted circular three-body problem and is
equal to the distance of the L1/L2 Lagrange points:

\begin{equation}
R_H = a_P \Big(\frac{M_P}{3 M_*}\Big)^{1/3}
\label{eqn:hillradius}
\end{equation}

Given that the focus of this thesis is detecting moons, and it is known that 
even detecting a Super-Earth mass planet is challenging with current 
instrumentation, one must acknowledge that the most favourable, and in all 
likelihood even feasible, case will be a single moon around a planet. This is 
because the total allowed mass will be concentrated into a single body. The 
Saturn-Titan system is perhaps the closest example of this whereas Jupiter has 
four large moons with roughly equivalent masses. Following the hypothesis of 
\citet{canup2006}, to reach an Earth-mass moon, one would therefore need a host 
planet of mass $\sim16$\,$M_J$,  which is really a brown dwarf rather than a 
``planet''.

The second formation mechanism is that of the irregular satellites. Classifying
an object as such does not really constrain the expected properties in
many ways. The mass of such an object could be anything from an asteroid to a
Super-Earth\footnote{Although, of course, no examples of this exist in the Solar
System}. The latter clearly would require a full-grown planet to get captured by
a gas giant planet. Similarly, this classification does not strongly constrain 
the origins of the satellite, which could be anything from an impact scenario 
(e.g. Earth-Moon \citep{taylor1992}) to a capture (e.g. Triton-Neptune 
\citep{agnor2006}).

This vast range of possible histories actually means that irregular satellites 
are quite favourable for an exomoon search. The permissible existence of Earth 
mass moons around ``normal'' gas giants (i.e. not brown dwarfs) certainly 
increases the chances of a detection. By this reasoning, I consider it much 
more likely that the first detected exomoon will be an irregular satellite, 
based upon our current understanding of the regular satellites.

I also mention that the existence of binary-planets is physically sound.
\citet{podsiadlowski2010} showed that one viable scattering history in the
formation of a planetary system is the tidal capture of two planets forming
a binary. Indeed, a Jupiter-Earth pair could be considered as an extreme binary,
much like Pluto-Charon.

\subsection{Evolution}
\label{sec:moonevolution}

Having identified that irregular satellites are the most probable bodies which 
can be detected in the near-future, let us consider the chances of a moon's 
survival. Two dominant effects are believed to cause the loss of moons for a 
single planet-moon pair\footnote{Multiple moons are more complex due to the 
dynamical interactions which occur between the moons}: i) tidal dissipation 
ii) inward planetary migration. 

Tidal dissipation between a planet and its companion cause a moon to either
spiral in or out. Which of these occurs depends on whether the tidal
bulge caused by the moon leads (spirals-in) or lags (spirals-out) relative to 
the rotation bulge of the planet. \citet{barnes2002} provide analytic
approximations for the maximum time a moon can survive and encouragingly predict
that Earth-mass moons are stable around Jupiter-like planets for 
habitable-zone periods for Sun-like stars. Planets on short periods have tighter
stable regions for a moon to reside and thus the planet-moon tides are greater 
leading to a rapid loss i.e. hot-Jupiters are unfavourable. 

The other effect is that of inward, disk-driven planetary migration, which 
occurs on a much faster time scale than tidal dissipation. Because of this fact,
the Hill radius of the planet shrinks very quickly. Thus, a moon 
initially well inside the Hill sphere can find itself outside and so be ejected. 
\citet{namouni2010} showed that once a migrating gas giant crosses 
$\sim0.1$\,AU, a moon is unlikely to survive. For this reason, hot-Jupiters are
again unfavourable hosts.

\subsection{Stable Region}
\label{sec:moonstable}

I earlier provided the distance at which a test particle feels a greater
gravitational attraction to the planet rather than the star- the Hill sphere
(see \S\ref{sec:moonformB} and Equation~(\ref{eqn:hillradius})).
However, this simple picture does not accurately represent the complete stable
region of an extrasolar moon. Other effects, such as the Yarkovsky effect 
\citep{yarkovsky1888} and three-body perturbations can eventually perturb a moon
to be ejected from the system.

\citet{domingos2006} showed that the actual region over which a moon is stable
depends upon the eccentricity of the orbit and the sense of orbital motion. For
prograde and retrograde moons respectively, the stable regions are:

\begin{align}
a_{S,\mathrm{max}}^{\mathrm{prograde}} &= 0.4895 R_H (1.0000 - 1.0305 e_P - 0.2738 e_S ) \\
a_{S,\mathrm{max}}^{\mathrm{retrograde}} &= 0.9309 R_H (1.0000 - 1.0764 e_P - 0.9812 e_S )
\label{eqn:domingos}
\end{align}

Where $e_P$ and $e_S$ are the orbital eccentricities of the planet and satellite
respectively. Therefore, retrograde moons can be found at far larger distances 
than their prograde counterparts. This means the moon has a larger distance over 
which to tidally spin-out or in and thus ultimately a longer lifetime. This is 
equivalent to allowing a larger moon mass to be stable for a given amount of 
time. One can also see that planets on eccentric orbits offer severely reduced 
regions of stability for a putative moon.

More recently, \citet{donnison2010} showed that moons on inclined orbits
also yield contracted regions of orbital stability and thus in general one
expects moons to be roughly co-aligned to the orbital plane of the planetary
orbit, which is supported by the properties of the massive satellites in our
own solar system.

\subsection{Composition}
\label{sec:mooncomp}

For the regular satellites of the Solar System, the composition of a moon tends 
to be ice-rich with the rest being silicates and iron \citep{consolmagno1983}. 
This is because gas giants are thought to have formed beyond the snow-line, 
where ices do not sublimate away and thus the moons which form in situ (i.e. 
regular moons) have plenty of ice to accumulate in their formation 
\citep{pollack1996}. However, given that this thesis has identified the best 
chance of detecting an exomoon to be a terrestrial planet which becomes captured 
around a gas giant, the composition of a detected exomoon is more likely to 
resemble Mercury, Venus, Earth and Mars. Useful mass-radius scalings for such 
bodies are presented in \citet{valencia2006}, who estimate 
$(R/R_{\oplus}) \sim (M/M_{\oplus})^{0.27}$.

\subsection{Conclusions}

The largest and most massive (and thus most detectable) possible satellite which
could exist is likely to be a terrestrial planet which was captured as an 
irregular moon around a gas giant planet. This statement merely indicates 
physically feasible origin for an exomoon which could be detected in the 
coming years. The consequences are that the moon is likely to have a similar
composition and mass scaling as the terrestrial planets of the Solar System and
may exist in a retrograde orbit (e.g. Triton). Further, the 
host planet is likely to be on an orbit with semi-major axis $\gtrsim 0.1$\,AU 
with low eccentricity.

\section{Proposed Detection Methods for Non-Transiting Systems}
\label{sec:moonmethods}

In this thesis, I will present methods of detecting moons using transit light
curves. However, other methods are of course possible for non-transiting systems
as well. In this section, I will discuss the different methods available, which
will provide a context for the method of transits.

\subsection{Direct Imaging}
\label{sec:moonimaging}

Direct imaging of an exomoon would face similar, but ultimately amplified,
problems as imaging an exoplanet. For the latter, one needs to spatially resolve
the star and planet which have an angular separation of $a_P/d$. However, for
a moon one needs to acquire an angular resolution of $a_S/d$. Given that the
maximum allowed $a_S \simeq R_H$ (i.e. the Hill radius), then a simple
calculation reveals the angular resolution requirement is at least a factor of 
$(3 M_*/M_P)^{1/3}$ lower. For a Jupiter around a Sun-like star this is a factor
of 14.4 i.e. comfortably an order-of-magnitude more challenging.

The glare of the host star also becomes more problematic as a 1\,$R_{\oplus}$
exomoon would have a contrast ratio of $\sim4.5\times10^{-10}$ to the host
star (using Equation~(\ref{eqn:reflectcontrast}) and assuming $a_P = 1$\,AU). 
This is a factor of 125 times more difficult than a Jovian planet i.e. two 
orders of magnitude.

\subsection{Microlensing}
\label{sec:moonlensing}

Microlensing does not offer the ability to characterize moons once the lensing
event is over and thus is more useful for understanding the statistical
distributions of moons rather than their nature.

\citet{liebig2010} recently calculated the detectability of an Earth mass moon
around a Saturn-like planet at 2.5\,AU for various configurations using the 
microlensing technique. The authors find that for fairly rapid sampling of once
every 15 minutes with an RMS photometric precision of 20\,mmag, there is a
30\% chance of detecting such a system. As the planet-star separation is
decreased, this probability rapidly falls and the authors conclude only ``cold''
planets are feasible for an exomoon detection in the near future.

Nevertheless, the case for a microlensing detection of an exomoon is strong when
considered in light of the planned space missions (EUCLID/WFIRST). However, due
to the fact that a habitable-zone exomoon is unlikely to be found and that
follow-up characterization is not possible, the microlensing prospect also has
a lot to be desired.

\subsection{Pulsar Timing}
\label{sec:pulsarmoons}

Is there any way that studying the motion of the host star could reveal the
presence of a moon? At the simplest level, the reflex motion of a star is 
exactly the same for a 1\,$M_J$ planet with a $0.003$\,$M_J$ moon as it is for a 
$1.003$\,$M_J$ planet without a moon. In other words, the star sees the 
planet-moon system as a composite point-mass and thus there is no way one
could detect a moon by watching the motion of the host star.

However, this is not strictly true and the presence of the moon does induce
slight deviations in the star's reflex motion. These tiny deviations arise from
a full analysis of the three body problem. In reality, the moon and planet are
not a single composite point-mass but two distinct masses which give rise to a
varying torque on the host star.

\citet{lewis2008} first considered the possibility of detecting moons in
this way, for pulsars with a time-of-arrival (TOA) analysis. The planet plus
moon signal has two components i) a dominant term which essentially acts like
$M_P + M_S$ in orbit of the host star and thus indistinguishable from a slightly
more massive planet ii) a residual term which may be used to detect the moon.
This residual signal is given by \citep{lewis2008}:

\begin{align}
\mathrm{TOA}_{\mathrm{residual}} = - \sin i_P \frac{M_P M_S}{(M_P+M_S)^2} \frac{a_P}{c} \Big(\frac{2 a_S}{a_P}\Big)^5 \Bigg[ \frac{3}{32}\cos(\phi_{P}-2\phi_{S})+\frac{15}{32}\cos(3\phi_{P}-2\phi_{S})\Bigg]
\label{eqn:lewis}
\end{align}

Where $\phi_{P}$ is the angular position of the planet-moon barycentre around
the host star and $\phi_{S}$ is the angular position of the satellite around 
the planet-moon barycentre. Thus, the residual signal has two frequency
components at $(n_{P}-2n_{S})$ and $(3n_{P}-2n_{S})$, where $n_{P}$ denotes
the mean motion of the planet-moon barycentre around the star and $n_{S}$
denotes the mean motion of the satellite around the planet-moon barycentre. 
Replacing $a_S$ with the Hill radius, and assuming $M_P \gg M_S$, the maximized 
amplitude of this signal can be expressed as:

\begin{align}
|\mathrm{TOA}_{\mathrm{residual}}| &= \frac{9 \sin i_P}{16} \frac{M_P M_S}{(M_P+M_S)^2} \frac{a_P}{c} \Big(\frac{2 a_S}{a_P}\Big)^5 \nonumber \\
\qquad &\simeq \frac{9 \sin i_P}{16} M_S \frac{a_P}{c} \frac{M_P^{2/3}}{3^{5/3} M_*^{5/3}} \nonumber \\
\qquad &= (41918\,\mathrm{ns}) \sin i_P \Big(\frac{M_S}{M_{\oplus}}\Big) \Big(\frac{a_P}{\mathrm{AU}}\Big) \Big(\frac{M_P}{M_J}\Big)^{2/3} \Big(\frac{M_{\odot}}{M_*}\Big)^{5/3} 
\end{align}

This may be compared to the residuals of one of the most stable pulsars, 
PSR J0437-4715, which has an RMS of 130\,ns \citep{straten2001}.
Therefore, the detection of an Earth-mass moon would seem highly feasible.
However, the ratio of the amplitude of this residual signal to the 
planet-only signal is $\simeq 0.347 M_P^{1/3} M_*^{2/3} M_S^{-1}$, or a factor 
of 11400 for a $1$\,$M_J$ planet at 1\,AU with an Earth mass moon. This
ratio perhaps gives a more realistic quantification of the challenge faced.

Another drawback with the proposed method is that only two examples of pulsars
with planets are known to exist, PSR B1257+12 and PSR B1620-26 and therefore
one does not have many planets to investigate. The first of these is a 
multi-planet system where the approximations of \citet{lewis2008} are invalid 
(i.e. a single planet with a single moon) \citep{wolszczan1992}. The second is 
actually a binary star system with a poorly characterized planetary companion 
($M_P = 2.5\pm 1.0 M_J$) \citep{backer1993}. Thus, success does not seem likely
to be forthcoming for the prospect of pulsar moons.

\subsection{Astrometry}
\label{sec:astromoons}

I briefly point out that the TOA signal from pulsar timing is an astrometric
effect and thus a very similar equation exists for the astrometric amplitude.
One can show that the astrometric signal is:

\begin{align}
|\alpha_{\mathrm{residual}}| &= \frac{c |\mathrm{TOA}_{\mathrm{residual}}|}{d} \nonumber \\
\qquad&= \frac{9 \sin i_P}{16} M_S \frac{a_P}{d} \frac{M_P^{2/3}}{3^{5/3} M_*^{5/3}} \nonumber \\
\qquad&= (0.147\,\mathrm{nas}) \Big(\frac{d}{10\,\mathrm{pc}}\Big) \Big(\frac{M_S}{M_{\oplus}}\Big) \Big(\frac{a_P}{\mathrm{AU}}\Big) \Big(\frac{M_P}{M_J}\Big)^{2/3} \Big(\frac{M_{\odot}}{M_*}\Big)^{5/3} 
\label{eqn:astromoons}
\end{align}

Where ``nas'' indicates nano-arcseconds. Whilst a dedicated astrometric space 
mission, such as SIM, may be launched in the next 10-20 years, a sensitivity
of at best $\mu$as would be achievable with such an enterprise 
\citep{goullioud2008}. Consequently, I conclude that achieving the 
nano-arcsecond precision needed to locate a moon is unfeasible in the near 
future.

\subsection{Radial Velocity}
\label{sec:rvmoons}

In a similar manner to the astrometric signature predicted by \citet{lewis2008},
the torque supplied by a binary should induce a residual radial velocity
signature over the bulk centre-of-mass signal. This torque was investigated
by \citet{morais2008}, who performed numerical simulations to compute the
predicted RV signals for a variety of possible configurations.

The authors found that a Jupiter-Jupiter binary at 1\,AU would induce a residual
RV signal of $10^{-5}$ to $10^{-6}$\,m/s and a Jupiter-Earth system is even more
challenging at $10^{-7}$ to $10^{-8}$\,m/s. These magnitudes are certainly far 
below the typical intrinsic noise level of even the quietest stars 
($\sim1$\,m/s, \citet{martinez2010}), the so-called stellar ``jitter'' which is 
due to spots, plages and magnetic activity. In conclusion, a RV detection 
appears quite unfeasible in the near-future.

\section{Summary}

Amongst the myriad of motivations for embarking on a search for exomoons, I have
identified three examples which outline the importance of this work. 
\S\ref{sec:motivations} discusses the potential of moons to be intrinsically 
habitable (\S\ref{sec:intrinhab}) as well as increasing the probability of 
complex life prospering on the host planet (\S\ref{sec:extrinhab}). Away from
astrobiology, moons would also shed new light onto the theories of planet/moon
formation (\S\ref{sec:moonformA}).

The expected properties of exomoons is discussed in \S\ref{sec:moonproperties},
including their formation (\S\ref{sec:moonformB}), evolution 
(\S\ref{sec:moonevolution}), dynamical stability (\S\ref{sec:moonstable}) and 
expected composition (\S\ref{sec:mooncomp}). I conclude the most feasible and
detectable case is a terrestrial planet which is captured by a Jupiter as an
irregular satellite. This therefore means retrograde orbits are possible for
the moon and it would likely exhibit a composition similar to the rocky planets
of our own Solar System.

Finally, I have explored the possible methods of detecting a non-transiting
planet. Direct imaging (\S\ref{sec:moonimaging}), astrometry 
(\S\ref{sec:astromoons}) and radial velocity methods (\S\ref{sec:rvmoons}) 
can be seen to be unfeasible as a method for detecting exomoons in the near 
future. Pulsar timing (\S\ref{sec:pulsarmoons}) has the necessary sensitivity 
but an insufficient catalogue of planets to survey.

Microlensing (\S\ref{sec:moonlensing}) has the greatest chance of detecting an 
exomoon in a non-transiting system in the near future. This seems to require 
space-based photometry to give a reasonable chance of success and indeed a 
mission such as EUCLID/WFIRST \citep{euclid2010} looks likely to fly within a 
decade. However, microlensing is unlikely to identify habitable-zone exomoons in
the near-future and would not be able to offer opportunities for intensive 
follow-up (e.g. to look for bio-signatures).

There is clearly a strong desideratum for a method which can detect exomoons, 
which have not only the potential to be habitable, but also be subsequently 
investigated to verify/reject such a hypothesis. Transiting systems, then,
may offer the only hope of a near-future achievement of such a goal. However,
before one can even entertain the notion of looking for exomoons with the 
transit method, one must first properly understand the transit light curve of a 
single planet, which I will discuss in the next chapter.

%% file: Chapt3.tex
\chapter{The Transiting Planet}
\label{ch:Chapt3}

\vspace{1mm}
\leftskip=4cm

{\it ``
From immemorial antiquity, men have dreamed of a royal road to success - leading
directly and easily to some goal that could be reached otherwise only be long
approaches and with weary toil. Times beyond number, this dream has proved to be
a delusion... Nevertheless, there are ways of approach to unknown territory
which lead surprisingly far, and repay their followers richly. There is probably
no better example of this than eclipses of heavenly bodies.''} 

\vspace{1mm}

\hfill {\bf --- Henry Norris Russell, 1946} 

\leftskip=0cm


\section{The Planetary Orbit}
\label{sec:planetaryorbit}

\subsection{Orbital Elements}
\label{sec:orbelements}

The first step in understanding the transit light curve is to define the 
appropriate coordinate system. In defining the so-called orbital elements, a 
Cartesian coordinate system is usually adopted. I begin with the simplest 
reference frame possible, where the planet orbits the star in the 
$\hat{x}$-$\hat{y}$ plane with the star at one focus, defined to be the origin, 
as shown in Figure~\ref{fig:simpleframe}. By working in the rest frame of the 
star, the reflex motion is inherently accounted for by the model. Concordantly, 
the Cartesian coordinate of the planet may be expressed as a function of the 
planet's true anomaly, $f_P$:

\begin{figure}
\begin{center}
\includegraphics[width=10.0 cm]{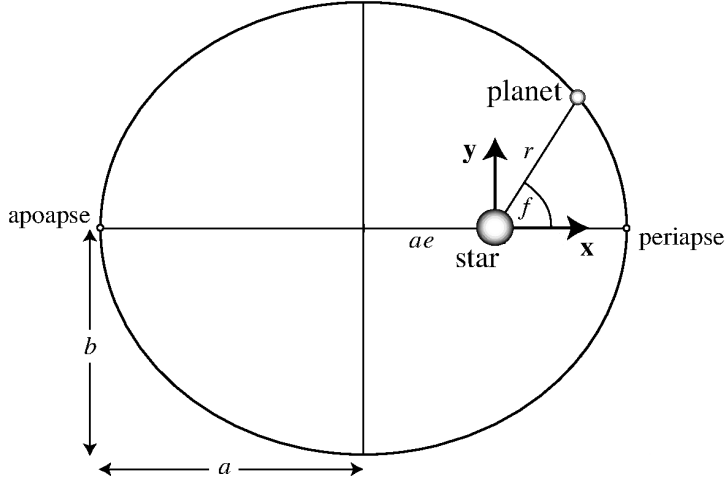}
\caption[The orbit of a planet around a star in the simplest reference frame]
{\emph{The orbit of a planet around a star in the simplest reference frame.
The orbit follows an elliptical path with the star at one focus, in accordance
with Kepler's First Law. Figure adapted from \citet{murray2010}.}}
\label{fig:simpleframe}
\end{center}
\end{figure}

\begin{align}
x_P &= r_P \cos f_P \nonumber \\
y_P &= r_P \sin f_P \nonumber \\
z_P &= 0
\label{eqn:simplecoords}
\end{align}

Where $r_P$ is the planet-star separation, given by:

\begin{align}
r_P &= a_P \frac{1-e_P^2}{1+e_P\cos f_P} = a_p \varrho_P(f_P)
\label{eqn:planetstarsep}
\end{align}

Where $\varrho_P(f_P)$ is used to absorb the effects of orbital eccentricity.
This simple 2D picture is, of course, impractical since real orbits exist in 3D. 
Say one executes $N$ rotations in three dimensions to account for the viewing 
angle of the target solar system in its local cluster, then the rotation of that
cluster to the Galactic plane and then the Solar System's rotation relative to 
the Galactic plane, etc, etc. One can very quickly obtain a large integer value 
for $N$ which would severely complicate the mathematics. However, 
\citet{euler1776} showed that any finite chain of rotations can be simplified to
just three rotations.

\begin{figure}
\begin{center}
\includegraphics[width=15.0 cm]{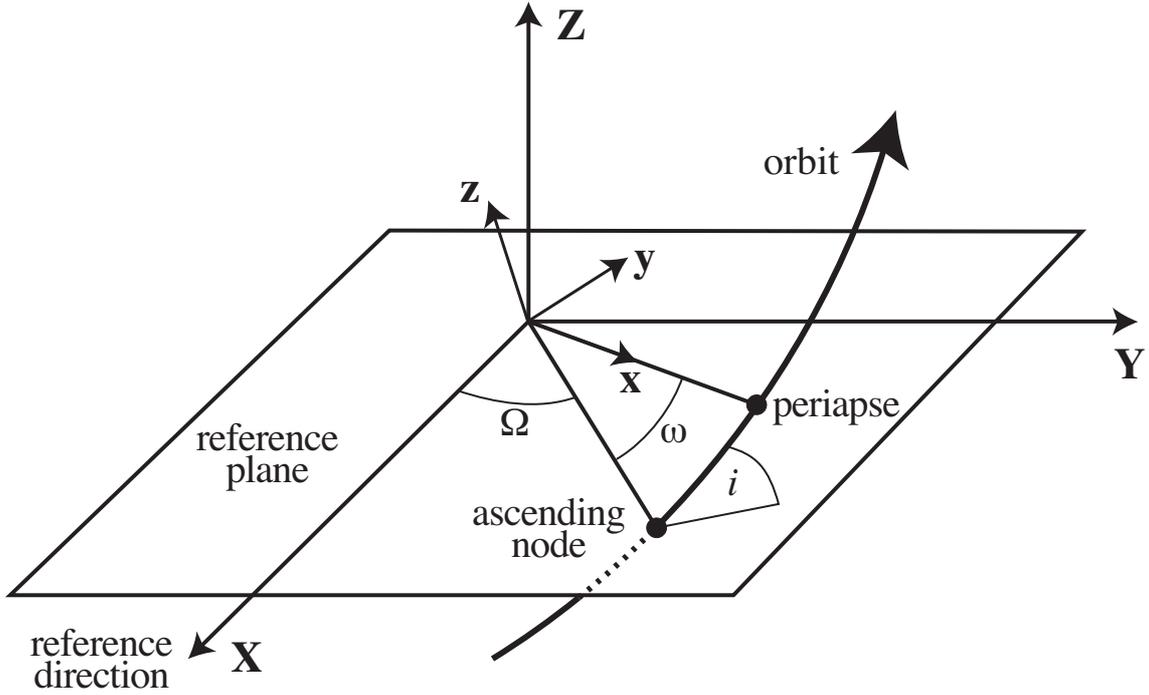}
\caption[Orbital elements of a planet orbiting a star]
{\emph{Orbital elements of a planet orbiting a star.
In the standard theory of transits, $i \simeq 90^{\circ}$ and the 
observer is located at $\{X,Y,Z\} = \{0,0,+\infty\}$. Figure from 
\citet{murray2010}.}}
\label{fig:complexframe}
\end{center}
\end{figure}

Standard celestial mechanics exploits Euler's theorem so that any planetary 
system can be described by just three angles: the argument of periapse, 
$\omega_P$, the orbital inclination angle, $i_P$, and the longitude of the 
ascending node, $\Omega_P$. The rotations are performed sequentially in a 
clockwise sense, about the axes $\hat{z}$-$\hat{x}$-$\hat{z}$. The rotations are
shown in Figure~\ref{fig:complexframe}, which illustrates the orbital elements.

As one is performing a series of three dimensional rotations, the easiest way to
represent the transformations mathematically is using matrices. Following 
\citet{murray2010}, I denote the following clockwise rotation matrices about 
the $\hat{x}$ and $\hat{z}$ axes respectively:

\begin{eqnarray}
{\bf P}_x (\phi) =\left(\begin{matrix}
1&0&0\cr
0&\cos\phi&-\sin\phi\cr
0&\sin\phi&\cos\phi\cr \end{matrix}\right)
\label{eqn:Px}
\end{eqnarray}

\begin{eqnarray}
{\bf P}_z (\phi) =\left(\begin{matrix}
\cos\phi&-\sin\phi&0\cr
\sin\phi&\cos\phi&0\cr
0&0&1\cr\end{matrix}\right)\,.
\label{eqn:Pz}
\end{eqnarray}

Armed with these definitions, the transformation from the simple frame of
$\{x,y,z\}$ to the observed frame of $\{X,Y,Z\}$ may be written as:

\begin{eqnarray}
\left(\begin{matrix}X_P\cr Y_P\cr Z_P\end{matrix}\right)=
{\bf P}_z (\Omega_P) {\bf P}_x (i_P) {\bf P}_z (\omega_P) \left(\begin{matrix}x_P\cr y_P\cr z_P\end{matrix}\right)
\label{eqn:matrixrotn}
\end{eqnarray}

After simplification, this yields the following final Cartesian coordinates:

\begin{align}
X_P &= r_P [ \cos\Omega_P\cos(\omega_P+f_P)-\sin\Omega_P\sin(\omega_P+f_P)\cos i_P] \nonumber \\
Y_P &= r_P [ \sin\Omega_P\cos(\omega_P+f_P)+\cos\Omega_P\sin(\omega_P+f_P)\cos i_P] \nonumber \\
Z_P &= r_P \sin(\omega_P+f_P)\sin i_P
\label{eqn:complexcoords}
\end{align}

In the final frame, plane of the sky is $\hat{X}$-$\hat{Y}$ and the observer is 
located far down the $+\hat{Z}$ axis at $\{0,0,+\infty\}$. From these simple 
equations, one is able to derive a large proportion of the equations governing 
exoplanetary transit light curves. However, a description of the planetary
motion is still missing.

\subsection{Kepler's Equation}
\label{sec:keplerseqn}

Equation~(\ref{eqn:complexcoords}) provides the instantaneous position of a 
planet as seen by the observer in three dimensions. However, one complication of
the equation is that it is expressed in terms of the true anomaly of the planet, 
$f_P$, rather than time, $t$. To convert from $f$ to $t$ is non-trivial and 
requires solving Kepler's Equation. A detailed discussion of the origins of the 
Kepler's Equation can be found in \citet{murray1999}, but for the purposes of 
this thesis, a statement of the relevant equations is sufficient.

In standard notation, a new parameter known as the mean anomaly, $\mathfrak{M}$,
is specified, which by definition scales linearly with time:

\begin{equation}
\mathfrak{M}(t) = n t + \mathfrak{M}_{0}
\label{eqn:meananomaly}
\end{equation}

The mean anomaly does not directly relate to the true anomaly but can be 
connected using an intermediary term known as the eccentric anomaly, $E$. The
relationship between $\mathfrak{M}$ and $E$ is given by:

\begin{equation}
\mathfrak{M} = E - e \sin E
\label{eqn:eccentricanomaly}
\end{equation}

Solving $E$ as a function of $\mathfrak{M}$ gives a transcendental equation, 
which is a source of perennial nuisance in computing orbits. This is typically 
done with a numerical iteration, starting from $E_1 = \mathfrak{M}$, and I opted
for the Newton-Rhapson iteration method \citep{danby1988}, given by:

\begin{equation}
E_{i+1} = E_i - \frac{E_i - e \sin E_i - \mathfrak{M}}{1-e \cos E_i}
\label{eqn:keplerseqn}
\end{equation}

Once $E$ has been found to the desired level of precision, the true anomaly is
computed using:

\begin{equation}
\tan\frac{f}{2} = \sqrt{\frac{1+e}{1-e}} \tan\frac{E}{2}
\label{eqn:trueanomaly}
\end{equation}

A very useful equation to introduce at this point is the rate of change of $f$,
with respect to time:

\begin{equation}
\frac{\mathrm{d}t}{\mathrm{d}f} = \frac{P}{2\pi} \frac{(1-e^2)^{3/2}}{(1+e\cos f)^2}
\label{eqn:angmom}
\end{equation}

\subsection{Kepler's Laws}
\label{eqn:keplerslaw}

Kepler's Laws of planetary motion are one of the bedrocks underlying the physics
of exoplanetary transits and it is appropriate to mention it here as part of the
fundamentals of transit theory. The three laws are:

\begin{enumerate}
\item[{$\mathbf{1.}$}] The orbit of every planet is an ellipse with the 
Sun at one of the two foci (i.e. Figure~\ref{fig:simpleframe})
\item[{$\mathbf{2.}$}] A line joining a planet and the Sun sweeps out 
equal areas during equal intervals of time
\item[{$\mathbf{3.}$}] The square of the orbital period of a planet is 
directly proportional to the cube of the semi-major axis of its orbit
\end{enumerate}

Kepler's Third Law is particularly cogent for exoplanet transits, as will be
seen later in \S\ref{sec:stellardensity}. It was originally stated as 
$P^2 \sim a^3$ but Newton's version of Kepler's Third Law (which I simply 
refer to Kepler's Third Law in the rest of this thesis) provides the constants 
of proportionality, derived from his Laws of Gravitation. For two bodies
labelled with subscripts 1 and 2, in the reference frame of body 1's rest frame,
Kepler's Third Law states:

\begin{equation}
\Big(\frac{P_1}{2\pi}\Big)^2 = \frac{a_1^3}{G (M_1+M_2)}
\label{eqn:kep3}
\end{equation}

\section{The Radial Velocity Signal}
\label{sec:therveqn}

A useful demonstration of the power of the equations obtained thus far is 
illustrated in the derivation of the radial velocity signal of a transiting 
planet. As discussed in \S\ref{sec:rv}, radial velocity measures the reflex 
motion of a star due to the presence of a companion along the line-of-sight. 
One can now see that the line-of-sight is $\hat{Z}$. Let us denote the motion of 
the separation between the star and the planet-star barycentre as $r_*$. Since 
$r_P$ is the planet-star separation then, $r_p = r_P M_*/(M_*+M_P)$ will be the 
planet-barycentre separation and finally $r_* = r_P M_P/(M_*+M_P)$ 
will be the star-barycentre separation. Therefore, we have:

\begin{equation}
r_* = a_P \varrho_P(f_P) \Big(\frac{M_P}{M_*+M_P}\Big)
\label{eqn:reflexmotion}
\end{equation}

In \S\ref{sec:orbelements}, I selected a reference frame with the star at
rest, but here one is interested in the star's reflex motion. To move from the
star's rest frame to that of the planet-moon barycentre is quite 
straightforward. I simply replace $r_P \rightarrow r_*$ in 
Equation~(\ref{eqn:complexcoords}) and then make use of 
Equation~(\ref{eqn:reflexmotion}) to obtain the correct description of the 
star's observed reflex motion in three dimensions. For radial velocity, the 
relevant term is $Z_*$:

\begin{equation}
Z_* = a_P \varrho_P(f_P) \Big(\frac{M_P}{M_*+M_P}\Big) \sin(\omega_P+f_P)\sin i_P
\label{eqn:Zstar}
\end{equation}

Note that $Z_*$ does not contain any dependency upon $\Omega_P$ and thus RV
measurements cannot constrain this parameter. The RV method measures 
d$Z_*$/d$t$, but a minus sign is usually introduced to give the historically 
standard definition of a positive RV corresponding to a receding object:

\begin{align}
\mathrm{RV} &= \gamma -\frac{\mathrm{d}Z_*}{\mathrm{d}t} \nonumber \\
\qquad &= \gamma -\frac{\mathrm{d}Z_*}{\mathrm{d}f_P} \frac{\mathrm{d}f_P}{\mathrm{d}t}
\label{eqn:RVformula}
\end{align}

Where $\gamma$ represents the constant systematic drift of the star relative to
the Sun, along our line of sight. Differentiating $Z_*$ with respect to $f_P$ 
and substituting Equation~(\ref{eqn:angmom}) into 
Equation~(\ref{eqn:RVformula}):

\begin{equation}
\mathrm{RV} = \gamma - K \sin i_P [e_P\cos\omega_P+\cos(f_P+\omega_P)]
\label{eqn:fullrveqn}
\end{equation}

Where I have substituted $K$, known as the ``radial velocity semi-amplitude'',
for:

\begin{align}
K &= a_P n_P \Bigg(\frac{M_P}{M_* + M_P}\Bigg) (1-e_P^2)^{-1/2} \nonumber \\
\qquad&= G^{1/3} n_P^{1/3} \Bigg(\frac{M_P}{(M_* + M_P)^{2/3}}\Bigg) (1-e_P^2)^{-1/2}
\label{eqn:rvK}
\end{align}

Where I have used Kepler's Third Law (Equation~(\ref{eqn:kep3})) on the second 
line. The mass terms may be grouped into a parameter sometimes referred to as
the ``mass function'', given by $M_P (M_* + M_P)^{-2/3}$. If one is able to 
determine $K$ from a set of radial velocity measurements, and the eccentricity, 
inclination, period and stellar mass are known, it should be possible to 
determine $M_P$. Solving Equation~(\ref{eqn:rvK}) for $M_P$ in this way yields a 
cubic with only one real root:

\begin{align}
M_P &= \Bigg[ -12 G K^3 M_* P_P \pi \xi_P^{3/2} - K^6 P_P^2 \xi_P^3 + K^3 P_P \xi_P^{3/2} \Bigg( -54 G^2 K^3 M_*^2 P_P \pi^2 \xi_P^{3/2} \nonumber \\
\qquad& - 18 G K^6 M_* P_P^2 \pi \xi_P^3 - K^9 P_P^3 \xi_P^{9/2} + 6\sqrt{3}\pi^{3/2} \sqrt{G^3 K^6 M_*^3 P_P^2 \xi_P^3 (27 G M_* \pi + 2 K^3 P_P \xi_P^{3/2})}\Bigg)^{1/3} \nonumber \\
\qquad& -\Bigg(-54 G^2 K^3 M_*^2 P_P \pi^2 \xi_P^{3/2} - 18 G K^6 M_* P_P^2 \pi \xi_P^3 - K^9 P_P^3 \xi_P^{9/2} \nonumber \\
\qquad& + 6\sqrt{3}\pi^{3/2} \sqrt{G^3 K^6 M_*^3 P_P^2 \xi_P^3 (27 G M_* \pi + 2 K^3 P_P \xi_P^{3/2})}\Bigg)^{2/3} \Bigg] \Bigg[6 G \pi \Bigg(-54 G^2 K^3 M_*^2 P_P \pi^2 \xi_P^{3/2} \nonumber \\
\qquad& - 18 G K^6 M_* P_P^2 \pi \xi_P^3 - K^9 P_P^3 \xi_P^{9/2} + 6\sqrt{3}\pi^{3/2} \sqrt{G^3 K^6 M_*^3 P_P^2 \xi_P^3 (27 G M_* \pi + 2 K^3 P_P \xi_P^{3/2})}\Bigg)^{1/3}\Bigg]^{-1}
\label{eqn:massfn}
\end{align}

Where I have used $\xi_P = (1-e_P^2)$. I have not found this equation
previously in the exoplanet literature but it is clearly preferable to use this 
exact solution as opposed to approximating $M_* \gg M_P$ in the mass function
given in Equation~(\ref{eqn:rvK}) and solving from there.

\section{The Transit Light Curve for a Uniform Source Star}
\label{sec:uniformtransit}

\subsection{Transit Basics}
\label{sec:transitbasics}

We have now accumulated sufficient background knowledge to describe the transit
light curve. For the purpose of introducing the topic, I will 
here only describe the case of a star of uniform brightness, which eliminates 
the complications of limb darkening effects. Although a detailed account of limb 
darkening is not given in this thesis for the sake of brevity, I will later
briefly discuss in \S\ref{sec:LDtransit} how it is incorporated into my 
modelling routines.

I start by assuming that A1) there are only two bodies in the system
A2) the motion of these bodies is completely described by Newton's laws only. 
The second assumption clearly excludes many other possible complications e.g.
relativistic effects, Newton's laws are not universal, tidal dissipation, disc
migration, etc. Armed with these two assumptions, the equations of motion
provided earlier in \S\ref{sec:planetaryorbit} are valid here. I also define
assumption A0 to be that the star has uniform brightness.

As was seen in \S\ref{sec:therveqn}, the star exhibits reflex motion, which 
means that in the inertial frame of the planet-star barycentre there are two 
bodies in motion. It is therefore simpler to choose a frame with the star at 
rest, as was done in \S\ref{sec:orbelements}, and so the position of the planet 
is given by Equation~(\ref{eqn:complexcoords}). Since the star is at rest then, 
this equation also describes the planet-star separation, which is of course the 
relevant issue for transits.

More specifically, it is not really the planet-star separation which matters, 
but the \textbf{sky-projected} planet-star separation. In other words, I
acknowledge that an observer has a certain perspective of the system.
The sky-projected planet-star separation, $S_{P*}'$, is given by the quadrature
sum of the two coordinates lying in the sky-plane\footnote{The subscripts ``P''
and ``*'' in $S_{P*}'$ are commutative}:

\begin{align}
S_{P*}' &= \sqrt{X_P^2 + Y_P^2} \nonumber \\
\qquad&= \frac{a_P (1-e_P^2)}{1+e_P\cos f_P} \sqrt{1-\sin^2(\omega_P+f_P)\sin^2i_P}
\label{eqn:Sdasheqn}
\end{align}

$S_{P*}'$ completely describes the transit light curve for a uniformly emitting
star. Note how there is no dependence on $\Omega_P$, just as we had for the
RV signal, and so once again this parameter cannot be determined from the 
transit light curve. For a planet crossing over the stellar disc, there are 
three, and only three, distinct cases:

\begin{itemize}
\item[{\tiny$\blacksquare$}] \emph{Out-of-transit}: No part of the sky-projected 
planetary disc overlaps with the sky-projected stellar disc
\item[{\tiny$\blacksquare$}] \emph{On-the-limb}: The sky-projected planetary 
disc lies partially within the sky-projected stellar disc
\item[{\tiny$\blacksquare$}] \emph{In-transit}: The sky-projected planetary disc 
lies fully inside the sky-projected stellar disc
\end{itemize}

So far I have only made three assumptions; A0, A1 and A2. Let us make the 
additional assumption, A3, that both the planet and star are perfect spheres. 
The 2D projection of a sphere viewed from any angle will always appear as a 
circle, which means that the three cases mentioned above can be completely 
described in terms of the radii of these circles (radius of the planet, $R_P$, 
and radius of the star, $R_*$) and $S_{P*}'$.

\begin{itemize}
\item[{\tiny$\blacksquare$}] \emph{Out-of-transit}: 
$R_*+R_P\leq S_{P*}' < \infty$
\item[{\tiny$\blacksquare$}] \emph{On-the-limb}: $R_*-R_P \leq S_{P*}' < R_*+R_P$
\item[{\tiny$\blacksquare$}] \emph{In-transit}: $0 \leq S_{P*}' < R_*-R_P$
\end{itemize}

With the three cases defined, let us ask what is the observed flux for each
case. To do this, one has to make several new assumptions which are often not 
explicitly stated. However, for completion, I will here state all of these 
assumptions (including those stated earlier):

\begin{itemize}
\item[{\textbf{A0}}] The star has uniform brightness
\item[{\textbf{A1}}] There are only two bodies in the system
\item[{\textbf{A2}}] The motion of these bodies is completely described by 
Newton's laws only
\item[{\textbf{A3}}] Both the planet and star are perfect spheres
\item[{\textbf{A4}}] The planet emits no flux and is completely opaque
\item[{\textbf{A5}}] The star's emission is of constant flux
\item[{\textbf{A6}}] There are no background/foreground luminous objects
\item[{\textbf{A7}}] The exoplanetary system is a constant distance $d$ from the
observer, such that $d\gg a_P$
\item[{\textbf{A8}}] The planet has no extended features, such as an atmosphere,
rings, etc
\item[{\textbf{A9}}] The planet does not cause gravitational microlensing of
the host star's light
\item[{\textbf{A10}}] The change in flux over a single integration is
much smaller than the flux measurement uncertainty
\end{itemize}

Now, that in-transit, the flux of a star is attenuated from 
$F_* \rightarrow F_* [1-(R_P^2/R_*^2)]$ i.e. the ratio-of-areas. Therefore,
the depth of a transit, $\delta$, is:

\begin{equation}
\delta = R_P^2/R_*^2 = p^2
\label{eqn:uniformdepth}
\end{equation}

Where I have used $p=R_P/R_*$ i.e. the ratio-of-radii. As $p$ is essentially a 
measurement of the planetary radius in units of the stellar
radius, it will be useful to translate $S_{P*}'$ to also be in units of the
stellar radius. I therefore define $S_{P*} = S_{P*}'/R_*$. I now normalize the
observed flux at any instant to that occurring during the out-of-transit times,
to give the so-called ``normalized flux'', $F_*^N$.

\begin{equation}
F_{*,\mathrm{transit}}^N(S_{P*}) = 
\begin{cases}
1 & 1+p \leq S_{P*} < \infty \\
1-\alpha(1,p;S_{P*}) & 1-p \leq S_{P*} < 1+p \\
1-p^2 & 0 \leq S_{P*} \leq 1-p
\end{cases}
\label{eqn:transitcases}
\end{equation}

Where $\alpha(R,r;S)$ is the area of overlap between two circles, of radii $R$ 
and $r$ with separation $S$. This simple scenario has a well-known solution:

\begin{align}
\alpha(R,r;S) &= r^2 \kappa_0(R,r;S) + R^2 \kappa_1(R,r;S) - \kappa_2(R,r;S) \\
\kappa_0(R,r;S) &= \arccos\Big[\frac{S^2+r^2-R^2}{2 S r}\Big] \\
\kappa_1(R,r;S) &= \arccos\Big[\frac{S^2+R^2-r^2}{2 S R}\Big] \\
\kappa_2(R,r;S) &= \sqrt{\frac{4 S^2 R^2 - (R^2+S^2-r^2)^2}{4}}
\label{eqn:alpha}
\end{align}

\subsection{Conditions for a Transit}
\label{sec:transitconditions}

One can see from Equation~(\ref{eqn:transitcases}) that if there is no instance 
in the orbit where $S_{P*}<1+p$, then $F_*^N=1$ at all times and so no transit 
features will be observable. Therefore, the condition for a transit to occur
is dependent upon the minimum value of $S_{P*}$. Let us denote this minimum as
$S_{P*,T}$ and the true anomaly at which it occurs as $f_{P,T}$ (the ``T'' 
subscript denotes transit, as opposed to ``O'' for occultation which is
discussed more in \S\ref{sec:occultations}).

$S_{P*,T}$ occurs approximately at the time of inferior conjunction, but the 
exact solution has a rather elaborate analytic form and is presented in 
\S\ref{sec:transitminima}. Proceeding with the approximate form for now, the 
minimum corresponding to the primary transit is therefore 
$f_{P,T} \simeq (\pi/2-\omega_P)$ . Substituting $f_{P,T}$ into 
Equation~(\ref{eqn:Sdasheqn}), one obtains $S_{P*,T}$:

\begin{equation}
S_{P*,T} \simeq \frac{a_P}{R_*} \frac{1-e_P^2}{1+e_P\sin \omega_P} \cos i_P = b_{P,T}
\label{eqn:impactparameter}
\end{equation}

Where $b_{P,T}$ is known as the ``impact parameter'' of the planetary transit 
event\footnote{Some of the subscripts present in the notation employed at this
stage of the thesis may seem superfluous, but once moons and occultations are 
introduced, the value of these subscripts will become apparent}. Therefore, for 
a full-transit to occur one requires $b_{P,T}\leq (1-p)$ and for a 
so-called ``grazing'' transit one requires $(1-p)<b_{P,T}<(1+p)$. For a circular 
orbit, one can see that two critical parameters will impinge on the light curve, 
$b_{P,T}$ and $a_P/R_*$. For an eccentric orbit, one has the additional 
parameter $\varrho_{P,T} = \varrho_{P}(f_P=\pi/2-\omega_P)$ (see 
Equation~(\ref{eqn:planetstarsep}) for the definition of $\varrho_P(f_P)$).
The importance of these parameters becomes apparent when fitting light curves.

\subsection{Anatomy of the Transit Light Curve}
\label{sec:transitanatomy}

\begin{figure}
\begin{center}
\includegraphics[width=10.0 cm]{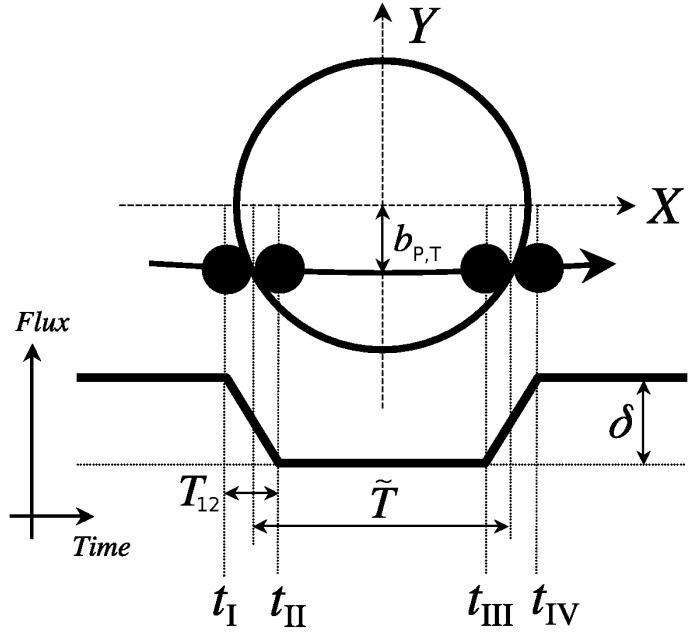}
\caption[Anatomy of the transit light curve]
{\emph{Anatomy of the transit light curve. The four key contact points are
labelled along the bottom and the corresponding position of the planet is
visible from the upper part of the figure. The transit durations $\tilde{T}$ and
$T_{12}$ are also marked, which will be discussed further in 
Chapter~\ref{ch:Chapt4}. The impact parameter, $b_{P,T}$, is defined as the
sky-projected planet-star separation in units of the stellar radius at the
instant of inferior conjunction. Figure adapted from \citet{winn2010}.}}
\label{fig:transitanatomy}
\end{center}
\end{figure}

As established in the previous subsection and stated explicitly in 
Equation~(\ref{eqn:transitcases}), there are three distinct cases for the 
transit. The three cases are buffeted by two boundaries i.e. $S_{P*} = (1+p)$ 
and $S_{P*} = (1-p)$. For a grazing transit, only the former of these two 
boundaries is crossed, but for a full-transit both are traversed. In either 
case, the planet must both enter and exit these boundaries and thus a 
full-transit has a total of four boundary crossings, which are usually labelled 
as the ``contact points'' of the transit. These contact points are labelled 
sequentially as times $t_I$, $t_{II}$, $t_{III}$ and $t_{IV}$, and the instant 
when $S_{P*}$ is minimized is $\tau_T$ (see Figure~\ref{fig:transitanatomy}). 
The durations between these contact points will be discussed in 
Chapter~\ref{ch:Chapt4}.

\subsection{Occultations}
\label{sec:occultations}

As well as the primary transit event, secondary eclipses or ``occultations'' can
occur too. Occultations cause a much shallower eclipse depth as one is
now seeing the flux decrease from $(F_*+F_P) \rightarrow F_*$ and, in general,
$F_P \ll F_*$ (where $F_P$ is the flux from the planet). $F_P$ can be caused by 
reflected light and thermal emission from the exoplanet and so in both cases it 
is largest for close-in planets. By definition, in order to acknowledge the 
presence of an occultation one must break assumption A4. This does affect the
transit event and will be discussed in \S\ref{sec:nightside}. Proceeding with
the occultation for now, one may combine the two possible sources for the 
eclipse, to calculate the occultation depth, $\epsilon$:

\begin{align}
\epsilon &= \Bigg[ \underbrace{\frac{\int_{\lambda=0}^{\infty} R_I(\lambda) \mathfrak{A}_G(\lambda) \,\mathrm{d}\lambda}{(a_P/R_*)^2}}_\text{reflection} + \underbrace{\frac{\int_{\lambda=0}^{\infty} R_I(\lambda) F_P(\lambda)\,\mathrm{d}\lambda}{\int_{\lambda=0}^{\infty} R_I(\lambda) F_*(\lambda)\,\mathrm{d}\lambda}}_\text{thermal emission}  \Bigg] p^2 \nonumber \\
\qquad&= \mathfrak{F}_{P*} p^2 
\label{eqn:seccy}
\end{align}

Where the left-hand side is for reflected light and the right-hand side for 
thermal emission. $R_I(\lambda)$ denotes the response function of the 
instrument used to observe the star. All of these effects can be absorbed into
the quantity $\mathfrak{F}_{P*}$, the ratio of the planet's flux per unit area to 
that of the star. This means in modelling the occultation one may use:

\begin{equation}
F_{*,\mathrm{eclipse}}^N(S_{P*}) = 
\begin{cases}
1 & 1+p \leq S_{P*} < \infty \\
1-\mathfrak{F}_{P*} \alpha(1,p;S_{P*}) & 1-p \leq S_{P*} < 1+p \\
1-\mathfrak{F}_{P*} p^2 & 0 \leq S_{P*} < 1-p
\end{cases}
\label{eqn:eclipsecases}
\end{equation}

Just as transits occur near the time of inferior conjunction, occultations occur
near the time of superior conjunction. 
For circular orbits, the presence of a primary transit guarantees that an
occultation of equal duration must occur at a time 
$\tau_O=\tau_T + 0.5P_P + \Delta t_{\mathrm{light}}$, where 
$\Delta t_{\mathrm{light}}$ is the light travel time across the system and is
typically less than a minute for hot-Jupiters \citep{loeb2005}. However, for 
eccentric orbits the presence of an occultation is not guaranteed and the 
duration and time of the event may also vary, which can be exploited to actually
measure the eccentricity \citep{deming2007}. These points will be addressed in 
\S\ref{sec:transitminima} and \S\ref{sec:pritosec}. 

\subsection{Phase Curves}
\label{sec:phasecurves}

\begin{figure}
\begin{center}
\includegraphics[width=15.0 cm]{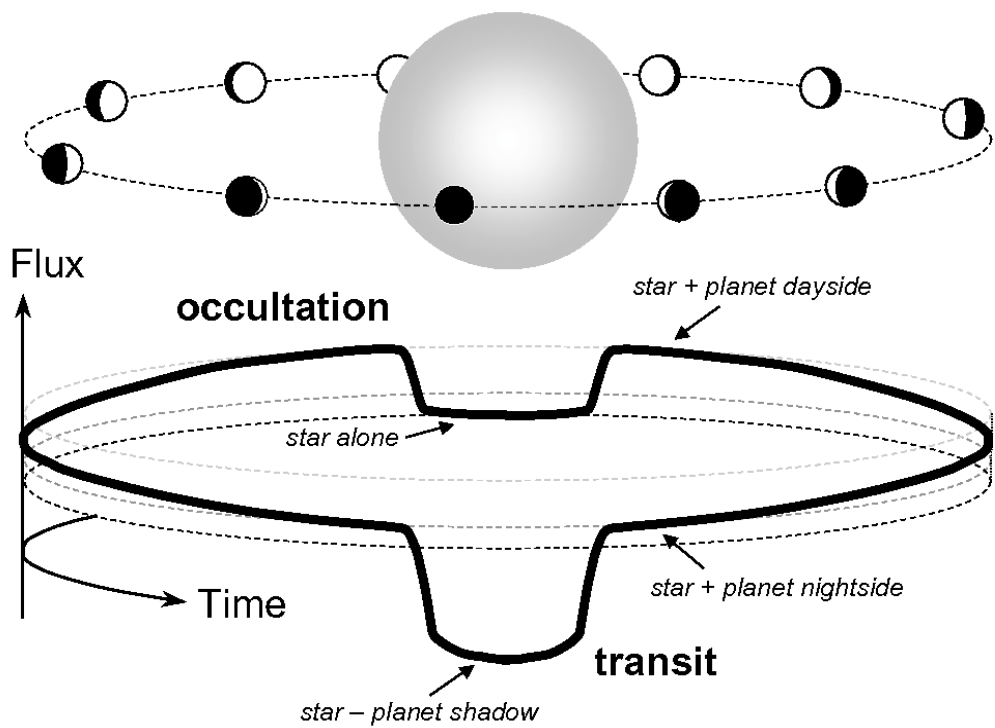}
\caption[Phase curve of a transiting exoplanet]
{\emph{Illustration of the phase curve of an exoplanet in the context of the
primary and secondary transits. As the day-side comes into view, the flux
from the planet appears to increase giving rise to a slight increase in the 
total flux observed from the star+planet pair. Figure from \citet{winn2010}.}}
\label{fig:phasecurve}
\end{center}
\end{figure}

For a hot-Jupiter system, the very short-period is thought to cause tidal 
locking of the planet's rotational period to the orbital period 
\citep{gladman1996}. Therefore, the planet has the same side facing the star at 
all times leading to potentially large changes in emitted flux between the day 
and night side. As the planet moves from the primary to secondary positions, one 
sees the day side come into view and thus continuous monitoring could reveal 
small changes in the total flux, known as the ``phase curve'', indicative of the
day-night contrast (the first successful observation of this effect was achieved
by \citet{knutson2007}). Figure~\ref{fig:phasecurve} provides an illustration of
this in the context of the primary and occultations.

\subsection{Extending to Limb Darkened Stars}
\label{sec:LDtransit}

The equations presented thus far allow one to fully model the primary transit
and occultations of an exoplanet. Whilst occultations are completely
unaffected by stellar limb darkening, the primary light curve is modified. As
limb darkening is circularly symmetric, the $S_{P*}$ parameter is still the only
term one needs to describe the light curve. This is contrast to gravity 
darkening, which breaks this symmetry and thus requires one to compute $X_P$ and 
$Y_P$ separately at every instance. Fortunately, gravity darkening is generally
ineffective and may be neglected, although it does become significant for 
rapidly rotating stars (spectral types $\sim$F6 and earlier) and the appropriate
treatment is described in \citet{barnes2009}.

Limb darkened stars exhibit a monotonically decreasing brightness profile from 
centre to limb. Numerous profile forms have been proposed including linear 
\citep{claret1995}, quadratic \citep{claret1995}, square-root 
\citep{claret1995}, four-coefficient polynomial \citep{claret2000} and 
logarithmic \citep{claret2000}. \citet{claret2000} argue that the 
four-coefficient model gives the most accurate representation but the quadratic 
model is sufficient for most applications. The use of just two coefficients 
rather than four is preferable from a parameter fitting perspective and I will 
adopt the quadratic law in this thesis. I treat the specific stellar intensity 
as:

\begin{equation}
\frac{I_{\mu}}{I_1} = 1 - u_1 (1-\mu) - u_2 (1-\mu)^2
\label{eqn:LDstar}
\end{equation}

Where $\mu\equiv\cos\theta$, $\theta$ is the angle between the stellar 
surface normal vector and the line-of-sight, and $u_1$ and $u_2$ are the linear
and quadratic limb darkening coefficients respectively.

\citet{mandel2002} presented equations for modelling the transit light curve
with linear, quadratic and four-coefficient polynomial limb darkening. Another
benefit of the quadratic model is that \citet{mandel2002} find the 
four-coefficient method requires computing hypergeometric functions, which slow 
down a fitting routine significantly. In this thesis, I employ the quadratic 
expressions from \citet{mandel2002} to generalize the modelling of transit light
curves and thus assumption A0 (a uniform source) is broken.

\subsection{Breaking the Assumptions}
\label{sec:breakassumptions}

Breaking any one of the assumptions listed in \S\ref{sec:transitbasics} will
invalidate the equations presented. As one may expect, some of the assumptions
hold in almost all situations whereas others are almost invariably broken. The
consequences and likelihood of each assumption being broken is discussed here.

\subsubsection{A0) A uniformly emitting star}

I have discussed how limb darkening \citep{mandel2002} strongly affects the 
transit light curve but has no effect on the occultation event. Limb darkening 
becomes less prominent for stars observed in the infrared wavelengths but in 
general is always present and so it is prudent to always account for limb 
darkening.

Gravity darkening \citep{barnes2009} is rarely prominent and currently no light
curves of a transiting planet have ever been shown to exhibit this effect.
However, rapidly rotating stars will have significant gravity darkening and
in these cases one should model the effect \citep{barnes2009}.

Other non-uniform effects, such as starspots and plages, can have a significant
effect on the transit light curve, especially at visible wavelengths. If a
planet transits across a starspot then an apparent flux increment will occur
which may be used to determine the spot's parameters \citep{pont2007}. Much
more troubling (and more probable) are out-of-transit spots which can induce
variations in the transit depth from epoch-to-epoch \citep{czesla2009}.

\subsubsection{A1) There are only two bodies in the system}

For just two bodies following Keplerian orbits, the orbital motions are
described with Equation~(\ref{eqn:angmom}). Accordingly, transits occur once
every orbital period, $P_P$. However, the introduction of a third body leads to 
an extremely challenging analytic problem. Indeed, the general solution to the
three-body problem has never been found and as such one is forced to use
either approximate solution for restricted cases or numerical methods.

Additional bodies, whether a nearby star, outer/inner planet or companion moon
will perturb the orbital motion of the transiting planet. More specifically,
the gravitational influence of the third body acts to perturb the position and 
velocity of the transiting body, which manifest as transit time and duration
variations (TTV and TDV). TTV means that the planet no longer transits once
every orbital period (the deviation can be seconds up to days) and TDV means 
that the duration of the transit events vary from orbit-to-orbit.

\citet{agol2005} and \citet{holman2005} discuss the TTV effects of a perturbing
planet and show that bodies in mean motion resonance (MMR) with the transiting
planet induce heavily amplified signals. \citet{borkovits2010} recently extended
these analyses to longer time scale perturbations too. Apsidal and nodal
precession which may be induced by a third body lead to TTV and TDV effects
as well (\citet{jordan2008};\citet{investigations2010}).
\citet{sartoretti1999} discuss the TTV effects caused by a companion moon and 
\citet{kipping2009a,kipping2009b} predict an accompanying TDV signal (also see 
Chapter~\ref{ch:Chapt6}). \citet{ford2007} discuss the TTV effect caused by a 
Trojan body.

Aside from TTV and TDV, the third body can also transit the host star as well.
If these transits are simultaneous (e.g. the case of a moon), then the transit
signal can modify the light curve shape in an elaborate manner. Whilst authors
such as \citet{sato2009} have touched on these points, a generalized model
which inherently includes both transit timing effects and the light curve
features remains absent in the exoplanet literature. A new framework to solve
this problem, \luna, is currently under development (see 
Chapter~\ref{ch:Chapt8}).

\subsubsection{A2) The motion are completely described by Newton's laws only}

Newton's laws produce very accurate predictions of the planetary motions in the
Solar System. However, we have already seen in Chapter~\ref{ch:Chapt1} how 
Mercury's observed precession rate cannot be explained by Newtonian mechanics 
alone \citep{clemence1947}. General relativistic effects perturb the precession 
rate and so in this case one of the assumptions is clearly invalid. 
\citet{jordan2008} provide a detailed discussion of the observability of
general relativistic precession and conclude that decades of duration 
measurements may be able to infer the rate. However, for time scales of years
or less these effects are not significant.

Other effects include tidal dissipation and disc migration. Migration generally
occurs on the time scale of Myr and thus is insignificant. Tidal dissipation is
usually a slow process as well, although it has been hypothesized that in some
extreme cases for planets on ultra-short periods it may result in TTV and TDV
effects as the planet spirals inwards \citep{hellier2009,investigations2010}. 
Finally, stellar quadrupole moments may also cause timing deviations through the 
Applegate effect, but are generally long-term deviations \citep{watson2010}.

\subsubsection{A3) Both the planet and star are perfect spheres}

Whilst a detailed discussion of stellar oblateness is absent in the exoplanet
literature, planetary oblateness has been investigated. \citet{seager2002}
suggested such planets would induce asymmetric light curves but requiring
photometric precisions at or beyond the ppm level. \citet{carter2010} suggested
that precession of the axial tilt of an oblate planet may offer a more 
detectable signal due to resultant depth changes, but predict decades of data
are required in most cases. In general, these effects do not need to be included
in a transit model.

\subsubsection{A4) The planet emits no flux and is completely opaque}

Although I could find no work considering the effects of a non-opaque planet,
it is perhaps safe to assume that such an event is quite unlikely. For the
planetary flux issue, occultations cannot be understood at all unless A4 is
discarded. With many observations of such events, particularly at infrared 
wavelengths, now existing in the literature (e.g. \citet{deming2006}, 
\citet{charbonneau2008}, \citet{swain2009}) the assumption must be seen as
frequently invalid.

\citet{kiptin2010} presented a method for correcting transit light curves for
the effects of nightside emission (or ``pollution'' as dubbed by the authors
due to the observational consequences) and this will be covered in this thesis
in \S\ref{sec:nightside}. As will be seen later, the overall effect is a
predictable dilution of the transit depth.

\subsubsection{A5) The star's emission is of constant flux}

Stellar activity is always present at various degrees of intensity and depends
heavily on the spectral type and rotation. Stars with very large activities
are generally not selected for transit surveys and thus few examples exist in
the literature. Those exhibiting pronounced activity induce photometric
variations at the level of a percent which can induce TTV \citep{alonso2008}.
TDV is also predicted to be induced by magnetic activity in the star over the
time scales of decades \citep{loeb2009}. Techniques such as Fourier filtering 
\citep{alapini2009} and wavelet analysis \citep{carter2009} can remove the bulk
of the activity but of course it is preferable to select a ``quiet'' star in the
first place (criterion C6).

Fortunately, 65-70\% of F7-K9 main-sequence stars in the are likely to have 
similar or lower intrinsic variability than the Sun \citep{batalha2002}, which
is 10\,ppm. Such a tiny noise level is generally inconsequential to light curve
analyses.

\subsubsection{A6) There are no background luminous objects}

Background luminous objects cause a dilution of transit depths, in a very
similar way to how nightside emission does \citep{kiptin2010}. These dilutions
are usually called ``blends'' and thus nightside emission is a self-blending
scenario. More generally, blends mean that instead of seeing just $F_*$ one has
$F_*+F_B$, where $F_B$ is the flux from the blended object. This causes the
observed transit depth to attenuate to:

\begin{align}
\delta' &= \frac{\delta}{B} \\
B &= \frac{F_* + F_B}{F_*}
\end{align}

Where $B$ is the blending factor. Due to the simple nature of the effects of 
blends, it is easy to account for them in transit modelling if $B$ is known. For
surveys like \emph{Kepler}, adaptive optics follow-up of transit candidates 
allows one to spatially resolve nearby stars from the target and thus determine
$B$ to high confidence (e.g. \citet{latham2010}). There is always a small chance
that a blended target is too close to spatially resolve but the low probability
of this occurrence is quantifiable and manageable.

The assumption of no luminous background objects also excludes gravitational
microlensing of a distant background source. Such an effect would be
extremely pronounced and easily identified but no examples of a transiting
system acting as a lens star are known to exist. \citet{lewis2001} estimate a
very low probability, $\sim10^{-6}$, of such an event being detected.

\subsubsection{A7) The exoplanetary system is a large and constant distance away}

By treating the distance of the system, $d$, to be at a large distance away
(i.e. $d\gg a_P$), one can neglect the angular size of the planet. The exact
size of the shadow of the planet cast onto the stellar surface has a radius 
$R_P'$ and causes a transit depth $\delta'$:

\begin{align}
R_P' &= \frac{r_{P,T} + d}{d} R_P \nonumber \\
\delta' &\simeq \Bigg[1 + 2.34\times10^{-13}\,\Bigg(\frac{a_P}{\mathrm{AU}}\Bigg)^2 \Bigg(\frac{10\,\mathrm{pc}}{d}\Bigg)^2 \Bigg] \delta 
\end{align}

Where $r_{P,T}$ is the planet-star separation at the transit minimum and on the
second line I have approximated the orbit to be circular. The change in the
transit depth is clearly negligible and thus assumption A7 seems justified.

Finally, one interesting effect which can occur for systems with high proper 
motion is a deviation in the times of transit due to the varying light travel 
time \citep{rafikov2009}.

\subsubsection{A8) The planet has no extended features}

Planetary rings will induce transit features before and after the transit event.
These features can be either brightenings due to forward scattering or dimmings
due to opacity, but no detection of such phenomena have been made. 
\citet{barnes2004} provide simulations of transit light curves featuring ring
systems and argue that \emph{Kepler} should be able to detect many 
feasible configurations.

An atmosphere is also neglected in this analysis. The opacity of an atmosphere
will vary as a function of wavelength due to absorption of light due by
different molecules. This has been exploited to detect or constrain the presence
of molecules in exo-atmospheres, in what is usually dubbed as ``transmission 
spectroscopy'' (e.g. \citet{charbonneau2002b}, \citet{tinetti2007}, 
\citet{beaulieu2009}). For a single bandpass observation, this actually
requires no modification to the model as $R_P/R_*$ simply increases or 
decreases.

One atmospheric effect which can cause a breakdown of this model is refraction
through the atmosphere, causing a lensing effect (\citet{seager2002}, 
\citet{sidis2010}). The most significant perturbation is the presence of 
out-of-transit features caused by the lensing, which may mimic rings in their 
shape and size. No observations of these effects have been reported though, and 
they do not need to be included in general.

\subsubsection{A9) The planet does not cause gravitational microlensing}

\citet{kasuya2010} discuss the possibility of a planet acting as a gravitational
lens to the host star and causing a degree of microlensing. Notably, spikes 
before and after the transit event may be observable, with amplitudes 
$\sim 100$\,ppm. However, typically these events are only observable for 
transiting planets at very large separation from their host star 
($\sim 200$\,AU), which means they are geometrically highly unlikely to transit 
in the first place ($\sim10^{-5}$). Therefore, I do no consider this effect to 
be necessary to model in most circumstances.

\subsubsection{A10) The change in flux over a single integration is
much smaller than the flux measurement uncertainty}

Taking long integrations for the photometric observations can result in a 
smearing of the transit signal. This effect was first discussed in 
\citet{kipbin2010} and I will come back to it in \S\ref{sec:binning}. The main
lesson is that it is possible to modify the light curve model to incorporate
the integration times and thus produce accurate parameter estimates for the
system. One minute integration typically exhibit negligible smearing whilst
anything above $\sim$15 minutes can cause significant effects. Given that the
\emph{Kepler Mission} performs the vast majority of its photometry using 30 
minute integrations, this effect can be quite severe and should be accounted
for.

\section{Summary}

I have presented the fundamental equations for the sky-projected motion 
(\S\ref{sec:planetaryorbit}), the associated radial velocity signal
(\S\ref{sec:therveqn}) and observed light curve of a transiting planet
(\S\ref{sec:uniformtransit}). The light curve for a uniform brightness star can
be completely understood in terms of the sky-projected planet-star separation,
$S_{P*}$, and this critical parameter is derived for two bodies in Keplerian
orbits. The derived equations are easily manipulated to give the associated RV
signal, which is critical in global modelling of the available data. Finally,
I have presented a simple model for the observed flux during a transit event
and clearly outlined my assumptions. The effect(s) of breaking the various
assumptions are overviewed in \S\ref{sec:breakassumptions} and I discuss the
frequency and significance of each assumption being broken. In particular, I 
find that three out of the eleven assumptions are likely to be frequently broken
with significant effects. These are A0) a uniform source star; A4/A5) the planet
emits no flux/there are no background luminous objects; A10) the integration
times are small. A4 and A5 are grouped together because they may be modelled
using the same method, the introduction of a blending factor, $B$, which will be
discussed in more detail in \S\ref{sec:nightside}. For A0, I discuss how the
expressions of \citet{mandel2002} are used in my fitting routines to account for
limb darkening. For A10, this assumption is broken by the long-cadence data of
the \emph{Kepler Mission} which uses 30\,minute integrations inducing smearing
of the light curve. In \S\ref{sec:binning}, I will discuss a method to
compensate for this effect.

With the basics established, there are numerous subtleties which I have skimmed
over and wish to return to. Notably, the ``transit time'' is defined as the
moment when $S_{P*}$ is minimized and so far I have only stated that these
moments occur near the times of inferior and superior conjunction for the
transit and occultation events respectively. Also, the duration of a transit
event has not been addressed. All of the time-related quantities will be
discussed in the next chapter, Chapter~\ref{ch:Chapt4}.

%% file: Chapt4.tex
\chapter{Timing the Transit}
\label{ch:Chapt4}

\vspace{1mm}
\leftskip=4cm

{\it ``
What then is time?  If no one asks me, I know what it is.  If I wish to explain 
it to him who asks, I do not know.''} 

\vspace{1mm}

\hfill {\bf --- Saint Augustine} 

\leftskip=0cm


\section{Transit Minima}
\label{sec:transitminima}

\subsection{The Exact Solution}
\label{sec:exactminima}

I discussed earlier how the minima of $S_{P*}$ occur approximately at the times
of inferior and superior conjunction for the transit and occultation 
respectively. In this section, I will derive the exact solution.

The instants of minima (and maxima) $S_{P*}$ occur when d$S_{P*}/$d$t=0$. This
may be expanded using the chain rule to:

\begin{align}
\frac{\mathrm{d}S_{P*}}{\mathrm{d}t} = \frac{\mathrm{d}S_{P*}}{\mathrm{d}f_P} \frac{\mathrm{d}f_P}{\mathrm{d}t}
\end{align}

Since $\dot{f_P} \neq 0$ under any circumstances, then the condition that
d$S_{P*}/$d$t=0$ is equivalent to d$S_{P*}/$d$f_P=0$, which is more manageable
mathematically. However, $S_{P*}$ involves a square root function, which I 
prefer to avoid as it tends to produce more elaborate forms when differentiated 
relative to functions of integer indices. $S_{P*}^2$ therefore makes a more 
useful starting point.

\begin{align}
\frac{\mathrm{d}(S_{P*}^2)}{\mathrm{d}f_P} &= 2 S_{P*} \frac{\mathrm{d}S_{P*}}{\mathrm{d}f_P}
\end{align}

From the above, one can see that the condition d$S_{P*}^2/$d$f_P=0$ is 
equivalent to the condition d$S_{P*}/$d$f_P=0$, except in the case where
$S_{P*} = 0$. However, $S_{P*} \geq 0$ at all times, then the instant when
$S_{P*}=0$ must correspond to a minimum anyway. I
therefore proceed to consider transit minima occurring at d$S_{P*}^2/$d$f_P=0$.

\begin{align}
S_{P*}^2 &= \frac{a_P^2 (1-e_P^2)^2}{(1+e_P \cos f_P)^2} [1-\sin^2i_P\sin^2(\omega_P+f_P)] \\
\frac{\mathrm{d}(S_{P*}^2)}{\mathrm{d}f_P} &= -\frac{2 a_P^2 (1-e_P^2)^2}{(1+e_P \cos f_P)^3} \Bigg( (1+e_P\cos f_P)\sin(f_P+\omega_P)\cos(f_P+\omega_P)\sin^2i_P \nonumber \\
\qquad& - e_P\sin f_P [1-\sin^2i_P\sin^2(f_P+\omega_P)] \Bigg)
\label{eqn:dSdf}
\end{align}

Inspection of d$S_{P*}^2/$d$f_P$ shows a common factor on the outside. This
factor has no bearing on the minima and can only reach an extrema if $e_P\geq1$,
which is forbidden for bounded orbits. Therefore, the condition under which
minima occur is:

\begin{align}
(1+e_P\cos f_P)\sin(f_P+\omega_P)\cos(f_P+\omega_P)\sin^2i_P = e_P\sin f_P [1-\sin^2i_P\sin^2(f_P+\omega_P)]
\end{align}

Rearranging and writing in terms of the coefficients of $\cos f_P$, one obtains
a quartic equation, of form 
$R_0 + R_1 \cos f_P + R_2 \cos^2f_P + R_3 \cos^3f_P + R_4 \cos^4f_P = 0$, where:

\begin{align}
R_0 &= -4 e_P^2 (1-\sin^2i_P\cos^2\omega_P)^2+\sin^4i_P\sin^2(2\omega_P) \\
R_1 &= 8 e_P \sin^2i_P (\cos2\omega_P - \cos^4\omega_P\sin^2i_P) \\
R_2 &= 4 e_P^2-\sin^2i_P\Big( 2 + 3e_P^2 - (2-e_P^2)\cos2i_P + e_P^2\cos2\omega_P(3+\cos2i_P)\Big) \\
R_3 &= -2e_P\cos2\omega_P\sin^2i_P(3+\cos2i_P) + 4 e_P \sin^4i_P \\
R_4 &= 4\sin^4i_P
\end{align}

The quartic solution has four roots, two of which correspond to the instants
where minima occur and two of which correspond to the maxima. The standard
analytic solutions for the roots of a quartic equation, known as Ferrari's 
solution, are unfortunately highly elaborate and take several pages to write 
out. Further, which root corresponds to a minima or maxima varies depending upon
$\omega_P$, $e_P$ and $i_P$ with complex boundary conditions. A final 
complication comes from the fact I have solved the equation in $\cos f_P$, 
which means there are twice as many roots (i.e. 8 roots) for $f_P$ (concordantly
I refer to the problem as a bi-quartic equation). 

For these reasons, solving the bi-quartic is clearly possible, but somewhat
impractical. One has to find all eight roots and then test each of them to see
which one is the true solution. The resulting algorithm is rather inefficient 
and an alternative approach is clearly desirable.

\subsection{The Series Expansion Solution}
\label{sec:approxminima}

A series expansion approach is attractive as one can continue up to any desired
level of precision and yet possess just one root i.e. avoid the root-selection 
problems encountered with the bi-quartic. The expansion can also be written as a
series of much simpler expressions than that encountered in 
\S\ref{sec:exactminima}. 

Many methods exist for iterating towards a root with the most commonly used
being the Newton-Rhapson method:

\begin{equation}
x_{i+1} = x_i \frac{f(x_i)}{f'(x_i)}
\end{equation}

A good starting point is required for the iteration and so I select the time of
inferior conjunction, $f_P = \pi/2-\omega_P$. I define the function, $f(x)$, to
be d$S_{P*}^2/$d$f_P$, as given in Equation~(\ref{eqn:dSdf}). The final solution
for $f_{P,T}$ may be written as:

\begin{align}
f_{P,T} &= \Big[ \frac{\pi}{2}-\omega_P \Big] - \sum_{i=1}^{n} \eta_i^T
\end{align}

The $\eta_i$ terms are used to absorb the perturbing terms from the time of
inferior conjunction. This notation comes from \citet{kopal1959}, who used a 
single $\eta$ parameter and only presented up to the first-order expansion,
whereas here I wish to go up to higher order. 

Newton's method produces quite elaborate formulas when used naively. However,
one useful trick is to perform a series expansion of $x_i$ using a Taylor 
series. This Taylor expansion should be performed in a parameter which is very
close to zero, so that only a few terms are needed to achieve the required
precision. An excellent candidate for this parameter is $\cos^2i_P$ which is
very close to zero for transits. Previous authors, such as \citet{irwin1952} and
\citet{kopal1959}, expanded in $\cot^2i_P$ rather than $\cos^2i_P$ though, so I
will pause to consider the value of such an approach. One wishes to choose the
parameter which is closest to zero for $i_P \rightarrow\pi/2$. Writing out
the first few terms of each one obtains:

\begin{align}
\cos(\pi/2 - x) &= x - \frac{x^3}{6} + \mathcal{O}[x^5] \nonumber \\
\cot(\pi/2 - x) &= x + \frac{x^3}{3} + \mathcal{O}[x^5] \nonumber
\end{align}

It can therefore be seen that $\cos^2i_P$ will approach zero faster than
$\cot^2i_P$. Another subtlety is that if one has performed $n$ iterations with
Newton's method, any parts of the series expansion in $\cos^2i_P$ above order
$n$ will change after the next iteration. In other words, they are unstable.
Therefore, one must be careful to only Taylor expand up to order $n$ in 
$\cos^2i_P$ if one has performed $n$ iterations with Newton's method. Taking 
into account all of these issues, the first six terms of the $\eta_i^T$ series 
are given by:

\begin{align}
\eta_1^T &=+\Big(\frac{k_P}{1+h_P}\Big) (\cos^2i_P)^1 \\
\eta_2^T &=+\Big(\frac{k_P}{1+h_P}\Big) \Big(\frac{1}{1+h_P}\Big) (\cos^2i_P)^2 \\
\eta_3^T &=-\Big(\frac{k_P}{1+h_P}\Big) \Big(\frac{-6 (1+h_P)+k_P^2 (-1+2h_P)}{6 (1+h_P)^3}\Big) (\cos^2i_P)^3 \\
\eta_4^T &=-\Big(\frac{k_P}{1+h_P}\Big) \Big(\frac{-2 (1+h_P)+k_P^2 (-1+3h_P)}{2 (1+h_P)^4}\Big) (\cos^2i_P)^4 \\
\eta_5^T &=+\Big(\frac{k_P}{1+h_P}\Big) \Big(\frac{40 (1+h_P)^2-40k_P^2(-1+3h_P+4h_P^2) + k_P^4(3-19h_P+8h_P^2)}{40 (1+h_P)^6}\Big) (\cos^2i_P)^5 \\
\eta_6^T &=+\Big(\frac{k_P}{1+h_P}\Big) \Big(\frac{24 (1+h_P)^2-40k_P^2(-1+4h_P+5h_P^2) +9k_P^4(1- 8h_P+5h_P^2)}{24 (1+h_P)^7}\Big) (\cos^2i_P)^6
\label{eqn:etasT}
\end{align}

Where $h_P = e_P\sin\omega_P$ and $k_P = e_p\cos\omega_P$.
For the occultation event, one may repeat the process but using the
initial guess $3\pi/2-\omega_P$ instead. The occultation minimum then occurs at:

\begin{align}
f_{P,O} &= \Big[ \frac{3\pi}{2}-\omega_P \Big] - \sum_{i=1}^{n} \eta_i^O
\end{align}

Where:

\begin{align}
\eta_1^O &= -\Big(\frac{k_P}{1-h_P}\Big) (\cos^2i_P)^1 \\
\eta_2^O &= -\Big(\frac{k_P}{1-h_P}\Big) \Big(\frac{1}{1-h_P}\Big) (\cos^2i_P)^2 \\
\eta_3^O &= -\Big(\frac{k_P}{1-h_P}\Big) \Big(\frac{6 (1-h_P)+k_P^2 (1+2h_P)}{6 (1-h_P)^3}\Big) (\cos^2i_P)^3 \\
\eta_4^O &= -\Big(\frac{k_P}{1-h_P}\Big) \Big(\frac{2 (1-h_P)+k_P^2 (1+3h_P)}{2 (1-h_P)^4}\Big) (\cos^2i_P)^4 \\
\eta_5^O &= -\Big(\frac{k_P}{1-h_P}\Big) \Big(\frac{40 (1-h_P)^2-40k_P^2(-1-3h_P+4h_P^2) + k_P^4(3+19h_P+8h_P^2)}{40 (1-h_P)^6}\Big) (\cos^2i_P)^5 \\
\eta_6^O &= -\Big(\frac{k_P}{1-h_P}\Big) \Big(\frac{24 (1-h_P)^2-40k_P^2(-1-4h_P+5h_P^2) +9k_P^4(1 +8h_P+5h_P^2)}{24 (1-h_P)^7}\Big) (\cos^2i_P)^6
\label{eqn:etasO}
\end{align}

I was unable to find the series presented in 
Equations~(\ref{eqn:etasT})\&(\ref{eqn:etasO}) in the previous exoplanet, or
eclipsing binary, literature above second order and thus conclude they are novel 
results. The equations have been verified against the
solution found from the bi-quartic equation and seem to converge very rapidly. I
found that it was not practical to derive the solutions above $6^\mathrm{th}$
order. The above equations were computed using \emph{Mathematica} with 
computation time increasing exponentially with the order number. On a 2GHz
\emph{Intel} processor, the final term requires around 20 minutes to generate
and I extrapolate that the next order will require 2\,days of continuous
processing. However, in practice, the calculation uses up all of the 
available memory before this point and this dramatically decreases the 
computation speed.

One may compare the $\eta^T$-series against the first-order solution presented 
in \citet{kopal1959}, who finds:

\begin{equation}
f_{P,T}^{\mathrm{Kopal}} = \frac{\pi}{2} - \omega_P - \Big(\frac{k_P}{1+h_P}\Big) \cot^2i_P
\end{equation}

Therefore, provided one acknowledges that $\cot^2i_P \simeq \cos^2i_P$ for 
$i_P\simeq(\pi/2)$, the \citet{kopal1959} solution is identical to the first
order term in the $\eta^T$-series.

It is useful to consider the error formulas of the $\eta$-series. If one goes up
to order $n$ in the series, the error in $f_{P,T}$ will be 
$\simeq|\eta_{n+1}^P|$. It is more useful, though, to consider the error in 
terms of time. To make the conversion from $f\rightarrow t$ requires solving 
Kepler's Equation, which is clearly not desirable due its transcendental 
nature. However, since the error in $f_P$ will be small, it is justified to 
approximate d$f_P\simeq\Delta f_P$ and d$t\simeq\Delta t$ in 
Equation~(\ref{eqn:angmom}). I will also approximate 
$f_{P,T} \simeq \pi/2-\omega_P$, which essentially is saying the speed of the 
planet at the time of transit minimum is approximately the same as that as at 
the time of inferior conjunction. This allows one to write that the error on the
$n^{\mathrm{th}}$-order $\eta^T$-series expansion for the time of transit
minimum, $\tau_T$, is\footnote{Note, the notation of using $\tau$ for the
transit minima is not present in the exoplanet literature, but is adopted in
this thesis in an effort to produce a self-consistent notation set spanning both
planets and moons (see Appendix~\ref{app:notation})}\footnote{Note, in virtually 
cases in the exoplanet literature this time is known as the mid-transit time, 
whereas in the eclipsing binary community is known as the eclipse minimum. The 
second definition is far more accurate as mid-time refers to simply the halfway
point of the transit, which is not the same as the minimum 
(see Appendix~\ref{app:notation})}:

\begin{equation}
(\Delta \tau_T)_n \simeq |\eta_{n+1}^T| \frac{P_P}{2\pi} \frac{(1-e_P^2)^{3/2}}{(1+h_P)^2}
\label{eqn:etaTerror}
\end{equation}

Similarly, for the occultation minimum:

\begin{equation}
(\Delta \tau_O)_n \simeq |\eta_{n+1}^O| \frac{P_P}{2\pi} \frac{(1-e_P^2)^{3/2}}{(1-h_P)^2}
\label{eqn:etaOerror}
\end{equation}

\begin{figure}
\begin{center}
\includegraphics[width=15.0 cm]{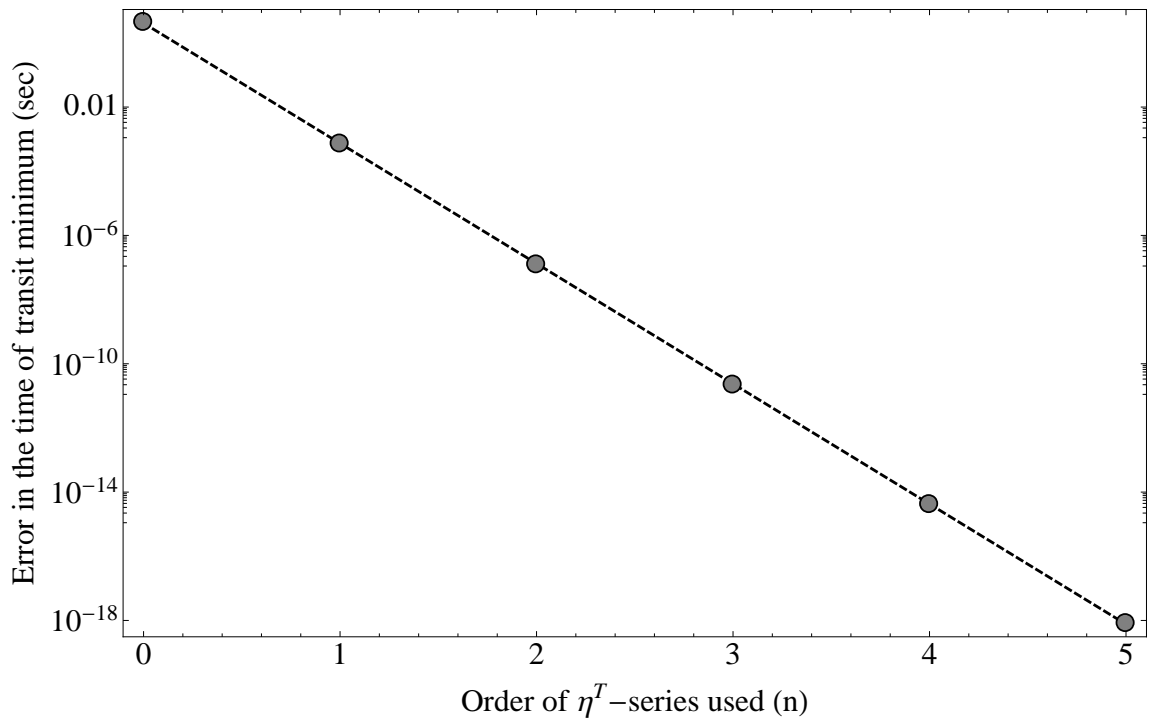}
\caption[Convergence of the $\eta^T$-series for predicting the time of the 
transit minimum]
{\emph{Convergence of the $\eta^T$-series for predicting the time of the transit 
minimum, for the example of the HAT-P-2b system. The very rapid convergence
is shown by the best-fit line, indicating a rate of convergence of 8.6 here.}}
\label{fig:etaconvergence}
\end{center}
\end{figure}

I will here illustrate the convergence of the $\eta^T$-series with an example of 
a known transiting system. To maximize the number of iterations needed, I 
require a short-period highly eccentric planet. HAT-P-2b \citep{bakos2007} gives
an ideal test with $P_P = 5.6334729$\,d, $e_P = 0.5171$, 
$\omega_P= 185.22^{\circ}$ and $i_P = 86.72^{\circ}$. In 
Figure~\ref{fig:etaconvergence}, I show the resulting errors in $\tau_T$ as a 
function of the order used from the $\eta^T$ series. Using a zeroth-order 
expansion gives an error of 4.5\,s, which is generally too large, but going to 
first-order reduces this to less than a millisecond as a result of the very 
rapid convergence. Although the general speed of convergence is variable, this 
typical example has a convergence of rate of 8.6, illustrating how rapidly the 
series works.

\section{Transit Duration for a Circular Orbit}
\label{sec:circduration}

\subsection{Fundamental Equations}
\label{sec:circfundamentals}

Calculating the duration of a transit is non-trivial and I begin the
discussion by considering the simple case of a circular orbit. The first point
to establish is that numerous definitions for the transit duration exist
in the literature. As was seen earlier in \S\ref{sec:transitanatomy} and 
Figure~\ref{fig:transitanatomy}, there are four principal contact points which
define the transit, $t_I$, $t_{II}$, $t_{III}$ and $t_{IV}$. $t_{I}$ and 
$t_{IV}$ occur when $S_{P*} = (1+p)$ and $t_{II}$ and $t_{III}$ occur when
$S_{P*} = (1-p)$. I define the time between any two contact points as $T_{xy}$,
where $x$ and $y$ are the contact points\footnote{Note that $x$ and $y$ are
commutable}. $T_{14}$ is sometimes referred to as the ``total duration'' and 
$T_{23}$ as the ``flat-bottomed duration'' or confusingly the 
``full duration''. $T_{12}$ and $T_{34}$ tend to have a negligible difference 
(even for very eccentric systems) and are therefore often grouped together and 
called the ``ingress/egress duration''.

The duration between any two instances is found by computing the time between
the planetary true anomalies at those points. For contact points $x$ and $y$, I 
define the associated true anomalies as $f_{P,x}$ and $f_{P,y}$ and thus the 
time between these points can be found by integrating 
Equation~(\ref{eqn:angmom}):

\begin{align}
T_{xy} &= \int_{f_P=f_{P,x}}^{f_{P,y}} \frac{\mathrm{d}t}{\mathrm{d}f_P}\,\mathrm{d}f_P \nonumber \\
\qquad &=  \frac{P_P}{2\pi} \int_{f_P=f_{P,x}}^{f_{P,y}} \frac{(1-e_P^2)^{3/2}}{(1+e_P\cos f_P)^2}\,\mathrm{d}f_P
\label{eqn:generalduration}
\end{align}

Recall earlier in \S\ref{sec:keplerseqn}, Equation~(\ref{eqn:meananomaly}),
that the mean anomaly is defined to scale linearly with time. Therefore, by 
virtue of its very definition, the mean anomaly is also directly related to the 
transit duration:

\begin{align}
T_{xy} &= \frac{P_P}{2\pi} (\mathfrak{M}_{P,y} - \mathfrak{M}_{P,x}) \nonumber \\
\Rightarrow \Delta \mathfrak{M}_P &= \int_{f_P=f_{P,x}}^{f_{P,y}} \frac{(1-e_P^2)^{3/2}}{(1+e_P\cos f_P)^2}\,\mathrm{d}f_P
\label{eqn:durationmeananomaly}
\end{align}

From the above, it can therefore be seen that $\Delta\mathfrak{M}_P=\Delta f_P$
if $e_P = 0$ i.e. a circular orbit. In such a case, the transit duration may
therefore be simply expressed as:

\begin{equation}
T_{xy}^{\mathrm{circ}} = \frac{P_P}{2\pi} (f_{P,y} - f_{P,x})
\label{eqn:circduration}
\end{equation}

The outstanding problem is now to solve for $f_{P,x}$ and $f_{P,y}$. Let us 
assume that contact points $x$ and $y$ have the same sky-projected
planet-star separation i.e. $S_{P*}(f_{P,x}) = S_{P*}(f_{P,y})$. 
Consequently, the true anomalies $f_{P,x}$ and $f_{P,y}$ may be found by 
solving the following for $f_P$:

\begin{align}
S_{P*} &= (a_P/R_*) \Bigg(\frac{1-e_P^2}{1+e_P \cos f_P}\Bigg) \sqrt{1-\sin^2i_P\sin^2(\omega_P+f_P)}
\label{eqn:solvingforfP}
\end{align}

For a circular orbit, the $e_P$ terms vanish, but an $\omega_P$ term would 
persist in the function $\sin^2(\omega_P+f_P)$. For a circular orbit, the 
transit minimum occurs exactly at the time of inferior conjunction (visible by 
setting $e_P=0$ in the $\eta^T$-series in Equation~(\ref{eqn:etasT})). I 
therefore define the difference in the true anomaly between the instant of 
inferior conjunction and the relevant contact points as $\breve{f}_P$. 
Therefore, $\breve{f}_{P,x} = \pi/2 - \omega_P - f_{P,x}$ and 
$\breve{f}_{P,y} = \pi/2 - \omega_P - f_{P,y}$ and consequently 
$\Delta \breve{f}_P = \Delta f_P$.  Equation~(\ref{eqn:solvingforfP}) now 
becomes:

\begin{align}
S_{P*} &= (a_P/R_*) \sqrt{1-\sin^2i_P\cos^2\breve{f}_P^{\mathrm{circ}}}
\label{eqn:circS}
\end{align}

Solving for $\breve{f}_P$ yields:

\begin{align}
\breve{f}_{P,x}^{\mathrm{circ}} &= -\arcsin\Bigg[\sqrt{ \frac{S_{P*}^2-b_{P,T}^2}{(a_P/R_*)^2 -b_{P,T}^2} }\Bigg] \nonumber \\
\breve{f}_{P,y}^{\mathrm{circ}} &=  \arcsin\Bigg[\sqrt{ \frac{S_{P*}^2-b_{P,T}^2}{(a_P/R_*)^2 -b_{P,T}^2} }\Bigg] \nonumber \\
\Delta f_P^{\mathrm{circ}} &= \Delta \breve{f}_P^{\mathrm{circ}} = 2 \arcsin\Bigg[\sqrt{ \frac{S_{P*}^2-b_{P,T}^2}{(a_P/R_*)^2 -b_{P,T}^2} }\Bigg]
\end{align}

Plugging this into Equation~(\ref{eqn:circduration}) gives an exact solution for
the duration of a transit, for circular orbits.

\begin{equation}
T_{xy}^{\mathrm{circ}} = \frac{P_P}{\pi} \arcsin\Bigg[ \sqrt{ \frac{S_{P*}^2-b_{P,T}^2}{(a_P/R_*)^2 -b_{P,T}^2} }\Bigg]
\label{eqn:Txycirc}
\end{equation}

Equation~(\ref{eqn:Txycirc}) is only valid if contact points $x$ and $y$ have 
the same $S_{P*}$. Two cases which are therefore applicable are $T_{14}$ and
$T_{23}$:

\begin{align}
T_{14}^{\mathrm{circ}} &= \frac{P_P}{\pi} \arcsin\Bigg[ \sqrt{ \frac{(1+p)^2-b_{P,T}^2}{(a_P/R_*)^2 -b_{P,T}^2} }\Bigg] \\
T_{23}^{\mathrm{circ}} &= \frac{P_P}{\pi} \arcsin\Bigg[ \sqrt{ \frac{(1-p)^2-b_{P,T}^2}{(a_P/R_*)^2 -b_{P,T}^2} }\Bigg]
\label{eqn:tTeqn}
\end{align}

For the ingress/egress duration, one cannot use Equation~(\ref{eqn:Txycirc})
since the contact points now have different values for the sky-projected 
separation. However, one may retrace the steps in the above derivation to obtain
such a result and find that the time difference between contact 
points 1\&2 and 3\&4 give an identical result:

\begin{align}
T_{12}^{\mathrm{circ}} &= T_{34}^{\mathrm{circ}} = \frac{P_P}{2\pi} \Bigg(\arcsin\Bigg[\sqrt{ \frac{(1+p)^2-b_{P,T}^2}{(a_P/R_*)^2 -b_{P,T}^2} }\Bigg] - \arcsin\Bigg[\sqrt{ \frac{(1-p)^2-b_{P,T}^2}{(a_P/R_*)^2 -b_{P,T}^2} }\Bigg]\Bigg) \nonumber \\
T_{12}^{\mathrm{circ}} &= T_{34}^{\mathrm{circ}} = \frac{T_{14}^{\mathrm{circ}} - T_{34}^{\mathrm{circ}}}{2}
\label{eqn:ingressdurcirc}
\end{align}

Where on the last line I have shown how the ingress/egress duration may be 
written as a linear combination of the two durations previously defined.

\subsection{Parameter Retrieval}
\label{sec:stellardensity}

The parameters $p$, $b_{P,T}$ and $(a_P/R_*)$ should now be clearly identifiable
as the main players in these equations. These three parameters combined with
the time of the transit minimum, $\tau_T$, provide the four-observables of 
transit light curves for both circular and eccentric orbits. It has already been
seen how the ingress/egress duration is simply a linear combination of $T_{14}$ 
and $T_{23}$ and thus it does not provide any new information. In other words, 
one only needs two quantities to define the transit durations. Since the transit 
depth gives $p$, and the timing gives $\tau_T$, the outstanding challenge one 
faces is whether these two durations can be converted into $b_{P,T}$ and 
$(a_P/R_*)$.

\citet{seager2003} were the first to appreciate this point and successfully
provide a solution. Through rearranging the expressions for 
$T_{14}^{\mathrm{circ}}$ and $T_{23}^{\mathrm{circ}}$, they showed:

\begin{align}
b_{P,T}^{\mathrm{circ}} &= \Bigg[ \frac{(1-p)^2-\frac{\sin^2(T_{23}^{\mathrm{circ}} \pi/P_P)}{\sin^2(T_{14}^{\mathrm{circ}} \pi/P_P)} (1+p)^2}{1-\frac{\sin^2(T_{23}^{\mathrm{circ}} \pi/P_P)}{\sin^2(T_{14}^{\mathrm{circ}}\pi/P_P)}} \Bigg]^{1/2} \\
\Big(\frac{a_P}{R_*}\Big)^{\mathrm{circ}} &= \Bigg[ \frac{(1+p)^2-[b_{P,T}^{\mathrm{circ}}]^2(1-\sin^2(T_{14}^{\mathrm{circ}}\pi/P_P)}{\sin^2(T_{23}^{\mathrm{circ}}\pi/P_P)}\Bigg]^{1/2}
\label{eqn:seagerparams}
\end{align}

If the semi-major axis of the orbit, $a_P$, and the period, $P_P$, was known
(through timing of consecutive transits) then it would be possible to determine
the mass of the star (or technically $M_*+M_P$) through Kepler's Third Law.
However, nature throws a spanner into the works because one cannot determine
$a_P$, only $(a_P/R_*)$. As a result, instead of being able to measure $M_*$,
one obtains the average stellar density, $\rho_*$:

\begin{align}
\frac{a_P^3}{G (M_* + M_P)} &= \Big(\frac{P_P}{2\pi}\Big)^2 \nonumber \\
\frac{a_P^3}{R_*^3} &= \Big(\frac{P_P}{2\pi}\Big)^2 G \Big(\frac{M_*}{R_*^3} + \frac{M_P}{R_*^3}\Big) \nonumber \\
\rho_* + p^3 \rho_P &= \frac{3\pi (a_P/R_*)^3}{G P_P^2}
\label{eqn:circdensity}
\end{align}

The importance of Equation~(\ref{eqn:circdensity}) cannot be overstated. This
pioneering insight by \citet{seager2003} now forms the standard method of
determining the properties of a transiting planet's host star. A typical
approach is to compute stellar evolution isochrones of a main sequence star
(e.g. \citet{baraffe1998}, \citet{yi2001}) and see which isochrones best-match 
the measured stellar density, as well as other stellar parameters determined 
spectroscopically (e.g. effective temperature of the star, $T_{\mathrm{eff}}$, 
metallicity, [Fe/H], and surface gravity, $\log g$).

Of course, the suggested method is actually somewhat restricted in that the
equations of \citet{seager2003} are only valid for circular orbits. A large
fraction of the known transiting planets have been found to maintain non-zero
eccentricities and thus this assumption is critically flawed. Clearly, what is
required is an updated version of the \citet{seager2003} equations but
accounting for eccentricity. To do this, one first needs to derive the transit
duration for such systems, which will be provided in \S\ref{sec:eccduration}.
First though, I will introduce an additional definition of the transit
duration, which will be very useful in the rest of this thesis.

\subsection{The $\tilde{T}$ Duration}
\label{sec:Tduration}

So far, I have considered the durations between the four contact points defined
in \S\ref{sec:transitanatomy} and Figure~\ref{fig:transitanatomy}. However, 
\citet{carter2008} proposed an additional, and very useful, definition of the 
transit duration which does not involve these four points. $\tilde{T}$ is 
defined as the time it takes for the planet's centre to cross the stellar limb 
and exit under the same condition\footnote{Since $\tilde{T}$ is not defined in
terms of the four contact points, it is not possible to use the $T_{xy}$ 
notation employed earlier, and thus the tilde symbol is used instead}. 
Mathematically, these points occur when $S_{P*} = 1$. Critically, this 
definition excludes the parameter $p$ and can be expressed in terms of $b_{P,T}$ 
and $(a_P/R_*)$ only (for a circular orbit). Naturally, if one has a measurement 
error on all three observables, then these will propagate into the estimates of 
the various durations, but $\tilde{T}$ is only exposed to two of these errors 
and thus tends to be found with the highest signal-to-noise (SNR). For this 
reason, $\tilde{T}$ is very useful for TDV (transit duration variation) studies.

On a brief aside, \citet{carter2008} showed that $\tilde{T}$ has the highest
SNR for a trapezoid-approximated light curve and thus limb darkening is notably
absent. If one uses the same bandpass for a set of TDV measurements, one may 
safely fix the quadratic limb darkening coefficients to some reasonable estimate
and $\tilde{T}$ will tend to give the lowest errors. However, if one compares 
multiple wavelengths, it must be acknowledged that the limb darkening will vary 
across each light curve and thus should be fitted for. This tends to increase 
the error on $\tilde{T}$ in particular and, often, it is no longer the best 
duration parameter. However, performing TDV across multiple wavelengths should 
be avoided anyway, based upon the detection criterion C5. For the 
\citet{seager2003} (SMO03) model, it is trivial to show that $\tilde{T}$ is 
given by using $S_{P*} = 1$ in Equation~(\ref{eqn:Txycirc}):

\begin{equation}
\tilde{T}^{\mathrm{circ}} = \frac{P_P}{\pi} \arcsin\Bigg[\sqrt{\frac{1-b_{P,T}^2}{ (a_P/R_*)^2-b_{P,T}^2 }}\Bigg]
\label{eqn:tildeTdurcirc}
\end{equation}

It should be stressed at this point that 
$\tilde{T}^{\mathrm{circ}} \neq (T_{14}^{\mathrm{circ}} + T_{23}^{\mathrm{circ}})/2$.

\section{Transit Duration for an Eccentric Orbit}
\label{sec:eccduration}

\subsection{The Exact Solution}
\label{sec:eccdurexact}

The duration between any two true anomalies is most concisely expressed in terms
of the change in the mean anomaly between those points, as first shown in
Equation~(\ref{eqn:durationmeananomaly}).

\begin{align}
T_{xy} &= \frac{P_P}{2\pi} \Delta \mathfrak{M}_P \nonumber \\
\Delta \mathfrak{M}_P &= \int_{f_P=f_{P,x}}^{f_{P,y}} \frac{(1-e_P^2)^{3/2}}{(1+e_P\cos f_P)^2}\,\mathrm{d}f_P
\end{align}

For circular orbits, $\Delta \mathfrak{M}_P = \Delta f_P$, but this is not the case for
eccentric ones. Integrating the above equation yields:

\begin{align}
\Delta \mathfrak{M}_P &= \Bigg( 2 (1-e_P^2) \arctan\Bigg[\sqrt{\frac{1-e_P}{1+e_P}}\tan\frac{f_{P,y}}{2}\Bigg] - \frac{e_P (1-e_P^2)^{3/2} \sin f_{P,y}}{1+e_P\cos f_{P,y}} \Bigg) \nonumber \\
\qquad& - \Bigg( 2 (1-e_P^2) \arctan\Bigg[\sqrt{\frac{1-e_P}{1+e_P}}\tan\frac{f_{P,x}}{2}\Bigg] - \frac{e_P (1-e_P^2)^{3/2} \sin f_{P,x}}{1+e_P\cos f_{P,x}} \Bigg)
\label{eqn:durationfn}
\end{align}

Now, the only remaining challenge is to find $f_{P,x}$ and $f_{P,y}$. As was
seen before, for $T_{14}$, one has:

\begin{align}
S_{P*} &= (a_P/R_*) \Bigg(\frac{1-e_P^2}{1+e_P \cos f_P}\Bigg) \sqrt{1-\sin^2i_P\sin^2(\omega_P+f_P)}
\end{align}

Solving the above for $f_P$ is a challenging problem. As was done in 
\S\ref{sec:transitminima}, the problem is easier if one changes all $f_P$ terms
to $\cos f_P$ parameters. This, again, yields a quartic equation
$Q_0 + Q_1 \cos f_P + Q_2 \cos^2f_P + Q_3 \cos^3f_P + Q_4\cos^4f_P$, where:

\begin{align}
Q_0 =& \Big( \csc^2i_P(S_{P*}^2-\Lambda_P^2) + \Lambda_P^2 \cos^2\omega_P \Big)^2 \\
Q_1 =& 2 e_P S_{P*}^2 \csc^2i_P \Big(2\csc^2i_P(S_{P*}^2-\Lambda_P^2) + \Lambda_P^2 + \Lambda_P^2\cos2\omega_P\Big) \\
Q_2 =& \Lambda_P^2 \cos2\omega_P \Big( S_{P*}^2 \csc^2i_P (-2+e_P^2) - \Lambda_P^2 + 2\Lambda_P^2\csc^2i_P\Big) \nonumber \\
\qquad&- \Lambda_P^4 + e_P^2 S_{P*}^2 \Lambda_P^2 \csc^2i_P + 2\csc^4i_P e_P^2 S_{P*}^2 (3 S_{P*}^2-\Lambda_P^2) \\
Q_3 =& 4 e_P S_{P*}^2 \csc^2i_P (e_P^2 S_{P*}^2 \csc^2i_P -\Lambda_P^2 \cos2\omega_P) \\
Q_4 =& \Lambda_P^4 - 2 e_P^2 S_{P*}^2 \Lambda_P^2 \cos2\omega_P \csc^2i_P + e_P^4 S_{P*}^4 \csc^4i_P \\
\Lambda_P =& (a_P/R_*) (1-e_P^2)
\label{eqn:durationquartic}
\end{align}

The quartic nature of the solution for this problem was first identified by
\citet{kipping2008}. As the solution is presented in $\cos f_P$ rather than 
$f_P$, Equation~(\ref{eqn:durationquartic}) actually has eight roots in the same
way as found for the transit minima, and thus is again a bi-quartic equation. 
The large number of solutions causes similar problems as before and a root 
selection routine is required. As before, I will seek methods to avoid the 
bi-quartic.

\subsection{Approximate Solutions}
\label{sec:eccdurapprox}

One way to avoid the bi-quartic is to make an approximation.  A useful 
approximation which can be made is that 
$\varrho_P(f_P) \simeq \varrho_{P,T} = \varrho_P(f_P=f_{P,T})$, i.e. the 
planet-star separation is approximately a constant value given by the 
planet-star separation at the transit minimum.  I will also make the 
approximation $f_{P,T} \simeq (\pi/2 - \omega_P)$ to simplify the mathematics, 
which can be justified at the level of a few seconds accuracy for even the most 
extreme systems (e.g. Figure~\ref{fig:etaconvergence}).

Modifying the definition of the transit impact parameter to account for the 
altered planet-star separation so that 
$b_{P,T} = (a_P/R_*) \varrho_{P,T} \cos i_P$,
the true anomalies satisfy Equation~(\ref{eqn:deltaf}), which is subsequently 
solved as was done before for the circular orbits 
(\S\ref{sec:circfundamentals}):

\begin{align}
S_{P*} &\simeq (a_P/R_*) \varrho_{P,T} \sqrt{1-\sin^2i_P\cos^2\breve{f}_P} \nonumber \\
\Delta f_P &= \Delta \breve{f}_P \simeq 2 \arcsin \Bigg[ \sqrt{ \frac{S_{P*}^2-b_{P,T}^2}{(a_P/R_*)^2 \varrho_{P,T}^2 - b_{P,T}^2} } \Bigg]
\label{eqn:deltaf}
\end{align}

Having assigned the true anomaly at the moment of the transit minimum and the 
change in the true anomaly over the transit event, the necessary tools to 
compute the duration have been found. Unlike the circular orbit case, there are 
actually two options as to how to proceed here. Firstly, one could use the 
solutions for $f_{P,x}$ and $f_{P,y}$ and feed them into the 
Equation~(\ref{eqn:durationfn}) to get $\Delta \mathfrak{M}_P$ and then use this
in Equation~(\ref{eqn:generalduration}) to get the duration. Given that 
Equation~(\ref{eqn:durationfn}) has two principal 
functions, this will yield a ``two-term'' expression for the duration. 
Alternatively, one could assume $\Delta \mathfrak{M}_P \simeq \Delta f_P$ and 
avoid Equation~(\ref{eqn:durationfn}) altogether, which would yield a ``one-term'' 
expression. I will investigate both approaches here, followed by alternative 
methods adopted by some previous authors.

\subsubsection{Two-Term Expression}

By combining Equation~(\ref{eqn:durationfn}) with 
Equation~(\ref{eqn:generalduration}), one may obtain a final expression for
the duration, which I label $T^{\mathrm{two}}$ (``two-term''):

\begin{align}
T_{xy}^{\mathrm{two}} &= \frac{P_P}{2\pi} \Bigg[ \arctan\Bigg[\frac{\cos(\Delta f_P/2)+e_P \cos(\langle f_P \rangle)}{\sqrt{1-e_P^2} \sin(\Delta f_P/2)}\Bigg] \nonumber \\ 
\qquad& +\frac{e_P (1-e_P^2)^{1/2} \sin(\Delta f_P/2) (e_P \cos(\Delta f_P/2) + \cos(\langle f_P \rangle))}{(1-e_P^2) \sin^2(\langle f_P \rangle) + (e_P \cos(\Delta f_P/2)+\cos(\langle f_P \rangle))^2} \Bigg]
\label{eqn:twoterm}
\end{align}

Where I have used $\Delta f_P$ from Equation~(\ref{eqn:deltaf}) and defined 
$\langle f_P \rangle = (f_{P,x}+f_{P,y})/2$. The approximate entry for
this latter term would be $\langle f_P \rangle \simeq \pi/2 - \omega_P$.
Testing $T_{xy}^{\mathrm{two}}$ for the exact solutions for $f_{P,T}$ and 
$\Delta f_P$ provided precisely the correct transit duration for all $e_P$, as 
expected. However, I found that using approximate entries for these terms 
severely limited the precision of the derived equation for large $e_P$ 
(the results of numerical tests will be discussed later in 
\S\ref{sec:numericaltests}).

The source of the problem is visible in Equation~(\ref{eqn:twoterm}), which 
consists of taking the difference between two terms. Both terms are of 
comparable magnitude for large $e_P$ and thus one is obtaining a small term by 
taking the difference between two large terms. These kinds of expressions are 
very sensitive to slight errors. In this case, the error is from using 
approximate entries for $\langle f_P \rangle$ and in particular $\Delta f_P$. 
Next, I will consider a possible ``one-term'' expression, which avoids the 
problem of taking the difference between two comparable-magnitude terms and thus
produces greater numerical stability.

\subsubsection{One-Term Expression}

There are numerous possible methods for finding ``one-term'' transit duration 
expressions. The first one I consider is to assume 
$\Delta \mathfrak{M}_P \simeq \Delta f_P$ thereby avoiding the two-term nature 
of Equation~(\ref{eqn:durationfn}):

\begin{align}
T_{xy}^{\mathrm{one}} &= \frac{P_P}{2 \pi} \frac{\varrho_{P,T}^2}{\sqrt{1-e_P^2}} \Delta f_P \nonumber \\
\qquad &= \frac{P_P}{\pi} \frac{\varrho_{P,T}^2}{\sqrt{1-e_P^2}} \arcsin\Bigg[\sqrt{ \frac{S_{P*}^2-b_{P,T}^2}{(a_P/R_*)^2 \varrho_{P,T}^2 - b_{P,T}^2} } \Bigg]
\label{eqn:T1}
\end{align}

Where Equation~(\ref{eqn:deltaf}) has been used for $\Delta f_P$.  Another 
derivation would be to assume the planet takes a tangential orbital velocity and
constant orbital separation from the planet, sweeping out an arc of length 
$r_{P,T} \Delta f_P$.  It is trivial to show that this argument will lead to 
precisely the same expression for $T_{xy}^{\mathrm{one}}$.

\subsubsection{\citet{tingley2005} Equation}

I will also describe the derivations of two other commonly used formulas in
the exoplanet literature. The \citet{seager2003} (SMO03) formula has already
been discussed in \S\ref{sec:circduration} and so will not be covered again.

The first alternative I discuss is that of \citet{tingley2005} (TS05), which 
has been used by numerous authors since (e.g. \citet{ford2008}; 
\citet{jordan2008}).  It is also forms the basis of a light curve parameter 
fitting set proposed by \citet{bakos2007}. There are two critical assumptions 
made in the derivation of the TS05 formula. The first of these is that:

\begin{itemize}
\item[{\tiny$\blacksquare$}] The planet-star separation, $r_P$, is constant 
during the planetary transit event and equals $r_{P,T}$
\end{itemize}

This is the same assumption made in the derivation of the 
$T_{xy}^{\mathrm{one}}$ equation.  Under this assumption, TS05 quote the 
following expression for $T_{xy}^{\mathrm{TS05}}$ (changing to consistent 
notation).

\begin{equation}
T_{xy}^{\mathrm{TS05}} = \frac{r_{P,T} \Delta \phi}{v_{P,T}}
\end{equation}

Where TS05 define $v_{P,T}$ as the planet's orbital velocity at the transit
minimum and $\Delta \phi$ as ``the eccentric angle'' between the contact 
points. In the standard notation, there is no such parameter 
defined strictly as the ``eccentric angle'' and thus I initially assumed that 
TS05 were referring to the eccentric anomaly.  However, substituting the 
relevant terms for $r_{P,T}$ and $v_{P,T}$ gives:

\begin{equation}
T_{xy}^{\mathrm{TS05}} = \frac{P_P}{2 \pi} \frac{\varrho_{P,T}^2}{\sqrt{1-e_P^2}} \Delta \phi
\label{eqn:tingley1}
\end{equation}

By comparing Equation~(\ref{eqn:tingley1}) to Equation~(\ref{eqn:T1}), it is 
clear that $\Delta \phi = \Delta f_P$ (also note that Equation~(\ref{eqn:T1}) 
was derived under precisely the same assumptions as that assumed by TS05 at this 
stage of the derivation).  I therefore conclude that the term TS05 refer to as 
the ``eccentric angle'' in fact refers to the true anomaly. This is an important 
point to make because the derivation of the TS05 equation would otherwise be 
very difficult to understand by those working outside of the field.  Continuing
the derivation from this point, the second assumption made by TS05 is:

\begin{itemize}
\item[{\tiny$\blacksquare$}] The planet-star separation is much greater than the 
stellar radius, $r_{P,T} \gg R_*$
\end{itemize}

Critically, this assumption was not made in the derivation of $T^{\mathrm{one}}$
or $T^{\mathrm{two}}$. Using this assumption, TS05 propose that (replacing 
$\Delta\phi \rightarrow \Delta f_P$ and $1+p$ to the more general form of 
$S_{P*}$, to remain consistent with the notations used in this thesis):

\begin{align}
\Delta f_P^{\mathrm{TS05}} &= \arcsin \Bigg[2 \frac{\sqrt{S_{P*}^2 - (a_P/R_*)^2 \varrho_{P,T}^2 \cos^2i_P}}{(a_P/R_*) \varrho_{P,T}}\Bigg]
\label{eqn:tingleyupper}
\end{align}
\begin{align}
\Delta f_P^{\mathrm{TS05}} &\simeq 2 \frac{\sqrt{S_{P*}^2 - (a_P/R_*)^2 \varrho_{P,T}^2 \cos^2i_P}}{(a_P/R_*) \varrho_{P,T}}
\label{eqn:tingleylower}
\end{align}

Where TS05 use Equation~(\ref{eqn:tingleylower}) rather than 
Equation~(\ref{eqn:tingleyupper}) in the final version of $T^{\mathrm{TS05}}$. 
Therefore, TS05 effectively make a small-angle approximation for $\Delta f_P$, 
which is a consequence of assuming $r_{P,T} \gg R_*$.  I argue here that
losing the $\arcsin$ function does not offer any great simplification of the 
transit duration equation but does lead to an unnecessary source of error in 
the resultant expression, in particular for close-in orbits, which is common 
for transits.  Also note that even Equation~(\ref{eqn:tingleyupper}) exhibits 
differences to Equation~(\ref{eqn:deltaf}).

Firstly, inside the $\arcsin$ function, the factor of $\csc i_P$ is missing,
which is present in both the derivation presented in Equation~(\ref{eqn:T1}) 
and the derivation of SMO03 for circular orbits, Equation~(\ref{eqn:Txycirc}). 
The absence of this term can be understood as a result of the $r_{P,T} \gg R_*$ 
assumption. As $r_{P,T} \rightarrow \infty$, in order to maintain a transit 
event, one must have $i_P \rightarrow (\pi/2)$.

Secondly, the expression I presented for $\Delta f_P$ earlier in 
Equation~(\ref{eqn:deltaf}) has the factor of 2 present outside of the arcsin 
function, whereas TS05 have this factor inside the function.  Further more, the 
SMO03 derivation also predicts that the factor of 2 should be outside of the 
$\arcsin$ function and this expression is known to be an exact solution for 
circular orbits.  In a small angle approximation, 
$\arcsin2 x \simeq 2 \arcsin x$, but moving the factor of 2 to
within the $\arcsin$ function seems to serve no purpose except to invite further
error into the expression.  As a result of these differences, the 
$T_{xy}^{\mathrm{TS05}}$ expression does not reduce down to the original SMO03 
equation and is given by:

\begin{equation}
T_{xy}^{\mathrm{TS05}} = \frac{P_P}{\pi} \frac{\varrho_{P,T}}{\sqrt{1-e_P^2}} \frac{\sqrt{S_{P*}^2 - (a_P/R_*)^2 \varrho_{P,T}^2 \cos^2i_P}}{(a_P/R_*)}
\label{eqn:tingleyeqn}
\end{equation}

\subsubsection{\citet{winn2010} Equation}

\citet{winn2010} (W10) proposed an expression for $T_{xy}$ based upon 
modification to the SMO03 equation.  The first change was to modify the impact 
parameter from $(a_P/R_*) \cos i_P \rightarrow \varrho_{P,T}(a_P/R_*) \cos i_P$,
i.e. to allow for the altered planet-star separation for eccentric orbits, as
was done in \S\ref{sec:eccdurapprox}. Secondly, the altered planetary velocity 
should also be incorporated. W10 propose that a reasonable approximation for the
transit duration is obtained by multiplying the SMO03 expressions by the 
following ratio:

\begin{equation}
\frac{\frac{\mathrm{d}X_P}{\mathrm{d}t}(f_{P,T})[e_P=0]}{\frac{\mathrm{d}X_P}{\mathrm{d}t}(f_{P,T})} = \frac{\varrho_{P,T}}{\sqrt{1-e_P^2}}
\end{equation}

Where $X_P$ is given in Equation~(\ref{eqn:complexcoords}).  This yields a new 
transit duration equation of:

\begin{equation}
T_{xy}^{\mathrm{W10}} = \frac{P_P}{\pi} \frac{\varrho_{P,T}}{\sqrt{1-e_P^2}} \arcsin\Bigg[\frac{1}{(a_P/R_*)} \Bigg(\frac{S_{P*}^2-(a_P/R_*)^2 \varrho_{P,T}^2 \cos^2 i_P}{\sin^2 i_P}\Bigg)^{1/2}\Bigg]
\end{equation}

Firstly, note an obvious improvement of the W10 expression is that one recovers
the original SMO03 equation for $e_P=0$.  Secondly, comparison to the equation 
for $T_{xy}^{\mathrm{one}}$ (Equation~(\ref{eqn:T1})) reveals that the two 
expressions are very similar except for the position of an extra $\varrho_{P,T}$
term. Indeed, the $T_{xy}^{\mathrm{one}}$ and $T_{xy}^{\mathrm{W10}}$ 
expressions are equivalent in the small-angle approximation.

\subsection{Numerical Testing of the Approximations}
\label{sec:numericaltests}

\subsubsection{Example Systems}

Insights into the robustness and accuracy of the various expressions may be 
obtained through numerical testing of the various approximate expressions. I 
here compare the accuracy of the four expressions: $\tilde{T}^{\mathrm{TS05}}$, 
$\tilde{T}^{\mathrm{W10}}$, $\tilde{T}^{\mathrm{one}}$ and 
$\tilde{T}^{\mathrm{two}}$. These expressions depend only on five 
parameters\footnote{There is no dependence on $p$ because I am using the
$\tilde{T}$ definition of the duration to deliberately avoid $p$}:
$P_P$, $(a_P/R_*)$, $b_{P,T}$, $e_P\sin\omega_P$ and $e_P\cos\omega_P$. One of 
the clearest ways of comparing the equations is to consider a typical 
transiting exoplanet example with system parameters for $(a_P/R_*)$ and 
$b_{P,T}$, and vary the eccentricity parameters. $P_P$ may be selected by simply 
assuming a star of Solar density and using Equation~(\ref{eqn:circdensity}). 

I created a 1000 by 1000 grid of $e_P\sin\omega_P$ and $e_P\cos\omega_P$ values 
from -1 to 1 in equal steps. Grid positions for hyperbolic orbits ($e_P > 1$) 
are excluded. I then calculated the transit duration through the exact solution 
of the bi-quartic equation, $\tilde{T}^{\mathrm{biquartic}}$, plus all four 
approximate formulas. Then, the fractional deviation of each 
equation from the true solution is found using:

\begin{equation}
\mathcal{D}^{\mathrm{candidate}} = \frac{\tilde{T}^{\mathrm{candidate}} - \tilde{T}^{\mathrm{biquartic}}}{\tilde{T}^{\mathrm{biquartic}}}
\end{equation}

\begin{figure*}
\begin{center}
\includegraphics[width=15.0 cm]{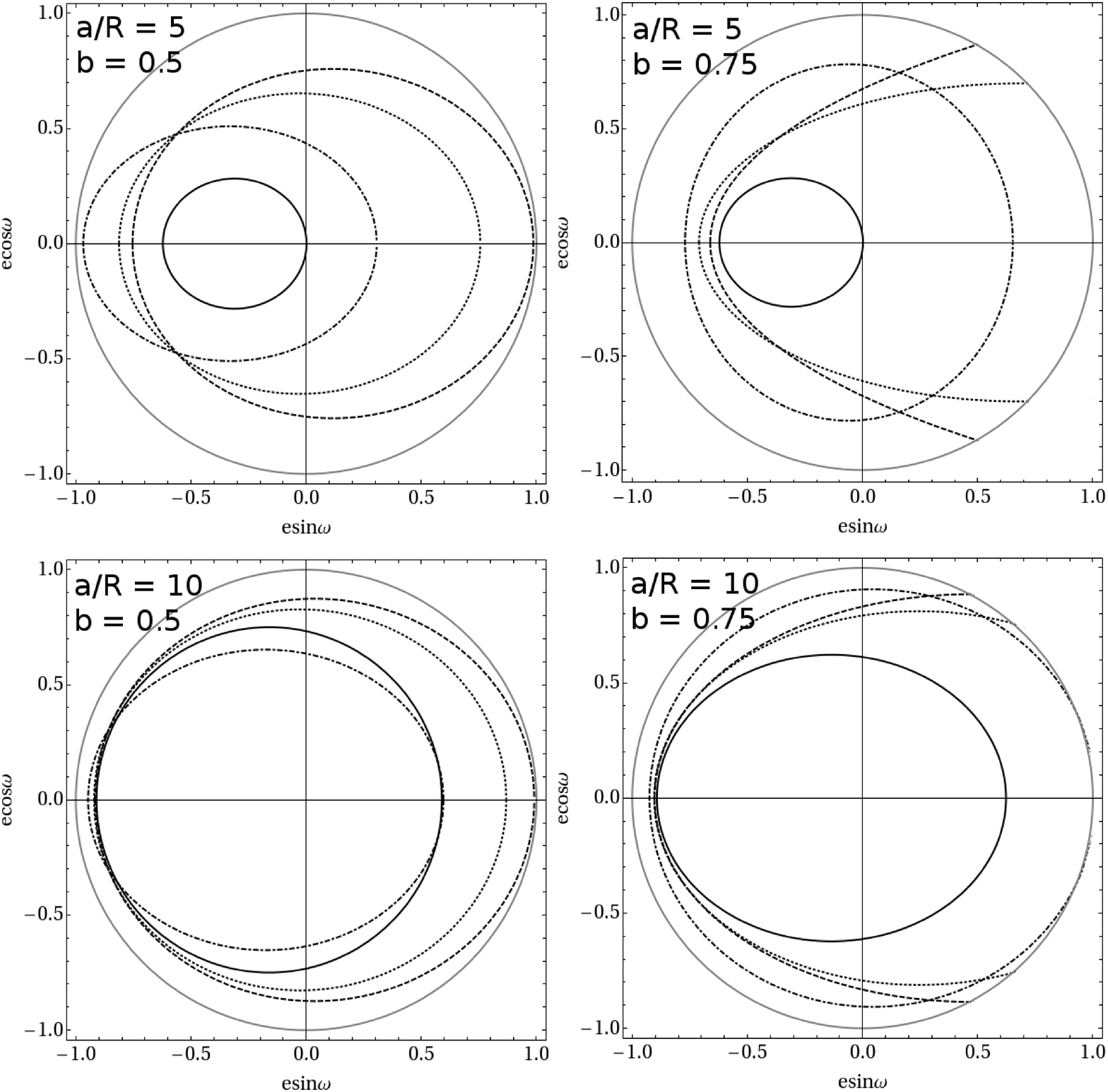}
\caption[Comparison of the accuracies of four approximate formulas for the
transit duration]
{\emph{Loci of points for which the accuracy is better than 99\% for all
four candidate expressions, as a function of eccentricity. The 
$\tilde{T}^{\mathrm{one}}$ expression offers both consistency and excellent 
accuracy. Other system parameters fixed to typical transit values. Black solid 
is for $\tilde{T}^{\mathrm{TS05}}$, dotted is for $\tilde{T}^{\mathrm{W10}}$, 
dashed is for $\tilde{T}^{\mathrm{one}}$ and dot-dashed for 
$\tilde{T}^{\mathrm{two}}$. The gray ellipse represents the allowed physical 
limits. Parameter subscripts removed for clarity.}}
\label{fig:durationloci}
\end{center}
\end{figure*}

The loci of points for which the deviation is less than 1\% (i.e. 
$\mathcal{D}^{\mathrm{candidate}} < 0.01$) may then be computed. In 
Figure~\ref{fig:durationloci}, I present four such plots for different choices 
of $(a_P/R_*)$ and $b_{P,T}$. The plot reveals several interesting features:

\begin{itemize}
\item[{\tiny$\blacksquare$}] $T^{\mathrm{one}}$ consistently yields the largest 
loci
\item[{\tiny$\blacksquare$}] $T^{\mathrm{two}}$ is sometimes accurate and 
sometimes not, demonstrating that the approximation is not stable
\item[{\tiny$\blacksquare$}] $T^{\mathrm{W10}}$ also yields consistently large 
loci
\item[{\tiny$\blacksquare$}] $T^{\mathrm{TS05}}$ consistently yields the 
smallest loci
\end{itemize}

\subsubsection{Additional Tests}

I also briefly discuss additional tests which I performed for two sets of $10^7$ 
different hypothetical transiting exoplanet systems; one for eccentricities 
$0.0<e_P<1$ and the other for $0.9<e_P<1$. In all cases, I randomly 
generated\footnote{With uniform probability distributions} the system parameters 
weighted by the transit probability and calculated the deviation of the various 
formulas.

I found that the $\tilde{T}^{\mathrm{one}}$ expression was consistently the most 
accurate, with the W10 of similar accuracy but higher asymmetry. I therefore 
find that the results yield a overall preference for the 
$\tilde{T}^{\mathrm{one}}$ approximation.  Note that authors using the 
$T^{\mathrm{W10}}$ formulation can also expect an extremely good approximation 
but I will only consider using $T^{\mathrm{one}}$ for the later derivations. I
define the ``improvement'' of the $\tilde{T}^{\mathrm{one}}$ expression relative 
to the $\tilde{T}^{\mathrm{TS05}}$ equation (since this is the previously most 
cited expression) as:

\begin{equation}
\mathcal{I}^{\mathrm{one}} = [ (\mathcal{D}^{\mathrm{TS05}}/\mathcal{D}^{\mathrm{one}} ) - 1 ]*100
\end{equation}

Where $\mathcal{I}$ is measured in \%.  One can see that if TS05 gives a lower 
deviation (i.e. more accurate solution), one will obtain $\mathcal{I} < 0$\% 
whereas if the candidate expression gives a closer solution one obtains 
$\mathcal{I} > 0$\% and is essentially the percentage improvement in accuracy. 
For the range $0<e_P<1$, I find that the mean value of this parameter is 
$\mathcal{I}^{\mathrm{one}} = 210$\% and for the range $0.9<e_P<1$ I find 
$\mathcal{I}^{\mathrm{one}} = 458$\%. Note that one caveat of these tests is 
that they are sensitive to the a-priori inputs. In conclusion, the qualitative,
rather than quantitative, aspects of these results offer the greatest insights.

For the case of spaced-based photometry, the typical measurement uncertainty on 
$\tilde{T}$ will be $\sim$0.1\% in most cases \citep{kippingetal2009}. I find,
on average, that $\tilde{T}^{\mathrm{one}}$ is accurate to 0.1\% or better over 
a range of $|e_P\sin\omega_P|<0.5$ and $|e_P\cos\omega_P|<0.85$.

\subsection{The Series Expansion Solution}
\label{sec:seriessoln}

I also investigated whether a series expansion solution could be obtained in a
similar way as was done for the transit minima in \S\ref{sec:approxminima}.
The equation which needs to be solved may be written as:

\begin{equation}
0 = \frac{(a_P/R_*)^2 (1-e_P^2)^2}{(1+e_P \cos f_P)^2} [1-\sin^2i_P\sin^2(\omega_P+f_P)] -  S_{P*}^2
\end{equation}

I tried using Newton's method again with initial starting points for $f_{P,x}$ 
and $f_{P,y}$ being $\pi/2-\omega_P-\Delta f_P/2$ and
$\pi/2-\omega_P+\Delta f_P/2$, where I define $\Delta f_P$ using 
Equation~(\ref{eqn:deltaf}). The resulting expressions do not yield elegant 
forms despite efforts to perform Taylor expansions in numerous terms. Further,
computational memory limits restrict the calculation in \emph{Mathematica} to
just first order. As a result of the limited use of these expressions, I do not 
present them here.

\subsection{The Consequences of Using Circular Expressions for Eccentric Orbits}
\label{sec:usingcircforecc}

SMO03 showed that the $1^{\mathrm{st}}$ to $4^{\mathrm{th}}$ contact duration, 
$T_{14}$, and the $2^{\mathrm{nd}}$ the $3^{\mathrm{rd}}$ contact duration, 
$T_{23}$, may be used to derive $(a_P/R_*)$ and $b_{P,T}$, and consequently the
derivative terms such as $i_P$ and $\rho_*$.  I here consider how biased these 
retrieved parameters would be if one used the circular equations for an 
eccentric orbit.  Such a circumstance may occur for newly discovered transiting
planets with only a few RV points and thus a simple circular model is 
erroneously adopted for the sake of simplicity. From here, I will employ the 
$T^{\mathrm{one}}$ expression for the transit duration, as this equation has 
been shown to provide the greatest accuracy in \S\ref{sec:numericaltests}. The 
transit durations $T_{14}^{\mathrm{circ}}$ and $T_{23}^{\mathrm{circ}}$ are 
given Equation~(\ref{eqn:tTeqn}), but for the $T_{xy}^{\mathrm{one}}$ solutions 
I utilize Equation~(\ref{eqn:T1}) to get:

\begin{align}
T_{14}^{\mathrm{one}} &= \frac{P_P}{\pi} \frac{\varrho_{P,T}^2}{\sqrt{1-e_P^2}} \arcsin\Bigg(\frac{\sqrt{(1+p)^2 - (a_P/R_*)^2 \varrho_{P,T}^2 \cos^2 i_P}}{(a_P/R_*) \varrho_{P,T} \sin i_P}\Bigg) \\
T_{23}^{\mathrm{one}} &= \frac{P_P}{\pi} \frac{\varrho_{P,T}^2}{\sqrt{1-e_P^2}} \arcsin\Bigg(\frac{\sqrt{(1-p)^2 - (a_P/R_*)^2 \varrho_{P,T}^2 \cos^2 i_P}}{(a_P/R_*) \varrho_{P,T} \sin i_P}\Bigg)
\end{align}

Using Equation~(\ref{eqn:tTeqn}), SMO03 show that the impact parameter of a
transiting planet on a circular orbit may be determined by using:

\begin{equation}
[b_{P,T}^{\mathrm{circ}}]^2 = \frac{(1-p)^2 - \frac{\sin^2(T_{14}^{\mathrm{circ}} \pi/P_P)}{\sin^2(T_{14}^{\mathrm{circ}} \pi/P_P)} (1+p)^2}{ 1-\frac{\sin^2(T_{23}^{\mathrm{circ}} \pi/P)}{\sin^2(T_{14}^{\mathrm{circ}} \pi/P_P)} }
\end{equation}

Now consider that an observer is unaware that the orbit is eccentric and makes
the simple, but false, assumption of a circular orbit. In this case, the
observer would assign 
$T_{xy}^{\mathrm{circ}} = T_{xy}^{\mathrm{biquartic}} \simeq T_{xy}^{\mathrm{one}}$.
As a result of this assumption, the observer would falsely calculate an impact
parameter:

\begin{align}
&[b_{P,T}^{\mathrm{circ}}]^2 = 1 + p^2 + 2p \nonumber \\
&\times \Bigg(\frac{ \sin^2[\frac{\varrho_{P,T}^2}{\sqrt{1-e_P^2}} \arcsin(\frac{\sqrt{(1-p)^2 - b_{P,T}^2}}{(a_P/R_*) \varrho_{P,T} \sin i_P})] + \sin^2[\frac{\varrho_{P,T}^2}{\sqrt{1-e_P^2}} \arcsin(\frac{\sqrt{(1+p)^2 - b_{P,T}^2}}{(a_P/R_*) \varrho_{P,T} \sin i_P})] }{ \sin^2[\frac{\varrho_{P,T}^2}{\sqrt{1-e_P^2}} \arcsin(\frac{\sqrt{(1-p)^2 - b_{P,T}^2}}{(a_P/R_*)( \varrho_{P,T} \sin i_P})] - \sin^2[\frac{\varrho_{P,T}^2}{\sqrt{1-e_P^2}} \arcsin(\frac{\sqrt{(1+p)^2 - b_{P,T}^2}}{(a_P/R_*) \varrho_{P,T} \sin i_P})] }\Bigg)
\label{eqn:bpwrong}
\end{align}

\begin{figure}
\begin{center}
\includegraphics[width=15.0 cm]{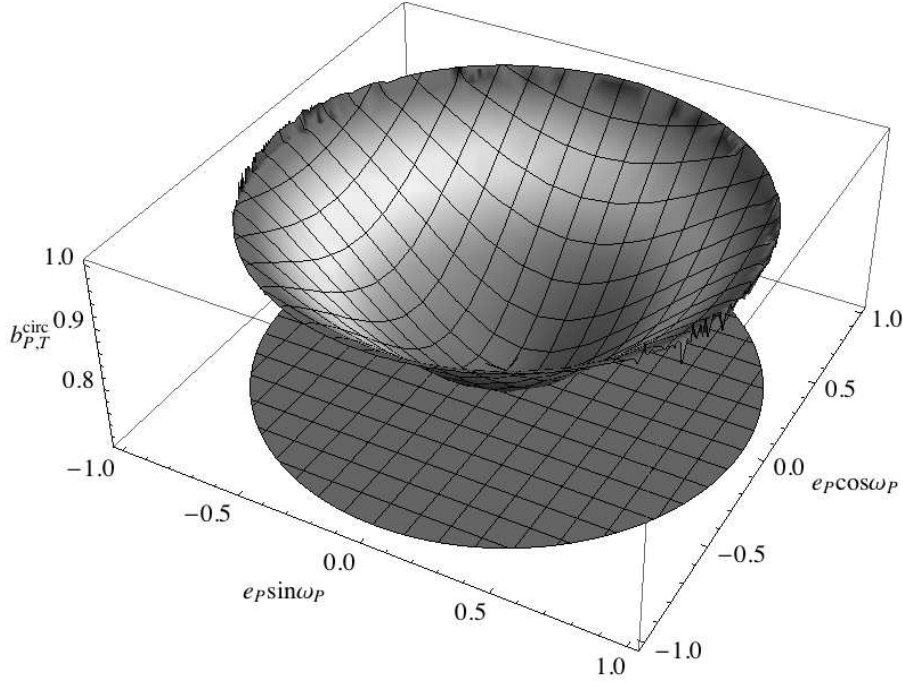}
\caption[The effect of eccentricity on the retrieved impact parameter, when
falsely using the circular equations]
{\emph{If one uses the circular expressions, the retrieved impact 
parameter is heavily biased by eccentricity.  In this example, the 
true value of $b_{P,T}$ is $\sqrt{0.5}$ but the introduction of eccentricity causes 
$b_{P,T}$ to be overestimated.}} 
\label{fig:bpwrong}
\end{center}
\end{figure}

Where it is understood that terms on the right-hand side with $b_{P,T}$ in them
refer to the true impact parameter, 
$b_{P,T} = (a_P/R_*) \varrho_{P,T} \cos i_P$.  This function is plotted in the 
case of $(a_P/R_*) = 10$, $b_{P,T}^2=0.5$ and $p=0.1$ in 
Figure~\ref{fig:bpwrong}. Making small-angle approximations, this yields 
$[b_{P,T}^{\mathrm{circ}}]^2 \simeq b_{P,T}^2$. However, for larger 
$e_P\sin\omega_P$ and $e_P\cos\omega_P$ values, the overall effect is to 
overestimate $b_{P,T}$ for eccentric orbits.

In addition to the impact parameter, SMO03 proposed that the parameter 
$(a_P/R_*)$ may be derived using:

\begin{equation}
[(a_P/R_*)^{\mathrm{circ}}]^2 = \frac{(1+p)^2 - [b_{P,T}^{\mathrm{circ}}]^2 }{\sin^2 (T_{14}^{\mathrm{circ}} \pi/P_P)} + [b_{P,T}^{\mathrm{circ}}]^2
\end{equation}

If one uses the assumption $b_{P,T}^{\mathrm{circ}} \simeq b_{P,T}$, then this 
equation yields:

\begin{align}
&[(a_P/R_*)^{\mathrm{circ}}]^2 = b_{P,T}^2 + [(1+p^2)-b_{P,T}^2] \csc^2\Bigg[\frac{\varrho_{P,T}^2}{\sqrt{1-e_P^2}} \arcsin\Bigg(\frac{\sqrt{(1+p)^2-b_{P,T}^2}}{(a_P/R_*) \varrho_{P,T} \sin i_P}\Bigg)\Bigg]
\end{align}

With small-angle approximations, this simplifies to:

\begin{equation}
(a_P/R_*)^{\mathrm{circ}} \simeq (a_P/R_*) \sqrt{\varrho_{P,T}^2 \cos^2 i_P + \frac{(1-e_P^2) \sin^2 i_P}{\varrho_{P,T}^2}}
\label{eqn:aRwrong}
\end{equation}

\begin{figure}
\begin{center}
\includegraphics[width=15.0 cm]{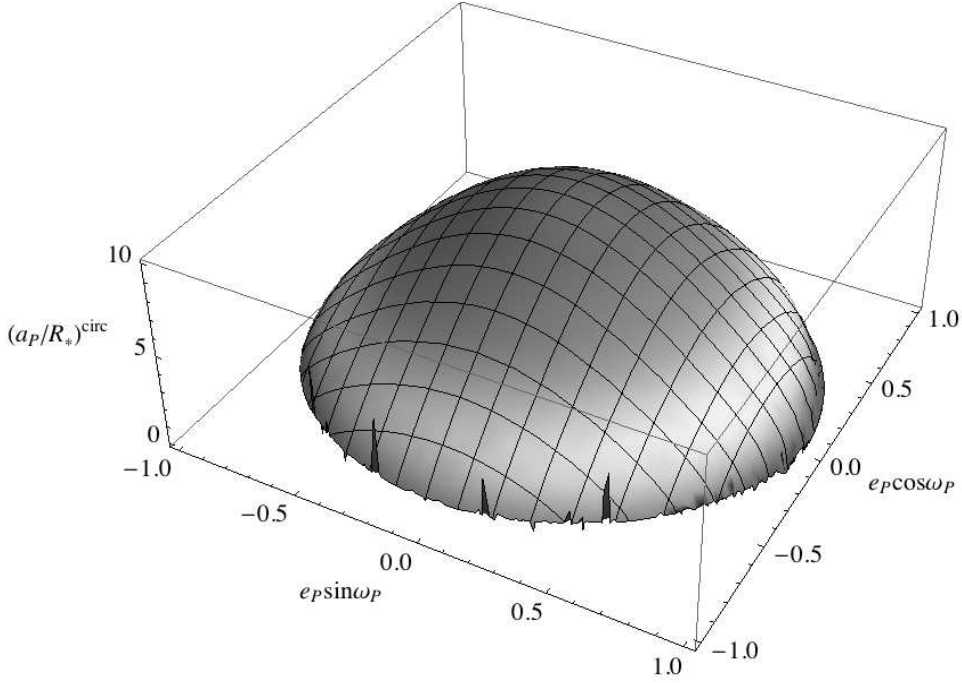}
\caption[The effect of eccentricity on the retrieved $(a_P/R_*)$, when
falsely using the circular equations]
{\emph{If one uses the circular expressions, the retrieved value of 
$(a_P/R_*)$ is heavily biased by eccentricity.  In this example, the true value 
of $(a_P/R_*)$ is 10 but the introduction of eccentricity causes $(a_P/R_*)$ to 
be underestimated.}} \label{fig:aRwrong}
\end{center}
\end{figure}

The term inside the square root goes to unity for circular orbits, as expected. 
The deviation in $(a_P/R_*)$ can be seen to become quite significant for 
eccentric orbits, as seen in Figure~\ref{fig:aRwrong} where the exact expression
for Equation~(\ref{eqn:aRwrong}) is plotted. This will have significant 
consequences for the next parameter, the stellar density.  As seen earlier, 
stellar density is related to $(a_P/R_*)$ by manipulation of Kepler's Laws
(see Equation~(\ref{eqn:circdensity})):

\begin{align}
\rho_* &= \frac{3 \pi}{G P_P^2} (a_P/R_*)^3 - p^3 \rho_P \nonumber \\
\qquad& \simeq \frac{3 \pi}{G P_P^2} (a_P/R_*)^3
\end{align}

Where the approximation is made using the assumption $p \ll 1$.  It can 
therefore be seen that:

\begin{align}
\rho_*^{\mathrm{circ}} &\simeq \rho_* \Bigg[\varrho_{P,T}^2 \cos^2 i_P + \frac{(1-e_P^2) \sin^2 i_P}{\varrho_{P,T}^2}\Bigg]^{3/2} \\
\rho_*^{\mathrm{circ}} &\simeq \rho_* \Psi_P = \rho_* \Bigg[\frac{(1+e_P \sin \omega_P)^3}{(1-e_P^2)^{3/2}}\Bigg]
\label{eqn:rhowrong}
\end{align}

Where in the second line I have assumed that $i_P \simeq \pi/2$.  A series 
expansion of $\Psi_P$ into first order of $e_P$ yields 
$\Psi_P \simeq 1 + 3 e_P \sin \omega_P + \mathcal{O}[e_P^2]$.  So observers 
neglecting an eccentricity of $e_P \sim 0.1$ may alter the stellar density by 
30\%.  As an example, if one decreased the density of a solar type G2V star by 
30\%, the biased average stellar density would be more consistent with a star 
of spectral type K0V. Indeed, asteroseismologically determined stellar densities
of transiting systems could be used to infer $\Psi_P$.

\begin{figure}
\begin{center}
\includegraphics[width=15.0 cm]{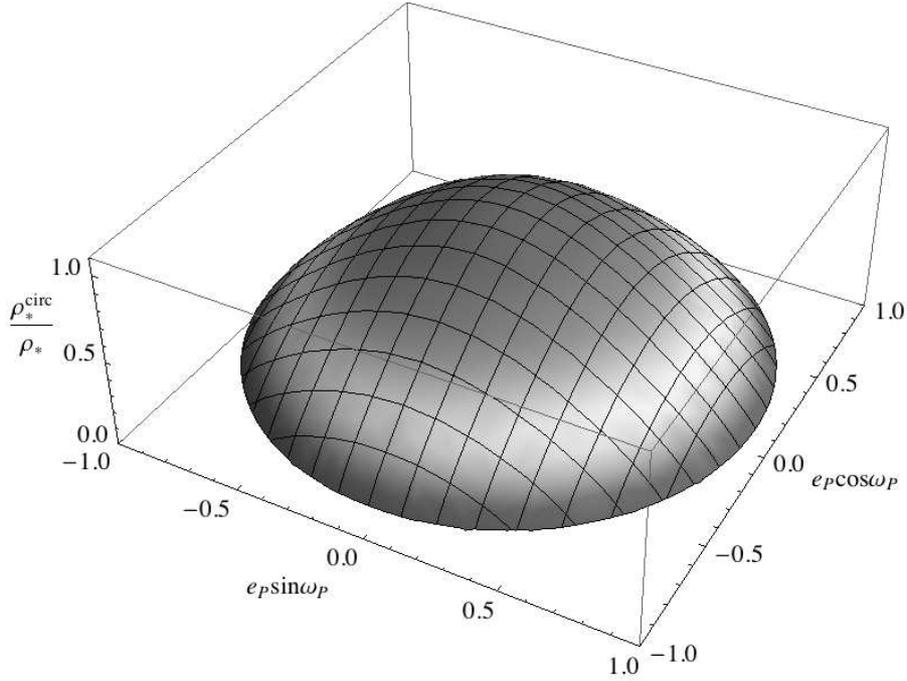}
\caption[The effect of eccentricity on the retrieved stellar density, when
falsely using the circular equations]
{\emph{If one uses the circular expressions, the retrieved value of 
$\rho_*$ is heavily biased by eccentricity.  In this example, the true value of 
$\rho_*$ is 1\,$\rho_{\odot}$ but the introduction of eccentricity causes 
$\rho_*$ to be underestimated.}} 
\label{fig:rhowrong}
\end{center}
\end{figure}

This density bias, which is plotted in Figure~\ref{fig:rhowrong}, could be 
extremely crucial in the search for transiting planets.  Many discovery papers 
of new transiting planets have only sparse radial velocity data and usually no 
secondary eclipse measurement.  As a result, the uncertainty on the 
eccentricity is very large.

Critically, planets are often accepted or rejected as being genuine or not on 
the basis of this light curve derived stellar density.  If the light curve 
derived stellar density is very different from the combination of stellar 
evolution and spectroscopic determination, these candidates are generally 
regarded as unphysical.  This method of discriminating between genuine planets 
and blends, which may mimic such objects, was proposed by \citet{seager2003} 
(see \S6.3 of SMO03) but crucially is only applicable for circular orbits.

Since the typical upper limit on $e_P$ is around 0.1 in discovery papers, then 
the light curve derived stellar density also has a maximum possible error of 
$\sim30$\%. In practice, the uncertainty on $e_P$ will result in a larger 
uncertainty in $\rho_*$.  Typical procedure is to fix $e_P=0$ if the radial 
velocity data is quite poor, despite the fact the upper limit on $e_P \sim 0.1$. 
As a result, the posterior distribution of $\rho_*$ would be artificially narrow
and erroneous if $e_P \neq 0$. I propose that global fits should allow $e_P$ to 
vary when analyzing radial velocity and transit photometry, as well as a fixed
$e_P=0$ fit for comparison.  This would allow the full range of possible 
eccentricities to be explored, which would result in a broader and more accurate
distribution for $b_{P,T}$, $(a_P/R_*)$ and critically $\rho_*$.

\subsection{Application to Multiple Transiting Planet Systems}
\label{sec:multitransit}

One interesting application of these ideas comes with the multi-planet 
transiting systems being found by \emph{Kepler}. Recently, \citet{steffen2010}
announced the discovery of five multiple transiting planet candidate systems. 
The planets have not been confirmed (and thus are ``candidate'' at the time of
writing) because the targets are too faint for RV follow-up. Nevertheless, in 
such cases, one can assess the ratio of $\Psi_P$ between the two planets, dubbed
with subscripts ``b'' and ``c'':

\begin{equation}
\frac{\rho_*^{\mathrm{circ}}(\mathrm{using\,\,planet\,\,b})}{\rho_*^{\mathrm{circ}}(\mathrm{using\,\,planet\,\,c})} = \frac{\Psi_b}{\Psi_c} = \Bigg[\frac{1+e_b \sin \omega_b}{1+e_c \sin \omega_c}\Bigg]^3 \Bigg[\frac{1-e_c^2}{1-e_b^2}\Bigg]^{3/2}
\end{equation}

The $\sin\omega_P$ functions must be in the range of -1 to +1 only, and so it is 
possible to consider the maximum and minimum ratios which bound the orbits:

\begin{align}
\Bigg(\frac{\Psi_b}{\Psi_c}\Bigg)_{\mathrm{min}} \leq &\Bigg(\frac{\Psi_b}{\Psi_c}\Bigg) \leq \Bigg(\frac{\Psi_b}{\Psi_c}\Bigg)_{\mathrm{max}} \nonumber \\
\Bigg(\frac{1-e_b}{1+e_b}\Bigg) \Bigg(\frac{1-e_c}{1+e_c}\Bigg) \leq &\Bigg(\frac{\Psi_b}{\Psi_c}\Bigg)^{2/3} \leq \Bigg(\frac{1+e_b}{1-e_b}\Bigg) \Bigg(\frac{1+e_c}{1-e_c}\Bigg)
\end{align}

Re-writing $e_c$ as $e_b (e_c/e_b)$ and then expanding to first 
order\footnote{A first order expansion in eccentricity can be justified by the 
fact multi-planet systems are rarely permitted to have dynamically stable highly
eccentric orbits for either body} in $e_b$ on both sides of the inequality, one
obtains a symmetric solution on either side which can be consequently
re-arranged to:

\begin{align}
e_b + e_c \geq \frac{1}{2} \Bigg|\Bigg(\frac{\Psi_b}{\Psi_c}\Bigg)^{2/3} - 1\Bigg|
\label{eqn:psiratio}
\end{align}

\begin{figure}
\begin{center}
\includegraphics[width=15.0 cm]{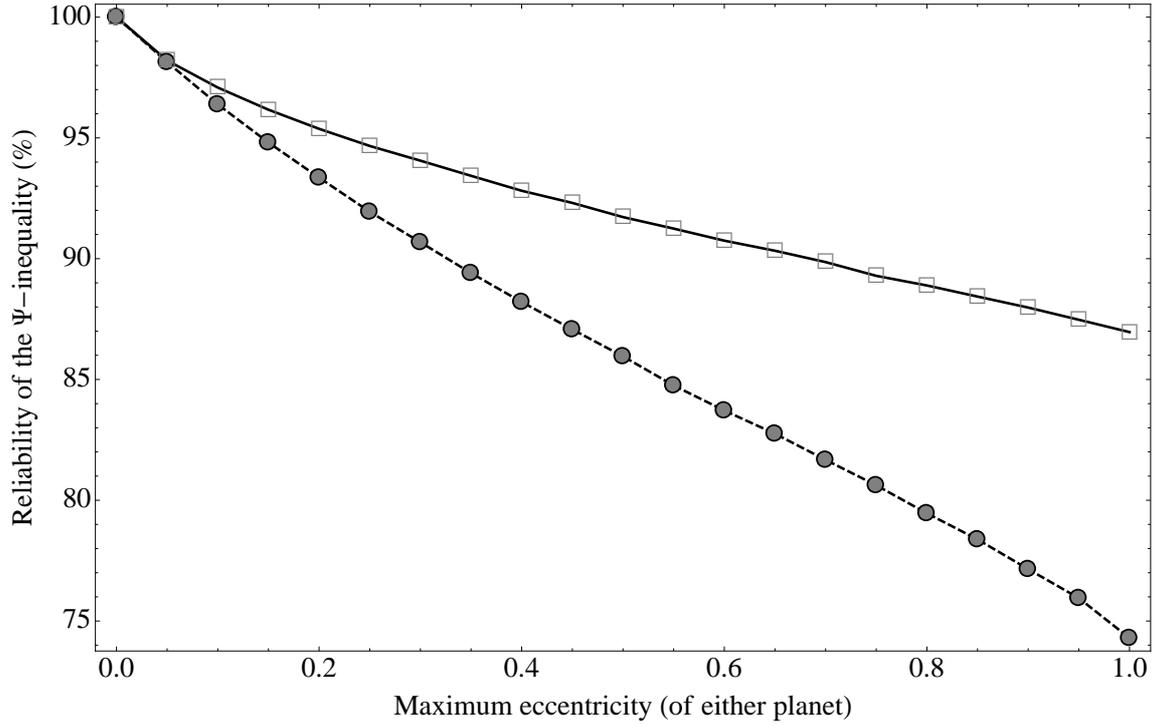}
\caption[Reliability of the photometric constraints on the eccentricities in
multiple transiting planet systems]
{\emph{Assuming circular orbits, one may derive two values for the stellar
density for a system with two transiting planets. The ratio of these densities
constrains the sum of the eccentricities via the $\Psi$-inequality given in
Equation~(\ref{eqn:psiratio}). The reliability of this inequality is plotted 
here as a function of the maximum allowed eccentricity, assuming a uniform 
distribution in $e_P$ (dashed line) and also a non-uniform physically motivated
distribution (solid line). The reliability is $\geq$90\% for $e_P\leq0.30$ for 
the uniform case, and $e_P\leq0.65$ for the non-uniform case.}} 
\label{fig:psis}
\end{center}
\end{figure}

Where it should be understood that the above is only an approximate formula
due to the first-order expansion in $e_b$. To test the accuracy of
Equation~(\ref{eqn:psiratio}), which I label as the $\Psi$-inequality, I 
generated some random values for $e_b$, $\omega_b$, $e_c$ and $\omega_c$. The 
$\omega_P$ values have uniform distributions between 0 and $2\pi$ and the $e_P$ 
values have uniform distributions between $0$ and $e_{\mathrm{max}}$. I 
generated these random values $10^6$ times and tested if the inequality in 
Equation~(\ref{eqn:psiratio}) was true or not each time. As an example, using 
$e_{\mathrm{max}} = 0.25$, the inequality is true in 91.9\% of all of the Monte 
Carlo simulations. In Figure~\ref{fig:psis}, I show the percentage of trials for
which the inequality is correct as a function of $e_{\mathrm{max}}$, which 
reveals that the $\Psi$-inequality provides useful eccentricity constraints in 
the absence of any other information and is $\geq$90\% reliable for 
$e_{\mathrm{max}} \leq 0.30$. 

I also tried using a potentially more realistic non-uniform distribution 
eccentricity distribution using a mixture of an exponential and a Rayleigh
distribution (see \citet{juric2008} and \citet{zakamska2010}):

\begin{equation}
e_P(x) = \alpha \lambda \exp(-\lambda x) + (1-\alpha)\frac{x}{\sigma_e^2} \exp(-x^2/2\sigma_e^2)
\end{equation}

The values of the constants were found by fitting the distribution of 
eccentricities in known multi-planet systems measured from radial velocity 
surveys using only systems with measured eccentricities, which find
$\alpha =0.38$, $\lambda=15$ and $\sigma_e = 0.17$ \citep{steffen2010}. Finally, 
this distribution can produce values of $e_P$ greater than unity, and so I 
ignored any simulations where $e_P > e_{\mathrm{max}}$ for either planet. Using 
$e_{\mathrm{max}} = 1$, I found that 87.0\% of simulations agreed with the 
inequality presented in Equation~(\ref{eqn:psiratio}), and $\geq$90\% agree for 
$e_{\mathrm{max}}\leq0.65$ (Figure~\ref{fig:psis} shows dependency of this 
percentage with $e_{\mathrm{max}}$).

Additional constraints may come from dynamical modelling of
the orbits to explore the stable regimes of $e_P$ and $\omega_P$ for each 
planet. Put together, these methods raise the chances of characterizing the 
eccentricity of such systems.

\subsection{Generalized Parameter Retrieval}

In the previous subsection, I discussed how using the circular expressions to 
derive $(a_P/R_*)$ and $\rho_*$ can lead to severe errors for even mildly 
eccentric systems.  Here, I present expressions which will recover excellent 
approximate values for $b_{P,T}$, $(a_P/R_*)$ and $\rho_*$.  The new equations 
(derived using the $\tilde{T}^{\mathrm{one}}$ model and thus given the
same superscript notation) are given by:

\begin{align} 
[b_{P,T}^{\mathrm{one}}]^2 &= \frac{(1-p)^2 - \frac{\sin^2[(T_{23}^{\mathrm{one}} \pi \sqrt{1-e_P^2})/(P_P \varrho_{P,T}^2)]}{\sin^2[(T_{14}^{\mathrm{one}} \pi \sqrt{1-e_P^2})/(P_P \varrho_{P,T}^2)]} (1+p)^2}{ 1-\frac{\sin^2[(T_{23}^{\mathrm{one}} \pi \sqrt{1-e_P^2})/(P_P \varrho_{P,T}^2)]}{\sin^2[(T_{14}^{\mathrm{one}} \pi \sqrt{1-e_P^2})/(P_P \varrho_{P,T}^2)]} } \\
[(a_P/R_*)^{\mathrm{one}}]^2 &= \frac{(1+p^2) - b_{P,T}^2}{\varrho_{P,T}^2 \sin^2[(T_{14}^{\mathrm{one}} \pi \sqrt{1-e_P^2})/(P_P \varrho_{P,T}^2)]} + \frac{b_{P,T}^2}{\varrho_{P,T}^2} \\
\rho_{*}^{\mathrm{one}} &= \frac{3 \pi}{G P_P^2} [(a_P/R_*)^{\mathrm{one}}]^3 - p^3 \rho_P
\label{eqn:oneretrieval}
\end{align}

These expressions can be shown to reduce down to the original SMO03 derivations 
if $e_P \rightarrow 0$ (Equations~(7)\&(8) of SMO03). The new stellar density 
parameter may be used with floating $e_P$ and $\omega_P$ values to correctly 
estimate the probability distribution of this critical parameter.

It is also possible to write down the retrieval of the parameters in terms of
the duration $\tilde{T}$, which was defined in \S\ref{sec:Tduration}, and the
ingress/egress duration. Due to the approximations made in the derivation of
$T_{xy}^{\mathrm{one}}$, the ingress and egress have an equivalent duration,
as was seen for circular orbits. Therefore, $T_{12}$ and $\tilde{T}$ form a
complete set which can be used instead of the $T_{14}$ and $T_{23}$ set. With 
this alternate set, the retrieved parameters may be shown to be:

\begin{align}
\tilde{\mathfrak{T}}^{\mathrm{one}} &= \frac{ 2\pi\sqrt{1-e_P^2} \tilde{T}^{\mathrm{one}} }{ P_P \varrho_{P,T}^2 } \\
\mathfrak{T}_{1,2}^{\mathrm{one}} &= \frac{2 \pi \sqrt{1-e_P^2} T_{12}^{\mathrm{one}} }{ P_P \varrho_{P,T}^2 } \\
(a_P/R_*)^2 &= \Bigg[ \varrho_{P,T}^2 \Bigg( 2p^2 \cos\tilde{\mathfrak{T}}^{\mathrm{one}}+(p^2+1)\cos2\tilde{\mathfrak{T}}^{\mathrm{one}} - \cos2\mathfrak{T}_{1,2}^{\mathrm{one}} + p^2\Bigg) \nonumber \\
\qquad& - \sqrt{2} p \varrho_{P,T}^2 \csc^2\mathfrak{T}_{1,2}^{\mathrm{one}}\cos^2\frac{\tilde{\mathfrak{T}}^{\mathrm{one}}}{2} \sin2\mathfrak{T}_{1,2}^{\mathrm{one}} \sqrt{p^2 - (p^2-4)\cos2\mathfrak{T}_{1,2}^{\mathrm{one}} - 4\cos4\mathfrak{T}_{1,2}^{\mathrm{one}}}\Bigg] \nonumber \\
\qquad& \times \Bigg[\varrho_{P,T}^4 \Bigg(\cos2\tilde{\mathfrak{T}}^{\mathrm{one}} - \cos2\mathfrak{T}_{1,2}^{\mathrm{one}}\Bigg)\Bigg]^{-1} \\
b_{P,T}^2 &= \Bigg[ \varrho_{P,T}^2 \Bigg( -2p^2 \cos\tilde{\mathfrak{T}}^{\mathrm{one}} + (p^2+1) \cos2\tilde{\mathfrak{T}}^{\mathrm{one}} + p^2 -1 \Bigg) \nonumber \\
\qquad& + \sqrt{2} p \varrho_{P,T}^2 \tan^2\frac{\tilde{\mathfrak{T}}^{\mathrm{one}}}{2}\csc^2\mathfrak{T}_{1,2}^{\mathrm{one}}\cos^2\frac{\tilde{\mathfrak{T}}^{\mathrm{one}}}{2}\sin2\mathfrak{T}_{1,2}^{\mathrm{one}} \nonumber \\
\qquad& \times \sqrt{p^2 - (p^2-4) \cos2\mathfrak{T}_{1,2}^{\mathrm{one}} - 4\cos2\tilde{\mathfrak{T}}^{\mathrm{one}}} + 2 \varrho_{P,T}^2 \sin^2\mathfrak{T}_{1,2}^{\mathrm{one}}\Bigg] \Bigg[ \varrho_{P,T}^2 \Bigg( \cos2\tilde{\mathfrak{T}}^{\mathrm{one}} - \cos2\mathfrak{T}_{1,2}^{\mathrm{one}} \Bigg) \Bigg]^{-1}
\end{align}

Where the first two lines give a necessary substitution.

\subsection{Application to Light Curve Fitting}

In fitting transit light curves, one is trying to calculate the parameters
$\tau_T$, $p$, $(a_P/R_*)$ and $b_{P,T}$. One can fit for these parameters 
directly but the strong correlation between $(a_P/R_*)$ and $b_{P,T}$ severely 
slows down such algorithms \citep{carter2008}. Consequently, it is preferable to 
use an alternative pair of parameters over $(a_P/R_*)$ and $b_{P,T}$.

To understand what alternative pair should be used, one must analyze the light
curve morphology which can be done by approximating the transit light curve as a
trapezoid and calculating the various covariances between the different 
parameters. \citet{carter2008} showed that this exercise reveals that $T_{12}$
and $\tilde{T}$ exhibit much lower correlations than $b_{P,T}$ and $(a_P/R_*)$. One 
must also consider the effect of whether the fitted parameters have uniform 
priors or not, which is also a desideratum.

Various sets exist in the literature e.g. \citet{winn2009} advocate $\tilde{T}$ 
and $b_{P,T}$; \citet{bakos2007} propose $(\zeta/R_*)=2/\tilde{T}$ and
$b_{P,T}^2$. The former set is generally preferable to the latter 
due to the non-uniform prior which exists for $b_{P,T}^2$ \citep{hatp24}, but 
nevertheless the \citet{bakos2007} set is far better than using $(a_P/R_*)$ and 
$b_{P,T}$. However, \citet{bakos2007} used the TS05 expression for $\tilde{T}$. 
By using an improved estimate for the duration, one would expect the 
correlations to further decrease, since the offset between the approximation and
the true value is lower.

To investigate this hypothesis, I generated an artificial transit and
refitted the data using the Markov Chain Monte Carlo (MCMC) method, which allows
one to see the various inter-parameter correlations. I used three parameter sets
to fit the data. Firstly, I use the physical parameter set 
$\{\tau_T,p,(a/R_*),b_{P,T}\}$ as a reference, or ``control'', model. In the 
second I use $\{\tau_T,p^2,(\zeta/R_*),b_{P,T}^2\}$ and in the third I use 
$\{\tau_T,p^2,(\Upsilon/R_*),b_{P,T}^2\}$, where I have defined
$(\Upsilon/R_*) = 2/\tilde{T}^{\mathrm{one}}$.

Since the TS05 formula does not reduce down to the SMO03 equation, the formula
will perform worse than $T^{\mathrm{one}}$ for even circular orbits (or 
near-circular), which are the most common type of planet in the exoplanet
transiting catalogue (see http://www.exoplanet.eu). I therefore consider a 
hot-Jupiter on a circular orbit with a planet-star separation of 
$(a_P/R_*) =3.5$ from a Sun-like star ($P_P=0.76$ days).  I choose a 
near-grazing transit with $b_{P,T}=0.9$, corresponding to an orbital inclination 
of $i_P=75.1^{\circ}$. The light curve is generated using the \citet{mandel2002} 
algorithm with no limb-darkening and $0.25$\,mmag Gaussian noise over a 30\,sec
cadence. The light curve is then passed onto an MCMC fitting algorithm. I also 
repeated the whole exercise for an eccentric orbit example as well, where I 
chose the same system parameters as that of the most eccentric transiting 
planet known at the time of writing, HD 80606b (\citet{fossey2009}, 
\citet{winn2009}).

In the MCMC runs, the jump sizes are set to be equal to $\sim 1$-$\sigma$ 
uncertainties from a preliminary short-run.  The MCMC is started from 
5-$\sigma$'s away from the solution for each parameter, utilizing 500,000 trials 
with a 100,000 burn-in time.  I then compute the cross-correlations between the
various parameters in trials which are within $\Delta \chi^2 = 1$ of 
$\chi^2|_{\mathrm{best}}$ (errors rescaled such that $\chi^2|_{\mathrm{best}}$ 
equals number of data points minus the degrees of freedom).  The 
inter-parameter correlations are computed and used to construct correlation 
matrices for each parameter fitting set.  As an example, the correlation matrix
for the $\{\tau_T, p, (a_P/R_*), b_{P,T}\}$ parameter set is given by:

\begin{eqnarray} \nonumber
&\mathrm{Corr}(\{\tau_T, p, (a_P/R_*), b_{P,T}\},\{\tau_T, p, (a_P/R_*), b_{P,T}\})= \nonumber \\
& \left(
\begin{array}{cccc}
         1              & \mathrm{Corr}(\tau_T,p) & \mathrm{Corr}(\tau_T,(a_P/R_*)) & \mathrm{Corr}(\tau_T,b_{P,T}) \\
 \mathrm{Corr}(p,\tau_T)   &            1         & \mathrm{Corr}(p,(a_P/R_*))   & \mathrm{Corr}(p,b_{P,T}) \\
 \mathrm{Corr}((a_P/R_*),\tau_T) & \mathrm{Corr}((a_P/R_*),p) &             1          & \mathrm{Corr}((a_P/R_*),b_{P,T}) \\
 \mathrm{Corr}(b_{P,T},\tau_T)   & \mathrm{Corr}(b_{P,T},p)   & \mathrm{Corr}(p,(a_P/R_*))   &           1
\end{array} \nonumber
\right)
\end{eqnarray}

I then calculated the semi-principal axes of correlation ellipsoid by 
diagnolizing the matrices.  For a completely optimal parameter set, this 
diagnolized matrix would be the identity matrix.  I quantify the departure of 
each proposed parameter set from the identity matrix by calculating 
$\sum_{i=1}^4 |M_{i,i}-1|$ where $\mathbf{M}$ is the diagnolized correlation 
matrix. The results are shown in upper half of Table~\ref{tab:durations}.

\begin{table}
\caption[Comparison of the inter-parameter correlations for three proposed
fitting sets]
{\emph{For several proposed light curve fitting parameter sets (left 
column), I calculate the inter-parameter correlation matrices in the examples of
i) a hypothetical near-grazing hot-Jupiter on a circular orbit ii) a system 
similar to the eccentric planet HD 80606b. I diagnolize the correlation 
matrices to give $M$ and then quantify the departure from a perfectly optimal 
parameter set (right column), where it is understood that 0 corresponds to 
optimal and larger values correspond to greater inter-parameter correlations.}} 
\centering 
\begin{tabular}{c c} 
\hline\hline 
Parameter Set & $\sum_{i=1}^4 |M_{i,i}-1|$ \\ [0.5ex] 
\hline
Circular orbit example \\
\{$\tau_T$, $p$, $(a_P/R_*)$, $b_{P,T}$\} & 2.19333 \\
\{$\tau_T$, $p^2$, $(\zeta/R_*)$, $b_{P,T}^2$\} & 1.71236 \\
\{$\tau_T$, $p^2$, $(\Upsilon/R_*)$, $b_{P,T}^2$\} & 1.32974 \\
\hline
Eccentric orbit example \\
\{$\tau_T$, $p$, $(a_P/R_*)$, $b_{P,T}$\} & 2.46676 \\
\{$\tau_T$, $p^2$, $(\zeta/R_*)$, $b_{P,T}^2$\} & 1.56816 \\
\{$\tau_T$, $p^2$, $(\Upsilon/R_*)$, $b_{P,T}^2$\} & 1.56799 \\
\hline
\end{tabular}
\label{tab:durations} 
\end{table}

\begin{figure}
\begin{center}
\includegraphics[width=15.0 cm]{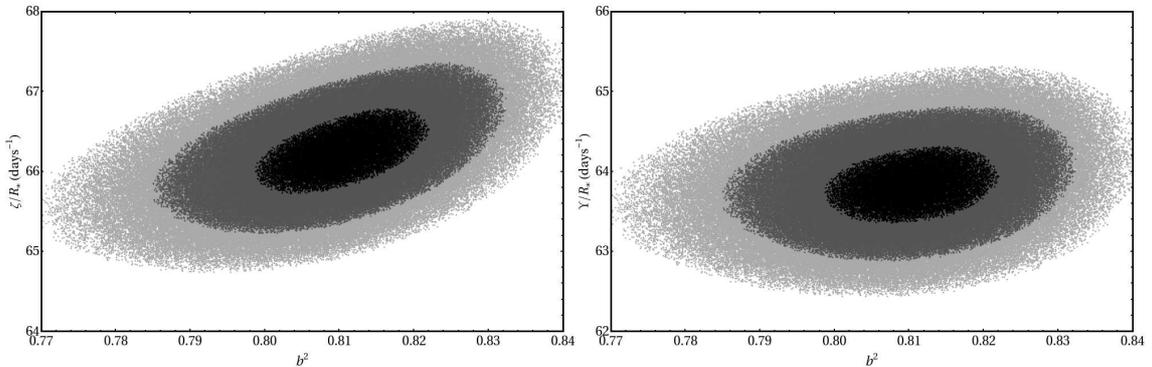}
\caption[Comparison of the correlations exhibited by $(\zeta/R_*)$ versus that
of $(\Upsilon/R_*)$]
{\emph{Comparison of the correlations between $(\zeta/R_*)$ against 
$b_{P,T}^2$ (left panel) and $(\Upsilon/R_*)$ against $b_{P,T}^2$ (right panel).  
Data comes from fitting a synthetic hot-Jupiter light curve on a circular, 
near-grazing orbit with an MCMC algorithm.  The $(\Upsilon/R_*)$ parameter 
provides two-fold lower correlation and preserves the ability to be inversely 
mapped to more physical parameters (see Equation~(\ref{eqn:oneretrieval})).  
The three different types of shading represent the 1-$\sigma$, 2-$\sigma$ and 
3-$\sigma$ confidence regions.}} 
\label{fig:upsilon}
\end{center}
\end{figure}

The correlations of the physical parameter set are predictably very large, in 
particular between $(a_P/R_*)$ and $b_{P,T}$, which approaches negative unity.   
The improvement of $T^{\mathrm{one}}$ over $T^{\mathrm{TS05}}$ is evident by 
examining the effect of modifying $(\zeta/R_*)$ to $(\Upsilon/R_*)$, which 
produces a clear improvement in the corresponding correlations, as seen in 
Figure~\ref{fig:upsilon}.

\subsection{Secular Transit Duration Variations}

Transit duration variation (TDV) can occur in two possible formats: i) periodic 
change and/or ii) secular change (or apparently secular).  Periodic changes in 
the duration of a transit will be discussed later in Chapter~\ref{ch:Chapt6}.

Secular changes in transit duration can be caused by numerous possible 
scenarios. For example, \citet{jordan2008} showed that apsidal precession would 
induce changes in $\tilde{T}$ and used the TS05 equation to predict the size of 
these changes.  As was already demonstrated a better formulation for 
$\tilde{T}$ is possible; I will here present an improved equation for the rate 
of change in $\tilde{T}$ due to apsidal precession, or essentially changes in 
$\omega_P$.

\citet{jordan2008} argued that apsidal precession can be caused by stellar 
oblateness, general relativistic effects and/or a perturbing planet.  
Additionally, \citet{murray1999} showed that, in general, nodal precession 
should always accompany apsidal precession, leading to changes in the orbital 
inclination angle, $i_P$.  These changes lead to another form of secular TDV.

Additionally, I consider here that falling planets, such as proposed for 
WASP-18b \citep{hellier2009}, would experience a changing semi-major axis, 
$a_P$, leading to another form of secular TDV.  Finally, I will consider the 
effect of varying the orbital eccentricity.  All four possible TDVs will be 
derived here using $\tilde{T}^{\mathrm{one}}$, since $T^{\mathrm{one}}$ has
demonstrated the greatest accuracy in numerical tests (see 
\S\ref{sec:numericaltests}) and $\tilde{T}$, in general, is the most useful 
definition of the duration for TDV studies \citep{carter2008}.

\subsubsection{Apsidal Precession}

Apsidal precession is the precession of the argument of periapse over time and 
it may be induced from several different effects including:

\begin{itemize}
\item[{\tiny$\blacksquare$}] General relativistic effects (\citet{einstein1915}; 
\citet{pal2008})
\item[{\tiny$\blacksquare$}] Rotational quadrupole bulges on the planet 
\citep{sterne1939}
\item[{\tiny$\blacksquare$}] Tides raised on the planet and the star 
\citep{sterne1939}
\item[{\tiny$\blacksquare$}] Stellar quadrupole moment \citep{murray1999}
\item[{\tiny$\blacksquare$}] Kozai mechanism \citep{kozai1962}
\item[{\tiny$\blacksquare$}] Perturbing planets (\citet{murray1999}, 
\citet{miralda2002}; \citet{heyl2007})
\end{itemize}

Out of these examples, planets on nearby orbits of masses $\geq M_{\oplus}$ are 
expected to produce the largest effect.  Thus the detection of apsidal 
precession could actually be used to infer the presence of Earth-mass planets.

As \citet{jordan2008} noted, apsidal precession should cause a change in the 
transit duration and in order to estimate the magnitude of this effect, the 
authors differentiated $T^{\mathrm{TS05}}$ with respect to $\omega_P$.  Having 
shown the $T^{\mathrm{one}}$ offers substantial improvement over the 
$T^{\mathrm{TS05}}$ in the previous section, I am here able to provide an 
improved estimate for the secular TDV caused by apsidal precession:

\begin{equation}
\frac{\partial \tilde{T}^{\mathrm{one}}}{\partial\omega_P} = \frac{P_P}{\pi} \frac{e_P \varrho_{P,T}^3 \cos \omega_P}{(1-e_P^2)^{3/2}} \Bigg(\frac{1}{\sqrt{1-b_{P,T}^2} \sqrt{(a_P/R_*)^2 \varrho_{P,T}^2 - 1}} - 2 \arcsin\Big(\frac{\sqrt{1-b_{P,T}^2}}{(a_P/R_*) \varrho_{P,T} \sin i_P}\Big) \Bigg)
\label{eqn:apsidal}
\end{equation}

One can see that there are two terms counter-acting in the derived quantity. The 
two terms can be understood to originate from the planet-star separation 
changing as a result of the precession, which has two effects: 1) decreasing the 
planet-star separation causes a near-grazing transit's impact parameter to 
decrease and thus increases $\tilde{T}$ (the first term) 2) decreasing the 
planet-star separation causes the tangential orbital velocity to increase and 
thus decreases $\tilde{T}$ (the second term).  The $\cos \omega_P$ term outside 
of the brackets determines the sign of which term causes an increase and which 
to decrease.

\citet{kopal1959} showed that the two effects approximately cancel out for 
$b_{P,T} \simeq 1/\sqrt{2}$.  The \citet{kopal1959} derivation is quite different 
for the ones produced in this paper.  It is done by first solving for the 
mid-eclipse time by a series expansion of the differential of the planet-star 
separation with respect to true anomaly, disregarding terms in $\sin^3i_P$ or 
higher.  The duration between the primary and secondary occultation is then 
solved for in another series expansion in first order of $e_P$.  Nevertheless, 
setting $b_{P,T}$ to this value, the terms inside the bracket of 
Equation~(\ref{eqn:apsidal}) become:

\begin{equation}
\frac{ \sqrt{2} }{ \sqrt{(a_P/R_*)^2 \varrho_{P,T}^2 - 1} } - 2 \arcsin\Bigg( \frac{1}{\sqrt{2} \sqrt{(a_P/R_*)^2 \varrho_{P,T}^2 - \frac{1}{2}}} \Bigg)
\end{equation}

Under the condition $a_P \gg R_*$, I find that Equation~(\ref{eqn:apsidal}) 
gives $\partial \tilde{T}^{\mathrm{one}}/\partial\omega_P = 0$, in agreement 
with \citet{kopal1959}. For very close-in orbits, this does not appear to hold.

One may compare this estimate of the apsidal precession to Equation~(15) of 
\citet{jordan2008}, which was found by differentiating the expression of TS05 
with respect to $\omega_P$.  The difference between the two expression is 
typically less than 1\% across a broad range of parameters.  However, if 
$b_{P,T}\simeq1/\sqrt{2}$, the difference between the two diverges and can reach 
10\%-100\%.  Given the sensitivity of both equations to this critical value of 
$b_{P,T}$, I recommend numerical calculations over analytic approximations if 
$b_{P,T}$ is known to be close to 0.707.

\subsubsection{Nodal Precession}

Nodal precession causes changes in the orbital inclination of the planetary 
orbit, which would be a source of secular TDV.  The secular theory of 
\citet{murray1999} predicts the rate of inclination change due to a perturbing 
planet as the nodes precess:

\begin{equation}
\frac{\partial i_P}{\partial t} = - \frac{\partial \omega_P}{\partial t} \Delta \Omega
\end{equation}

Where $\Delta \Omega$ is the ascending node of the perturbing planet relative 
to the ascending node of the transiting planet, measured clockwise on the plane 
of the sky.  Therefore, any occurrence of apsidal precession from a perturbing 
planet will, in general, be coupled with nodal precession.  One may derive the 
rate of secular TDV from inclination change as before:

\begin{equation}
\frac{\partial \tilde{T}^{\mathrm{one}}}{\partial i_P} = \frac{P_P}{\pi} \frac{\varrho_{P,T}^2 \sqrt{(a_P/R_*)^2 \varrho_{P,T}^2 - 1}}{\tan i_P \sqrt{1-e_P^2} \sqrt{1-b_{P,T}^2}}
\end{equation}

This expression has only one term and therefore one can see that decreasing the 
inclination towards a more grazing transit always yields a shorter transit 
duration, and vice versa.

\subsubsection{Falling Exoplanets}

Planetary bodies experience infall towards the host star through tidal 
dissipation and to a much lesser degree gravitational radiation.  The effects 
increase as the orbit becomes smaller leading to runaway fall-in.  Therefore, 
for very close-in exoplanets, the change in semi-major axis may be detectable. 
The transit duration will vary as:

\begin{align}
&\frac{\partial \tilde{T}^{\mathrm{one}}}{\partial a_P} = \frac{P_P}{\pi} \frac{\varrho_{P,T}^2}{a_P \sqrt{1-e_P^2}} \Bigg(\frac{3}{2} \arcsin\Big(\frac{\sqrt{1-b_{P,T}^2}}{(a_P/R_*) \varrho_{P,T} \sin i_P}\Big) - \frac{1}{\sqrt{1-b_{P,T}^2} \sqrt{(a_P/R_*)^2 \varrho_{P,T}^2 - 1}}\Bigg)
\end{align}

As for apsidal precession, there are two countering components, which are the 
same as before except for a slightly different constant in front of the first 
term.  This different constant means that the impact parameter at which both 
effects cancel has now changed to $b_{P,T} \simeq 1/\sqrt{3} = 0.577$.  This result 
could not be found in the previous literature and is of particular interest 
given the recent discovery of exoplanets on periods of around a day or less, 
for example WASP-18b \citep{hellier2009} with period of 0.94\,days and 
$b_{P,T} = 0.25$.

\subsubsection{Eccentricity Variation}

Irregular satellites are known to exchange orbital inclination and eccentricity 
through the Kozai mechanism, which roughly conserves the value 
$\cos I \sqrt{1-e^2}$, where $I$ is the angle to the ecliptic. Variations in 
$e_P$ may lead to a form of secular TDV as well:

\begin{align}
&\frac{\partial \tilde{T}^{\mathrm{one}}}{\partial e_P} = \frac{P_P}{\pi} \frac{\varrho_{P,T}^3}{(1-e_P^2)^{5/2}} \Bigg[\frac{2 e_P+(1+e_P^2) \sin \omega_P}{\sqrt{1-b_{P,T}^2} \sqrt{(a_P/R_*)^2 \varrho_{P,T}^2 -1}} - [3e_P+(2+e_P^2)\sin\omega_P] \arcsin\Big(\frac{\sqrt{1-b_{P,T}^2}}{(a_P/R_*) \varrho_{P,T} \sin i_P}\Big)\Bigg]
\label{eqn:eccvar}
\end{align}

The two terms here seem to exhibit a more complicated inter-dependency, which is
physically based on the same idea of varying the planet-star separation.  The 
balance-point between the two effects occurs for:

\begin{align}
b_{P,T} &\simeq \sqrt{\frac{e_P+\sin\omega_P}{(3e_P+(2+e_P^2) \sin\omega_P}} \nonumber \\
\qquad &= \frac{1}{\sqrt{2}} - \frac{e_P}{4 \sqrt{2} \sin\omega_P} + \mathcal{O}[e_P]^2
\end{align}

Thus to zeroth-order in $e_P$, the balance point is the same as for apsidal 
precession.

\section{Other Important Timings}

\subsection{Transit to Occultation Time}
\label{sec:pritosec}

Before moving on, I will discuss other important timings related to transits,
aside from the transit minima and the durations. The first one will I consider
is the time it takes for a planet to move from the transit minimum to the
occultation minimum i.e. $\tau_O - \tau_P$. As was discussed in 
\S\ref{sec:occultations}, this time difference is given by 
$0.5P_P + \Delta t_{\mathrm{light}}$ for circular orbits, where 
$\Delta t_{\mathrm{light}}$ is the light travel time across the system. For an 
eccentric orbit, things are more complicated and the exact time is given by:

\begin{align}
(\tau_O - \tau_T) &= \frac{P_P}{2\pi} \Delta \mathfrak{M}_{T\rightarrow O} + \Delta t_{\mathrm{light}} \\
\Delta \mathfrak{M}_{T\rightarrow O} &= \int_{f_P=f_{P,T}}^{f_{P,O}} \frac{(1-e_P^2)^{3/2}}{(1+e_P\cos f_P)^2}\,\mathrm{d}f_P
\end{align}

Dealing with the light travel time first, if one assumes a nearly coplanar 
orbit, then the transit minima occur at the times of the conjunctions, so:

\begin{align}
\Delta t_{\mathrm{light}} &= \frac{Z_P(f_{P,O}) - Z_P(f_{P,T})}{c} \simeq \frac{a_P (\varrho_{P,T} + \varrho_{O,T})}{c} \nonumber \\
\lim_{i_P \rightarrow \pi/2} \Delta t_{\mathrm{light}} &= \frac{2 a_P (1-e_P^2)}{c (1-e_P^2 \sin^2\omega_P)}
\label{eqn:lighttime}
\end{align}

For the dynamical time difference, one must deal with the integral once again.
In practice, my light curve fitting routine computes the transit and occultation
minima using the $\eta$-series and then uses Equation~(\ref{eqn:durationfn}) to
compute $\Delta \mathfrak{M}_{T\rightarrow O}$. If one must insist on using an 
approximation though, then $f_{P,T} \simeq (\pi/2-\omega_P)$ and 
$f_{P,O} \simeq (3\pi/2 - \omega_P)$ may be used to give:

\begin{align}
\lim_{i_P \rightarrow \pi/2} (\tau_O - \tau_T) &= \frac{P_P}{2} + \frac{P_P}{\pi} \Bigg( \arctan\Big[\frac{e_P\cos\omega_P}{\sqrt{1-e_P^2}}\Big] + \frac{e_P \sqrt{1-e_P^2} \cos\omega_P}{1-e_P^2\sin^2\omega_P}\Bigg) + \frac{2 a_P (1-e_P^2)}{c (1-e_P^2 \sin^2\omega_P)}
\label{eqn:secpri}
\end{align}

Another insightful approximation comes from expanding to first
order in $e_P$ and ignoring the light travel time, which is usually very small:

\begin{equation}
\lim_{i_P \rightarrow \pi/2} (\tau_O - \tau_T) \simeq \frac{P_P}{2} + \frac{2 P_P}{\pi} \Big[e_P \cos\omega_P + \mathcal{O}[e_P^2]\Big]
\end{equation}

Therefore, the measurement of the time between the transit and the occultation
strongly constrains $e_P \cos \omega_P$.

\subsection{Transit to Occultation Duration Ratio}

To estimate of the ratio between the two eclipse durations, I will
proceed with the $T^{\mathrm{one}}$ approximations. I will label the 
occultation duration between the $x$ and $y$ contact points as $O_{xy}$ (thus
$T$ corresponds to the transit and $O$ for the occultation). It is trivial to 
show that:

\begin{align}
O_{xy}^{\mathrm{one}} &= \frac{P_P}{\pi} \frac{\varrho_{P,O}^2}{\sqrt{1-e_P^2}} \arcsin\Bigg[\sqrt{ \frac{S_{P*}^2-b_{P,O}^2}{(a_P/R_*)^2 \varrho_{P,O}^2 - b_{P,O}^2} } \Bigg]
\label{eqn:occultationdur}
\end{align}

Where $b_{P,O}$ is the impact parameter of the planet for the occultation event,
given by $b_{P,O} = (a_P/R_*) \varrho_{P,O} \cos i_P$. Taking the ratio of the
tilde definitions for the duration gives an elaborate expression. However,
I expand to zeroth-order in $\cos^2 i_P$ and then to first order in $e_P$,
simplifying the result with the assumption $a_P\gg R_*$:

\begin{align}
\frac{\tilde{T}^{\mathrm{one}}}{\tilde{O}^{\mathrm{one}}} &\simeq 1-2 e_P \sin\omega_P + \mathcal{O}[e_P^2]
\end{align}

Due to the approximations made, the above expression is only useful for
providing a sense of the scaling of the ratio. However, they do allow one to see
that the dominant term is $e_P \sin\omega_P$. Therefore, the transit to
occultation timing gives $e_P \cos\omega_P$ and the duration ratios give
$e_P\sin\omega_P$ suggesting that eccentricity could be measured 
photometrically with high quality data.

\subsection{Transit to RV Time}

As a final piece to the puzzle, I will discuss the time difference between
the transit minimum and the time of the stellar reflex velocity null (determined
by radial velocity measurements). The RV-null occurs at exactly the same instant
as the transit minimum for circular orbits. However, for eccentric orbits a
slight offset is introduced. This may be found by first finding the true
anomaly at which the RV-null occurs, by solving:

\begin{align}
\mathrm{RV}(f_P) - \gamma &= -K \sin i_P [e_P\cos\omega_P+\cos(f_P+\omega_P)] = 0 \nonumber \\
\Rightarrow f_{P,\mathrm{null}} &= -\omega_P + \arccos[-e_P \cos \omega_P]
\end{align}

Thus for $e_P = 0$ one has $f_{P,\mathrm{null}} = \pi/2-\omega_P$, as expected.
The disturbance of the RV-null away from the point of inferior conjunction is
much greater than that of the transit minimum. Therefore, a good approximation
is given by assuming $f_{P,T} \simeq \pi/2 - \omega_P$ for the purposes of
computing the time difference between the RV-null and the transit minimum. I
will also use the approximation $\Delta \mathfrak{M}_P \sim \Delta f_P$ for
simplicity:

\begin{align}
(\tau_{\mathrm{null}} - \tau_{T}) &\simeq \frac{P_P}{2\pi} \Big[ \arccos[-e_P \cos \omega_P] - \pi/2 \Big] \nonumber \\
\qquad &= \frac{P}{2\pi} \Big[ e_P \cos \omega_P + \mathcal{O}[e_P^2] \Big]
\end{align}

The first-order expansion in $e_P$, which is presented on the second-line,
illustrates how the RV-null time has a strong dependence on $e_P \cos\omega_P$.
Indeed, at the first-order level, 
$(\tau_{\mathrm{null}} - \tau_{T}) \simeq (\tau_O - \tau_T + 0.5 P_P)/4$.
Put together, these additional timings permit the joint modelling of transit,
occultation and RV data to strongly constrain $e_P$ and $\omega_P$. With these
quantities known, reliable estimates for the other parameters, such as $i_P$
and $\rho_*$, are enabled.


%% file: Chapt5.tex
\chapter{Transit Distortions}
\label{ch:Chapt5}

\vspace{1mm}
\leftskip=4cm

{\it ``
Nothing is as simple as we hope it will be''} 

\vspace{1mm}

\hfill {\bf --- Jim Horning} 

\leftskip=0cm


\section{Introduction}
\label{sec:intro5}

The properties and morphology of the transit light curve have now been 
established, provided assumptions A0 to A10 are in effect. In 
\S\ref{sec:breakassumptions}, I discussed the consequences of breaking each of 
the assumptions and how frequent and significant the invalidation of each 
assumption was. In that section, it was concluded that three assumptions, in 
particular, were likely to be frequently broken and lead to significant
distortion to the transit light curve. These are A0) a uniform source star; 
A4/A5) the planet emits no flux/there are no background luminous objects; 
A10) the integration times are small. A0 can be compensated for by using the 
\citet{mandel2002} code to account for limb darkening and generally avoid
observing spotty stars. In this section, I will discuss the consequences
of breaking A4/A5 (\S\ref{sec:nightside}) and A10 (\S\ref{sec:binning}) in 
more detail and present methods for compensating for each distortion. These
two sections are based upon the papers \citet{kiptin2010} and \citet{kipbin2010}
respectively.

\section{Blending and Nightside Pollution}
\label{sec:nightside}

\subsection{Introduction}

During the early days of exoplanetary science, the dedicated transit space 
missions now available did not exist and the pioneers of those early times
sought the first transit using off-the-shelf instrumentation and small-aperture
telescopes.  It was therefore not surprising that the first transit light curve
was obtained in the visible wavelength range (\citet{charbonneau2000}; 
\citet{henry2000}) and for the subsequent few years this was established as the 
normal practice in later observations and surveys (e.g. \citet{brown2001}; 
\citet{bakos2004}; \citet{pollacco2006}), mostly down to the cost efficiency of 
the available CCDs (see \S\ref{sec:transits} for a more detailed account of the 
first discovery).

As transit measurements became routine and staggering photometric quality 
became available with space-based observatories like HST \citep{brown2001}, 
it became clear that improved models of the transit light curve were required.
In response to the growing need for accurate parameterization of light curves, 
several authors produced equations modelling the light curve behaviour, such as
\citet{mandel2002} and \citet{seager2003}, both of which have been discussed
extensively earlier in this thesis (see \S\ref{sec:LDtransit} and 
\S\ref{sec:circduration} respectively).

In the last few years, the value of infrared measurements of transiting systems 
has become apparent with numerous pioneering detections; emission from a 
transiting planet \citep{deming2005}, emission from a non-transiting planet 
\citep{harrington2006}, an exoplanetary spectrum \citep{grillmair2007}, 
detection of water in the atmosphere of a hot-Jupiter \citep{tinetti2007}, 
methane \citep{swain2008} and more recently carbon dioxide \citep{swain2009}. 
More details on the use of transmission spectroscopy as a tool for detecting 
molecular species can be found in \citet{seager2000} and \citet{tinetti2009}. 
With JWST set to replace HST in the next decade, one can expect an abundance of 
high-precision infrared transits to be observed in order to detect more 
molecular species, perhaps including biosignatures \citep{seager2005}. In this 
section, I discuss the consequences of significant nightside planetary emission 
on precise infrared transit light curves.  Nightside emission invalidates one of 
the original assumptions made both earlier in this thesis (assumption A4, 
\S\ref{sec:transitbasics}) and in the models of \citet{seager2003} and 
\citet{mandel2002}: ``The planet emits no flux''.  In the case of hot-Jupiter 
systems, the nightside of the planet is hot and thus flux-emitting. This 
additional flux can be considered as a blend, but from the planet itself, i.e. a 
\emph{self-blend}.

Conceptually, it is very easy to see that this will cause mid-infrared transit 
depth measurements (and to a lesser degree in the visible and near-infrared 
range) to become underestimates of the true depth.  The reason is that there are
two sources of flux, the star and the planet, and only one of these is being 
occulted, whilst the other is the blend source. This is highly analogous to the
case of a nearby companion star which is not spatially resolved in the 
point-spread-function (PSF) and thus induces blending.  In this section, I will 
derive expressions estimating the amplitude of the effect, propose methods for 
correcting the light curves and apply them to two cases where the nightside 
emission of an exoplanet has been determined in the mid-infrared.

Although the fundamental effects of both blending and self-blending (or 
nightside pollution) are the same (i.e. a dilution of the depth) I will here
focus more on the nightside pollution effect. This is because this was not
previously known until pointed out in \citet{kiptin2010} (the paper upon which
this section is based) whereas blending is a well-known occurrence and generally 
corrected for by resolving the companion's light using high resolution imaging 
\citep{latham2010}\footnote{Obviously this is not possible for self-blending}.

\subsection{Derivation}
\subsubsection{Depth Dilution}

Let us define the total out-of-transit flux surrounding a transit event, as 
shown in Figure~\ref{fig:night2}a, to be given by:

\begin{equation}
F_{\mathrm{out,tra}} = F_{*} + F_{P}
\end{equation}

Where $F_{*}$ is the total flux received from the star and $F_{P}$ 
is the total flux received from the nightside of the planet, over a time 
interval of d$t$.  Let us assume that the star is a uniform emitter (A0) and 
that both the stellar and planetary total flux are invariable over the timescale
of the transit event (A5).  One may then write down the flux 
during a transit as a function of the ratio of the radii, $p$:

\begin{equation}
F_{\mathrm{in,tra}} = (1-p^2) F_{*} + F_{P}
\end{equation}

Note how the flux of the star has been attenuated as a result of the eclipse but
the planetary flux is still present.  The observed transit depth in the flux 
domain, $\delta'$, is defined by:

\begin{align}
\delta' &= \frac{F_{\mathrm{out,tra}} - F_{\mathrm{in,tra}}}{F_{\mathrm{out,tra}}} \\
\qquad &= \Big(\frac{F_*}{F_* + F_{P}}\Big) \delta
\label{eqn:dilution1}
\end{align}

Where $\delta$ is the undiluted transit depth.  In the case of 
$F_{\mathrm{P}} \rightarrow 0$, the standard equation for the 
depth is recovered, i.e. $\delta' = \delta = p^2$.  For cases 
where the nightside flux of the planet is non-negligible, the transit depth will
therefore be affected.  One may re-write Equation~(\ref{eqn:dilution1}) as 
$\delta = B_{P} \delta'$ where I define:

\begin{align}
B_{x} &= \frac{F_* + F_{x}}{F_*} \nonumber \\
\Rightarrow B_{P} &= \frac{F_* + F_{P}}{F_*}
\end{align}

Where the above equation emphasizes how the equations presented here are
generally applicable to any kind of blend from a source $x$. For the nightside
emission case, the blend source is generally at a much cooler temperature than 
the host star and thus the contrast between the two bodies is greater at 
infrared wavelengths. Consequently, the self-blending has a significant effect 
on infrared measurements (e.g. see \S\ref{sec:appliedexamples}) but a much 
lower impact on visible wavelengths (e.g. see \S\ref{sec:compeffects}). This 
makes the inclusion of such an effect paramount since visible and infrared 
measurements must be considered incommensurable unless this systematic is 
corrected for.

\subsubsection{The Consequences for Other Parameters}

The largest effect of a blend is to underestimate the transit depth. However, I
evaluate here the effect of the nightside blend, or indeed any kind of blend, on
the other light curve parameters. For simplicity, I consider here a circular 
orbit and so I may use the expressions of \citet{seager2003}.

Earlier in \S\ref{sec:usingcircforecc}, I showed how using the circular
expressions when the orbit is eccentric leads to the erroneous retrieval
of the impact parameter, the scaled orbital semi-major axis and the light curve
derived stellar density (see Equations~(\ref{eqn:bpwrong},\ref{eqn:aRwrong} \&
\ref{eqn:rhowrong})). In a similar manner, I will here show how erroneous
the retrieval of these parameters would be if one did not know that a blend
was present:

\begin{align}
[b_{P,T}^{\mathrm{undiluted}}]^2 &= \frac{p^2 - (1 + p^2 -b_{P,T}^2) \sqrt{B_x} + B_x}{B_x} \nonumber \\
\qquad &= b_{P,T}^2 + \frac{1}{2} (1-b_{P,T}^2 - p^2) (B_x-1) + \mathcal{O}[(B_x-1)^2]
\label{eqn:bpundiluted}
\end{align}

\begin{align}
[(a_P/R_*)^{\mathrm{undiluted}}]^2 &= (a_P/R_*)^2 - \frac{(1-b_{P,T}^2-p^2) ((a_P/R_*)^2-(1+p)^2)}{2 ((1+p)^2-b_{P,T}^2)} (B_x-1) + \mathcal{O}[(B_x-1)^2]
\label{eqn:aRundiluted}
\end{align}

Equations~(\ref{eqn:bpundiluted}) and (\ref{eqn:aRundiluted}) imply that 
negating the blending factor causes one to overestimate the impact parameter and 
underestimate $(a_P/R_*)$. It is important to recall that the light curve 
derived stellar density is found by taking the cube of $(a_P/R_*)$ and therefore 
will exacerbate any errors at this stage. Note that both equations give the 
expected results for $B_x=1$, i.e. no blend source present. 

These expressions have been derived assuming no limb darkening is present, which
is typically a very good approximation for the wavelength range we are 
interested in. However, in reality the incorporation of limb darkening is easily
implemented and demonstrated later in \S\ref{sec:appliedexamples}. Therefore, 
Equations~(\ref{eqn:bpundiluted}) and (\ref{eqn:aRundiluted}) should not be used
to attempt to correct parameters derived from fits not accounting for nightside 
pollution, rather they offer an approximate quantification of the direction and 
magnitude of the expected errors\footnote{For this reason, I considered it
not worthwhile to derive the full versions of Equations~(\ref{eqn:bpundiluted}) 
and (\ref{eqn:aRundiluted}) which account for eccentricity}.

\subsection{Compensating for the Effect}
\label{sec:compensating}

\subsubsection{Empirical Method}

In the previous section, it was seen how the undiluted transit depth, 
$\delta$, and the observed transit depth, $\delta'$, are 
related by the factor $B_{x}$. If the blend source was due to a nearby star
which was hidden within the PSF (point spread function) during the observations, 
then it usually possible to evaluate $B_{x}$ through higher resolution imaging 
e.g. using adaptic optics \citep{latham2010}. However, if the blend source is 
from the planet itself, then no feasible improvements in spatial resolution will
ever aid in determining $B_{P}$.

Fortunately, $B_{P}$ can be obtained in other ways, specifically through
measuring the phase curve of an extrasolar planet (for the first measured 
example, see \citet{knutson2007}). With such a measurement, the difference 
between the day and nightside fluxes may be determined and thus $(F_{P}/F_{*})$ 
can be calculated.

Another possible method is to measure the occultation and transit without 
the intermediate phase curve information, which would require instruments with 
extremely stable calibration.  For a very inactive star, a highly calibrated 
instrument could, in principle, measure the nightside flux by measuring just the
transits and occultations.  This would equate to an absolute calibration 
accurate to a fraction of the difference between the day and nightside fluxes, 
estimated to be $\sim 10^{-3}$ for HD 189733b, over one half of the orbital 
period.  I therefore estimate calibration requirements to be at least 
$\sim 10^{-4}$ during the full $\sim 30$ hour period.  Staring 
telescopes like \emph{Kepler} and \emph{CoRoT} have been shown to reach
these precisions \citep{kippingbakos2010b}, but it is the 
infrared telescopes of \emph{Spitzer} and JWST that are most heavily affected 
by nightside pollution and these observatories frequently slew around looking at 
different patches of the sky.  After the slewing one requires the target to be 
at the same centroid position to within a fraction of a pixel.  Regardless as to
whether the whole phase curve or simply the eclipse-only observations are made, 
the same method may be used to correct both for the effects of nightside 
pollution.

\begin{figure*}
\begin{center}
\includegraphics[height=22.5 cm]{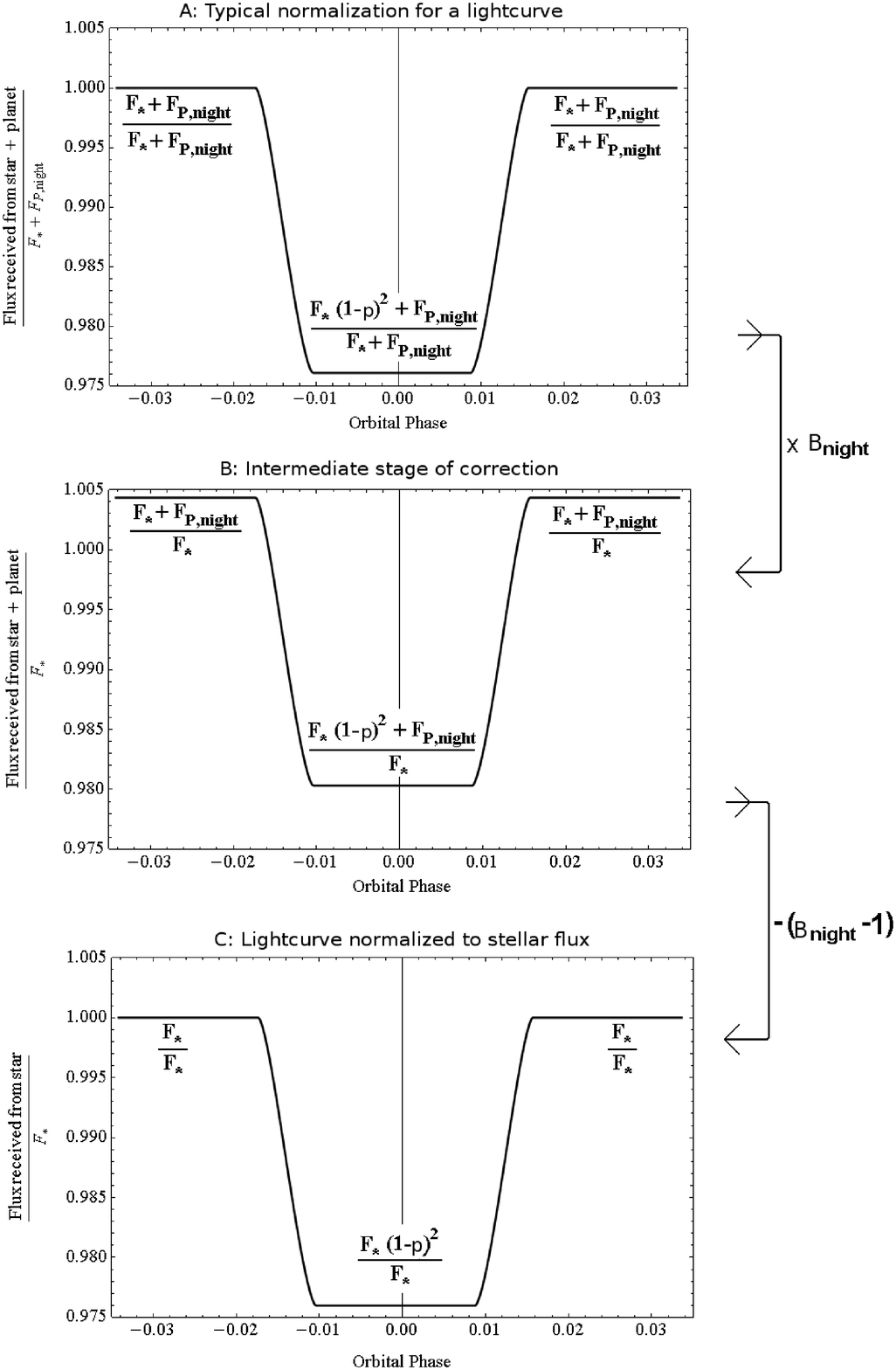}
\caption[Three-stage process to correct for blending]
{\emph{Illustration of the three-stages involved in the corrective
procedure, to compensate for the effects of nightside pollution.}} 
\label{fig:night2}
\end{center}
\end{figure*}

A phase curve time series is typically normalized to $F_*$ (see 
Figure~\ref{fig:phasecurve} for illustration of the phase curve), which is in 
contrast to a normal transit measurement, which is normalized to $(F_{*}+F_{P})$ 
i.e. the local out-of-transit baseline (see Figure~\ref{fig:night2}a). 
Therefore, for a phase curve, the stellar normalized flux immediately before and 
after the transit event is equal to $B_{P}$, as shown in 
Figure~\ref{fig:night2}b. It is possible that shifted hot 
spots on the planetary surface could cause an inequality between the pre and 
post transit baselines, but in practice the net effect of nightside pollution 
is very well accounted for by averaging over this time range.

In order to correct a transit light curve, one needs to modify the 
normalization. In Figure~\ref{fig:night2}, I show the two-step transformation 
which can achieve this.  I consider initially normalizing a light curve using 
the local baseline as usual for such measurements, as shown in 
Figure~\ref{fig:night2}a. After this, the two-step correction may be performed, 
provided the observer has knowledge of $B_{x}$.  The whole process may be 
summarized by the following (also illustrated in Figure~\ref{fig:night2}):

\begin{enumerate}
\item[1.] Normalize fluxes to local out-of-transit baseline, as usual.
\item[2.] Multiply all flux values by $B_{x}$.
\item[3.] Subtract $(B_{x}-1)$ from all data points.
\end{enumerate}

In practice, these steps are incorporated into the light curve fitting algorithm 
directly. In the case of using Monte Carlo based routines for error estimation,
$B_{x}$ may be floated around its best-fit value and corresponding
uncertainty. An example of this method is shown in \S\ref{sec:appliedexamples} 
for the planet HD 189733b.  Defining $I_{j,\mathrm{uncorr}}$ as the locally 
normalized flux measurement of the $j^{\mathrm{th}}$ data point, one may 
explicitly write down the corrected data point using 
Equation~(\ref{eqn:correctioneqn}).

\begin{equation}
I_{j,\mathrm{corr}} = B_{x} I_{j,\mathrm{uncorr}} - (B_{x} - 1)
\label{eqn:correctioneqn}
\end{equation}

I also briefly mention here that the transformation operations on the 
light curve time series will not only change the transit depth but also provides
a more physical transit signal and thus one should expect a slightly lower 
$\chi^2$ in the final fitting, as indeed is seen later in 
Tables~\ref{tab:night1} and \ref{tab:night2}.

One caveat with the described method is the possible presence of ellipsoidal 
variations of the star, which would mix the phase curve signature. For example, 
\citet{welsh2010} detected ellipsoidal variations in HAT-P-7. Such signals peak 
at orbital phases of 0.25 and 0.75, whereas a phase curve should peak close to 
orbital phase 0.5, but can be offset by a small factor due to hot spots. 
\citet{welsh2010} provide a detailed discussion of modelling both signals and 
although ellipsoidal variations complicate the analysis, they certainly do not 
undermine it.

\subsubsection{Semi-empirical Method}

To correct for the effect of nightside pollution, in an accurate way, I have
proposed using phase curve information to obtain $B_{P}$, which
requires many hours of telescope time.  Further, the phase curve should be 
obtained at every wavelength simultaneously and for every epoch\footnote{Each
epoch should be done in case temporal variability exists in the system} one
wishes to measure the transit event at, in order to be sure of a completely 
reliable correction.  I label this resource-intensive method of correcting for 
the effect as the ``empirical method''.  However, it is appreciated that 
obtaining phase curves at every wavelength for each epoch is somewhat 
unrealistic and I propose a ``semi-empirical method'' of achieving the same goal 
with far fewer resources.

I propose that observations of the phase curve are made at several wavelengths
in the infrared; for example the IRAC and MIPS wavelengths of \emph{Spitzer} 
are very suitable but most of these channels are unfortunately no longer 
available. The resulting measurements of the nightside flux may be used to
compile a spectrum for the purposes of interpolating/extrapolating the emission 
to other wavelengths which are missing phase curve observations (an example of 
this will be provided in \S\ref{sec:extrapolating}).  Should higher resolution 
spectra of the nightside emission become available in the future, one may 
construct a more sophisticated model as appropriate.  Transit observations at 
different wavelengths may then interpolate/extrapolate the model template to 
estimate the magnitude of the effect and apply the required correction at any 
wavelength. This allows one to estimate the blending factor at all wavelengths 
and times.

Long-term monitoring of the planet may also be necessary in order to ascertain 
the presence or absence of temporal variability in the system.  Ideally, this
may be achieved by obtaining phase curves of the exoplanet to measure the 
nightside flux at regular times.  More practically, it could be done by 
measuring only the occultation, which is the dayside of the exoplanet, at 
regular times (for example \citet{agol2009}).  Any large changes in the 
nightside flux are likely to be correlated to large changes in the dayside flux 
too, assuming a constant energy budget for the planet.  This second approach 
would reduce the demands on telescope time by an order of magnitude or more.

\subsubsection{Non-empirical Method}

The final method I propose here is the least accurate, but requires the fewest 
resources to implement.  If a planet has recently been discovered, no phase 
curves or even occultations may have been obtained yet.  Accordingly, 
the only avenue available is to estimate the temperature of the nightside 
through either a simple analytic estimation or a dynamical model of the 
atmosphere, although the latter may be excessive given the absence of any 
observational constraints.  I illustrate here how the temperature may be 
quickly estimated in such a case.

The nightside temperature may be estimated by assuming the dominant source of
heating is from the incident stellar flux.  In this case, the only unknown 
factors affecting the nightside temperature are the re-distribution of heat 
factor, $\mathfrak{F}$ and the Bond albedo of the planet, $\mathfrak{A}_B$. 
First-order estimations of these values can be made based upon empirical upper 
limits and measurements of other planets and atmospheric models.  The following 
expression may be used as a first-order estimate of the planet's brightness 
temperature:

\begin{equation}
T_{P,\mathrm{hemisphere}}(\lambda) = \Bigg(T_{*}^4 \frac{R_{*}^2}{4 a_{P}^2} [\mathfrak{F} \frac{(1-\mathfrak{A}_B)}{\epsilon}] + \frac{ L_{P}^{\mathrm{internal}} }{4 \pi R_{P}^2 \sigma_B \epsilon}\Bigg)^{1/4}
\label{eqn:Themi}
\end{equation}

Where $T_{*}$ is the effective temperature of the host star, $\mathfrak{F}$ is 
the distribution of energy to the hemisphere in question, $\epsilon$ is the
emissivity of the planet ($0\leq\epsilon\leq1$), $L_{P}^{\mathrm{internal}}$ is 
the luminosity of the planet from internal heat 
generation (e.g. tidal heating, radioactivity), $\sigma_B$ is the 
Stefan-Boltzmann constant.

For this calculation, $L_{P}^{\mathrm{internal}}$ is generally assumed to be 
negligible unless large tidal forces are expected as a result of a highly 
eccentric orbit, for example.  All of the other quantities are typically 
measured except for $\mathfrak{F}$ and $\mathfrak{A}_B$.  Choices for these 
values may come from atmospheric models or experience with other exoplanets.

\subsection{Comparison to Other Effects}
\label{sec:compeffects}

\subsubsection{Starspots}

Starspots have been observed within the transit events in several cases; e.g. 
\citet{pont2007}; \citet{dittmann2009}. They typically have been observed to 
have a radius of less than or equal to a few Earth radii and are estimated to 
have temperatures from 100\,K to 1000\,K cooler than the rest of the stellar 
surface. When a planet passes over a starspot, it results in an increase in 
relative flux within the transit signal which is easily identified. If one 
assumes that the only starspot is the starspot which has been crossed, then 
accounting for the effect is quite trivial and may be incorporated in the light 
curve modelling.

What is much more troublesome are out-of-transit starspots for which one has no 
direct evidence. The presence of out-of-transit starspots will cause the transit
depth to appear larger, in general. As a typical example, \citet{czesla2009} 
estimate that the effect can cause underestimations of the planetary radius by a 
fraction of $\sim 3$\% for CoRoT-2b, which is a 1.6\% change in the transit 
depth. This effect is larger than the nightside pollution effect by an order
of magnitude.

However, these effects can only be present for spotty stars whereas nightside 
pollution simply requires a hot planet. Also, the spot coverage of a stellar 
surface varies periodically giving rise to an ultimately regular pattern which 
may therefore be corrected for. In contrast, the nightside pollution effect is 
not periodic: it is a constant offset in a single direction. Further, prior 
information such as a phase curve or a dayside eclipse places strong constraints
on $B_{P}$ and concordantly the estimation of this parameter is not an
issue. Therefore, even for spotty stars, no-one would propose ignoring the 
effects of blending induced by a nearby companion star and so it can be seen 
that negating the self-blend of the planet would also be folly.

\subsubsection{Temporal Variability of $F_P$ and $F_*$}

If the stellar flux or the nightside emission of the planet experiences temporal
variations, then one would expect $\delta_{\mathrm{obs}}$ to also change over 
time. I will here estimate the magnitude of this effect.  The expected changes 
in nightside emission has not been studied in as much detail as that for the 
dayside, but one expects the magnitude of the variations to be very similar, 
based upon an energy budget argument. \citet{rauscher2007} have used 
shallow-layer circulation models to estimate variations in the dayside emission 
at the 1\%-10\% level. This is consistent with the observations of HD 189733b's 
dayside flux by \citet{agol2009}, who measure variations in the dayside below 
10\%. Typical stellar flux changes are at the $\lesssim$1\% level and so the 
ratio $(F_P/F_*)$ is more likely to vary due to the planet than the star.

For the case of HD 189733b, I later show that $B_{P} = 1.002571$ at 8.0\,$\mu$m. 
If $F_{P}$ increased by $\pm 10$\%, this would correspond to $B_{P} = 1.002828$ 
causing the transit depth to vary by 6\,ppm.  This would be around an order of 
magnitude below \emph{Spitzer}'s sensitivity but could be potentially close to a 
1-$\sigma$ effect for JWST. Nevertheless, the effect is sufficiently small that 
it is unlikely to be significant in most cases.

\subsubsection{Limb Darkening}

The nightside pollution effect is generally only relevant for hot-Jupiters at 
infrared wavelengths. As one moves towards the 10-30\,$\mu$m wavelength range, 
the effects of limb darkening become negligible. The curvature of the transit 
trough is essentially flat. However, the limb of the star will possess a more 
complicated profile (\citet{orosz2000}; \citet{jeffers2006}) and this could 
potentially introduce errors into the fitting procedure. It is generally prudent 
to include even the very weak limb darkening effects when modelling such 
transits.

As a result, Equations~(\ref{eqn:bpundiluted}) and (\ref{eqn:aRundiluted}) 
should not be used to attempt to correct light curves which were fitted without 
accounting for nightside pollution. They do, however, offer a useful approximate 
quantification as to the direction and magnitude of any errors. A comparison 
between the predictions of Equations~(\ref{eqn:bpundiluted}) and 
(\ref{eqn:aRundiluted}) and the exact limb-darkened nightside-polluted light 
curve fits is given later in \S\ref{sec:appliedexamples} for the example of 
HD 189733b. A discussion of this method required to produce this exact modelling 
is given in \S\ref{sec:compensating}.

\subsubsection{Significance at Visible Wavelengths}

I briefly consider the value of including the nightside pollution effect at 
visible wavelengths, in particular for the \emph{Kepler Mission}. 
\citet{borucki2009} recently reported visible-wavelength photometry for HAT-P-7b
which exhibits a combination of ellipsoidal variations and a phase curve 
\citep{welsh2010}, as well as an occultation of depth $(130 \pm 11)$\,ppm. 
HAT-P-7b is one of the very hottest transiting exoplanets discovered, so it 
offers a useful upper-limit example. For the purposes of nightside pollution, 
the maximum possible effect would occur if eclipse was both due to thermal 
emission alone and efficient day-night circulation. In this hypothetical 
scenario which maximizes the nightside pollution, one would have 
$B_{P} = 1.00013$. In the case of HAT-P-7b, the transit depth was reported to be 
$\delta_{\mathrm{obs}} = (6056 \pm 47)$\,ppm by \citet{welsh2010}, implying that
the undiluted transit depth is larger by $0.79$\,ppm, or $0.017$-$\sigma$. 
Therefore, as expected, visible wavelength transits will not, in general, be 
significantly affected by nightside pollution for even \emph{Kepler} class 
photometry.

\subsection{Applied Example - HD 189733b}
\label{sec:appliedexamples}

\subsubsection{\emph{Spitzer} IRAC 8.0\,$\mu$m measurement}

I will here provide an example of the empirical method of compensating for 
nightside pollution. I used the corrected data of HD 189733b's phase curve at 
8.0\,$\mu$m, as measured by \citet{knutson2007} (obtained through personal 
communication). I applied a median-stack smoothing function to the light curve 
with a one-minute window in order to identify the eclipse contact points.  I 
find the flux of the star by taking the mean of fluxes between the 
$2^{\mathrm{nd}}$ and $3^{\mathrm{rd}}$ contact points during the occultation, 
weighting each point by the reported error.  The standard deviation 
within this region is divided by the square root of the number of data points to 
provide the error on the mean.  All fluxes are then divided by the derived 
stellar flux and the error on each flux stamp is propagated through, 
incorporating the error on the stellar flux estimate.

In order to determine the nightside flux, which is not the same as the minimum
flux, I adopt a baseline defined as 30 minutes before $1^{\mathrm{st}}$ contact 
and 30 minutes after $4^{\mathrm{th}}$ contact and find a mean of 
$B_{P} = 1.002571 \pm 0.000048$.  The average RMS in this baseline
is 0.65\,mmag/minute.  If it were possible for the nightside of the planet to 
induce an occultation, as the dayside does, one would measure an occultation 
nightside depth of $(0.256 \pm 0.023)$\%, whereas \citet{charbonneau2008} report
a dayside occultation depth of $(0.391 \pm 0.022)$\%.

As discussed in \S\ref{sec:compensating}, ellipsoidal variations can also be 
responsible for out-of-transit flux variations and can be potentially confused 
with the phase curve. For HD 189733b, I use Equation~(1) of \citet{pfahl2008} to 
estimate an ellipsoidal variation amplitude of $2.2$\,ppm. Given that the phase 
curve exhibits a variation $1350$\,ppm amplitude, ellipsoidal variations can be 
neglected for the rest of this analysis.

I now produce two fits of the light curve: 1) no blending factor 2) blending 
factor $B_{\mathrm{night}}$ included.  Each light curve is fitted independently
assuming a fixed period of $P_P = 2.2185733$ days and zero orbital eccentricity
\footnote{This is done for simplicity, as our goal is merely to compare
corrected versus uncorrected photometry, not re-derive more accurate system
parameters}. The results of the fits are displayed in Table~\ref{tab:night1}.

For the fitting, I used a Markov Chain Monte Carlo (MCMC) algorithm which 
employs the limb darkening model of \citet{mandel2002} and I chose the 
light curve fitting parameter set: 
$\{\tau_T, p^2, \tilde{T}^{\mathrm{one}}, b_{P,T}, OOT\}$.  I used 125,000 
trials with the first 25,000 discarded for burn-in. Employing the code of I. 
Ribas, a Kurucz (2006) style atmosphere is used to interpolate the four 
non-linear limb darkening coefficients \citep{claret2000}, following the same 
methodology of \citet{beaulieu2009}, giving us $c_1 = 0.2790207$, 
$c_2 = -0.1506885$, $c_3 = 0.0779481$ and $c_4 = -0.0087653$.  I use the same 
local baseline as defined earlier, constituting 22382 data points and assume a 
circular orbit\footnote{A circular orbit is used for the purposes of fair 
comparison with Equations~(\ref{eqn:bpundiluted}) and (\ref{eqn:bpundiluted}), 
and justified by the fact the goal here is not to produce refined parameters, 
merely illustrate the significance of nightside pollution}. At this stage, no 
outliers have yet been rejected but I proceed to fit the unbinned light curve. 
I take the best-fit light curve and subtract it from the data to obtain the 
residues and then look for outlier points.  I then use the 
median-absolute-deviation (MAD) \citep{gauss1817} to provide a robust estimate 
of the standard deviation of the data, as this parameter is highly resistant to 
outliers, and find $\textrm{MAD} = 2.92062 \times 10^{-3}$.  Since there are 
$22382$ points, then the maximum expectant departure from a normal distribution 
is 4.08 standard deviations.  Any points above this level are rejected, where 
the evaluation of the standard deviation comes from the MAD value multiplied by 
1.4826, as appropriate for a normal distribution\footnote{Although strictly a 
Poisson distribution, for $22382$ data points, the distribution is very well 
approximated by a Gaussian}.  This procedure rejects any points with a residual 
deviation greater or equal to 0.0176749, corresponding to 10 
points\footnote{Note that the data has already been cleaned of outlier 
measurements, which is why the MAD rejection criteria only identifies 10 
outliers from $22382$ points}.

\begin{table*}
\caption[Comparison of nightside pollution corrected versus uncorrected transit 
parameters of HD 189733b at 8.0\,$\mu$m]
{\emph{Best-fit transit parameters for the HD 189733b 8.0\,$\mu$m 
primary transit light curve; data obtained by \citet{knutson2007}. Fits 
performed for the case of 1) typical normalization of the local baseline 
2) correction for the effects of nightside pollution. The number of data points 
is 22372.}} 
\centering 
\begin{tabular}{l c c c c c c} 
\hline\hline 
Method & Depth, $p^2$,\% & $\tilde{T}$/s & $b_{P,T}$ & $(a_P/R_*)$ & $i_P$/$^{\circ}$ & $\chi^2$ \\ [0.5ex] 
\hline 
(1) Local baseline & 2.3824 & 5127.45 & 0.66264 & 8.9121 & 85.7360 & 22412.4691 \\
(2) Nightside correction & 2.3884 & 5127.55 & 0.66204 & 8.9183 & 85.7428 & 22412.4605 \\
Uncertainty & 0.0061 & 8.1 & 0.0061 & 0.050 & 0.054 & - \\
\hline
(2) - (1) & +0.0060 & +0.10 & -0.00060 & +0.0062 & +0.0068 & -0.0086  \\ [1ex]
\hline\hline 
\end{tabular}
\label{tab:night1}
\end{table*}

The new light curve is then refitted in the normal way and I present the 
best-fit value in Table~\ref{tab:night1}.  I find the uncorrected light curve 
has a depth of $\delta_{\mathrm{obs}} = 2.3824 \pm 0.0061$\%.  For comparison, 
\citet{knutson2009} report the fitted 8.0\,$\mu$m depth to be 
$2.387 \pm 0.006$\%, which is consistent with this value\footnote{The slight
difference likely comes from a different limb darkening treatment}. Applying 
the correction due to nightside pollution, I find the undiluted transit depth to 
be $\delta = 2.3884 \pm 0.0061$\% meaning that the depth has 
increased by 60\,ppm corresponding to 1-$\sigma$.  From this example, it is 
clear that negating an effect which systematically biases transit depth 
measurements by $\sim1$-$\sigma$ would be imprudent.

Using Equation~(\ref{eqn:bpundiluted}), to first-order in $(B_x-1)$, I estimate 
that the impact parameter should be overestimated by $6.9\times10^{-4}$. The 
fits reveal a very similar value of $6.0\times10^{-4}$. Similarly, for 
$(a_P/R_*)$, Equation~(\ref{eqn:aRundiluted}) predicts an underestimation of 
$3.4\times10^{-3}$ whereas the light curve fit finds $6.2\times10^{-3}$. As 
expected, the effect of a blend is less pronounced on the other parameters.

Based on the difference in collecting area, it is expected JWST will achieve a
precision $\sim 6.6$ times greater than \emph{Spitzer}, suggesting this 
nightside pollution effect will become significant at the 
$\sim 5$-$10$\,$\sigma$ level for future infrared transit observations of 
hot-Jupiters.  Additionally, the binning of multiple \emph{Spitzer} transits 
would raise the significance of the effect. For example, \citet{agol2009} 
reported seven 8.0\,$\mu$m transits of HD 189733b which, if globally fitted 
would increase the significance of the nightside pollution effect to 
2.6-$\sigma$. Such a large effect cannot be justifiably neglected.

\subsubsection{\emph{Spitzer} MIPS 24\,$\mu$m}

\citet{knutson2009} measured the phase curve of HD 189733b with the MIPS 
instrument onboard \emph{Spitzer} about a year after the observations of the 
8.0\,$\mu$m phase curve for the same system.  Using the original 
normalized-to-stellar-flux unbinned data (personal correspondence with H. 
Knutson), I took the mean of data points $\simeq 1$\,hour either side of the 
transit event, which exhibit an RMS of 2.1 mmag/minute.  I combined the two 
baseline estimates to calculate a nightside relative flux of 
$B_{P} = 1.00438 \pm 0.00025$.

\citet{knutson2009} reported a 24\,$\mu$m transit depth of $(2.396 \pm 0.027)$\% 
and this re-analysis of the data yields 
$\delta_{\mathrm{obs}} = (2.398 \pm 0.019)$\%, where the fit has been performed 
using the same methodology as for 8.0\,$\mu$m, except that I assume no limb 
darkening at 24\,$\mu$m. As before, I apply the correction due to nightside 
pollution and estimate a new 24\,$\mu$m transit depth of 
$\delta = (2.409 \pm 0.020)$\%, which increases the depth by 
$\sim 0.5$-$\sigma$. Despite the absolute effect being larger than that at 
8.0\,$\mu$m, the difference is fewer standard deviations away due to the much
poorer signal-to-noise of the transit event itself.

\begin{table*}
\caption[Comparison of nightside pollution corrected versus uncorrected transit 
parameters of HD 189733b at 24.0\,$\mu$m]
{\emph{Best-fit transit parameters of the HD 189733b 24.0\,$\mu$m 
primary transit light curve; data obtained by \citet{knutson2009}.  Fits 
performed for the case of 1) typical normalization of the local baseline 2) 
correction for the effects of nightside pollution.  Number of data points is 
1198.}} 
\centering 
\begin{tabular}{l c c c c c c} 
\hline\hline 
Method & Depth, $p^2$,\% & $\tilde{T}$/s & $b_{P,T}$ & $(a_P/R_*)$ & $i_P$/$^{\circ}$ & $\chi^2$ \\ [0.5ex] 
\hline 
(1) Local baseline & 2.3980 & 5072.527 & 0.61425 & 9.492 & 86.290 & 1260.9125 \\
(2) Nightside correction & 2.4085 & 5072.505 & 0.6131 & 9.502 & 86.300 & 1260.9092 \\
Uncertainty & 0.019 & 19.7 & 0.003 & 0.033 & 0.022 & - \\
\hline
(2) - (1) & +0.011 & -0.022 & -0.0012 & +0.010 & +0.010 & -0.0033 \\ [1ex]
\hline\hline 
\end{tabular}
\label{tab:night2}
\end{table*}

\subsection{Pollution of the Transmission Spectrum}

In the standard theory of transmission spectroscopy, planetary nightside 
emission is assumed to be negligible and thus disregarded (e.g. as explicitly 
stated in the foundational theory of \citet{brown2001b}).  However, it has been 
shown here that the effect noticeably changes the transit depth for high quality 
photometry.  Essentially, it is posited here that the ``traditional'' 
transmission spectrum is in fact a combination of transmission through the 
terminator and the self-blending caused by emission from the nightside. 

One subtle point is that the nightside pollution effect is not something which 
can be accounted for in the modelling of the transmission spectrum. It is 
generally useful to think of the nightside pollution effect as an astrophysical 
blend which happens to be related to the planetary properties. A transmission 
spectrum is typically found by fitting a transit light curve at multiple 
wavelengths and then fitting a spectrum through the retrieved transit depths 
which models the planetary atmosphere. These routines usually make use of 
radiative transfer, chemical equilibrium, molecular line lists, etc to estimate
the opacity of the atmosphere at each wavelength. However, attempting to 
increase the sophistication of these models would not accurately account for the
self-blending scenario. Recall that each transit depth is obtained by fitting 
an eclipse model through the light curve time series. Critically, it is at this 
stage where blending needs to be accounted for. The transit signal plus blend 
should be modelled as such from the outset due to the subtle, and quite 
intricate, inter-dependencies between $b_{P,T}$, $(a_P/R_*)$, $p^2$ and the limb 
darkening. Thus it can be seen that attempting to incorporate the effects later 
on is far more challenging and completely unnecessary than simply fitting each 
transit light curve with a physically accurate model in the first place.

In this section, I will estimate how different an exoplanet's transmission 
spectrum would appear with and without nightside pollution. The spectra
generated in this section were computed by Giovanna Tinetti, who was the
co-author on the paper on which this thesis chapter is based \citep{kiptin2010}.
All other parts of this paper including the original concept, derivation and
light curve analyses were conducted by myself.

In order to evaluate the magnitude of the spectral pollution, we will here 
consider a planet of similar type to HD 189733b.  It is important to stress that 
the effects of nightside pollution will vary from case to case and the example 
we give here is indeed just for one example which is somewhat typical for an 
observed hot-Jupiter.  Therefore the results here are only for a hypothetical, 
but typical, example.

\begin{figure*}
\begin{center}
\includegraphics[width= 15.0 cm]{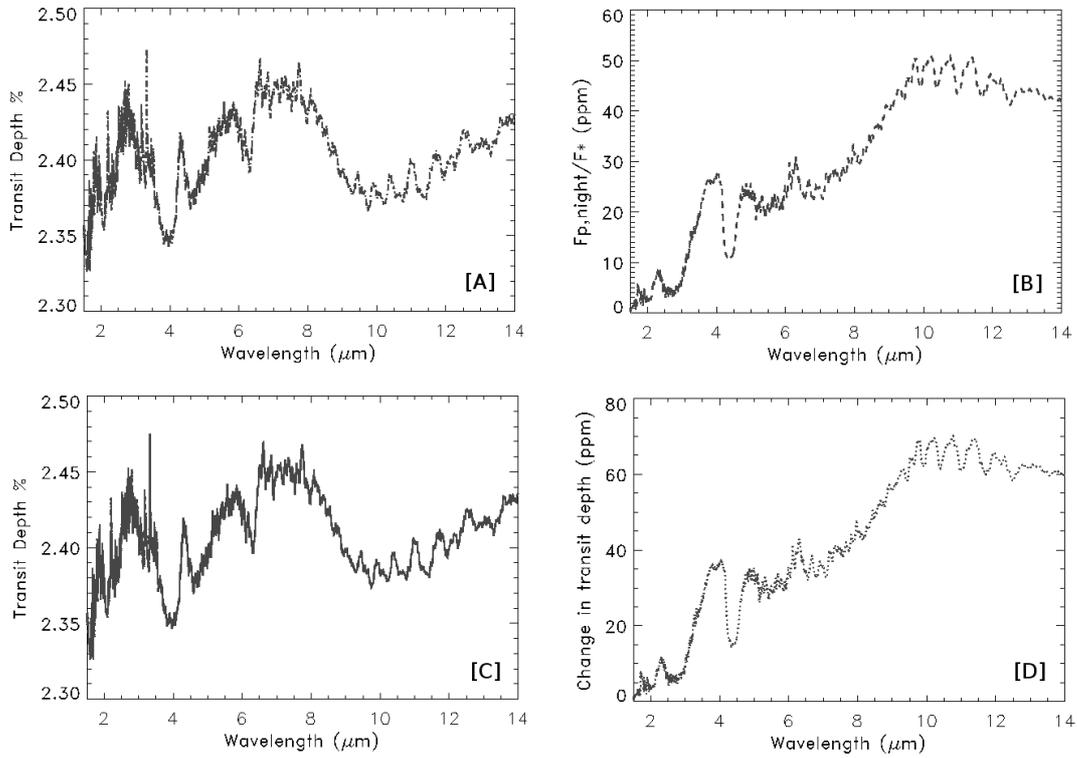}
\caption[Pollution of transmission spectra due to nightside emission]
{\emph{Top left:} \textbf{A}- Transmission spectrum of a hypothetical 
exoplanet similar to HD 189733b, generated considering the transmission through
the terminator only.  \emph{Top right:} \textbf{B}- Emission spectrum from the 
nightside of the hypothetical planet.  \emph{Bottom left:} \textbf{C}- 
Transmission spectrum of the planet incorporating the pollution of the nightside
emission.  \emph{Bottom right:} \textbf{D}- Residual between two transmission 
spectra.  We conclude that not accounting for nightside emission would result in
a 60-80\,ppm error in the transit depth.} \label{fig:night3}
\end{center}
\end{figure*}

The real question we need to answer is how much does a planetary transmission 
spectra change due to nightside self-blending?  We therefore need to generate 
two versions of the planetary transmission spectra, one including 
(Figure~\ref{fig:night3}a) and one excluding the effects of nightside pollution 
(Figure~\ref{fig:night3}c), and then take the difference between the two 
(Figure~\ref{fig:night3}d).

For a description of the models used to generate the spectra, details may be 
found in \citet{tinetti2007} for the transmission spectrum and 
\citet{swain2009} for the emission.  Planet and star properties are set to be 
that of the HD 189733 system.  The model contains water, carbon dioxide and 
methane to provide us with the effects of molecular species on nightside 
pollution. No carbon monoxide or hazes/clouds are included in our example. 
We note that the transmission and emission models are good fits to the current 
available spectroscopy/photometry data of HD 189733b in the NIR/MIR 
\citep{swain2009}. The effects of water absorption are quantified with the BT2 
water line list \citep{barber2006}, which characterises water absorption at the 
range of temperatures probed in HD 189733b. Methane was simulated by using a 
combination of HITRAN 2008 \citep{rothman1995} and PNNL data-lists.  Carbon 
dioxide absorption coefficients were estimated with a combination of HITEMP and 
CDSD-1000 (Carbon Dioxide Spectroscopic Databank version for high temperature 
applications; \citet{tashkun2003}). The continuum was computed using $H_2-H_2$ 
absorption data \citep{borysow2001}.

Generated spectra are always plotted in terms of the quantities determined with 
the lowest measurement uncertainty, namely $(R_P/R_*)^2$ and $(F_P/F_*)$, for 
the transits and occultations respectively.  Transmission spectra which are 
plotted in units of $R_P$ (e.g. as done by \citet{fortney2010}) will cause the 
measurement uncertainties to be much larger since the error on $R_*$ must 
necessarily be propagated in such a recipe. In fact, the measurement 
uncertainties on a spectra plotted in units of $R_P$ will be dominated by the 
error on $R_*$ since this property is typically determined to much lower 
precision.  Consequently, statistically significant molecular features would be 
overwhelmed by the artificially large error bars.

Using the model described above, we first compute the transit depth from 
transmission absorption effects only (i.e. excluding nightside pollution) as 
visible in Figure~\ref{fig:night3}a.  We then generate the dayside emission spectra 
for the same planet and make the assumption that the dayside and nightside 
emission spectra are identical (Figure~\ref{fig:night3}b).  This assumption is 
unlikely to be true for the exact case of HD 189733b and really constitutes an 
upper limit, but we again stress that we are here only considering a planet 
similar to that of HD 189733b and thus we are free to make this assumption for 
our hypothetical example.

\begin{figure}
\begin{center}
\includegraphics[width= 15.0 cm]{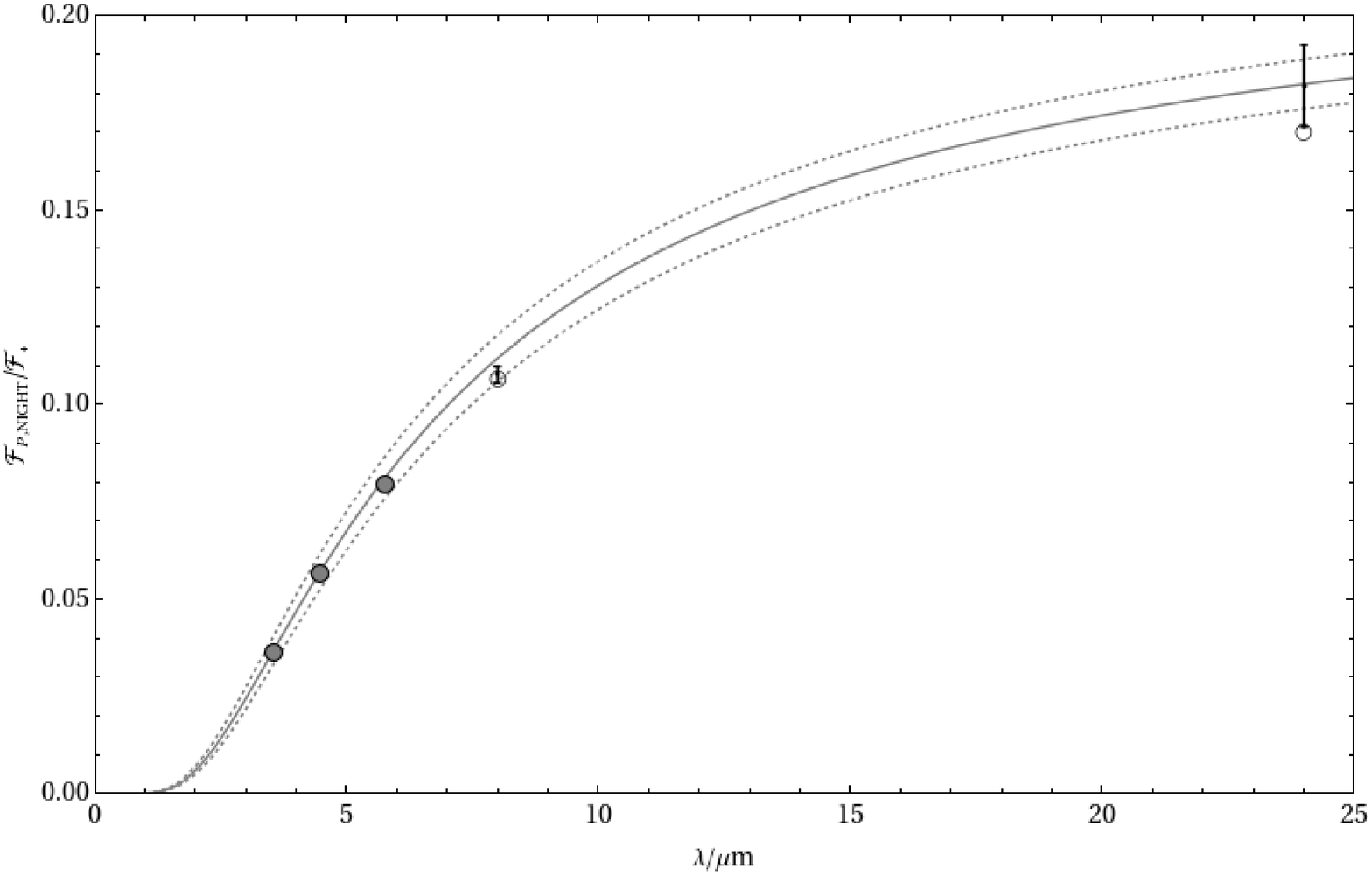}
\caption[Observed nightside emission spectrum of HD 189733b]
{\emph{Flux per unit area (i.e. not total flux) of the planetary 
\textbf{nightside} emission divided by that of the star, plotted as a function 
of wavelength.  Using the two measurements at 8.0\,$\mu$m and 24.0\,$\mu$m 
(black dots with error bars), we fit a blackbody curve through the points 
(gray lines) with a nightside planetary temperature of $T = 1148 \pm 32$K. Open 
circles represent the integrated blackbody function across the instrument 
bandpasses. Filled circles represent the same but extrapolated to the other IRAC 
wavelengths, which allow us to conclude the nightside effect will be much less 
at lower wavelengths.}} \label{fig:night4}
\end{center}
\end{figure}

We combine the nightside emission and transmission spectra to produce a 
transmission spectra which includes the effects of the nightside, as seen in 
Figure~\ref{fig:night3}c. We then take the difference between the corrected 
spectrum and the one which excludes nightside emission. The resultant residual 
spectrum is plotted in Figure~\ref{fig:night3}d.  The residual spectrum reveals 
nightside emission affects the transmission spectra at the level of 
$6 \times 10^{-5}$ for $\lambda \geq 10$\,$\mu$m and very closely matches the 
behaviour of the emission features, as expected.  As we saw earlier, the 
magnitude of the effect is equal to the typical measurement uncertainty for a 
target like HD 189733b with \emph{Spitzer}. This supports our hypothesis that 
the nightside pollution effect is a $\sim$1-$\sigma$ effect for 8.0\,$\mu$m 
\emph{Spitzer} photometry.

\subsection{Extrapolating the Nightside Correction for HD 189733b}
\label{sec:extrapolating}

Only two measurements exist for the nightside flux for HD 189733b (or indeed any
other exoplanet) at 8.0\,$\mu$m and 24\,$\mu$m, but several other primary 
transit light curves exist in the mid-infrared wavelengths. \citet{beaulieu2008} 
presented 3.6\,$\mu$m and 5.8\,$\mu$m measurements and \citet{desert2009} 
obtained photometry at 4.5\,$\mu$m and 8.0\,$\mu$m.  I will here estimate the 
nightside effect on the 3.6\,$\mu$m, 4.5\,$\mu$m and 5.8\,$\mu$m channels. 
Currently, only two data points exist and this does not warrant modelling the
nightside emission spectrum in any more complexity than that of a blackbody, as 
a first-order approximation. Should more measurements become available in the
future, molecular species could be included to improve the accuracy of our 
model.

I first assume that the star is blackbody emitter with $T_* = (5040 \pm 50)$\,K 
\citep{torres2008}.  At this point is advantageous to consider only the emission 
per unit area from both the planet and the star, in order to avoid the effects 
of transmission through the planetary atmosphere.  I therefore define the flux 
per unit area of each object using:

\begin{align}
\mathcal{F}_{P} &= F_{P}/(\pi R_P^2) \\
\mathcal{F}_* &= F_*/(\pi R_*^2) \\
\mathcal{R}_{P} &= \frac{\mathcal{F}_P}{\mathcal{F}_*} = \frac{F_P}{F_*} \delta
\end{align}

I then convert the $B_{P}$ measurements for 8.0\,$\mu$m and 
24\,$\mu$m into $\mathcal{R}_{P}$ by using the two values of the 
corrected ratio-of-radii shown in Tables~\ref{tab:night1} and 
\ref{tab:night2} and propagating the uncertainties. Using the Planck function
as a blackbody model for each body and the relevant \emph{Spitzer} bandpass 
response functions, one may estimate $\mathcal{R}_{\mathrm{night}}$ for 
any given value of $T_{P,\mathrm{night}}$. I allow this temperature to vary 
from 700\,K to 1700\,K in 1\,K steps and numerically integrate the bandpasses 
to find $\chi^2$ at each temperature, which I define as:

\begin{align}
\chi^2 &= \Bigg(\frac{\mathcal{F}_{\mathrm{obs}} - \mathcal{F}_{\mathrm{calc}}}{\Delta(\mathcal{F}_{\mathrm{obs}})}\Bigg)^2\Bigg|_{8.0\mu\mathrm{m}} + \Bigg(\frac{\mathcal{F}_{\mathrm{obs}} - \mathcal{F}_{\mathrm{calc}}}{\Delta(\mathcal{F}_{\mathrm{obs}})}\Bigg)^2\Bigg|_{24\mu\mathrm{m}} + \Bigg(\frac{T_{*,obs} - T_{*,calc}}{\Delta(T_*)}\Bigg)^2
\end{align}

Where I additionally define $T_{*,\mathrm{calc}}$ as the temperature used in 
the integration and $T_{*,\mathrm{obs}}$ as being equal to the value determined 
by \citet{torres2008}. Note that I do not fit for $T_*$ but do allow the value 
to float in order to correctly estimate the uncertainty of 
$T_{P,\mathrm{night}}$. The final analysis reveals a best-fit planetary 
nightside temperature of $T_{P,\mathrm{night}} = (1148 \pm 32)$\,K with 
$\chi^2 = 1.12$, suggesting the blackbody model gives a satisfactory fit for 
these two measurements.  Note that not accounting for the response function 
of the instruments would yield a erroneous result of 
$T_{P,\mathrm{night}} = 1120$\,K.

Using our derived planetary temperature, one may now use the blackbody function 
to extrapolate $\mathcal{R}_{\mathrm{night}}$ to other wavelengths and thus the 
nightside corrected transit depths for 3.6\,$\mu$m, 4.5\,$\mu$ and 5.8\,$\mu$m.
The undiluted transit depth will be given by Equation~(\ref{eqn:truedepth}) and 
the results of this analysis are summarized in Table~\ref{tab:night3} and 
Figure~\ref{fig:night4}.

\begin{equation}
\delta = \frac{\delta_{\mathrm{obs}}}{1-\mathcal{R}_{P}\delta_{\mathrm{obs}}}
\label{eqn:truedepth}
\end{equation}

\begin{table*}
\caption[Extrapolation of the nightside pollution effect to other \emph{Spitzer}
channels for HD 189733b]
{\emph{Using a fitted blackbody function for the nightside emission of 
HD 189733b, we calculate the nightside corrections to \emph{Spitzer} channels 
for which no phase curve information currently exists. Values with a $\dagger$ 
superscript cannot have their uncertainties estimated since they are 
extrapolated parameters. 8.0\,$\mu$m data comes from \citet{knutson2007}, 
24.0\,$\mu$m from \citet{knutson2009}, 3.6\,$\mu$m \& 5.8\,$\mu$m from 
\citet{beaulieu2008} and 4.5\,$\mu$m from \citet{desert2009}.}} 
\centering 
\begin{tabular}{l c c c c} 
\hline\hline 
Channel & Observed depth $\delta_{\mathrm{obs}}$,\% & $\mathcal{R}_{P}$ & Corrected depth $\delta$,\% & $\frac{\delta - \delta_{\mathrm{obs}}}{\sigma_{\delta}}$ \\ [0.5ex] 
\hline
\emph{Measured} \\
\hline
8.0\,$\mu$m & $2.3824 \pm 0.0060$ & $0.1076 \pm 0.0020$ & $2.3884 \pm 0.0061$ & 1.0 \\
24.0\,$\mu$m & $2.398 \pm 0.019$ & $0.1818 \pm 0.0105$ & $2.4085 \pm 0.020$ & 0.5 \\ 
\hline 
\emph{Extrapolated} \\
\hline
3.6\,$\mu$m & $2.356 \pm 0.019$ & $0.0365^{\dagger}$ & $2.358 \pm 0.019^{\dagger}$ & $\leq 0.1$ \\
4.5\,$\mu$m & $2.424 \pm 0.010$ & $0.0570^{\dagger}$ & $2.427 \pm 0.010^{\dagger}$ & $\leq 0.4$ \\
5.8\,$\mu$m & $2.436 \pm 0.020$ & $0.0800^{\dagger}$ & $2.441 \pm 0.020^{\dagger}$ & $\leq 0.25$ \\[1ex]
\hline\hline 
\end{tabular}
\label{tab:night3}
\end{table*}

The maximal deviation occurs for 8.0\,$\mu$m and is less than 1-$\sigma$ for all 
other wavelengths.  Consequently, the deduction of which molecules are evident 
from the spectrum of HD 189733b will not significantly affected by the nightside
effect, but derived abundances will change. 
\subsection{Conclusions}

Both transit photometry and transmission spectroscopy are expected to be
affected by self-blending of the planet's nightside emission at infrared
wavelengths. This self-blending, or ``nightside pollution'', causes a dilution
of transit depths and invalidates assumptions A4 and A5. However, I have
shown here how either high resolution imaging, for extrinsic blend sources,
or phase curve observations, for self-blends, can be used to determine the
extra flux contribution and thus compensate for the effect. In a worked example,
it is shown that HD 189733b's 8.0\,$\mu$m \emph{Spitzer} transit light curve 
experiences a 1-$\sigma$ bias if this effect is not accounted for. With A4/A5
now addressed, the outstanding assumption which is frequently broken in A10,
which I move onto next in \S\ref{sec:binning}.

\clearpage

\section{Binning is Sinning}
\label{sec:binning}

\subsection{Introduction}

The final assumption to be considered is A10) - ``The change in flux over a 
single integration is much smaller than the flux measurement uncertainty''. This
essentially assumes a very short integration time but the quantification of just
how short this really is has not been provided. This may be evaluated by 
breaking the assumption and then computing when the consequences surpass the 
noise. Assumption A10 was never stated in \citet{mandel2002} or 
\citet{seager2003}, but it can be seen to be implicit in both pioneering papers. 
Much like how nightside pollution was never conceived until precise 
infrared measurements became available with \emph{Spitzer}, the issue of A10 
was never considered until a space-based telescope combining long-integrations
with extremely high precision photometry was launched i.e. \emph{Kepler}. 

The \emph{Kepler Mission} surveys more than 150,000 stars for transiting
planets and thus produces large amounts of data. Due to telemetry limits,
almost all of these measurements are performed in long-cadence (LC) mode to
preserve bandwidth. The LC mode utilizes 30\,minute integrations and so one
has the combination of integration times of comparable size to the transit
features plus high precision photometry. This combination maximizes the
conditions under which A10 is broken. In this section, I will consider the
consequences of this invalidation, and how one can compensate for the effect.

\subsection{The Effects of Finite Integration Time}
\subsubsection{Ingress/Egress Durations}

As discussed in \S\ref{sec:transitanatomy} there are four critical contact 
points which define the transit light curve morphology, which represent the 
points where the time derivative is discontinuous. Physically speaking, contact 
points I and IV occur when the sky-projected planet-star separation is equal to 
the stellar radius plus the planetary radius.  Contact points II and III occur 
when this parameter equals the stellar radius minus the planetary radius. 
Defining $W$ as the mean of the durations between the 
1$^{\mathrm{st}}$-to-4$^{\mathrm{th}}$ and 
2$^{\mathrm{nd}}$-to-3$^{\mathrm{rd}}$ contacts, one
may write:

\begin{align}
t_{I} &= \tau_T - W/2 - T_{12}/2 \\
t_{II} &= \tau_T - W/2 + T_{12}/2 \\
t_{III} &= \tau_T + W/2 - T_{12}/2 \\
t_{IV} &= \tau_T + W/2 + T_{12}/2
\end{align}

\begin{figure*}
\begin{center}
\includegraphics[width=15.0 cm]{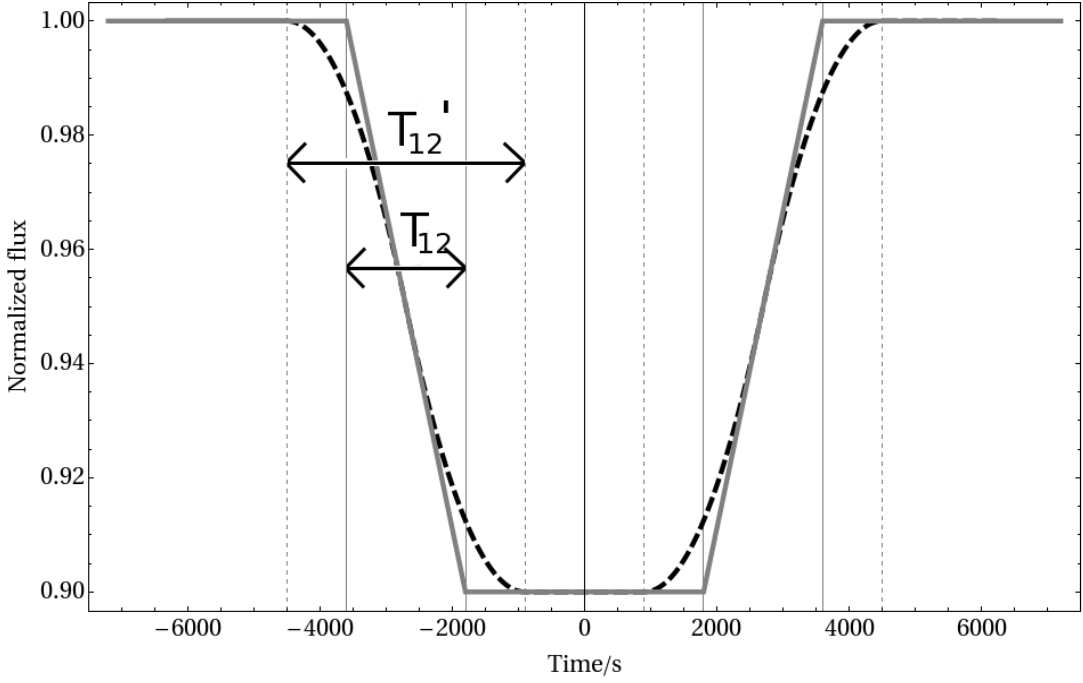}
\caption[Illustration of transit light curve smearing]
{\emph{A trapezoid approximated light curve with a one hour flat-bottom 
duration and 30 minute ingress/egress duration, $T_{12}$, is shown in solid. 
The dashed line shows the light curve morphology for an integration time of 
30\,minutes.  The apparent ingress/egress duration, $T_{12}'$, can be seen to 
have doubled purely as a consequence of the integration time.}}
\label{fig:binning1}
\end{center}
\end{figure*}

The principal effect of finite integration time is to smear out the light curve 
into a broader shape (see Figure~\ref{fig:binning1}).  The apparent ingress and 
egress durations will increase and additional curvature will be introduced into 
the light curve wings.  The ingress/egress stretching can be considered in terms 
of the apparent positions of the contact points being temporally shifted from 
their true value. The magnitude of this time shift is dependent on the relative 
phase difference between the sampling and the transit signal. If one assumes 
that a large number of transits observed with LC photometry are folded about the 
orbital period, as is typical in transit detection, then the effect becomes much 
more predictable with the deviation averaging out to $\mathcal{I}/2$.

Under these conditions, contact points I and IV will appear to move outwards 
from $\tau_T$ by one half of the integration time each, $\mathcal{I}/2$.  
Conversely, contact points II and III will appear to move inwards by the same 
amount.  Let us define the apparent contact points as $t_{x}'$:

\begin{align}
t_{I}' = t_{I} - \mathcal{I}/2 \\
t_{II}' = t_{II} + \mathcal{I}/2 \\
t_{III}' = t_{III} - \mathcal{I}/2 \\
t_{IV}' = t_{IV} + \mathcal{I}/2
\end{align}

These shifted apparent contact points will cause the apparent transit durations
$T_{14}'$ and $T_{23}'$ to differ from the true values. Specifically, an
observer would find $T_{14}' = (T_{14} + \mathcal{I})$ and 
$T_{23}' = (T_{23} - \mathcal{I})$. If the observer was unaware that this
smearing had occurred, they would naively use Equation~(\ref{eqn:oneretrieval}) 
(or alternatively the \citet{seager2003} equations if one assumes $e_P=0$) to
retrieve erroneous values of $b_{P,T}$, $(a_P/R_*)$, $i_P$ and $\rho_*$. I
denote these erroneous retrieved parameters with the superscript ``instant'',
in reference to the assumption of instantaneous integration times. They may
be computed using the same method I used in \S\ref{sec:nightside}, and
utilizing Equation~\ref{eqn:oneretrieval} I find:

\begin{align}
[b_{P,T}^{\mathrm{instant}}]^2 &= \frac{\Bigg[(1+p)^2 \frac{\sin^2\Big[\frac{\sqrt{1-e_P^2} \mathcal{I} \pi}{P_P \varrho_{P,T}^2} - \arcsin\Big(\frac{\sqrt{(1-p)^2 - b_{P,T}^2}}{(a_P/R_*) \varrho_{P,T} \sin i_P}\Big)\Big]}{\sin^2\Big[\frac{\sqrt{1-e_P^2} \mathcal{I} \pi}{P_P \varrho_{P,T}^2} + \arcsin\Big(\frac{\sqrt{(1+p)^2 - b_{P,T}^2}}{(a_P/R_*) \varrho_{P,T} \sin i_P}\Big)\Big]}- (1-p)^2\Bigg]}{\Bigg[\frac{\sin^2\Big[\frac{\sqrt{1-e_P^2} \mathcal{I} \pi}{P_P \varrho_{P,T}^2} - \arcsin\Big(\frac{\sqrt{(1-p)^2 - b_{P,T}^2}}{(a_P/R_*) \varrho_{P,T} \sin i_P}\Big)\Big]}{\sin^2\Big[\frac{\sqrt{1-e_P^2} \mathcal{I} \pi}{P_P \varrho_{P,T}^2} + \arcsin\Big(\frac{\sqrt{(1+p)^2 - b_{P,T}^2}}{(a_P/R_*) \varrho_{P,T} \sin i_P}\Big)\Big]}- 1\Bigg]} \\
[(a_P/R_*)^{\mathrm{instant}}]^2 &= \frac{\frac{\Bigg[(1+p)^2 \sin^2\Bigg(\frac{\sqrt{1-e_P^2} \pi \mathcal{I}}{\varrho_{P,T}^2 P_P} - \arcsin\Big(\frac{\sqrt{(1-p)^2 - b_{P,T}^2}}{(a_P/R_*) \varrho_{P,T} \sin i_P}\Big)\Bigg)-4p\Bigg]}{\sin^2\Bigg(\frac{\sqrt{1-e^2}\pi \mathcal{I}}{\varrho_c^2 P} + \arcsin\Big(\frac{\sqrt{(1+p)^2-b_{P,T}^2}}{(a_P/R_*) \varrho_c \sin i_P}\Big)\Bigg)}-(1-p)^2}{\varrho_{P,T}^2 \frac{\sin^2\Bigg(\frac{\sqrt{1-e_P^2}\pi \mathcal{I}}{\varrho_{P,T}^2 P_P} - \arcsin\Big(\frac{\sqrt{(1-p)^2 - b_{P,T}^2}}{(a_P/R_*) \varrho_{P,T} \sin i_P}\Big)\Bigg)}{\sin^2\Bigg(\frac{\sqrt{1-e_P^2}\pi \mathcal{I}}{\varrho_{P,T}^2 P_P} + \arcsin\Big(\frac{\sqrt{(1+p)^2 - b_{P,T}^2}}{(a_P/R_*) \varrho_{P,T} \sin i_P}\Big)\Bigg)}-\varrho_{P,T}^2} \\
\rho_{*}^{\mathrm{instant}} &\simeq \frac{3 \pi}{G P_P^2} [(a_P/R_*)^{\mathrm{instant}}]^3
\end{align}

Setting $\mathcal{I} = 0$ returns the original results as expected.  
Unfortunately, these equations are somewhat overly complex for one to draw any 
physical intuition.  To proceed, I will consider a typical case example by using 
the system parameters from one of the \emph{Kepler} planets, since these are 
discovered using LC data.  The following example is for the 
assumption of zero limb darkening, which is a very poor one for the 
\emph{Kepler} bandpass.  The effects of limb darkening will be discussed later.

\begin{figure*}
\begin{center}
\includegraphics[width=15.0 cm]{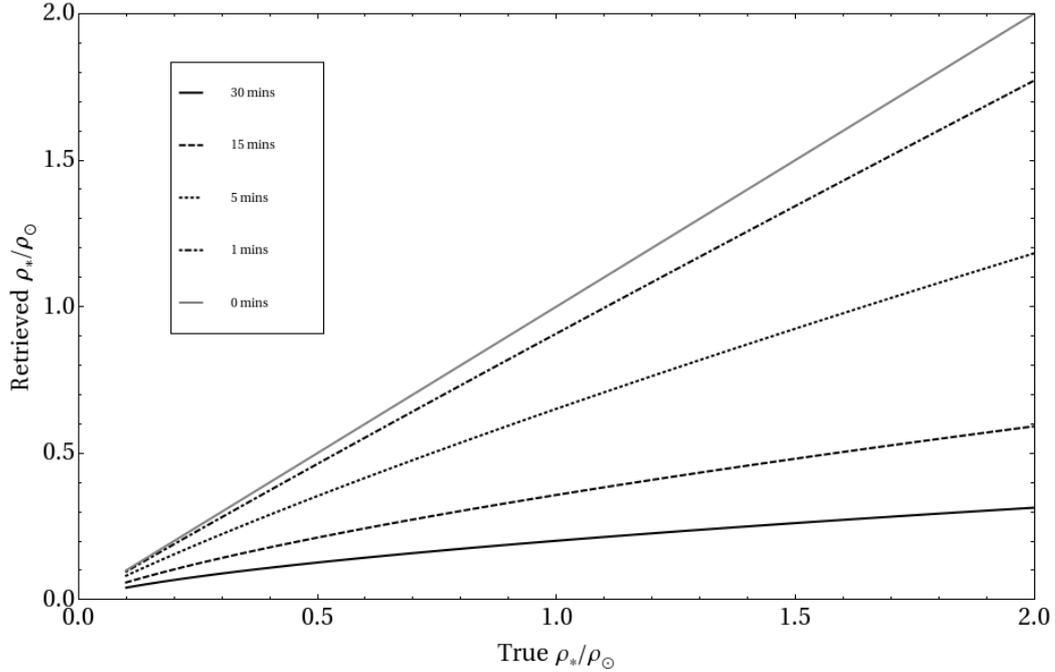}
\caption[Effect of transit smearing on the light curve derived stellar density]
{\emph{As an example, I use the system parameters of Kepler-5b to show 
the effect on the retrieved stellar density as a result of long integration 
times, in the case of no limb darkening.  The 1 min line appears to produce 
results within the typical uncertainties of the derived stellar density.}} 
\label{fig:binning2}
\end{center}
\end{figure*}

In Figure~\ref{fig:binning2}, I plot the retrieved stellar density, 
$\rho_*^{\mathrm{instant}}$, as a function of the true stellar density, 
$\rho_*$. All other parameters are fixed to be that of Kepler-5b, as reported by 
\citet{koch2010}. The effect can be seen to be highly 
significant, causing the retrieved stellar density to be underestimated by a 
factor which borders on an order-of-magnitude.  This scale of underestimation is 
sufficient to completely reject some planetary candidates as unphysical. 
However, note that in reality the underestimation of $\rho_*$ will not be 
this severe due to countering effects of limb darkening suppression discussed 
later.

Looking at Figure~\ref{fig:binning2} again, let us explore the physical reasons 
for this underestimation.  I begin this line of thought by considering the 
impact parameter, $b_{P,T}$.  If the ingress/egress duration is elongated, what 
does one expect to happen to the derived impact parameter?  Consider two extreme
cases.  When a planet transits with a very low impact parameter, one has 
essentially an equatorial transit.  This means the vector describing the 
sky-projected planetary velocity is nearly perpendicular to the stellar limb. 
As a result, the planet crosses the limb expediently.  In contrast, for a 
near-grazing transit, the mutual angle between the stellar limb and the 
sky-projected planetary velocity vector has become more acute, which has the 
effect of making the limb-crossing time much longer.  Therefore, stretching the 
ingress/egress duration causes $b_{P,T}^{\mathrm{instant}} > b_{P,T}$.

Having established this point, consider the effect on $(a_P/R_*)$.  The simplest
way to understand the effect on this parameter is to appreciate that $b_{P,T}$ 
and $(a_P/R_*)$ exhibit an extremely strong negative correlation, as 
demonstrated by \citet{carter2008} and seen earlier in 
Table~\ref{tab:durations}.  So the act of increasing the ingress/egress duration 
will increase $b_{P,T}$ and therefore decrease $(a_P/R_*)$.  Finally, one knows 
that $\rho_* \propto (a_P/R_*)^3$ meaning that decreasing $(a_P/R_*)$ by a 
factor of 2 would decrease $\rho_*$ by a factor of 8.

\subsubsection{Limb Darkening Effects}

For transit observations at visible wavelengths, limb darkening is quite 
pronounced producing a well-known curvature in the flat-bottom part of the 
light curve.  For large integration times, the curvature is smeared out, 
producing a flatter transit trough morphology.

However, for stars exhibiting limb darkening, the boundary between the end of
ingress and start of the trough is less obvious and can form an essentially
continuous curve. As a result, an algorithm which is fitting this light curve
may ``borrow'' some of the ingress portion to increase the net curvature in
what the algorithm considers to be the trough. The reason an algorithm would
do this is because usually the limb darkening parameters are known and often
fixed and so the algorithm is trying to find a solution which produces the
expected curvature a-priori.

By ``borrowing'' some of the ingress, the overall effect of limb darkening
is to actually decrease the systematic overestimation of the ingress discussed
in the last subsubsection. Therefore, LC data pushes 
$b_{P,T}^{\mathrm{instant}}$ towards a more equatorial value due to limb 
darkening.

\subsubsection{Two Countering Effects}

In conclusion, finite integration times have two countering effects on 
$b_{P,T}^{\mathrm{instant}}$ and thus $(a_P/R_*)^{\mathrm{instant}}$ and 
$\rho_*^{\mathrm{instant}}$ as well.  The ingress and egress smearing causes the 
ingress to appear larger, which occurs for more grazing transits.  In contrast, 
the limb darkening borrows some of the ingress and thuds attenuates the previous
effect, causing an apparently more equatorial transit.  
Whilst limb darkening is important, especially for transit surveys like 
\emph{CoRoT} and \emph{Kepler} which operate at visible wavelengths, the 
fundamental change in the transit morphology is sufficiently large that it will 
tend to dominate.  This is because the amplitude of the curvature in the 
light curve trough is usually at least an order-of-magnitude less than the 
amplitude of the transit signal itself.  Therefore, the consequences of 
smearing the overall transit signal will tend to dominate over the effects 
of limb darkening.

This general rule of thumb that $b_{P,T}^{\mathrm{instant}}$ is overestimated 
will break-down when one has $T_{12} \gg \mathcal{I}$ but 
$T_{23} \sim \mathcal{O}[\mathcal{I}]$, which occurs for near-grazing transits.
In such a case, the fractional change to the ingress durations is minimized but 
the change in the transit trough curvature is maximized.

To exactly calculate the net consequence of these two effects though, the 
integration time should be included when one generates the model light curves, 
rather than attempting ad-hoc corrections post-analysis.  This appears to be the 
only way to completely account for the effect in a reliable manner.

\subsubsection{Consequences for the Transit Depth}

I will briefly comment on the effect of finite integration time on the transit 
depth.  Assuming no limb darkening is present and $T_{23} > \mathcal{I}$, the 
transit depth is completely unaffected by the long integration time.  This has 
important consequences for occultations where the light curve is unaffected 
by stellar limb darkening.

For cases where limb darkening is present, the net effect on the retrieved 
$p^{\mathrm{instant}}$ will depend on whether 
$b_{P,T}^{\mathrm{instant}} >b_{P,T}$ or $b_{P,T} < b_{P,T}^{\mathrm{instant}}$.
Additionally, the effect will be a function of what assumptions were used in the
fitting algorithm (e.g. the fixing of various parameters), the true impact 
parameter and the limb darkening coefficients.  Given the large number of 
correlated factors, predicting the effect of integration time becomes less 
reliable and one must fit the transit light curve with a model which accounts
for integration time in the first place.

\subsubsection{Observed Effects with \emph{Kepler}}

I point out that the effects of long integration times have already been 
observed by the \emph{Kepler Mission}. Figure~4 of \citet{gilliland2010}, shows 
the transit light curve of long-cadence and short-cadence data for the same 
planet, TreS-2b.  The long-cadence light curve exhibits a broader shape with the
apparent position of the contact points shifted by $\sim \mathcal{I}/2$, as 
predicted in this work. Notice also that the curvature in the transit trough, 
due to limb darkening, has also been attenuated.

\subsection{Accurate Transit Light Curve Modelling}
\subsubsection{Analytic Integration}

The critical problem I have outlined can be simply summarized by the following:
\emph{Don't fit an unbinned model to binned data}.  The model usually used to 
generate a transit light curve is provided by \citet{mandel2002}, which 
includes the effects of stellar limb darkening.  To generate the 
\citet{mandel2002} light curve, one usually has a set of time stamps forming a 
time vector $\mathbf{t}$.  This time vector represents instantaneous moments 
rather than integrated time.  The $\mathbf{t}$ vector is converted into a 
vector of instantaneous true anomalies, $\mathbf{f_P}$, by solving Kepler's 
Equation numerically (see \S\ref{sec:keplerseqn}).  $\mathbf{f_P}$ is 
converted to a $\mathbf{S_{P*}}$ array, where $\mathbf{S_{P*}}$ is an array of
the instantaneous sky-projected planet-star separations.  Finally, the 
\citet{mandel2002} equations provide us with $\mathbf{F}$, where $F$ is the 
instantaneous flux.  The sequence of events may be summarized as:

\begin{equation}
\mathbf{t} \rightarrow \mathbf{f_P} \rightarrow \mathbf{S_{P*}} \rightarrow \mathbf{F}(\mathbf{t}) \nonumber
\end{equation}

Now that the mechanism of generating of a transit light 
curve for instantaneous time stamps, $\mathbf{F}(\mathbf{t})$, has been
established, I will consider what the transit light curve for integrated time 
stamps would be, $\mathbf{\check{F}}(\mathbf{\check{t}})$.  In this case, 
the integrated flux of the $i^{\mathrm{th}}$ data point would be given by:

\begin{equation}
\check{F_i}(\check{t_i}) = \frac{\int_{t=\check{t_i}-\mathcal{I}/2}^{\check{t_i}+\mathcal{I}/2} F(t) \mathrm{d}t}{\int_{t=\check{t}_i-\mathcal{I}/2}^{\check{t}_i+\mathcal{I}/2} \mathrm{d}t} \\
\end{equation}

This equation suffers from the problem that $F$ cannot be written 
as a function of $t$ analytically, since such a solution would require a 
closed-form solution to Kepler's Equation, which is transcendental.  Evaluating 
this expression for $F$ as a function of $S_{P*}$ is also not possible 
since one would find the following:

\begin{equation}
\int_{t=\check{t_i}-\mathcal{I}/2}^{\check{t_i}+\mathcal{I}/2} F(t) \mathrm{d}t = \int_{S_{P*}(\check{t_i}-\mathcal{I}/2)}^{S_{P*}(\check{t_i}+\mathcal{I}/2)} F(t(S_{P*})) \Big[ \frac{\mathrm{d}t}{\mathrm{d}S_{P*}}(S_{P*}) \Big] \mathrm{d}S_{P*}
\end{equation}

Whilst d$S_{P*}$/d$t$ may be evaluated analytically through a chain rule 
expansion of $($d$S_{P*}/$d$f_P) \times ($d$f_P/$d$t)$, the resultant expression 
will be as a function of $f_P$, rather than $S_{P*}$.  If one knew 
$f_P(S_{P*})$, then one would be able to write out the integrand in a 
closed-form, but $f_P(S_{P*})$ can only be found by solving a bi-quartic 
equation, as was shown earlier in \S\ref{sec:eccdurexact}. Unfortunately, as was 
discussed, there is no currently proposed method to find which roots
correspond to which orbital conjunction which makes a closed-form expression 
elusive.

The only remaining hope for a simple analytic expression would be to express 
the integral in terms of true or eccentric anomaly, which are inter-changeable. 
This would yield the following integral:

\begin{equation}
\int_{t=\check{t_i}-\mathcal{I}/2}^{\check{t_i}+\mathcal{I}/2} F(t) \mathrm{d}t = \int_{f_P(\check{t_i}-\mathcal{I}/2)}^{f_P(\check{t_i}+\mathcal{I}/2)} F(t(f_P)) \frac{\mathrm{d}t}{\mathrm{d}f_P}(f_P) \mathrm{d}f_P
\end{equation}

The integrand of this expression may be written out in a closed-form, using 
Equation~(\ref{eqn:angmom}):

\begin{equation}
\int_{t=\check{t_i}-\mathcal{I}/2}^{\check{t_i}+\mathcal{I}/2} F(t) \mathrm{d}t = \int_{f_P(\check{t_i}-\mathcal{I}/2)}^{f_P(\check{t_i}+\mathcal{I}/2)} F(S_{P*}(f_P)) \frac{P_P}{2\pi} \frac{(1-e_P^2)^{3/2}}{(1+e_P\cos f_P)^2} \mathrm{d}f_P
\end{equation}

Where $F(S_{P*})$ is given by \citet{mandel2002} and $S_{P*}(f_P)$ is given in
Equation~(\ref{eqn:Sdasheqn}).  The integral limits do not possess a closed-form
solution since once again one must solve Kepler's Equation, but in principle the
indefinite integral could be analytically evaluated and then the relevant limits
applied after a subroutine provides numerical solutions to Kepler's Equation. 
I believe that this strategy would be the most computationally efficient since 
this would obviate the need for any numerical integration.  However, I was 
unable to find a solution for the indefinite integral for even a uniform-source 
case and will therefore focus the remainder of this discussion onto numerical 
techniques.

\subsubsection{Numerical Integration}

Having established the significant challenges regarding analytic integration, 
I now consider numerical integration techniques. The functions one needs to 
integrate over are in fact very well-behaved and well-approximated by 
compositions of polynomials and thus I anticipate that even a low-resolution 
numerical integration technique should provide satisfactory accuracy.

In this subsection, I first consider the merit of Simpson's Rule or other 
Newton-Cotes based methods. I aim to avoid using nested quadrature methods 
like the Gauss-Kronrod or Clenshaw-Curtis, as the number of integrations 
required is large and it is preferable to avoid nested methods.  For the 
simplest case of Simpson's rule, the integrated flux is given by:

\begin{equation}
\check{F}_i(N=3) = \frac{F(t_i-\mathcal{I}/2) + 4 F(t_i) + F(t_i+\mathcal{I}/2)}{6}
\end{equation}

Where $N$ denotes the number of calls needed to the \citet{mandel2002} code and 
essentially is a measure of the resolution of the numerical integration. This 
method may be extended to higher intervals by using Simpson's composite rule. 
Alternatively, one can extend to cubic, quartic, etc interpolations by using the
Newton-Cotes formulas.  Each time one evaluates $F(t)$, one requires another 
call to the \citet{mandel2002} subroutine, and thus one wishes to minimize the 
number of calls, but maximize the accuracy of the employed technique.

Simpson's composite rule works by splitting up our integration range into $2m$ 
subintervals, therefore requiring $N = 2m + 1$ calls to the \citet{mandel2002} 
code. The error on the composite Simpson's rule is given by:

\begin{equation}
\underset{\mathrm{Comp. Simpson}}{\sigma_{\check{F}}(N)} = F^{(4)}(\epsilon) \frac{\mathcal{I}}{180} \Big(\frac{\mathcal{I}}{N-1}\Big)^4
\end{equation}

Where $\epsilon$ lies in the range 
$t_i-\mathcal{I}/2<\epsilon<t_i+\mathcal{I}/2$.  In contrast, the Newton-Cotes 
formulas move through increasing orders by increasing the interpolation order. 
For the $N=4$ case (which is the cubic interpolation scenario, known as 
Simpson's 3/8 rule), the equivalent errors between the two methods are:

\begin{align}
\underset{\mathrm{Newton-Cotes}}{\sigma_{\check{F}}(N=4)} &= F^{(4)}(\epsilon) \frac{3}{80} \mathcal{I}^5 \\
\underset{\mathrm{Comp. Simpson}}{\sigma_{\check{F}}(N=4)} &= F^{(4)}(\epsilon) \frac{1}{2880} \mathcal{I}^5
\end{align}

Thus for $N=4$, Simpson's composite rule offers greater accuracy than the 
Newton-Cotes based equation.  Moving through the higher orders in the 
Newton-Cotes family causes the error to have a functional dependence on 
$F^{(N)}$, i.e. the $N^{\mathrm{th}}$ differential of $F$.  So for $N>4$ it is 
not possible to give an exact comparison between the two methods since 
$F^{(N)}(t)$ is not known for any $N>1$. Therefore, our only reliable comparison 
is for the $N=4$ case, from which I conclude the composite Simpson's rule is 
superior in terms of accuracy versus computational requirement.

\subsubsection{Error in Numerical Integration}

I now consider what value of $N$ should be used.  There are essentially 
three segments of the light curve which exhibit curvature and thus would produce 
the maximum error in the numerical integrations, which employ linear piece-wise 
approximations.  

\begin{enumerate}
\item[{\tiny$\blacksquare$}] Curvature of the ingress/egress
\item[{\tiny$\blacksquare$}] Curvature of the limb-darkened light curve trough
\item[{\tiny$\blacksquare$}] Discontinuities at the contact points
\end{enumerate}

The last of these is due to a discontinuous function and the former two are due
to curvatures within continuous functions. I will treat these two different 
sources of ``curvature'' separately, although from the arguments made earlier, 
one expects the last of these effects to be the largest source of numerical 
error.
\\
\emph{\textbf{Ingress/egress curvature}}

The transit light curve has a depth $\delta$ and an ingress duration $T_{12}$. 
For most of the ingress, the curvature is close to zero and essentially mimics a 
linear slope.  However, near the contact points, the slope rapidly changes to a
flat line of zero gradient.  Therefore, near the contact points, the 
ingress/egress morphology causes large amounts of curvature.  These points will
exhibit the largest numerical errors in using a technique like Simpson's 
composite rule.

A suitable choice of resolution can be made by increasing $N$ until 
$\sigma_{\check{F}}|_{\mathrm{max}} \leq \sigma_{\check{F},\mathrm{obs}}$, i.e. 
the calculation should produce a flux which has a maximum systematic error which
is less than the observational uncertainty.  One should set the resolution to a 
point where it provides satisfactory accuracy even at the point of highest 
numerical error, i.e. within the ingress/egress near the contact points.

Another approach would be to use an adaptive composite Simpson's rule, for 
example like that proposed by McKeeman (1962).  However, it is preferable
to avoid using adaptive routines since they would require a new adaptation for 
every single data point and fitting trial, which would be time consuming.  The 
costs versus benefits of using such a method could warrant further investigation
in the future.  Instead, I choose to use the adaptation required for the most 
troublesome points, which have already been identified.  The required interval 
size in each element of the Simpson's composition should be decreased until one 
reaches:

\begin{equation}
\Big|\mathcal{S}\Big(a,\frac{a+b}{2}\Big) + \mathcal{S}\Big(\frac{a+b}{2},b\Big) - \mathcal{S}\Big(a,b\Big)\Big|/15 < \sigma_{\check{F},\mathrm{obs}}
\end{equation}

Where $\mathcal{S}(\alpha,\beta)$ is Simpson's rule evaluated over the interval 
$\alpha$ to $\beta$.  In this case, the integral is over time and $a=t_I$ and 
$b=t_I + (\mathcal{I}_0/m)$, where $2m$ is the number of subintervals which 
the integral has been split into and $2m = N-1$, where $N$ is the required 
factor by which the number of calls to the \citet{mandel2002} code increases.
The reason for the subscript of $0$ by the $\mathcal{I}$ term will be explained 
shortly. This requirement may be written as:

\begin{align}
\frac{|\mathcal{S}(t_I,t_I+\frac{\mathcal{I}_0}{2m}) + \mathcal{S}(t_I+\frac{\mathcal{I}_0}{2m},t_I+\frac{\mathcal{I}_0}{m}) - \mathcal{S}(t_I,t_I+\frac{\mathcal{I}_0}{m})|}{15} < \sigma_{\check{F},\mathrm{obs}}
\label{eqn:simperror}
\end{align}

\begin{align}
\mathcal{S}(\alpha,\beta) =\Big(\frac{\beta-\alpha}{6}\Big) \Big[F(\alpha) + 4F\Big(\frac{\alpha+\beta}{2}\Big) + F(\beta)\Big]
\end{align}

In order to continue, one needs to evaluate $F(t)$ in a closed-form, which 
cannot be achieved due to Kepler's Equation.  However, there exists a special 
case where Kepler's Equation does yield an exact closed-form solution and this 
occurs for circular orbits since $\mathfrak{M}_P = E_P = f_P$.  In such a case, 
one may use Equation~(\ref{eqn:circS}), which I remind the reader of here:

\begin{equation}
S_{P*}(t) = (a_P/R_*) \sqrt{\sin^2\Big(\frac{2 \pi t}{P_P}\Big) + \cos^2i_P\cos^2\Big(\frac{2 \pi t}{P_P}\Big)}
\end{equation}

The ingress/egress morphology is dominated by the expressions pertaining to a 
uniform source.  Limb darkening does affect the ingress/egress curvature but 
this is much less than the amplitude of the uniform source transit signal.  In 
the small-planet limit, \citet{mandel2002} provided the following approximation 
for the ingress/egress flux:

\begin{equation}
F(\mathcal{X}) = 1+\mathcal{X}\sqrt{p^2 - \mathcal{X}^2}-p^2 \arccos\Big[\frac{\mathcal{X}}{p}\Big]
\end{equation}

Where I have defined $S_{P*}=1+\mathcal{X}$ and it is understood that 
$-p < \mathcal{X} < p$ for the ingress/egress.  For the purposes of
evaluating the maximum error, one may bear in mind $\mathcal{X} \simeq p$ and 
thus one may expand the cosine term into second order using a Taylor series. 
I assume the simple case of $b_{P,T}=0$ so that $i_P = \pi/2$. I make further 
small-angle approximations to simplify the resultant expression for the error, 
which is justified since $2\pi t_I \ll P_P$.  The other adjustment one needs to 
account for is that I have approximated $b_{P,T}=0$ and $e_P=0$.  To 
generalize the result, I consider that the effect of $b_{P,T}>0$ and $e_P>0$ is 
to stretch or shrink the ingress/egress duration by a factor $\tau/\tau_0$. 
Therefore the expressions here are actually for $\mathcal{I}_0$, which may be 
written as $\mathcal{I}_0 = \mathcal{I} (\tau_0/\tau)$.  One may now rewrite 
Equation~(\ref{eqn:simperror}) as:

\begin{align}
\sigma_{\check{F},\mathrm{obs}} &> \Big|\frac{\psi^{5/2}}{108m^3} \Big[3 \Big(\sqrt{24 m p-9 \psi}- 4 \sqrt{2 m p -\psi} - 6\sqrt{4m p-\psi} + \sqrt{8 m p-\psi}\Big)\Big]\Big|
\label{eqn:Ferror}
\end{align}

Where I have used:

\begin{align}
\psi &= \frac{2 \pi (a_P/R_*)}{P_P} \frac{\kappa_0}{\kappa} \mathcal{I} \\
\frac{\kappa_0}{\kappa} &\simeq \frac{\sqrt{1-b_{P,T}^2} \sqrt{1-e_P^2}}{\varrho_{P,T}}
\end{align}

Due to the approximations made, I find that this equation is only stable for 
$m\geq 2$.  For any given data set, one simply needs to solve 
Equation~(\ref{eqn:Ferror}) for $m$ with some sensible estimates of $p$, 
$b_{P,T}$, $e_P$, $\omega_P$, $(a_P/R_*)$ and $P_P$. As an example, for 
Kepler-5b, taking the quoted parameters from the \citet{koch2010} paper, I find 
that even using $m=2$ provides an error of $0.1$\,ppm, which is well below the 
typical measurement uncertainty of $130$\,ppm.
\\
\emph{\textbf{Limb-darkened trough-curvature}}

Another part of the light curve where there is significant curvature, and thus 
one expects the maximum numerical integration error, is the limb-darkened 
light curve trough.  However, the peak-to-peak size of the changes in flux 
induced by the limb darkening are much lower than the transit signal itself 
(i.e. $\delta$); typically an order-of-magnitude. Further, the time scale over 
which these changes act is greater than that of the ingress/egress curvatures 
(i.e. $T_{14} \gg T_{12}$) except for grazing transits.  So one can see that, 
in general, the errors in these numerical integration techniques will be 
dominated by the ingress/egress curvatures rather than the 
limb-darkening-induced light curve trough curvatures.
\\
\emph{\textbf{Contact point discontinuities}}

The final source of variation in the light curve gradient is that of the 
discontinuous change located at the contact points. Estimating the error due to 
this discontinuity may be achieved by assuming a trapezoid approximated
light curve and considering the location where maximal error is induced. The 
largest error (and in fact only error) will occur for measurements close to 
contact points, or more specifically $|t_i - t_x| < \mathcal{I}/2$ where $t_x$ 
is the time of one of the contact points.

Before the first contact point, the flux is a flat line at $F=1$ and after this 
point there is a linear slope with a gradient $-(\delta/T_{12})$. The error in 
Simpson's composite rule will depend upon the relative phasing between the 
centre of the integration and the contact point, i.e. $(t_i - t_I)$. Generalized
to any phase, the true integrated flux of the trapezoid approximated light curve 
for the $i^{\mathrm{th}}$ time stamp is given by:

\begin{equation}
\check{F}_{\mathrm{true},i} = 1 - \frac{\delta}{T_{12}} \frac{(2 t_i + \mathcal{I})^2}{8 \mathcal{I}}
\end{equation}

For each value of $m=1,2,3...$ I choose to set the phase to be such that the 
difference between the true integrated flux and that from Simpson's method is 
maximized. Under such a condition, it may be shown that the maximum error is 
given by:

\begin{align}
\underset{\mathrm{Comp. Simp.}}{\sigma_{\check{F}}} &= \frac{\mathcal{I} \delta}{24 m^2 T_{12}} \\
\qquad&= \frac{\mathcal{I} \delta}{6 (N-1)^2 T_{12}}
\end{align}

For the system parameters of Kepler-5b, I find that using $m=1,2,3$ induces a 
maximal error of $371$\,ppm, $93$\,ppm and $41$\,ppm respectively. Given that 
the measurement uncertainties are $130$\,ppm \citep{koch2010}, a suitable choice
for the resolution would be $m=2$ since this means the maximum possible error of
a data point in the least-favourable phasing would be below the measurement 
error.

It is interesting to see that for $m=2$ the error was $0.1$\,ppm for the 
ingress/egress curvature of the same system, suggesting the discontinuity error 
dominates the error budget.  Actually, this is expected from the arguments made 
earlier in this section. Therefore, in most applications, a selection for $m$ 
based on the error induced by the contact point discontinuities will provide a 
robust integration resolution.

\subsubsection{Resampling}

An additional method for numerically integrating the light curve is discussed 
here. Consider a set of observations with integrated time stamps 
given by the vector $\mathbf{\check{t}}$.  A second way of calculating 
$\mathbf{\check{F}}(\mathbf{\check{t}})$ is to resample the time vector 
into a very fine cadence, at which point one may assume 
$\mathbf{\check{F}} \simeq \mathbf{F}$. I define the temporary resampled time 
vector as $\mathbf{\check{t}'}$.  As an example, for the \emph{Kepler} data, one 
may choose to resample the 30\,minute integrations into 1\,minute integrations 
by expanding each time stamp, $\check{t}_i$ into a sub-vector of 30 equally 
spaced time stamps with a mean value given by $\check{t}_i$.  The new temporary 
time array is used to generate a light curve using the normal \citet{mandel2002} 
expressions giving $\mathbf{F'}(\mathbf{\check{t}'})$ (note that $F$ here has no 
check sign because the \citet{mandel2002} equations can only generate 
instantaneous flux, not integrated flux).  The next step is to rebin the model 
light curve back to the original cadence to give 
$\mathbf{F}(\mathbf{\check{t}})$.  Finally, I make the assumption 
$\mathbf{\check{F}}(\mathbf{\check{t}'}) \simeq \mathbf{F}(\mathbf{t})$,
i.e. the high cadence resampled time vector yields a light curve model 
consistent with a time vector of infinite cadence.

\begin{equation}
\mathbf{\check{t}} \underset{\mathrm{resample}}{\rightarrow} \mathbf{\check{t}'} \underset{\mathrm{MA02}}{\rightarrow} \mathbf{F'}(\mathbf{\check{t}'}) \underset{\mathrm{rebin}}{\rightarrow} \mathbf{F}(\mathbf{\check{t}}) \simeq \mathbf{\check{F}}(\mathbf{\check{t}})
\end{equation}

It can be seen that resampling into $N$ sub-time stamps will increase the 
computation time by a factor of $\sim N$, since typically the \citet{mandel2002}
subroutine uses the majority of a light curve fitting algorithm's resources 
(especially for non-linear limb darkening). In the next subsubsection, I will 
show that the computation times can be decreased by ``selective resampling''.

One advantage of resampling is that one can choose to resample in such a way as 
to account for read-out and dead-times, which may be important if the 
instrument's duty cycle is quite poor\footnote{Note that this is not the case 
for \emph{Kepler} which has a duty cycle of 91.4\%.}. The resampling of the 
$i^{\mathrm{th}}$ time stamp into $N$ sub-time stamps with labels 
$j=1,2,...N-1,N$ can be expressed as:

\begin{equation}
t_{i,j}' = t_i + \Big(j - \frac{N+1}{2}\Big) \frac{\mathcal{I}}{N}
\end{equation}

The flux of the $i^{\mathrm{th}}$ time stamp is found by rebinning all $N$ flux 
stamps from $j=1$ to $j=N$.

\begin{equation}
\check{F}_i = \frac{\sum_{j=1}^N F_{i,j}'}{N}
\end{equation}

Thus for the first few values of $N=2$, $N=3$ and $N=4$ one would have:

\begin{align}
\check{F}_i(N=2) &= \frac{1}{2} \Big[ F(t_i - \mathcal{I}/4) + F(t_i + \mathcal{I}/4) \Big] \\
\check{F}_i(N=3) &= \frac{1}{3} \Big[ F(t_i - \mathcal{I}/3) + F(t_i) + F(t_i + \mathcal{I}/3) \Big] \\
\check{F}_i(N=4) &= \frac{1}{4} \Big[ F(t_i - 3\mathcal{I}/8) + F(t_i - \mathcal{I}/8) + F(t_i + \mathcal{I}/8)+F(t_i + 3\mathcal{I}/8) \Big]
\end{align}

For a trapezoid approximated light curve, it can be easily shown that the error 
in these expressions, as a function of $N$, is given by:

\begin{equation}
\underset{\mathrm{Resampling}}{\sigma_{\check{F}}} = \frac{\delta}{T_{12}} \frac{\mathcal{I}}{8 N^2}
\end{equation}

Therefore, the resampling method yields greater accuracy than the composite 
Simpson's method. In \citet{kippingbakos2010b}, both the resampling and 
Simpson's composite rule were employed in completely independent analyses and 
the obtained results were consistent. Therefore, whilst one is free to use 
either method discussed here, the most efficient approach out of the two is 
resampling.

\citet{gilliland2010} reported that they used a method for fitting the 
long-cadence light curve of TreS-2b which I interpret to be equivalent to the 
resampling method. The authors split the LC intervals into 30 contributing 
sub-intervals corresponding to the SC cadence i.e. $N=30$. For the reported LC 
RMS noise of $66$\,ppm and the system parameters of TreS-2b taken from 
\citet{kippingbakos2010b}, I estimate that using $N=5$ would produce a maximum 
possible error in the most unfavourably phased data point of $59$\,ppm and thus 
using $N=30$ is excessive for this light curve. These equations therefore permit 
for a reduction in computational time of 600\%. Such a saving is highly 
advantageous in MCMC fitting, which is inherently expensive on the CPU.

\subsubsection{Selective Resampling}

Resampling time stamps which satisfy $|t_i-\tau_T|>(T_{14} + \mathcal{I})/2$ 
and $|t_i - \tau_O| > (O_{14} + \mathcal{I})/2$ is unnecessary since 
$\check{F}_i = F_i = 1$ in such cases (assuming one has folded multiple transits
about the orbital period).  I label this method of optimization as 
``selective resampling''.

Since $(T_{14} + O_{14})/P_P \sim 2/(\pi (a_P/R_*))$, this can reduce the 
number of time stamps which require resampling by an order of magnitude for 
continuous staring telescopes like \emph{Kepler} and \emph{CoRoT}. It should be 
noted that selective resampling will not be possible if the light curve model 
includes phase variations of the planet e.g. HAT-P-7b, \citet{borucki2009}.

\subsection{Conclusions}

I have explored how long-cadence data, with particular focus on \emph{Kepler}, 
invalidates assumption A10 and consequently causes severe systematic errors 
in the retrieved physical parameters, unless accounted for.  The effect is valid 
for any finite exposure time but increases with longer cadences. Long-cadence 
data smears out the light curve morphology, which acts to stretch out the 
ingress/egress duration and suppress limb darkening in the light curve trough. 
These two effects act to increase and decrease the retrieved impact parameter 
respectively.  Critically, overestimating the impact parameter is shown to lead 
to severe underestimations of the stellar density which could lead to 
planetary candidates being rejected on the basis of being unphysical, in a
similar manner as to what was found for eccentric orbits in 
\S\ref{sec:usingcircforecc}.
 
Numerical integration techniques permit improved modelling of the transit 
light curve. I have discussed two particular methods, the composite Simpson's 
method and resampling. Expressions for estimating the errors of these 
techniques have been provided and I find that both methods produce an error 
which scales as $N^{-2}$, where $N$ is the numerical resolution of the 
techniques. Out of these two discussed methods, the resampling approach yields a 
greater efficiency.

With corrective procedures for the invalidation of assumptions A0, A4/A5 and 
A10 now established, it is now possible to produce highly accurate transit
light curve models. With the discussion and nuances of modelling the transit
light curve of a single exoplanet now complete, it is possible to progress
to the original aim of this thesis - \emph{methods to detect extrasolar moons in
transiting systems}, which will be explored in Chapter~\ref{ch:Chapt6}.

%% file: Chapt6.tex
\chapter{Transit Timing Effects due to an Exomoon}
\label{ch:Chapt6}

\vspace{1mm}
\leftskip=4cm

{\it ``
The moving moon went up to the sky,\\
And nowhere did abide;\\
Softly she was going up,\\ 
And a star or two beside. ''} 

\vspace{1mm}

\hfill {\bf --- Samuel Taylor Coleridge, \emph{The Ancient Mariner (pt. IV)}} 

\leftskip=0cm


\section{Introduction}
\label{sec:intro6}

The methods and intricacies of modelling the light curve of a single transiting
planet have now been covered. This necessary understanding enables the
investigation of more complex situations involving three-bodies, rather than
two. This therefore generalizes the model for cases where assumption A1 is
invalid (see \S\ref{sec:transitbasics}).

As discussed in Chapter~\ref{ch:Chapt2}, the moons of extrasolar planets 
present a unique and exciting challenge in modern astronomy. The motivations
for such a search have been clearly established (\S\ref{sec:motivations}) and
with the transit now well understood (see Chapters~\ref{ch:Chapt3}, 
\ref{ch:Chapt4} \& \ref{ch:Chapt5}), I begin here the discussion of what
forms the ultimate goal of this thesis (\S\ref{sec:thesisoutline}) - a method 
to detect an exomoon in a transiting system.

\section{Background Theory of Timing Deviations}
\label{sec:exomoonhistory}

\subsection{The \citet{sartoretti1999} Method}
\label{sec:sar99method}

The first serious scientific investigation into a method for detecting exomoons 
in transiting systems comes from \citet{sartoretti1999}. This groundbreaking 
paper, published in the supplementary series of A\&A, did not receive wide 
attention until very recently, where it has gone from an average of $\sim$4 
citations per year from 2000-2009 up to 16 citations in 2010.

Perhaps inspired by the astrometric method of detecting exoplanets, which in
1999 was considered to have more promise than it has since delivered,
\citet{sartoretti1999} argued that even if an exomoon is too small to detect
in transit, the gravitational perturbation which it induces on the host planet
may not be. The very simple premise was that the planet and moon orbited a
common centre-of-mass which itself orbited the star causing the planet to
exhibit composite motion. This composite motion was discernible from a simple
Keplerian orbit by virtue of the fact the position of the planet was offset
from the barycentre and thus the times of transit minimum would also be 
displaced. Although not labelled as such in the paper, the phenomenon of changes 
in the times of transit minimum would soon be dubbed ``transit time variation'' 
(TTV).

\begin{figure}
\begin{center}
\includegraphics[width=10.0 cm]{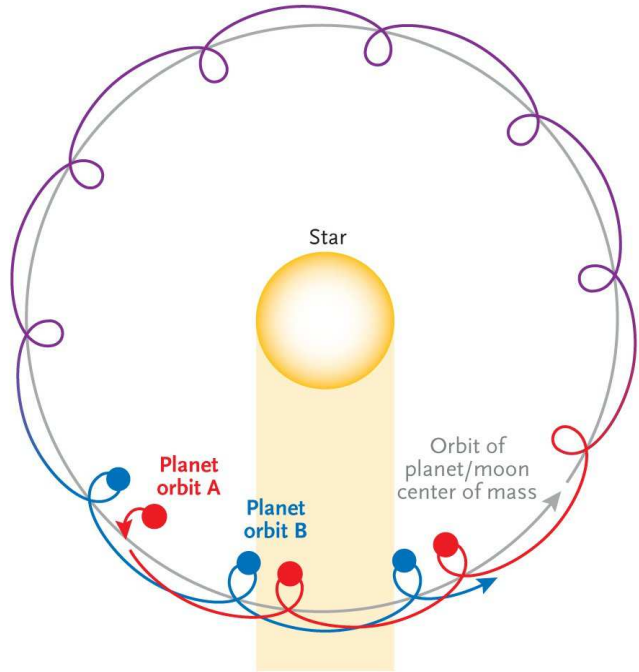}
\caption[Cartoon illustrating the composite motion of a planet in orbit of a
star with an unseen exomoon companion]
{\emph{Cartoon illustrating the composite motion of a planet in orbit of a
star with an unseen exomoon companion. Unless the period of the moon happens
to be resonant to the planet's orbital period (which there is no reason
to expect) then orbit B will exhibit a different shift in position than
orbit A at the time of inferior conjunction. These changes in position give
rise to the TTV effect.}} 
\label{fig:wobble}
\end{center}
\end{figure}

\subsection{Exomoon Transit Features}
\label{sec:luna1}

Before I provide Sartoretti \& Schneider's mathematical formulation of TTV, I 
first mention that the authors also considered detecting moons through their
transit features. These transit features could be due to either the moon eclipsing
the planet (causing a flux increase) or the star (causing a flux decrease). This
method is certainly highly viable for moons of large radius but modelling the
light curves of planet-moon systems in an efficient manner is a very challenging
problem. 

To illustrate this complexity, consider that for a planet-star system it is
possible to describe the light curve using just one parameter, $S_{P*}$, which
can be in three possible states (see \S\ref{sec:transitbasics}): i) in-transit 
ii) on-the-limb iii) out-of-transit. Therefore, there exists $3$ distinct
cases, each with its own unique, analytic solution. In contrast, a
planet-moon-star system requires three distances to characterize the light
curve $S_{P*}$, $S_{PS}$ and $S_{S*}$ (where the ``S'' subscript denotes
satellite). Since each one can be in three states again, there exists
$3^3=27$ distinct cases, each with its own unique, analytic solution. Things
are further exacerbated when one considers the effects of limb darkening and
the three-body problem, which has no exact closed-form solution. As a result,
modelling these light curves in an analytic manner is highly challenging and
no proposed method currently exists in the literature (one could model
the light curve numerically by pixelating the star, but such methods use
inordinate amounts of CPU and memory making a fitting routine impractical).

As will be discussed in Chapter~\ref{ch:Chapt8}, part of the future work in
exomoon detection theory will revolve around meeting this challenge. To this
end, I have been developing a new analytic framework and computer code called
\luna, which accomplishes the goal of complete analytic modelling of the 
planet-moon-star transit light curve. However, this is currently a work in 
progress and does not constitute the main focus of this thesis. Therefore,
for the remaining derivations I assume $(R_S/R_*)^2 \ll 1$ i.e. the moon does
not induce a detectable transit signal. I label this assumption as $\alpha$1.
I also point out here that assumption $\alpha$0 will be to consider the
system comprised of a star with a single planet with a single moon.

\subsection{The TTV of \citet{sartoretti1999}}

The authors explicitly assumed a coplanar configuration and implicitly assumed
that the orbit of both the planet and satellite are circular, plus assumption
$\alpha$1\footnote{This last assumption is required or the moon's transit
features would impair the timing of the planetary transit signal}. There are
other implicit assumptions which will be discussed in 
\S\ref{sec:ttvassumptions}, and it is unclear whether the authors were aware
of these or not. \citet{sartoretti1999} argued that the positional displacement 
of the planet away from the planet-moon barycentre is given by:

\begin{align}
a_{PB} &= (M_S/M_P) a_{SB} 
\end{align}

Where $a_{PB}$ denotes the semi-major axis of the planet (``P'') around the
planet-moon barycentre (``B'') and $a_{SB}$ denotes the semi-major axis of the 
satellite (``S'') around the same barycentre. The authors then argue that
the velocity of the planet at the time of transit minimum is given by
$v_{P,T} = 2\pi a_P/P_P$. Therefore, the maximum time shift of the planet away
from that expected from a strictly linear ephemeris would be given by
$(a_{PB}/v_{P,T})$ and the maximum possible observed difference between
consecutive transits would be twice this value:

\begin{align}
\Delta t &\sim 2 a_{SB} M_S M_P^{-1} \times P_P (2\pi a_P)^{-1}
\label{eqn:sar99eqn}
\end{align}

Equation~(\ref{eqn:sar99eqn}) is given in the exact same format as that
presented in the original paper, except for a change in notation to be 
consistent with that of this thesis. Equation~(\ref{eqn:sar99eqn}) may be
re-expressed in terms some typical astrophysical units and replacing
$a_{SB} = \mathfrak{D}' R_{\mathrm{Hill}}$, where $R_{\mathrm{Hill}}$ is the
Hill radius, I find:

\begin{align}
\Delta t &\sim \frac{ \mathfrak{D}' M_S P_P }{ 3^{1/3} \pi M_P^{2/3} M_*^{1/3} } \nonumber \\
\qquad &= 36.0\,\mathfrak{D}'\,\Big(\frac{M_S}{M_{\oplus}}\Big)\,\Big(\frac{P_P}{\mathrm{years}}\Big) \Big(\frac{M_J}{M_P}\Big)^{2/3} \Big(\frac{M_{\odot}}{M_*}\Big)^{1/3}\,\mathrm{minutes}
\label{eqn:sar99scaled}
\end{align}

This 36\,minute deviation illustrates the exciting potential of this technique 
to detect an exomoon. After this pioneering work, it seemed as though the 
methodology for detecting an exomoon was solved, but unfortunately problems
with this TTV technique soon became apparent, as discussed next in
\S\ref{sec:ttvproblems}.

\subsection{The Problems with Detecting Exomoons Using TTV}
\label{sec:ttvproblems}

\subsubsection{Problem 1 - A Multitude of Phenomena}

As was mentioned earlier, \citet{sartoretti1999} devised the TTV method for
detecting moons, although it was not called TTV at the time. Six years later,
\citet{agol2005} and \citet{holman2005} almost simultaneously predicted
that if a system has two or more planets then TTV would be induced due to
planet-planet gravitational interactions (this appears to be the first instance
of referring to the effect as TTV). These effects are particularly
pronounced for planets in mean motion resonance (MMR), where timing deviations 
of hours can be induced \citep{holman2010}.

Unfortunately for exomoons, this new prediction of the phenomenon meant that if 
one observed TTVs then it could be due to either a moon or a perturbing planet.
This was exacerbated later when a barrage of papers predicted
a plethora of possible sources for TTV, including:

\begin{itemize}
\item[{\tiny$\blacksquare$}] Trojan bodies \citep{ford2007}
\item[{\tiny$\blacksquare$}] Parallax effects \citep{scharf2007}
\item[{\tiny$\blacksquare$}] Kozai mechanism induced eccentricity variation 
\citep{kipping2008}
\item[{\tiny$\blacksquare$}] Apsidal precession \citep{jordan2008}
\item[{\tiny$\blacksquare$}] Star spots \citep{alonso2008}
\item[{\tiny$\blacksquare$}] Stellar proper motion \citep{rafikov2009}
\item[{\tiny$\blacksquare$}] Planetary in-fall \citep{hellier2009}
\item[{\tiny$\blacksquare$}] The Applegate effect \citep{watson2010}
\item[{\tiny$\blacksquare$}] Stellar binarity \citep{montalto2010}
\end{itemize}

With so many effects potentially responsible for TTV, the detection of TTVs
no longer seemed viable as an unambiguous detection tool for exomoons. This
can be summarized by the desideratum for some kind of unique signature of the
exomoon's timing deviations, which would distinguish it from the other
effects.

\subsubsection{Problem 2 - Undersampling}

The second major problem with the TTV due to an exomoon is undersampling. I
begin this discussion by considering the Hill radius. $R_{\mathrm{Hill}}$
represents the maximum possible stable orbital distance of an exomoon away
from its host planet. As discussed earlier in \S\ref{sec:moonstable}, the
true limit is actually less than $R_{\mathrm{Hill}}$ \citep{domingos2006}.
Let us therefore denote the semi-major axis of the satellite around the 
planet as some fraction, $\mathfrak{D}$, of the Hill radius, in a similar way
as to what was done in Equation~(\ref{eqn:sar99scaled}):

\begin{equation}
a_{SP} = \mathfrak{D} a_P \Big(\frac{M_P}{3 M_*}\Big)^{1/3}
\end{equation}

Now I use Kepler's Third Law to replace the distances with orbital periods:

\begin{align}
G^{1/3} (M_P + M_S)^{1/3} \Big(\frac{P_S}{2\pi}\Big)^{2/3} &= \mathfrak{D} G^{1/3} (M_* + M_P + M_S)^{1/3} \Big(\frac{P_P}{2\pi}\Big)^{2/3} \Big(\frac{M_P}{3 M_*}\Big)^{1/3}
\end{align}

Now I assume $M_* \gg M_P \gg M_S$, giving:

\begin{align}
P_S &= P_P \mathfrak{D}^{3/2} \Big(\frac{1}{3}\Big)^{1/2}
\label{eqn:kiprelation}
\end{align}

Equation~(\ref{eqn:kiprelation}) is very powerful, constraining a satellite's
orbital period purely in terms of $\mathfrak{D}$ and $P_P$. I have been
unable to find this expression in the previous literature and believe it may be 
a novel result, first pointed out in \citet{kipping2009a}. Since 
$\mathfrak{D}<1$ for all bounded moons, then $P_S < 0.57735 P_P$. 

The sampling rate of the exomoon TTV signal is once every transit i.e. $P_P$.
In order to avoid aliasing, one is required to sample the signal at above
the Nyquist rate i.e. $2/P_S = 3.46/P_P$. However, it is physically impossible
to sample the signal 3.46 times per $P_P$, the best one can ever do is 
once\footnote{Hypothetically, one could actually sample the signal twice per
$P_P$ using the occultation, but the error on this timing would be much higher.
In any case, even sampling of twice per $P_P$ would still cause aliasing.}
per $P_P$. Therefore, the observed TTVs caused by an exomoon \emph{will always
be undersampled}, in most cases heavily so. As a result, a Fourier periodogram 
of the TTVs would reveal a rich forest of possible harmonic frequencies with no 
way of distinguishing which signal is the correct one.

Despite not being able to retrieve the correct period, the deviations themselves
will still be present and cause an excess variance in the residuals of the
transit times from a linear ephemeris. This excess variance is detectable if
the standard deviation of the scatter is much greater than the errors bars
of the measurements themselves. 

Since it is possible to measure the amplitude through excess scatter, then
one can evaluate:

\begin{align}
M_{S} a_{SB} = \frac{\pi a_P M_P}{2 P_P} \sqrt{2} \delta_{\mathrm{TTV}}
\end{align}

Where $\delta_{\mathrm{TTV}}$ is the RMS amplitude of the TTV signal. Therefore,
the final problem of undersampling becomes evident - one can only measure
$M_S a_{SB}$ i.e. mass multiplied by distance, a.k.a. ``the moment'' of the 
exomoon.

\section{An Updated Model for TTV}
\label{sec:kipttv}

\subsection{Introduction}

Two major problems exist with the proposed TTV method of \citet{sartoretti1999}
and a hypothetical wish list of required breakthroughs would read as:
1) the ability to distinguish between an exomoon's TTV and other sources
2) the ability to measure both the exomoon mass and orbital distance. In the
remainder of this chapter, I provide a new method of detecting exomoons which 
achieves both of these goals and thus makes transit timing a viable method for 
detecting exomoons and thus coming full circle to satisfy the original goal of 
this thesis. This remainder of this chapter is based upon the papers 
\citet{kipping2009a} and \citet{kipping2009b}, with some new updates.

Before I provide this new method, I will first provide an updated model for the 
transit timing variations (TTV) of an exomoon, including the effects of orbital 
eccentricity and inclination of both the planet and moon plus the mutual 
longitude of the ascending node. I will then go on to predict and introduce a 
new observable, transit duration variations (TDV) and compare it to TTV.

I will begin by outlining the assumptions of the model presented here. 
So far, only one assumption has been made, $\alpha$1): the satellite induces 
no transit features $\Rightarrow (R_S/R_*)^2 \ll 1$. In order to proceed, the 
first requirement is a coordinate system for the positions of the planet and the
moon which will require an additional approximation, given that no general
analytic solution exists for the three-body problem.

\subsection{The Nested Two-Body Model}
\label{sec:nested}

Since there is a planet, star and moon all in play, one must deal with the
three-body problem. The three-body problem is quite daunting given that no
analytic, general solution has ever been found. In practice, one is forced to
either use numerical methods or some restricted case through an approximation. 
In this work, I will adopt what I label as ``the nested two-body'' model. In 
this framework, one has the planet-moon forming a close-binary which orbits the 
star at a much greater distance. The basic premise is that in the frame of 
reference of the planet-moon barycentre, both bodies follow Keplerian orbits. 
Then, in the reference frame of the star, the motions can be considered as a 
composite of these local Keplerian orbits on top of the global Keplerian 
motion around the star. 

The advantages of the nested two-body model are a) it is simple to
describe analytically since one only has Keplerian orbits b) more general than
the restricted circular three-body problem, which insists eccentricities are
zero c) particularly applicable for the problem of interest- motion of a
planet over short timescales (i.e. we do not care about long term precessions,
secular changes, etc). I will here derive the conditions under which the nested 
two-body model is a good approximation for the three-body problem.

For a planet-moon pair at infinite orbital separation, the validity of the
nested model can be understood intuitionally. As the planet-moon pair 
moves in, the disturbance due to the star becomes greater and the approximation 
will break down. One may follow the standard treatment used in lunar theory 
\citep{murray1999} to calculate the disturbance to this approximation. I select 
a non-rotating reference frame with the star at rest at the origin, for 
which one may treat the frame as inertial since the mass of the star is much 
greater than the planet or moon. The satellite's equation of motion is given by:

\begin{align}
\ddot{\mathbf{r}_S} &= - n_S^2 a_S^3 \frac{(\mathbf{r}_S - \mathbf{r}_P)}{|\mathbf{r}_S - \mathbf{r}_P|^3} - n_P^2 a_P^3 \frac{\mathbf{r}_S}{|\mathbf{r}_S|^3}
\end{align}

Where $\mathbf{r}_S$ and $\mathbf{r}_P$ are the position vectors of the 
satellite and planet relative to the star, $n_P$ is the mean-motion of the 
planet around the star and $a_P$ is the corresponding semi-major axis. 
Similarly, $n_S$ is the mean-motion of the moon around the planet and $a_S$ is 
the corresponding semi-major axis. To translate to the reference frame of the 
planet, one may substitute:

\begin{align}
\mathbf{r}  &= \mathbf{r}_S - \mathbf{r}_P \\
\mathbf{r}' &= -\mathbf{r}_P
\end{align}

These vectors now describe the positions of the satellite and the star, 
respectively, relative to the planet. It follows that in this non-inertial 
frame where the planet is at rest, but the coordinate axes point in 
fixed directions, the equation of motion for the satellite becomes:

\begin{align}
\ddot{\mathbf{r}} &= - n_S^2 a_S^3 \frac{\mathbf{r}}{|\mathbf{r}|^3} + n_P^2 a_P^3 \Bigg[ \frac{(\mathbf{r}'-\mathbf{r})}{|\mathbf{r}'-\mathbf{r}|^3} - \frac{\mathbf{r}'}{|\mathbf{r}'|^3}\Bigg]
\end{align}

The term to the right of the addition sign represents the disturbance to 
Kepler's Third Law from a simple nested two-body approximation. The disturbing 
function can be written as:

\begin{align}
\phi &= n_P^2 a_P^3 \Bigg[ \frac{\mathbf{r}_S}{|\mathbf{r}_S|^3} - \frac{\mathbf{r}_P}{|\mathbf{r}_P|^3}\Bigg] \nonumber \\
|\phi| &\simeq n_P^2 a_P \Big( [1 + (a_S/a_P)]^{-2} - 1\Big) \nonumber \\
|\phi| &\simeq 2 n_P^2 a_S
\end{align}

Where I have assumed a low eccentricity system and expanded to first order for 
$(a_S/a_P) \ll 1$. For the disturbing function to be small, one therefore 
requires:

\begin{align}
\Big| n_S^2 a_S^3 \frac{\mathbf{r}}{|\mathbf{r}|^3} \Big| \gg |\phi| \nonumber \\
n_S^2 \gg 2 n_P^2
\end{align}

Where the final line gives the condition under which this derivation is 
ultimately valid and is equivalent to:

\begin{equation}
P_S^2 \ll P_P^2/2
\end{equation}

Using Equation~(\ref{eqn:kiprelation}), 
$(P_S/P_P)\simeq \sqrt{\mathfrak{D}^3/3}$, this constraint becomes assumption 
$\alpha$2:

\begin{equation}
\mathfrak{D}^3 \ll 3/2
\label{eqn:alpha2}
\end{equation}

Defining $\ll$ to indicate an order-of-magnitude difference, i.e. a factor of 
10, this constrains $\mathfrak{D}\leq0.531$. Note that this distance is larger 
than that predicted as the maximum stable separation for a prograde satellite of
$\mathfrak{D}=0.4895$ \citep{domingos2006}. However, a retrograde moon can be 
stable at up to $\mathfrak{D}=0.9309$, at which point $\mathfrak{D}^3$ is only
about one half of 3/2 and thus Keplerian motion is not guaranteed.

\subsection{Coordinate System}

In the nested two body problem, the planet-moon barycentre orbits the star in
a Keplerian orbit and thus one can see that the planet-moon barycentre, denoted
by the subscript ``B'', replaces the planet (``P'') in 
Equation~(\ref{eqn:complexcoords}). The coordinates of the barycentre are 
therefore given by:

\begin{align}
X_B &= r_P \cos(\omega_P+f_P) \nonumber \\
Y_B &= r_P \sin(\omega_P+f_P)\cos i_P \nonumber \\
Z_B &= r_P \sin(\omega_P+f_P)\sin i_P
\label{eqn:complexcoords2}
\end{align}

Furthermore, the satellite orbits the barycentre of the planet-moon system with 
a semi-major axis of $a_{SB}$ in a Keplerian orbit. The motion is therefore 
highly analogous to the case derived earlier for a planet-star system. Indeed, 
in the reference frame of the planet-moon barycentre, one may simply adapt 
Equation~(\ref{eqn:complexcoords}), replacing the subscripts for the planet to 
that of the satellite:

\begin{align}
X_{SB} &= r_{SB} [ \cos\Omega_S\cos(\omega_S+f_S) - \sin i_S\sin\Omega_S\sin(\omega_P+f_S) ] \nonumber \\
Y_{SB} &= r_{SB} [ \sin\Omega_S\cos(\omega_S+f_S) + \sin i_S\cos\Omega_S\sin(\omega_P+f_S) ] \nonumber \\
Z_{SB} &= r_{SB} \cos i_S \sin(\omega_S+f_S)
\end{align}

Where $r_{SB} = a_{SB} (1-e_S^2)/(1+e_S\cos f_S)$. The planet's reflex motion 
is simply given by:

\begin{align}
X_{PB} &= -(M_S/M_P) X_{SB} = -r_{PB} [ \cos\Omega_S\cos(\omega_S+f_S) - \sin i_S\sin\Omega_S\sin(\omega_P+f_S) ] \nonumber \\
Y_{PB} &= -(M_S/M_P) Y_{SB} = -r_{PB} [ \sin\Omega_S\cos(\omega_S+f_S) + \sin i_S\cos\Omega_S\sin(\omega_P+f_S) ] \nonumber \\
Z_{PB} &= -(M_S/M_P) Z_{SB} = -r_{PB} \cos i_S \sin(\omega_S+f_S)
\end{align}

Where $r_{PB} = (M_S/M_P) r_{SB}$. These coordinates define the motion of
the planet in the reference frame of the planet-moon barycentre. However,
the reference frame of interest is that of the star at the origin. Proceeding
through the rotations for $\omega_P$ and $i_P$ ($\Omega_P$ rotation is not
performed as the transit is insensitive to this parameter) as was done in
\S\ref{sec:transitbasics}, the final sky-projected coordinates of the planet
are given by:

\begin{align}
X_P &= r_P \cos(f_P+\omega_P) + r_{PB} \Big[ -\cos(f_S+\omega_S) \cos(\omega_P+\Omega_S) \nonumber \\
\qquad& + \sin i_S \sin(f_S+\omega_S) \sin(\omega_P+\Omega_S) ] \\
Y_P &= r_P \sin(f_P+\omega_P) \cos i_P + r_{PB} \Big[ \sin(f_S+\omega_S) [ \cos i_S \sin i_P \nonumber \\
\qquad& - \cos i_P \sin i_S \cos(\omega_P+\Omega_S) ] - \cos i_P \sin (\omega_P+\Omega_S) \cos(f_S+\omega_S) \Big] \\
Z_P &= r_P \sin(f_P+\omega_P) \sin i_P - r_{PB} \Big[ \sin(f_S+\omega_S) [ \cos i_S \cos i_P \nonumber \\
\qquad& + \sin i_P \sin i_S \cos(\omega_P+\Omega_S) ] + \sin i_P \sin(\omega_P+\Omega_S) \cos(f_S+\omega_S) \Big]
\label{eqn:wobblecoords}
\end{align}

There are two things to note from these equations. First of all, 
$X_{PB} \neq (X_P - X_B)$, and equivalently for $Y_{PB}$ and $Z_{PB}$. I will
therefore define $\Delta X_P = (X_P - X_B)$, $\Delta Y_P = (Y_P - Y_B)$ and
$\Delta Z_P = (Z_P - Z_B)$. Second, all instances of $\Omega_S$ occur in the 
form $(\omega_P+\Omega_S)$. I will define this quantity as 
$\varpi_S = \omega_P+\Omega_S$, the ``longitude of the periapsis'', in the rest 
of this work.

\subsection{Model Assumptions}
\label{sec:ttvassumptions}

\subsubsection{The Importance of Declaring the Model Assumptions}

Curiously, the TTV of an exomoon is very easy to write down with some simple
intuitive feel but actually characterizing what the assumptions are and under
what conditions this toy model is valid is more subtle. In this subsection, I
will provide a strict outline of the model and the required assumptions, no
matter how seemingly obvious. This is extremely important as it has already
been seen how breaking model assumptions for the transit model causes 
significant departures (\S\ref{sec:breakassumptions}). Further, identifying
the assumptions allows one to calculate the range of moons and planets for which
it is actually legitimate to conduct TTV analyses on.

\subsubsection{RMS Amplitude Definition}

The first thing to establish is that whatever form is derived for the TTV 
effect, it is only ever possible to infer the RMS amplitude, as discussed in 
\S\ref{sec:ttvproblems}. Therefore, regardless as to the functional form,
the following integral must be performed to calculate the RMS amplitude:

\begin{align}
\delta_{\mathrm{TTV}} &= \sqrt{\frac{1}{2\pi} \int_{f_S=0}^{2\pi} [\mathrm{TTV}(f_S)]^2\,\mathrm{d}f_S }
\label{eqn:RMSdefinition}
\end{align}

I outline this point early on, because naturally if the TTV has an elaborate
functional form, the integral will become increasingly challenging to compute
analytically, and perhaps not even possible. Therefore, it is required to seek a 
relatively simple but nevertheless physically accurate and justified form for 
the TTV.

\subsubsection{Two Definitions of the TTV}
\label{sec:twoTTVs}

One subtlety with TTV is that there actually exists two definitions of what
constitutes the TTV. The first definition could be considered as the pure
theoretical definition whilst the second is the observable or measurable TTV.
I will discuss the first definition initially, which will make it clear why
a second definition exists.
\\
\emph{\textbf{Theoretical definition of the TTV}}

In the nested two-body problem, the planet-moon barycentre orbits the star in
a Keplerian orbit. Therefore the instant when the barycentre satisfies
d$S_{B*}$/d$t=0$ (i.e. the equivalent of the transit minimum) occurs at
approximately the time of inferior conjunction, as before. These events are
therefore separated in time by exactly $P_P$ and have a strictly linear 
ephemeris.

In contrast, the instant when the transit minimum of the planet occurs will be
given by when d$S_{P*}$/d$t=0$, which will be a function of $f_S$. In other
words, the transit minima do not occur at the same instances as the barycentric
minima. Therefore, the transit minima do not follow a strictly linear ephemeris.
The temporal deviation of the transit minima away from the barycentric
minima is the source of the TTV (for both definitions).

The theoretical definition for the TTV of a transit event would therefore be to 
calculate the difference between the times $t(\dot{S_{B*}}=0)$ and 
$t(\dot{S_{P*}}=0)$:

\begin{align}
\mathrm{TTV} = t(\dot{S_{P*}}=0) - t(\dot{S_{B*}}=0)
\label{eqn:theoryTTV}
\end{align}
\\
\emph{\textbf{Observable definition of the TTV}}


The theoretical definition classes the TTV as the time difference between the
instants of the barycentric and planetary transit minima. However, consider the
case of a moon moving on a very short orbital period, say less than the transit
duration. Under such a circumstance, the planet-moon separation would be very
different between the start and the end of the transit. When one fits such a
light curve, the timing deviation would be essentially the averaged-out 
temporal deviation across the whole duration. In other words, the motion of the
moon smears out the TTV.

It can be seen that a derivation of the theoretical TTV is required before it is
possible to account for the practicalities of smearing effects. I therefore
proceed here to consider what model assumptions are required for the theoretical 
TTV. The observational TTV will be discussed in \S\ref{sec:obsTTV}.

\subsubsection{Assumptions $\alpha$3 \& $\alpha$4: Slow-moving moon}

The theoretical definition of the TTV is based upon solving the instants when 
the sky-projected separations hit minima. However, it has been shown earlier 
(see \S\ref{sec:transitminima}) that solving the exact time when 
$\dot{S_{B*}}=0$ is non-trivial and possesses no simple analytic form. The 
mathematics will be even more complex for the case of $\dot{S_{P*}}=0$. 
Therefore, one can easily appreciate that approximations will be necessary to 
make this calculation possible.

The first two approximations I therefore make are $\alpha$3) the sky-projected 
distance between the barycentre and the planet does not change appreciably over 
the timescale of the TTV amplitude $\alpha$4) the sky-projected velocity of the
planet does not change appreciably over the timescale of the TTV amplitude.
Under such approximations, the TTV definition becomes:

\begin{align}
\mathrm{TTV} = \frac{[S_{P*}-S_{B*}](t(f_P=f_{P,T}))}{\frac{\mathrm{d}[S_{P*}]}{\mathrm{d}t}(t(f_P=f_{P,T}))}
\label{eqn:alpha34}
\end{align}

Where the parentheses after each term mean that the function is evaluated at the
time, $t$, when $f_P = f_{P,T}$ i.e. the transit minimum. The approximations 
$\alpha$3 and $\alpha$4 will be written out mathematically shortly, but first 
it is useful to adopt additional approximations. Firstly, when trying to solve
the true anomaly at the instant of the transit minimum for the simple case
of a moonless planet and a star, it was seen that a simple, closed-form solution
is elusive (\S\ref{sec:transitminima}). In light of this, one can easily
appreciate that solving for $f_{P,T}$ with the moon included
will be even worse. Therefore, a useful approximation to make is that 
the numerator and denominator in Equation~(\ref{eqn:alpha34}) experience 
negligible change when evaluated at $f_P = (\pi/2-\omega_P)$ rather than 
$f_{P,T}$. 

Since I have assumed the distance is constant over the timescale of the TTV
amplitude, assumption $\alpha$3, then assuming 
$[S_{P*}-S_{B*}](t(f_P=f_{P,T})) \simeq [S_{P*}-S_{B*}](t(f_P=\pi/2-\omega_P))$
is actually already implicit by assumption $\alpha$3 and no further assumption
is needed. This is because the timescale of the TTV amplitude is approximately
equal to that for the planet to move between $f_P=f_{P,T}$ and 
$f_P=(\pi/2-\omega_P)$.

The same is also true for the denominator, if one wishes to assume the velocity 
at $f_P=f_{P,T}$ and $f_P=(\pi/2-\omega_P)$ are approximately equal, then this
is equivalent to assuming that the velocity does not change appreciably over
the timescale of the TTV effect i.e. assumption $\alpha$4. This line of
argument leads to the following definition for the TTV effect:

\begin{align}
\mathrm{TTV} = \frac{[S_{P*}-S_{B*}](t(f_P=\pi/2-\omega_P))}{\frac{\mathrm{d}[S_{P*}]}{\mathrm{d}t}(t(\pi/2-\omega_P))}
\label{eqn:alpha342}
\end{align}

\subsubsection{Assumptions $\alpha$5 and $\alpha$6: Motion is in the $\hat{X}$-direction}

Now, it has been established that the key term of interest is the sky-projected 
distance between the planet and the planet-moon barycentre i.e. 
$(S_{P*} - S_{B*})$. I will write out this distance in terms of the Cartesian 
elements:

\begin{align}
S_{P*} - S_{B*} &= \frac{\sqrt{X_P^2+Y_P^2}}{R_*} - \frac{\sqrt{X_B^2+Y_B^2}}{R_*} \nonumber \\
\qquad& = \frac{\sqrt{(X_B+\Delta X_P)^2+(Y_P+\Delta Y_{P})^2}}{R_*} - \frac{\sqrt{X_B^2+Y_B^2}}{R_*}
\end{align}

The presence of terms with ``B'' subscripts is non-preferable as one can
intuitionally understand that it is the distance between the barycentre and
the planet (i.e. the ``$\Delta$'' terms) which ultimately dominate the TTV. 
Inspecting the above, it is clear that if one assumes the motion is 
predominantly in the $\hat{X}$-direction, then the $Y$-terms can be ignored and 
the following simplification would be possible:

\begin{align}
S_{P*} - S_{B*} &\simeq \frac{\sqrt{(X_B+\Delta X_{P})^2}}{R_*} - \frac{\sqrt{X_B^2}}{R_*} \nonumber \\
S_{P*} - S_{B*} &\simeq \frac{\Delta X_{P}}{R_*} = \frac{X_P - X_B}{R_*}
\label{eqn:alpha56}
\end{align}

These simplifications can be seen to greatly aid in the analysis and thus are
adopted in this work and denoted as assumptions $\alpha$5 and $\alpha$6. The
TTV equation now becomes:

\begin{align}
\mathrm{TTV} = \frac{[X_{P}-X_{B}](t(f_P=\pi/2-\omega_P))}{\frac{\mathrm{d}[X_{P}]}{\mathrm{d}t}(t(\pi/2-\omega_P))}
\end{align}

The mathematical form of assumption $\alpha$5 yields:

\begin{align}
|Y_B| &\ll |X_B| \nonumber \\
\Rightarrow |\cos i_P| &\ll 1
\label{eqn:alpha5}
\end{align}

Assumption $\alpha$6 can be seen to be equivalent to this assumption if one
assumes $r_{PB} \ll r_P$, which gives:

\begin{align}
a_{PB} &\ll a_P \nonumber \\
\mathfrak{D} &\ll \frac{3^{1/3} M_*^{1/3} M_P^{2/3}}{M_S} \nonumber \\
\mathfrak{D} &\ll 4655.19 \Big(\frac{M_*}{M_{\odot}}\Big)^{1/3} \Big(\frac{M_P}{M_J}\Big)^{2/3} \Big(\frac{M_{\oplus}}{M_S}\Big)
\end{align}

The above assumption can be seen to be essentially valid in virtually all 
conditions. Therefore since $a_{PB} \ll a_P$ in almost all cases and 
$\cos i_P \ll 1$ by definition for a transit to occur, then assumptions
$\alpha$5 and $\alpha$6 are excellent approximations.

\subsubsection{Assumption $\alpha$7: Barycentric dominated velocity}

The next step is to compute the velocity-like term. The 
$\dot{X_P}(t(f_P=\pi/2-\omega_P))$ term contains an $f_S$ dependency
and this causes problems later when it is necessary to perform the integration 
over $f_S$, as required by Equation~(\ref{eqn:RMSdefinition}). I have not been 
able to solve said integral if the velocity-like term has an $f_S$ dependency. 
Therefore, a required approximation is that $\dot{X_{P}} \simeq \dot{X_B}$. This 
is equivalent to assuming that 
$|\dot{X_{B}}(t(f_P=\pi/2-\omega_P))| \gg |\dot{\Delta X_{P}}|(t(f_P=\pi/2-\omega_P))$ 
i.e. the barycentric velocity is much greater than the planet's reflex motion,
at the instant of inferior conjunction. This forms assumption
$\alpha$7. Differentiating, and setting $f_P = (\pi/2 - \omega_P)$, I find:

\begin{align}
&|\dot{X_B}(t(f_P=\pi/2-\omega_P))| \gg |\dot{\Delta X_{P}}(t(f_P=\pi/2-\omega_P))| \nonumber \\
&\frac{a_P (1+e_P \sin \omega_P)}{\sqrt{1-e_P^2} P_P} \gg \frac{a_{PB}}{\sqrt{1-e_S^2} P_S} \nonumber \\
\qquad& \times \Bigg| \cos\varpi_S [e_S\sin\omega_S+\sin(f_S+\omega_S) + \sin i_S \sin\varpi_S [e_S\cos \omega_S + \cos(f_S+\omega_S)] \Bigg|
\label{eqn:alpha7}
\end{align}

The terms inside the large modulus contain all of the $f_S$ dependency.
In order to evaluate the limits under which the above inequality is valid, one
is interested in setting the terms inside the large modulus to be the
maximum possible value. I differentiated the expression with respect to $f_S$
and then solved for when the result is equal to zero. This yields four solutions
for $f_{S,\mathrm{max}}$. I then generated $10^6$ Monte Carlo simulations where
all of the parameters are randomly varied and four versions of the term in the
modulus are calculated each time, for each solution for $f_{S,\mathrm{max}}$. I 
then compute the maximum of each of these four lists of possible maxima which
gives $\{1.99424, 1.98929, 1.99485, 1.99275\}$. Therefore, it is reasonable
to conclude that the maximum possible value is equal to 2. Accordingly, the
assumption $\alpha$7 becomes:

\begin{align}
\frac{a_P (1+e_P \sin \omega_P)}{\sqrt{1-e_P^2} P_P} \gg \frac{2 a_{PB}}{\sqrt{1-e_S^2} P_S}
\end{align}

Now using Kepler's Third Law and Equation~\ref{eqn:kiprelation}, one obtains:

\begin{align}
\mathfrak{D}^{1/2} \gg 2 \times 3^{1/6} \Bigg(\frac{M_S}{M_*^{1/3} M_P^{2/3}}\Bigg) \Bigg( \frac{\sqrt{1-e_P^2}}{\sqrt{1-e_S^2} (1+e_P\sin\omega_P)}\Bigg)
\end{align}

For near circular orbits, and assuming $\gg$ to be an order-of-magnitude, this
becomes:

\begin{align}
\mathfrak{D} \geq 0.000055 \Big(\frac{M_S}{M_{\oplus}}\Big)^2 \Big(\frac{M_{\odot}}{M_*}\Big)^{2/3} \Big(\frac{M_J}{M_P}\Big)^{4/3}
\label{eqn:alpha7simple}
\end{align}

So this constraint will be satisfied in essentially all cases as well.

\subsection{Valid Range for the Theory TTV Model}
\label{sec:TTVvalidrange}

I have now established an upper constraint on $\mathfrak{D}$ using assumption 
$\alpha$2 and a lower limit from $\alpha$7. Therefore there is a permitted
bounded range of $\mathfrak{D}$ which can be investigated. Before I do so,
I will calculate the conditions which assumptions $\alpha$3 and $\alpha$4
imply. For $\alpha$3, the condition is that $(S_{P*}-S_{B*})$ does not change
appreciably over the timescale of the TTV amplitude. Under assumption $\alpha$5
this becomes that $(X_{P}-X_{B})=\Delta X_{P}$ does not change appreciably over 
the same time. $\Delta X_{P}$ is characterized by the distance $a_{PB}$ and 
the TTV amplitude is approximately given by the \citet{sartoretti1999} model,
given in Equation~(\ref{eqn:sar99eqn}):

\begin{align}
\frac{\mathrm{d}\Delta X_{P}}{\mathrm{d}t} &\ll a_{PB} \Big[2 a_{SB} M_S M_P^{-1} \times P_P (2\pi a_P)^{-1}\Big]^{-1} \nonumber \\
\mathfrak{D}^{1/2} &\gg \frac{4\times3^{1/6}}{\sqrt{1-e_S^2}} \frac{M_S}{M_P^{2/3} M_*{1/3}} \nonumber \\
\mathfrak{D}^{1/2} &\gg 0.00149 \Big(\frac{M_S}{M_{\oplus}}\Big) \Big(\frac{M_J}{M_P}\Big)^{2/3} \Big(\frac{M_{\odot}}{M_*}\Big)^{1/3}
\end{align}

Where on the last line I have have assumed $e_S\ll1$. If one defines
$\gg$ to indicate an order of magnitude then the above becomes:

\begin{align}
\mathfrak{D} \geq 0.00012 \Big(\frac{M_S}{M_{\oplus}}\Big)^2 \Big(\frac{M_J}{M_P}\Big)^{4/3} \Big(\frac{M_{\odot}}{M_*}\Big)^{2/3}
\label{eqn:alpha3}
\end{align}

Therefore, just like $\alpha$7, assumptions $\alpha$3 places a lower limit
on $\mathfrak{D}$ for TTV studies. The two have the same functional dependency
but $\alpha$3 places a more stringent lower constraint on $\mathfrak{D}$ (by a 
factor of 2) and so represents the more realistic bounding limit.

For $\alpha$4, the constraint bounds $\dot{X_B}$ to exhibit negligible 
variation over the timescale of the TTV amplitude.

\begin{align}
\frac{\mathrm{d}^2X_B}{\mathrm{d}t^2}\Big|_{f_P=\pi-2\omega_P} \delta_{\mathrm{TTV}} \ll \frac{\mathrm{d}X_B}{\mathrm{d}t}\Big|_{f_P=\pi-2\omega_P}
\end{align}

However, since $\ddot{X_B}(f_P=\pi/2-\omega_P) = 0$, then the above can be seen
to be valid in all cases.
It has now been established that $\alpha$7 places a lower limit and $\alpha$2
places an upper limit on $\mathfrak{D}$, which together imply:

\begin{align}
0.00012 \Big(\frac{M_S}{M_{\oplus}}\Big)^2 \Big(\frac{M_J}{M_P}\Big)^{4/3} \Big(\frac{M_{\odot}}{M_*}\Big)^{2/3} \leq \mathfrak{D} \leq \Big(\frac{3}{20}\Big)^{1/3}
\end{align}

It can therefore be seen that the theoretical TTV model presented in this thesis 
is invalid if the LHS exceeds the RHS of this inequality. This condition gives 
rise to the following condition, which must be satisfied for a TTV search using 
the model of this thesis to be possible:

\begin{align}
M_S < \Big( 48.98\,M_{\oplus}\Big) \Big(\frac{M_P}{M_J}\Big)^{1/3} \Big(\frac{M_{\odot}}{M_*}\Big)^{1/3} \frac{1}{\sqrt{1-e_S^2}}
\label{eqn:TTVconstraint}
\end{align}

Therefore, the model for the theory TTV presented here is generally valid for
all prograde satellites with terrestrial masses.
The final list of assumptions which are used, and their respective consequential
conditions which must be satisfied, are (conditions given for low-eccentricity 
systems):

\begin{itemize}
\item[{\tiny$\blacksquare$}] $\alpha$0) There is only one moon and one planet in 
the system
\item[{\tiny$\blacksquare$}] $\alpha$1) The moon does not impart any transit 
features onto the light curve $\Rightarrow R_S^2 \ll R_*^2$
\item[{\tiny$\blacksquare$}] $\alpha$2) The three-body problem may be described 
with the nested two-body model $\Rightarrow \mathfrak{D}^3 \ll 3/2$
\item[{\tiny$\blacksquare$}] $\alpha$3) The change in the sky-projected 
planet-moon separation is negligible during the transit event
$\Rightarrow \mathfrak{D}^{1/2} \gg 4 \times 3^{1/6} M_S M_*^{-1/3} M_P^{-2/3}$
\item[{\tiny$\blacksquare$}] $\alpha$4) The change in the sky-projected velocity 
of the planet is negligible during the transit event $\Rightarrow$ no further 
constraints
\item[{\tiny$\blacksquare$}] $\alpha$5) The sky-projected planet moves 
predominantly in the $\hat{X}$-direction $\Rightarrow a_{PB} \ll a_P$
\item[{\tiny$\blacksquare$}] $\alpha$6) The sky-projected planet-moon barycentre 
moves predominantly in the $\hat{X}$-direction $\Rightarrow |\cos i_P| \ll 1$
\item[{\tiny$\blacksquare$}] $\alpha$7) The sky-projected reflex motion of the 
planet is much less than the sky-projected barycentric motion 
$\Rightarrow \mathfrak{D}^{1/2} \gg 2 \times 3^{1/6} M_S M_*^{-1/3} M_P^{-2/3}$
\end{itemize}

\subsection{TTV Waveform and Amplitude}
\label{sec:ttvamplitude}

Under assumptions $\alpha$0$\rightarrow \alpha$7, the temporal deviation of 
the planet away from the planet-moon barycentre's position at the time
of inferior conjunction is given by:

\begin{align}
\mathrm{TTV} = \frac{[X_{P}-X_{B}](t(f_P=\pi/2-\omega_P))}{\frac{\mathrm{d}[X_{B}]}{\mathrm{d}t}(t(\pi/2-\omega_P))}
\end{align}

Now one requires evaluating $[X_{P}-X_{B}](t(f_P=\pi/2-\omega_P))$:

\begin{align}
[X_{P}-X_{B}](t(f_P=\pi/2-\omega_P)) &= r_{PB} \Big[ -\cos(f_S+\omega_S) \cos\varpi_S + \sin i_S \sin(f_S+\omega_S) \sin\varpi_S \Big]
\end{align}

And the sky-projected velocity of the planet-moon barycentre, in the 
$\hat{X}$-direction, is given by:

\begin{align}
\frac{\mathrm{d}[X_{B}]}{\mathrm{d}t}(t(\pi/2-\omega_P)) &= \frac{\mathrm{d}X_B}{\mathrm{d}f_P}\Big|_{f_P = \pi/2-\omega_P} \frac{\mathrm{d}f_P}{\mathrm{d}t}\Big|_{f_P = \pi/2-\omega_P} \nonumber \\
\qquad&= \Bigg[ a_P \frac{1-e_P^2}{(1+e_P \cos f_P)^2} [e_P\sin\omega_P + \sin(f_P + \omega_P)] \frac{2\pi}{P_P} \frac{(1+e_P\cos f_P)^2}{(1-e_P^2)^{3/2}} \Bigg]_{f_P = \pi/2-\omega_P} \nonumber \\
\qquad&= \frac{2 \pi a_P}{P_P} \frac{ 1+e_P\sin\omega_P }{ (1-e_P^2)^{1/2} }
\end{align}

The final TTV equation is therefore:

\begin{align}
\mathrm{TTV}(f_S) &\simeq \Bigg[\frac{a_S \sqrt{1-e_P^2} (1-e_S^2) M_S P_P}{2 \pi a_P M_P (1+e_P\sin\omega_P)}\Bigg] \Lambda_{\mathrm{TTV}}(f_S) \\
\Lambda_{\mathrm{TTV}}(f_S) &= \Bigg[ \frac{\sin i_S \sin \varpi_S \sin(f_S+\omega_S) - \cos\varpi_S \cos(f_S+\omega_S)}{1+e_S\cos f_S}\Bigg]
\label{eqn:TTVwaveform}
\end{align}

The TTV effect is therefore a function of the true anomaly of the satellite, as
expected. In general, the satellite will take a different true anomaly every 
time one measures a transit. Therefore, one can obtain the RMS amplitude of the
TTV effect using Equation~(\ref{eqn:RMSdefinition}). Integrating this equation, 
I obtain:

\begin{align}
\delta_{\mathrm{TTV}} &= \frac{1}{2\pi} \frac{a_S M_S P_P}{a_P M_P} \frac{(1-e_S^2) \sqrt{1-e_P^2}}{(1+e_P\sin\omega_P)} \sqrt{\frac{\Phi_{\mathrm{TTV}}}{2\pi}} \\
\Phi_{\mathrm{TTV}} &= \frac{2\pi}{(1-e_S^2)^2 [1+\sqrt{1-e_S^2}]} \Bigg[ \cos^2\omega_S \Bigg( \sin^2i_S\sin^2\varpi_S(1-e_S^2)^{3/2} + \cos^2\varpi_S \Big(1+e_S^2[\sqrt{1-e_S^2}-1]\Big) \Bigg) \nonumber \\
\qquad& + \sin^2\omega_S \Bigg( \sin^2i_S\sin^2\varpi_S\Big( 1+ e_S^2 [\sqrt{1-e_S^2}-1] \Big) + (1-e_S^2)^{3/2}\cos^2\varpi_S\Bigg) \nonumber \\
\qquad& + \frac{1}{2} \sin i_S \sin2\omega_S\sin2\varpi_S\Bigg( e_S^2[1-2\sqrt{1-e_S^2}]+\sqrt{1-e_S^2} - 1 \Bigg) \Bigg]
\label{eqn:TTVrms}
\end{align}

\subsection{Properties of the TTV RMS Amplitude}
\label{sec:ttvproperties}

\begin{figure}
\begin{center}
\includegraphics[width=10.0 cm]{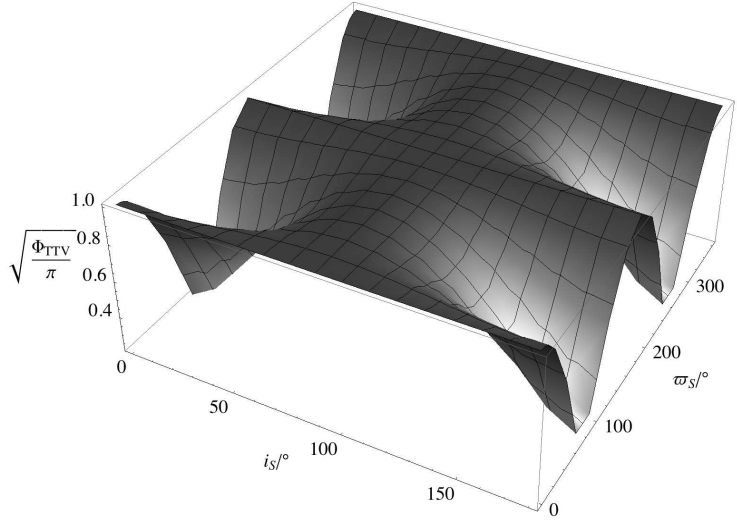}
\caption[Dependency of $\Phi_{\mathrm{TTV}}$, the TTV enhancement factor, on an 
exomoon's inclination and longitude of the periapsis]
{\emph{Dependency of $\Phi_{\mathrm{TTV}}$, the TTV enhancement factor, on an 
exomoon's orbital inclination, $i_S$, and longitude of the periapsis, 
$\varpi_S$, as measured from the plane of the planetary orbit. Maximum TTVs
occur for co-aligned systems. In this plot, I have assumed a circular orbit for 
both the exomoon and the host planet.}} 
\label{fig:phittvinc}
\end{center}
\end{figure}

In the limit of circular and co-aligned orbits, as was considered by
\citet{sartoretti1999}, $\Phi_{\mathrm{TTV}} \rightarrow \pi$ and thus the RMS 
amplitude agrees with that found by the original paper. However, the advantages 
of this much more general expression for the TTV waveform and RMS amplitude are
manifold. In Figure~\ref{fig:phittvinc}, I show the dependency of 
$\sqrt{\Phi_{\mathrm{TTV}}/\pi}$, which is essentially the scaling factor for 
the TTV amplitude, in the case of a circular lunar orbit. In this case, a very 
simple expression for $\Phi_{\mathrm{TTV}}$ is possible:

\begin{align}
\lim_{e_S \to 0} \Phi_{\mathrm{TTV}} = \pi \Big( 1 - \cos^2i_S\sin^2\varpi_S\Big)
\label{eqn:phittvinc}
\end{align}

Equation~(\ref{eqn:phittvinc}) and Figure~\ref{fig:phittvinc} reveal that 
exomoons possessing $|i_S- \pi/2|>0$ or $\varpi_S>0$ only act to decrease the 
TTV amplitude and are thus less favourable for detection.

Note, that the equations presented here are also improvements upon the model 
presented in \citet{kipping2009a} and \citet{kipping2009b}, where longitude of 
the ascending node was fixed to zero and it was approximated that the velocity
was equal to the tangential barycentric orbital velocity, rather than the 
sky-projected velocity  of the barycentre in the $\hat{X}$-direction. These 
improvements mean that the above equations should be used rather than those 
previously presented, for the greatest accuracy.

\begin{figure}
\begin{center}
\includegraphics[width=10.0 cm]{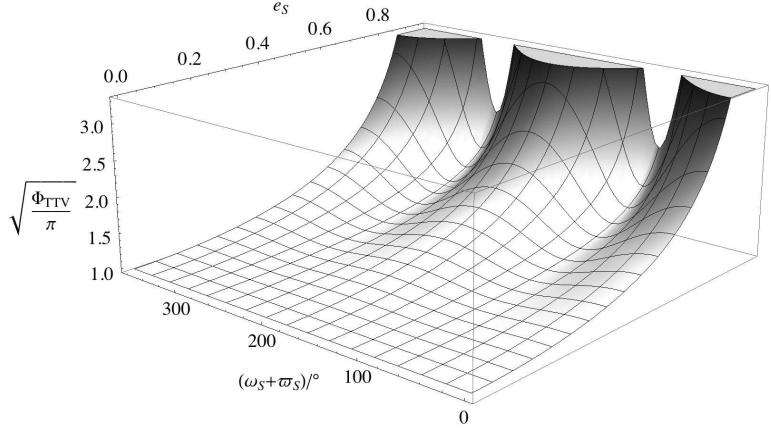}
\caption[Dependency of $\Phi_{\mathrm{TTV}}$, the TTV enhancement factor, on an 
exomoon's eccentricity and the sum of the argument of periapsis and the
longitude of periapsis]
{\emph{Dependency of $\Phi_{\mathrm{TTV}}$, the TTV enhancement factor, on an 
exomoon's eccentricity and the sum of the argument of periapsis and the
longitude of periapsis, $\omega_S+\varpi_S$. In this plot, I 
have assumed $i_S=\pi/2$.}} 
\label{fig:phittvecc}
\end{center}
\end{figure}

The effects of eccentricity, $e_S$, are illustrated in 
Figure~\ref{fig:phittvecc}. Here, I assume $i_S=\pi/2$ which gives:

\begin{align}
\lim_{i_S \to \pi/2} \Phi_{\mathrm{TTV}} = \pi \Bigg[ \frac{2 (1-e_S^2) \cos^2(\omega_S+\varpi_S) + \sqrt{1-e_S^2} [1+(-1+2e_S^2)\cos2(\omega_S+\varpi_S)]}{(1-e_S^2)^2 [1+\sqrt{1-e_S^2}]} \Bigg]
\label{eqn:phittvecc}
\end{align}

Equation~(\ref{eqn:phittvecc}) is plotted in Figure~\ref{fig:phittvecc} and
reveals how eccentric moons can cause strong enhancements to the TTV amplitude.
However, in general, one does not expect highly eccentric systems exomoons to be
dynamically stable \citep{domingos2006}.

\subsection{Observational TTV}
\label{sec:obsTTV}

The model presented thus far has been for the theoretical TTV, which ignores
the practicalities of fitting light curves. As discussed in \S\ref{sec:twoTTVs},
a real determination of the timing deviation of the transit would differ from
the theory definition if the moon was moving quickly enough. This motion would
cause a blurring or smearing of the signal.

At the simplest level, this could be considered by extending assumptions
$\alpha$3 and $\alpha$4 to hold over the timescale of the transit duration
rather than the TTV amplitude. However, this would cause the equations to hold
at a much higher accuracy than the TTV is ever likely to be measured to. One has
to acknowledge that moons will induce very small observable effects and so
the measurement of a TTV signal to, say, signal-to-noise SNR=10 would have to be
considered quite optimistic, in general. Therefore, if the smearing effect
induces an error much less than 10\%, then it would be safe to assume the
observational TTV is equivalent to the theory TTV.

Out of the two assumptions, $\alpha$4 is less constraining than 
$\alpha$3. This is because the sky-projected acceleration of the barycentre is 
zero at the time of inferior conjunction and thus the assumption of constant
velocity is excellent for both the timescale of the TTV amplitude and the
timescale of the transit duration. I will denote the extension of $\alpha$4
to the larger timescale as $\alpha$4*, which as discussed here has the same 
conditions as $\alpha$4. With this point established, I will turn
the focus onto $\alpha$3 and its extension to $\alpha$3*.

In light of the acknowledgement that there exists a measurement error in real
observations, I will consider the extension of $\alpha$3, which I denote as
$\alpha$3*, to be that \emph{the smearing effect of the lunar motion causes
a change in the TTV amplitude which is much less than the measurement error}.
If one had a measurement error of $\sigma_{\mathrm{TTV}}$, then the
acceptable tolerance level for the error in the TTV equation would be given
by $\sigma_{\mathrm{theory}}$:

\begin{align}
\sqrt{\sigma_{\mathrm{theory}}^2 + \sigma_{\mathrm{TTV}}^2} - \sigma_{\mathrm{TTV}} \ll \sigma_{\mathrm{TTV}} \nonumber \\
\Rightarrow \sigma_{\mathrm{theory}} \ll 2 \sigma_{\mathrm{TTV}}
\label{eqn:alpha3s}
\end{align}

So if for example $\sigma_{\mathrm{TTV}}$=10\%, then an acceptable tolerance
level in the smearing effect would be $\sigma_{\mathrm{theory}} \simeq$20\%.
The error in the theory TTV is controlled by the error in assuming
$[X_P - X_B] = [X_P-X_B](t(f_P=\pi/2-\omega_P))$. This may be evaluated by
computing 
$\Delta X_{P}(t(f_P=\pi/2-\omega_P-\Delta f_P/2)) - \Delta X_{P}(t(f_P=\pi/2-\omega_P))$:

\begin{align}
\sigma_{\mathrm{theory}} \simeq \frac{\Delta X_{P}(t(f_P=\pi/2-\omega_P-\Delta f_P/2)) - \Delta X_{P}(t(f_P=\pi/2-\omega_P))}{a_{PB}}
\end{align}

The two instances in time correspond to the ingress up to the moment of inferior
conjunction and so are separated by $\tilde{T}/2$. If one assumes the satellite
as a nearly circular orbit, then $\Delta \mathfrak{M}_S \simeq \Delta f_S$
and so $\Delta f_S \simeq \pi \tilde{T}/P_S$. The maximum smearing occurs
when the velocity is largest which occurs when $\Delta X_{P}=0$. This, in turn,
occurs when $f_S + \omega_S = \tan^{-1}[1/\sin i_S \tan\varpi_S]$ for
a low eccentricity orbit. If one assumes $\sin i_S\simeq 1$ (i.e. a coplanar
moon) then this becomes $(f_S + \omega_S) = (\pi/2 - \varpi_S)$. It is now 
possible to evaluate $\sigma_{\mathrm{theory}}$:

\begin{align}
\sigma_{\mathrm{theory}} \simeq \sin\Bigg[\frac{\pi \tilde{T}}{P_S}\Bigg]
\end{align}

The requirement of $\alpha$3* then becomes:

\begin{align}
\sin\Big[\frac{\pi \tilde{T}}{P_S}\Big] \ll 2 \sigma_{\mathrm{TTV}}
\label{eqn:alpha3s2}
\end{align}

Further expansion is possible by writing out $\tilde{T}$ assuming that
$(a_P/R_*)^2\varrho_{P,T}^2\gg b_{P,T}^2$ and employing a small-angle 
approximation and finally using Equation~(\ref{eqn:kiprelation}) plus
that $\gg$ indicates an order-of-magnitude:

\begin{align}
\mathfrak{D} \geq 0.196 \Bigg(\frac{10\%}{\sigma_{\mathrm{TTV}} } \Bigg)^{2/3} \Bigg(\frac{10}{(a_P/R_*)}\Bigg)^{2/3} (1-b_{P,T}^2)^{1/3} \Bigg(\frac{\sqrt{1-e_P^2}}{1+e_P\sin\omega_P}\Bigg)^{2/3}
\label{eqn:alpha3s3}
\end{align}

Equation~(\ref{eqn:alpha3s3}) reveals that for typical transiters with 
$(a_P/R_*) > 10$, the constraint will be satisfied across the majority of the
Hill sphere. For $(a_P/R_*) > 100$, which is the likely requirement for a moon
to be dynamically stable, this translates to $\mathfrak{D} > 0.042$. Having
established that the observational TTV is equal to the theory TTV under
assumption $\alpha$3*, the model for the TTV due to an exomoon is complete.

\subsection{Valid Range for the Observational TTV}

Just as was done in \S\ref{sec:TTVvalidrange}, the new lower bound from
$\alpha$3* may be combined with the upper bound from $\alpha$2 to estimate
the feasible range for observational TTV (where I here assume near-circular
orbits):

\begin{align}
\frac{a_P}{R_*} > 0.224 \frac{\sqrt{1-b_{P,T}^2}}{\sigma_{\mathrm{TTV}}}
\end{align}


\section{Velocity-Induced Transit Duration Variation (TDV-V)}
\label{sec:TDVV}

\subsection{The TDV-V Waveform and Amplitude}

The theory of TTV is now complete, but as pointed out in 
\S\ref{sec:ttvproblems}, it is not sufficient as a method for the unambiguous
detection of an exomoon. TTV is conceptually analogous to astrometry in that
one is looking for a companion to a massive body by searching for the reflex
motion and consequent variations in position. Astrometry is well-known to be
highly complementary to the radial velocity technique and thus this line of
thought brings one to consider whether variations in the velocity of a planet
could betray the presence of a moon. Such variations would induce 
transit duration variations (TDV).

TDV has been previously been discussed as a possible test of general 
relativity by \citet{jordan2008}. In this case, the changes were secular in 
nature whereas one would expect the reflex motion due to a moon to cause
periodic variations of the same frequency as the TTV. In this section, I 
consider the TDV due to an exomoon and conclude that it should produce a 
detectable signal.

An exomoon's TDV may be found by considering that the duration of a transit is 
inversely proportional to the sky-projected velocity of the planet across the 
star. Since I am only considering TDV to be due to velocity variations, I label
this TDV effect as TDV-V, where the ``V'' stands for velocity. The required
assumptions are $\alpha$0, $\alpha$1, $\alpha$2, $\alpha$4*, $\alpha$5, 
$\alpha$6 and $\alpha$7. The planetary transit duration scales as:

\begin{equation}
\tilde{T}_P \propto \Bigg[ \frac{\mathrm{d}X_P}{\mathrm{d}t}\Big|_{f_P=\pi/2-\omega_P} \Bigg]^{-1}
\label{eqn:alpha8}
\end{equation}

The presence of a moon means that d$X_P$/d$t \neq $d$X_B$/d$t$. The deviation of
the duration away from the barycentric motion is given by:

\begin{align}
\mathrm{TDV-V} &= \tilde{T}_P - \tilde{T}_B \nonumber \\
\qquad&= \tilde{T}_B \Bigg[ \frac{\tilde{T}_P}{\tilde{T}_B} - 1 \Bigg] \nonumber \\
\qquad&= \tilde{T}_B \Bigg[ \frac{\frac{\mathrm{d}X_B}{\mathrm{d}t}\Big|_{f_P=\pi/2-\omega_P} }{\frac{\mathrm{d}X_P}{\mathrm{d}t}\Big|_{f_P=\pi/2-\omega_P} } - 1 \Bigg]
\end{align}

The differentiated position of the planet-moon barycentre is given by:

\begin{align}
\frac{\mathrm{d}X_B}{\mathrm{d}t}\Big|_{f_P=\pi/2-\omega_P} &= \Bigg(\frac{2 \pi a_P}{P_P}\Bigg) \Bigg(\frac{1+e_P\sin\omega_P}{\sqrt{1-e_P^2}}\Bigg)
\end{align}

From Equation~(\ref{eqn:wobblecoords}), one may write 
$X_P = X_B + \Delta X_{P}$ where $\Delta X_{P}$ is the perturbation term due to 
the moon and may be expressed independent of $f_P$. Therefore:

\begin{align}
\frac{\mathrm{d}X_P}{\mathrm{d}t}\Big|_{f_P=\pi/2-\omega_P} &= \frac{\mathrm{d}X_B}{\mathrm{d}t}\Big|_{f_P=\pi/2-\omega_P} + \frac{\mathrm{d}\Delta X_{P}}{\mathrm{d}t}\Big|_{f_P=\pi/2-\omega_P} 
\end{align}

Where the first term on the RHS is already known. The second term is:

\begin{align}
\frac{\mathrm{d}\Delta X_{P}}{\mathrm{d}t} &= \frac{\mathrm{d}\Delta X_{P}}{\mathrm{d}f_S} \frac{\mathrm{d}f_S}{\mathrm{d}t} \nonumber \\
\qquad& = \frac{2 \pi a_S}{\sqrt{1-e_S^2} P_S} \Bigg[ \cos\varpi_S [e_S\sin\omega_S+\sin(f_S+\omega_S)] + \sin i_S \sin\varpi_S [e_S\cos\omega_S + \cos(f_S+\omega_S)] \Bigg]
\end{align}

Note, that since $X_{PB}$ has no dependency on $f_P$, then evaluating the
above at the instant of inferior conjunction does not actually change the 
equation at all. One may now bring assumption $\alpha$7 into play:

\begin{align}
\frac{\mathrm{d}\Delta X_{P}}{\mathrm{d}t} \ll \frac{\mathrm{d}X_B}{\mathrm{d}t}\Big|_{f_P=\pi/2-\omega_P}
\end{align}

With this approximation, a first-order expansion of the TDV-V waveform is 
given by:

\begin{align}
\mathrm{TDV-V} &= - \frac{ \frac{\mathrm{d}(X_P-X_B)}{\mathrm{d}t} }{ \frac{\mathrm{d}X_B}{\mathrm{d}t}\Big|_{f_P=\pi/2-\omega_P} } \tilde{T}_B \\
\qquad&= \tilde{T}_B \Bigg(\frac{a_S M_S P_P}{a_P M_P P_S}\Bigg) \Bigg(\frac{\sqrt{1-e_P^2}}{\sqrt{1-e_S^2} (1+e_P\sin\omega_P)}\Bigg) \Lambda_{\mathrm{TDV-V}}(f_S) \\
\Lambda_{\mathrm{TDV-V}}(f_S) &= \cos\varpi_S [e_S\sin\omega_S+\sin(f_S+\omega_S)] + \sin i_S \sin\varpi_S [e_S\cos\omega_S+\cos(f_S+\omega_S)] 
\label{eqn:TDVwaveform}
\end{align}

Integrating over $f_S$, as was done for Equation~(\ref{eqn:TTVrms}), one obtains
the RMS TDV-V amplitude:

\begin{align}
\delta_{\mathrm{TDV-V}} &= \sqrt{\frac{1}{2\pi} \int_{f_S=0}^{2\pi} [\mathrm{TDV-V}(f_S)]^2\,\mathrm{d}f_S } \nonumber \\
\delta_{\mathrm{TDV-V}} &= \tilde{T}_B \Bigg(\frac{a_S M_S P_P}{a_P M_P P_S}\Bigg) \Bigg(\frac{\sqrt{1-e_P^2}}{\sqrt{1-e_S^2} (1+e_P\sin\omega_P)}\Bigg) \sqrt{\frac{\Phi_{\mathrm{TDV-V}}}{2\pi}} \\
\Phi_{\mathrm{TDV-V}} &= \frac{\pi}{8} \Bigg[ -2e_S^2\cos2\omega_S [1+3\cos2\varpi_S] \nonumber \\ 
\qquad& + (1+e_S^2) [ 6+\cos2(i_S-\varpi_S)+2\cos2\varpi_S+\cos2(i_S+\varpi_S) ] \nonumber \\
\qquad& - 2\cos2i_S [1+e_S^2+2e_S^2\cos2\omega_S\sin^2\varpi_S)] + 8e_S^2\sin i_S \sin2\omega_S\sin2\varpi_S \Bigg]
\label{eqn:TDVrms}
\end{align}

As with $\Phi_{\mathrm{TTV}}$, $\Phi_{\mathrm{TDV-V}} \rightarrow \pi$ for
circular, co-aligned orbits.
As was provided for the TTV effect, it is useful to write the TDV-V amplitude
in astrophysical units:

\begin{align}
\delta_{\mathrm{TDV-V}} &= \Big(13.4\,\mathrm{s}\Big) \Bigg(\frac{1}{\mathfrak{D}}\Bigg)^{1/2} \Bigg(\frac{\tilde{T}_B}{10\,\mathrm{hrs}}\Bigg) \Bigg(\frac{M_S}{M_{\oplus}}\Bigg) \Bigg(\frac{M_J}{M_P}\Bigg)^{2/3} \Bigg(\frac{M_{\odot}}{M_*}\Bigg)^{1/3} \Bigg(\frac{\sqrt{1-e_P^2}}{\sqrt{1-e_S^2} (1+e_P\sin\omega_P)}\Bigg) \sqrt{\frac{\Phi_{\mathrm{TDV-V}}}{2\pi}}
\end{align}

\subsection{Properties of the TDV-V RMS Amplitude}
\label{sec:tdvvproperties}

As was done in \S\ref{sec:ttvproperties}, I will here investigate the
properties of the derived TDV-V RMS amplitude.
In the case of a moon on a circular orbit, a very simple
expression for $\Phi_{\mathrm{TDV-V}}$ is possible, which happens to be
equivalent to $\Phi_{\mathrm{TTV}}$ for the same conditions.

\begin{align}
\lim_{e_S \to 0} \Phi_{\mathrm{TDV-V}} &= \pi \Big( 1 - \cos^2i_S\sin^2\varpi_S\Big) \nonumber \\
\qquad&= \lim_{e_S \to 0} \Phi_{\mathrm{TTV}}
\label{eqn:phitdvvinc}
\end{align}

\begin{figure}
\begin{center}
\includegraphics[width=10.0 cm]{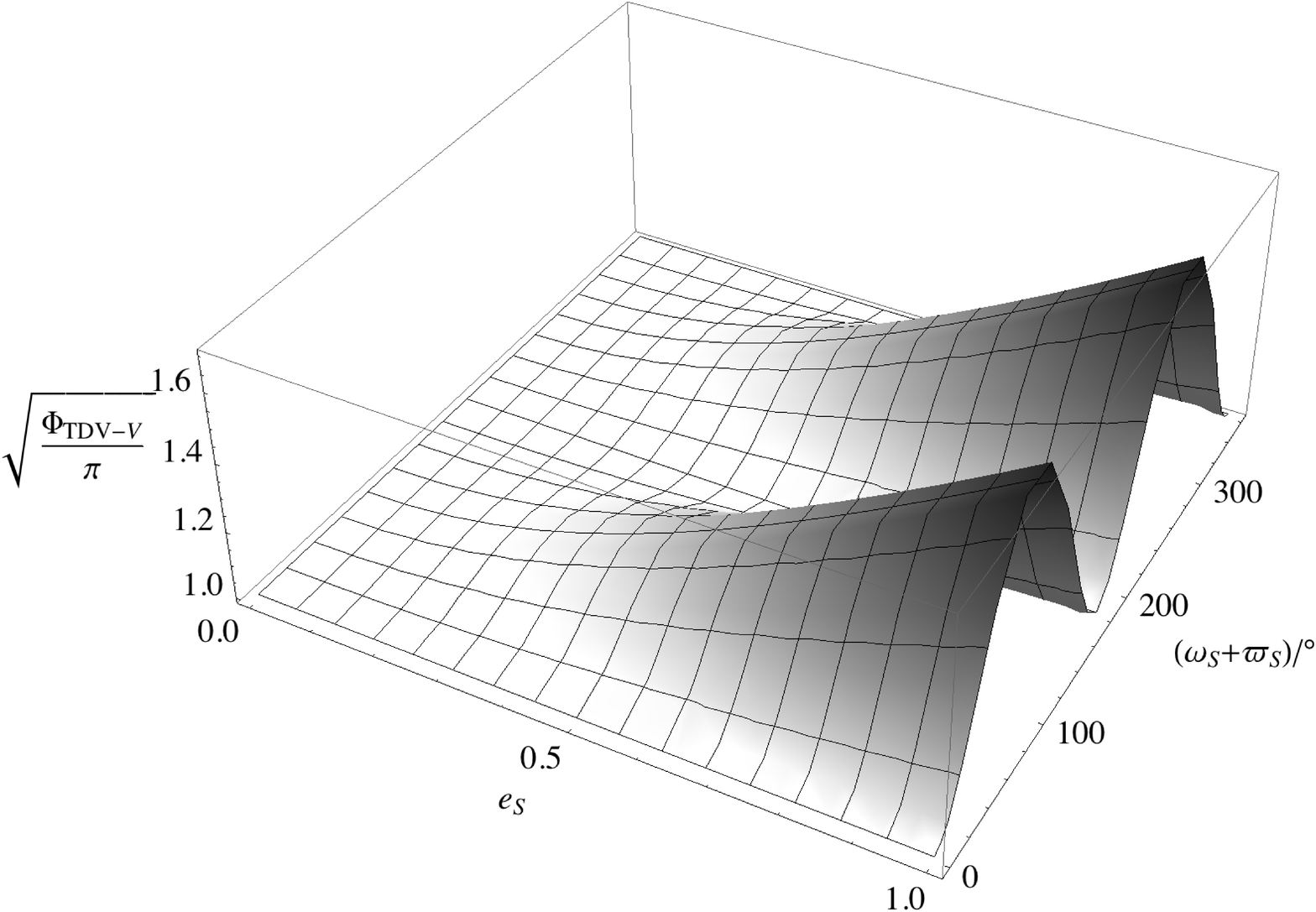}
\caption[Dependency of $\Phi_{\mathrm{TDV-V}}$, the TDV-V enhancement factor, on 
an exomoon's eccentricity and the sum of the argument of periapsis and the 
longitude of the periapsis]
{\emph{Dependency of $\Phi_{\mathrm{TDV-V}}$, the TDV-V enhancement factor, on 
an exomoon's eccentricity, $e_S$, and the sum of the argument of periapsis and 
the longitude of the periapsis, $(\omega_S+\varpi_S)$,
In this plot, I have assumed $i_S=\pi/2$.}} 
\label{fig:phitdvvecc}
\end{center}
\end{figure}

Since this function is plotted in Figure~\ref{fig:phittvinc}, it will not be
repeated here.
The effects of eccentricity, $e_S$ are illustrated in
Figure~\ref{fig:phitdvvecc}. As was done for the TTV case, I set $i_S=\pi/2$
which simplifies $\Phi_{\mathrm{TDV-V}}$ to:

\begin{align}
\lim_{i_S \to \pi/2} \Phi_{\mathrm{TDV-V}} = \pi \Big( 1 + 2 e_S^2 \sin^2(\omega_S+\varpi_S)\Big)
\label{eqn:phitdvvecc}
\end{align}

Figure~\ref{fig:phitdvvecc} plots Equation~(\ref{eqn:phitdvvecc}), which shows
how eccentricity always leads to an enhancement of the TDV-V amplitude. As for
the TTV case, the function is total controlled by one single angle, given
by the sum of the argument of periapsis and the longitude of the periapsis.

\section{TTV \& TDV-V as Complementary Methods}

\subsection{Mass and Period Determination}
\label{sec:moonmassdetermination}

I here define $\eta$ as the ratio of the TDV to the TTV RMS amplitudes. Each
of these amplitudes are derivable through searches for excess variance 
and so $\eta$ is readily measureable.

\begin{align}
\eta &= \frac{\delta_{\mathrm{TDV}}}{\delta_{\mathrm{TTV}}} \nonumber \\
\qquad&= \frac{\tilde{T}_B}{2\pi P_S} \frac{1}{(1-e_S^2)^{3/2}} \sqrt{\frac{\Phi_{\mathrm{TDV-V}}}{\Phi_{\mathrm{TTV}}}}
\label{eqn:etasimple}
\end{align}

If one makes the reasonable approximation that $e_S \ll 1$, which is expected
for dynamically stable exomoons \citep{domingos2006}, then one may make use of
the fact:

\begin{align}
\lim_{e_S \to 0} \Phi_{\mathrm{TTV}} = \lim_{e_S \to 0} \Phi_{\mathrm{TDV-V}} = \pi \Big(1-\cos^2i_S\sin^2\varpi_S\Big)
\end{align}

Therefore, for $e_S\ll1$, Equation~(\ref{eqn:etasimple}) becomes:

\begin{align}
\lim_{e_S \to 0} \eta &= \frac{\tilde{T}_B}{2\pi P_S}
\label{eqn:etasimplecirc}
\end{align}

The elegant Equation~(\ref{eqn:etasimplecirc}) shows that the measurement of a 
TTV and TDV amplitude due to an exomoon allows one to directly determine $P_S$. 
Armed with $P_S$ and Kepler's Third Law, one may use either 
$\delta_{\mathrm{TTV}}$ or $\delta_{\mathrm{TDV-V}}$ to obtain $M_S$ as well.

A further possibility to use the derived period to look for the closest
harmonic frequency in the periodogram which would further constrain $P_S$
and consequently $M_S$. Alternatively, one could use the harmonic period to
estimate $\sqrt{(1-e_S^2)^{-3}\Phi_{\mathrm{TDV-V}}/\Phi_{\mathrm{TTV}}}$,
which essentially characterizes how non-circular and non-coplanar the moon is.

Therefore, combining TTV and TDV together allows one to determine both the mass
and period of the exomoon, which satisfies the first critical desideratum
outlined in \S\ref{sec:ttvproblems}. Now just one problem remains - the
requirement for a unique signature of an exomoon.

\subsection{Phase Difference}
\label{sec:phasedifference}

Consider the simple case of $i_S\simeq\pi/2$ and $\Omega_S \simeq 0$
and $e_S = e_P = 0$. These conditions may seem limited but actually are
dynamically expected as the Hill sphere of stable moon orbits shrink as the
eccentricity and inclination angles move away from coplanarity 
\citep{donnison2010}. Under this simple circumstance, the waveform components of 
the two signals become:

\begin{align}
\lim_{e_P,e_S\to0} \lim_{\Omega_S\to0} \lim_{i_S\to\pi/2} \Lambda_{\mathrm{TTV}} &= \cos f_S \nonumber \\
\lim_{e_P,e_S\to0} \lim_{\Omega_S\to0} \lim_{i_S\to\pi/2} \Lambda_{\mathrm{TDV-V}} &= \sin f_S
\end{align}

Therefore, the TDV-V leads TTV by a $\pi/2$ phase shift. This
phase difference is paramount - it is the key to unlocking an exomoon detection
through timing effects. By detecting both signals and observing this phase
shift, the detection would be unambiguous against all other phenomenon, since
presently no other effect is predicted to induce such a phase shift. In fact,
no other phenomena are predicted to even induce both periodic (and detectable) 
TTVs \textbf{and} TDVs.

The origin of the phase shift can be understood by considering that the planet's
local Keplerian orbit appears as simple harmonic motion (SHM) on the sky-plane. 
Since TTV is a position effect and TDV-V is a velocity effect, then  
just like in SHM with a swinging pendulum, the velocity and position will always
be $90^{\circ}$ out of phase, and so too are the TTV and TDV-V effects.

In practice, the effects of orbital eccentricity and non-coalignment distort
the phase difference away from $\pi/2$. However, one expects such systems to be
relatively rare, due to the contracted regions of stability 
\citep{donnison2010}. Therefore, the existence of this phase shift acts as a 
unique signature for an exomoon and discriminates the timing signals from other 
phenomena.

For a retrograde orbit, $i_S\simeq3\pi/2$, but the phase shift is preserved. 
This tell us that the ratio of the TDV-V to the TTV effect provides no 
information about the sense of orbital motion of the moon:

\begin{align}
\lim_{e_P,e_S\to0} \lim_{\Omega_S\to0} \lim_{i_S\to3\pi/2} \Lambda_{\mathrm{TTV}} &= \cos f_S \nonumber \\
\lim_{e_P,e_S\to0} \lim_{\Omega_S\to0} \lim_{i_S\to3\pi/2} \Lambda_{\mathrm{TDV-V}} &= \sin f_S
\end{align}

With the ability to determine both the exomoon mass and discriminate it against 
other phenomena, both of the problems described with TTV alone have been solved 
(\S\ref{sec:ttvproblems}). TDV-V can therefore be seen to be the key to solving
the problem of detecting exomoons through timing effects. With this, I have now
met the original goal of this thesis - a method to detect exomoons in transiting
systems. However, in the next section, I will discuss a additional complication
which arises for non-coplanar configurations.

\section{Transit Impact Parameter induced Transit Duration Variations (TDV-TIP)}
\label{sec:tip}

\subsection{Definition}
\label{sec:tipdefinition}

\begin{figure}
\begin{center}
\includegraphics[width=10.0 cm]{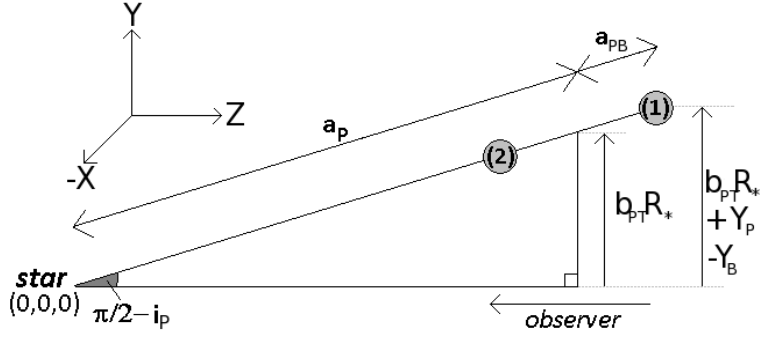}
\caption[Cartoon illustrating the TDV-TIP effect]
{\emph{Cartoon of the TDV-TIP effect. Here, the moon is relaxed into the
same orbital plane as the planet's orbit and causes the planet to experience
reflex motion illustrated by the two positions of the planet, (1) and (2). This
motion can be seen to cause a change in the apparent impact parameter, which 
causes a change in the transit duration.}} 
\label{fig:tipexplan}
\end{center}
\end{figure}

Not long after the prediction of TDV-V in \citet{kipping2009a}, a follow-on
paper soon predicted that an additional component to TDV should also exist
for non-coplanar systems \citep{kipping2009b}. The premise was that if the plane
of the planet-moon orbit was inclined to the line-of-sight, then the planet
would have a non-zero component of motion in the $\hat{Y}$-direction i.e.
$|\Delta Y_{P}|>0$ (illustrated in Figure~\ref{fig:tipexplan}). This motion 
would change the transit impact parameter and since the transit duration is 
highly sensitive to this term, then another form of TDV should occur.

TDV-TIP can be completely understood in terms of $\Delta Y_{P}$ and how it 
modifies the apparent transit impact parameter. The modified transit duration, 
using the $\tilde{T}^{\mathrm{one}}$ approximation, would be given by:

\begin{align}
\tilde{T}_P^{\mathrm{TIP}} &= \frac{P_P}{\pi} \frac{\varrho_{P,T}^2}{\sqrt{1-e_P^2}} \arcsin\sqrt{ \frac{1 - [b_{p,T}+(\Delta Y_{P}/R_*)]^2}{(a_P/R_*)^2 \varrho_{P,T}^2 - [b_{P,T}+(\Delta Y_{P}/R_*)]^2} }
\end{align}

Where the ``TIP'' superscript indicates that this definition of the duration
only accounts for the TDV-TIP effect and not the TDV-V. Since the V component
is just a scaling factor, it can be applied \emph{after} the TIP component
has been incorporated. This two-step process allows one
to write the transit duration accounting for both V and TIP components as:

\begin{align}
\tilde{T}_P &= \frac{ \dot{X_{B}} }{ \dot{X_{B}} + \dot{\Delta X_{P}} } \tilde{T}_P^{\mathrm{TIP}}
\end{align}

At this point, it is useful to define $\tilde{T}_P^{\mathrm{TIP}}$ as some
factor multiplied by the barycentric transit duration, $\tilde{T}_B$:

\begin{align}
\epsilon &= \frac{\tilde{T}_P^{\mathrm{TIP}}}{\tilde{T}_B}
\end{align}

Note that $\tilde{T}$ increases as the impact parameter approaches zero. Since
$Y_B$ is generally positive, then increases in the transit duration occur when
$\Delta Y_{P}$ is negative and decreases occur when $\Delta Y_{P}$
is positive. Mathematically, $\epsilon>1$ for $\Delta Y_{P}<0$ and $\epsilon<1$ 
for $\Delta Y_{P}>0$. The total TDV effect may be now written as:

\begin{align}
\mathrm{TDV} &= \Bigg[ \frac{ \epsilon \dot{X_{B}} }{ \dot{X_{B}} + \dot{\Delta X_{P}} } -1 \Bigg] \tilde{T}_B
\end{align}

Using assumption $\alpha$7, this simplifies to:

\begin{align}
\mathrm{TDV} &= \Bigg[ (\epsilon-1) - \frac{ \epsilon \dot{\Delta X_{P}} }{ \dot{X_{B}} } \Bigg] \tilde{T}_B
\end{align}

One can now see that the TDV signal has two clear components. In the absence of
any TIP-component, $\epsilon\rightarrow1$ and one recovers the TDV-V effect
seen earlier. Given that $\epsilon$ is of order unity, it is useful to write it
as:

\begin{equation}
\epsilon = 1 + \varphi
\end{equation}

Where it is understood that $\varphi$ is small compared to unity. Expanding out,
the TDV effect now becomes:

\begin{align}
\mathrm{TDV} &= \Bigg[ \varphi - \frac{ \dot{\Delta X_{P}} }{ \dot{X_{B}} } - \frac{ \varphi \dot{\Delta X_{P}} }{ \dot{X_{B}} } \Bigg] \tilde{T}_B
\end{align}

Both $(\dot{\Delta X_{P}}/\dot{X_{B}})$ and $\varphi$ are small terms and thus
the cross-product of the two gives an even smaller term i.e. it is second order.
Therefore, the total TTV effect may be expressed as:

\begin{align}
\mathrm{TDV} &= \Bigg[ \varphi - \frac{ \dot{\Delta X_{P}} }{ \dot{X_{B}} } \Bigg] \tilde{T}_B \nonumber \\
\mathrm{TDV} &= (\mathrm{TDV-TIP}) + (\mathrm{TDV-V}) \\
\mathrm{TDV-TIP} &= \varphi \tilde{T}_B
\end{align}

Critically, TDV may be treated as a linear combination of the TDV-V and TDV-TIP
components, which grossly simplifies the subsequent analysis.

\subsection{Derivation}

To provide the TDV-TIP waveform and RMS amplitude, it is necessary to derive
$\varphi$, which is a function of $\Delta Y_{P}$. $\epsilon$ is given by:

\begin{align}
\epsilon &= \frac{1}{\tilde{T}_B} \frac{P_P}{\pi} \frac{\varrho_{P,T}^2}{\sqrt{1-e_P^2}} \arcsin\sqrt{ \frac{1 - [b_{P,T}+(\Delta Y_{P}/R_*)]^2}{(a_P/R_*)^2 \varrho_{P,T}^2 - [b_{P,T}+(\Delta Y_{P}/R_*)]^2} }
\end{align}

Assuming $(a_P/R_*)^2\varrho_{P,T}^2\gg b_{P,T}^2$ and using a small-angle
approximation, $\epsilon$ may be written as:

\begin{align}
\epsilon^2 &= \frac{1 - [b_{P,T} + (\Delta Y_{P}/R_*)]^2}{1 - [b_{P,T}]^2}
\end{align}

Note that the above is true even for eccentric orbits. Expanding out the 
squares:

\begin{align}
\epsilon^2 &= \frac{1 - b_{P,T}^2 - 2 b_{P,T} (\Delta Y_{P}/R_*) - \mathcal{O}[(\Delta Y_{P}/R_*)^2]}{1 - b_{P,T}^2} \nonumber \\
\epsilon &= \sqrt{1 - \frac{2 b_{P,T} (\Delta Y_{P}/R_*)}{1-b_{P,T}^2}}
\end{align}

Where the second line has made the reasonable approximation that 
$\mathcal{O}[(\Delta Y_{P}/R_*)^2]$ is small, since it is second order. Since one 
knows that $\epsilon$ is of order unity due to its definition as the ratio of 
the transit duration with and without the TDV effect, then the term subtracted 
from unity inside the square root must be small i.e. $\varphi \ll 1$. Therefore, 
a Taylor expansion to first-order gives:

\begin{align}
\varphi \simeq -\frac{b_{P,T} (\Delta Y_{P}/R_*)}{1-b_{P,T}^2}
\end{align}

This tells us that for $\Delta Y_{P}>0$, $\varphi<0$ indicating $\epsilon<1$ 
which in turn means the duration has decreased, which follows the expected 
behaviour. The TDV-TIP waveform is therefore:

\begin{align}
\mathrm{TDV-TIP} &= \tilde{T}_B \Bigg( \frac{b_{P,T}}{1-b_{P,T}^2} \Bigg) \Bigg( \frac{a_S M_S (1-e_S^2)}{R_* M_P}\Bigg) \Lambda_{\mathrm{TDV-TIP}} \nonumber \\
\Lambda_{\mathrm{TDV-TIP}} &= \frac{\sin(f_S+\omega_S) [-\cos i_S \sin i_P + \sin i_S \cos i_P \cos\varpi_S] + \cos i_P \sin\varpi_S\cos(\omega_S+f_S)}{1+e_S\cos f_S}
\label{eqn:TIPwaveform}
\end{align}

In \citet{kipping2009b}, I had been unable to solve the integral of 
$\Lambda_{\mathrm{TDV-TIP}}^2$, but I here present a solution found through
a decomposition of the various elements, integrated and then re-combined:

\begin{align}
\delta_{\mathrm{TDV-TIP}} &= \tilde{T}_B \Bigg( \frac{b_{P,T}}{1-b_{P,T}^2} \Bigg) \Bigg( \frac{a_S M_S (1-e_S^2)}{R_* M_P}\Bigg) \sqrt{\frac{\Phi_{\mathrm{TDV-TIP}}}{2\pi}} \nonumber \\
\Phi_{\mathrm{TDV-TIP}} &= \pi e_S^{-2} (1-e_S^2)^{-3/2} \Bigg[ -\Bigg(-e_S^2 + \Big( e_S^2 (3 - 2\sqrt{1-e_S^2}) + 2 (-1+\sqrt{1-e_S^2}) \Big) \cos2\omega_S \Bigg) \nonumber \\
\qquad&  \times [\cos i_S\sin i_P-\cos i_P\cos\varpi_S\sin i_S]^2 \nonumber \\
\qquad& + 2 \Bigg( 2 - 2\sqrt{1-e_S^2} + e_S^2(-3+2\sqrt{1-e_S^2}) \Bigg)\cos i_P \sin2\omega_S\sin\varpi_S \nonumber \\
\qquad& \times [-\cos i_S\sin i_P+\cos i_P \cos\varpi_S\sin i_S] \nonumber \\
\qquad& -\cos^2i_P \sin^2\varpi_S \Bigg(-e_S^2+\cos2\omega_S\Big(2 - 2\sqrt{1-e_S^2}+e_S^2(-3+2\sqrt{1-e_S^2})\Big) \Bigg) \Bigg]
\label{eqn:TIPrms}
\end{align}

In astrophysical units, this becomes:

\begin{align}
\delta_{\mathrm{TDV-TIP}} &= \Big(7.73\,\mathrm{s}\Big) \mathfrak{D}^{1/2} \Bigg(\frac{\tilde{T}_B}{10\,\mathrm{hrs}}\Bigg) \Bigg( \frac{b_{P,T}}{1-b_{P,T}^2} \Bigg) \Bigg(\frac{a_P}{R_*}\Bigg) \Bigg(\frac{M_S}{M_{\oplus}}\Bigg) \Bigg(\frac{M_J}{M_P}\Bigg)^{2/3} \Bigg(\frac{M_{\odot}}{M_*}\Bigg)^{1/3} (1-e_S^2) \sqrt{\frac{\Phi_{\mathrm{TDV-TIP}}}{2\pi}}
\end{align}

\subsection{Properties of the TDV-TIP RMS Amplitude}

As with the other $\Phi$ terms, $\Phi_{\mathrm{TDV-TIP}} \rightarrow \pi$ in
the nominal case. Unlike, the other $\Phi$ terms though, this nominal case
occurs for $i_S=0$ rather than $i_S=\pi/2$ since this effect is non-coplanar, 
whereas the others were coplanar.
Taking the limit for circular orbits and $\varpi_S=0$, $\Phi_{\mathrm{TDV-TIP}}$
becomes:

\begin{align}
\lim_{e_S \to 0} \lim_{\varpi_S \to 0} \Phi_{\mathrm{TDV-TIP}} &= \pi \sin^2(i_P-i_S)
\end{align}

\begin{figure}
\begin{center}
\includegraphics[width=10.0 cm]{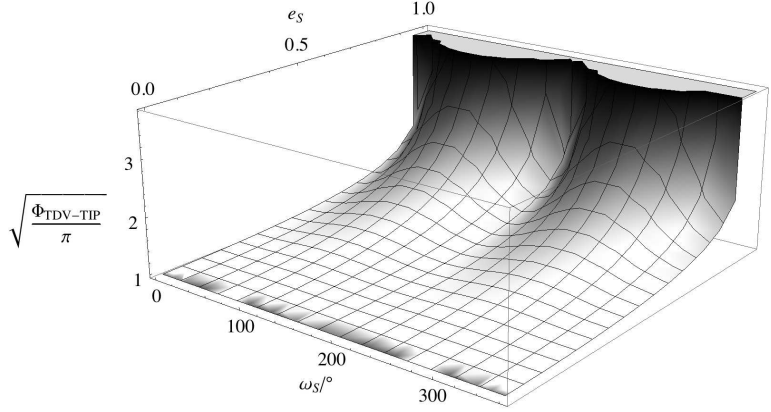}
\caption[Dependency of $\Phi_{\mathrm{TDV-TIP}}$, the TDV-TIP scaling factor,
as a function of the satellite's eccentricity and argument of periapsis for
$i_P=\pi/2$ and $i_S=0$.]
{\emph{Dependency of $\Phi_{\mathrm{TDV-TIP}}$, the TDV-TIP scaling factor,
as a function of the satellite's eccentricity and argument of periapsis for
$i_P=\pi/2$ and $i_S=0$.}} 
\label{fig:phitipecc}
\end{center}
\end{figure}

This indicates that maximum TIP effects occur for $i_P-i_S=n (\pi/2)$, where $n$
is any real integer. This prediction is exactly what one would expect 
intuitionally, as it corresponds to the planet's reflex motion being orthogonal
to the $\hat{X}$-direction. For the eccentricity dependency,
I take the limit for $i_S=0$, $\varpi_S=0$ and $i_P=\pi/2$ and then one may plot 
$\Phi_{\mathrm{TDV-TIP}}$ purely as a function of $e_S$ and $\omega_S$, 
as shown in Figure~\ref{fig:phitipecc}.

Another interesting point to bear in mind with TDV-TIP, is that one does not
require the moon to be inclined relative to the planet's orbital plane to still
have a TIP effect. The only requirement is that $i_P - i_S \neq n \pi $.
Therefore, if $i_P \neq \pi/2$, then a TIP effect is guaranteed unless one has
the coincidence that $i_S = n \pi+i_P$. Consider the case of 
$e_S\ll1$, and $i_S = \pi/2$, the dependency on $\varpi_S$ naturally drops out
leaving one with:

\begin{align}
\lim_{e_S \to 0} \lim_{i_S \to \pi/2} \Phi_{\mathrm{TDV-TIP}} &= \pi \cos^2i_P
\end{align}

This last equation is particularly useful because one expects moons to exhibit 
circular, coplanar configurations based upon the contracted region of orbital 
stability \citep{donnison2010} but there is no reason why the planet's orbital
plane should coincidentally align to the observer's line-of-sight.

\subsection{The Total TDV Effect}

As shown in \S\ref{sec:tipdefinition}, the total TDV effect may be treated as
a linear combination of the two components. This yields:

\begin{align}
\mathrm{TDV} &= \tilde{T}_B \Bigg(\frac{a_S M_S P_P}{a_P M_P P_S}\Bigg) \Bigg(\frac{\sqrt{1-e_P^2}}{(1+e_P\sin\omega_P) \sqrt{1-e_S^2}} \Bigg) \Lambda_{\mathrm{TDV-V}} \nonumber \\
\qquad& + \tilde{T}_B \Bigg(\frac{b_{P,T}}{1-b_{P,T}^2}\Bigg) \Bigg( \frac{a_S M_S (1-e_S^2)}{R_* M_P}\Bigg) \Lambda_{\mathrm{TDV-TIP}}
\end{align}

The RMS amplitude integral will be given by:

\begin{align}
2\pi \delta_{\mathrm{TDV}}^2 &= \int_{f_S=0}^{2\pi} \tilde{T}_B^2 \Bigg(\frac{a_S M_S P_P}{a_P M_P P_S}\Bigg)^2 \Bigg(\frac{\sqrt{1-e_P^2}}{(1+e_P\sin\omega_P) \sqrt{1-e_S^2}} \Bigg)^2 \Lambda_{\mathrm{TDV-V}}^2 \nonumber \\
\qquad& +2\tilde{T}_B^2 \Bigg(\frac{a_S M_S P_P}{a_P M_P P_S}\Bigg) \Bigg(\frac{\sqrt{1-e_P^2}}{(1+e_P\sin\omega_P) \sqrt{1-e_S^2}} \Bigg) \Bigg(\frac{b_{P,T}}{1-b_{P,T}^2}\Bigg) \Bigg( \frac{a_S M_S (1-e_S^2)}{R_* M_P}\Bigg) \Lambda_{\mathrm{TDV-V}} \Lambda_{\mathrm{TDV-TIP}} \nonumber \\
\qquad& + \tilde{T}_B^2 \Bigg(\frac{b_{P,T}}{1-b_{P,T}^2}\Bigg)^2 \Bigg( \frac{a_S M_S (1-e_S^2)}{R_* M_P}\Bigg)^2 \Lambda_{\mathrm{TDV-TIP}}^2 \,\mathrm{d}f_S
\end{align}

The cross-term is new, but the two squared terms were integrated previously:

\begin{align}
2\pi \delta_{\mathrm{TDV}}^2 &= \tilde{T}_B^2 \Bigg(\frac{a_S M_S P_P}{a_P M_P P_S}\Bigg)^2 \Bigg(\frac{\sqrt{1-e_P^2}}{(1+e_P\sin\omega_P) \sqrt{1-e_S^2}} \Bigg)^2 \Phi_{\mathrm{TDV-V}} \nonumber \\
\qquad& + 2\tilde{T}_B^2 \Bigg(\frac{a_S^2 M_S^2 P_P}{R_* a_P M_P^2 P_S}\Bigg) \Bigg(\frac{\sqrt{1-e_P^2}\sqrt{1-e_S^2}}{(1+e_P\sin\omega_P)} \Bigg) \Bigg(\frac{b_{P,T}}{1-b_{P,T}^2}\Bigg) \Phi_{\mathrm{cross}} \nonumber \\
\qquad& + \tilde{T}_B^2 \Bigg(\frac{b_{P,T}}{1-b_{P,T}^2}\Bigg)^2 \Bigg( \frac{a_S M_S (1-e_S^2)}{R_* M_P}\Bigg)^2 \Phi_{\mathrm{TDV-TIP}} 
\label{eqn:tottdv1}
\end{align}

Where the $\Phi_{\mathrm{cross}}$ term is given by:

\begin{align}
\Phi_{\mathrm{cross}} &= \int_{f_S=0}^{2\pi} \Lambda_{\mathrm{TDV-V}} \Lambda_{\mathrm{TDV-TIP}}\,\mathrm{d}f_S \nonumber \\
\Phi_{\mathrm{cross}} &= -\frac{\pi}{e_S^2} \Bigg[ \frac{\cos i_P \sin i_S \sin^2\varpi_S [2e_S^2\cos^2\omega_S (\sqrt{1-e_S^2}-1) - \cos2\omega_S(e_S^2+2\sqrt{1-e_S^2}-2) + e_S^2 ]}{\sqrt{1-e_S^2}} \nonumber \\
\qquad& + \sin i_S \sin2\omega_S\sin\varpi_S [\cos i_P\sin i_S \cos \varpi_S - \sin i_P\cos i_S] [e_S^2+2\sqrt{1-e_S^2}-2] \nonumber \\
\qquad& + \cos2\varpi_S [\cos i_P\sin i_S\cos\varpi_S-\sin i_P\cos i_S] [e_S^2-\cos2\omega_S(e_S^2+2\sqrt{1-e_S^2}-2)] \nonumber \\
\qquad& +\frac{1}{2} \cos i_P \sin2\omega_S\sin2\varpi_S [e_S^2+2\sqrt{1-e_S^2}-2] \Bigg]
\end{align}

The cross-term, whose ``parents'' could be considered to be 
$\Phi_{\mathrm{TDV-V}}$ and $\Phi_{\mathrm{TDV-TIP}}$, mimics the properties
of the $\Phi_{\mathrm{TDV-TIP}}$ parent in the sense of its nominal case.
Like $\Phi_{\mathrm{TDV-TIP}}$, $\Phi_{\mathrm{cross}}\rightarrow\pi$ when
$i_S\rightarrow0$ rather than $\pi/2$. Therefore, the cross-term describes an
out-of-the-plane effect. This is supported by the fact 
$\Phi_{\mathrm{TDV-TIP}}\rightarrow0$ when $i_S\rightarrow0$ for a coplanar 
planet.
One useful limiting case to consider is that of a circular moon, which yields:

\begin{align}
\lim_{e_S \to 0} \Phi_{\mathrm{cross}} = \pi \Big( \cos i_S \sin i_P \cos\varpi_S - \cos i_P \sin i_S \Big)
\end{align}

\subsection{The Consequences for $\eta$}

$\eta$, as defined in Equation~(\ref{eqn:etasimple}), is the ratio of the TDV
to TTV RMS amplitudes and was shown earlier to allow one to directly determine
$P_S$ and then $M_S$, thus solving Problem~1 in \S\ref{sec:ttvproblems}. With
the TDV-TIP effect now recognized, the expression for $\eta$ will become more
complicated.

Equation~(\ref{eqn:tottdv1}) presented the total TDV RMS amplitude as the 
quadrature sum of the two squares of the individual components plus a cross-term.
It is useful to complete the square in Equation~(\ref{eqn:tottdv1}). If one
defines 
$\mathcal{W}_{\mathrm{TDV-V}} = \mathrm{TDV-V}/\Lambda_{\mathrm{TDV-V}}$
and similarly for the TDV-TIP effect, then one may write:

\begin{align}
2 \pi \delta_{\mathrm{TDV}}^2 &= \mathcal{W}_{\mathrm{TDV-V}}^2 \Phi_{\mathrm{TDV-V}}^2 + \mathcal{W}_{\mathrm{TDV-TIP}}^2 \Phi_{\mathrm{TDV-TIP}}^2 + 2 \mathcal{W}_{\mathrm{TDV-V}} \mathcal{W}_{\mathrm{TDV-TIP}} \Phi_{\mathrm{cross}} \nonumber \\
\qquad&= \Big[\mathcal{W}_{\mathrm{TDV-V}} \Phi_{\mathrm{TDV-V}} + \mathcal{W}_{\mathrm{TDV-TIP}} \Phi_{\mathrm{TDV-TIP}}\Big]^2 + 2 \mathcal{W}_{\mathrm{TDV-V}} \mathcal{W}_{\mathrm{TDV-TIP}} \Big[ \Phi_{\mathrm{cross}} - \Phi_{\mathrm{TDV-V}} \Phi_{\mathrm{TDV-TIP}} \Big] \nonumber \\
\qquad&= \Big[\mathcal{W}_{\mathrm{TDV-V}} \Phi_{\mathrm{TDV-V}} + \mathcal{W}_{\mathrm{TDV-TIP}} \Phi_{\mathrm{TDV-TIP}}\Big]^2 + 2 \mathcal{W}_{\mathrm{TDV-V}} \mathcal{W}_{\mathrm{TDV-TIP}} \Delta\Phi
\end{align}

Where I have defined 
$\Delta \Phi = \Phi_{\mathrm{cross}} - \Phi_{\mathrm{TDV-V}} \Phi_{\mathrm{TDV-TIP}}$.
In order for the squares and square roots to cancel, one ideally wants 
$\Delta\Phi$ to be small. Assuming $e_S \ll 1$ and $i_S \simeq \pi/2$ in line
with the dynamical expectations \citep{donnison2010}, $\Delta \Phi$ is given by: 

\begin{align}
\lim_{e_S \to 0} \lim_{i_S \to \pi/2} \lim_{\varpi_S \to 0} \Delta \Phi = \pi \cos i_P + \mathcal{O}[\cos^2i_P]
\end{align}

Since the other $\Phi$ terms are of order unity and $\cos i_P \ll 1$, then in 
general, one can see that the $\Delta\Phi$ term will be negligible. In such a 
circumstance, the total TDV amplitude becomes:

\begin{align}
\delta_{\mathrm{TDV}} &\simeq \tilde{T}_B \Bigg(\frac{a_S M_S P_P}{a_P M_P P_S}\Bigg) \Bigg(\frac{\sqrt{1-e_P^2}}{(1+e_P\sin\omega_P) \sqrt{1-e_S^2}} \Bigg) \sqrt{\frac{\Phi_{\mathrm{TDV-V}}}{2\pi}} \nonumber \\
\qquad& + \tilde{T}_B \Bigg(\frac{b_{P,T}}{1-b_{P,T}^2}\Bigg) \Bigg( \frac{a_S M_S (1-e_S^2)}{R_* M_P}\Bigg) \sqrt{\frac{\Phi_{\mathrm{TDV-TIP}}}{2\pi}}
\label{eqn:tottdv2}
\end{align}

The expression for $\eta$ now becomes:

\begin{align}
\eta &= \frac{\delta_{\mathrm{TDV}}}{\delta_{\mathrm{TTV}}} \nonumber \\
\qquad&= \frac{\tilde{T}_B}{2\pi P_S} \frac{1}{(1-e_S^2)^{3/2}} \sqrt{\frac{\Phi_{\mathrm{TDV-V}}}{\Phi_{\mathrm{TTV}}}} \nonumber \\
\qquad& + \frac{\tilde{T}_B}{2\pi P_B} \frac{a_P}{R_*} \Bigg(\frac{b_{P,T}}{1-b_{P,T}^2}\Bigg) \Bigg(\frac{1+e_P\sin\omega_P}{\sqrt{1-e_P^2}}\Bigg)  \sqrt{\frac{\Phi_{\mathrm{TDV-TIP}}}{\Phi_{\mathrm{TTV}}}}
\label{eqn:etacomplex}
\end{align}

Now consider the case of $e_S\ll1$ and $i_S\simeq\pi/2$, as was done previously:

\begin{align}
\lim_{e_S \to 0} \lim_{i_S \to \pi/2} \eta &= \frac{\tilde{T}_B}{2\pi P_S} \sqrt{\frac{\Phi_{\mathrm{TDV-V}}}{\Phi_{\mathrm{TTV}}}} + \frac{\tilde{T}_B}{2\pi P_B} \frac{a_P}{R_*} \Bigg(\frac{b_{P,T}}{1-b_{P,T}^2}\Bigg) \sqrt{\frac{\Phi_{\mathrm{TDV-TIP}}}{\Phi_{\mathrm{TTV}}}} \nonumber \\
\qquad&= \frac{\tilde{T}_B}{2\pi P_S} + \frac{\tilde{T}_B}{2\pi P_B} \Bigg(\frac{b_{P,T}^2}{1-b_{P,T}^2}\Bigg)
\label{eqn:etacomplex2}
\end{align}

Therefore, the contribution of the TIP effect is just to add a constant onto
the $\eta$ term. This constant is purely a function of the properties of the
planet, and so can be estimated reliably.

A summary of the properties of TTV, TDV-V and TDV-TIP is provided in 
Table~\ref{tab:ttvsummary} and I provide some estimates of the TTV and TDV 
amplitudes for several transiting exoplanet systems in 
Table~\ref{tab:ttvexamples}.

\begin{table}
\caption[Key properties of various transit timing effects due to exomoons]
{\emph{Summary of key properties of the three known transit timing effects due to an exomoon.}} 
\centering 
\begin{tabular}{c c c c} 
\hline\hline 
& TTV & TDV-V & TDV-TIP \\ [0.5ex] 
\hline 
Type of effect & Positional & Velocity & Positional \\
Direction & $\hat{X}$ & $\hat{X}$ & $\hat{Y}$ \\
Proportionality & $M_S a_S$ & $M_S a_S^{-1/2}$ & $M_S a_S$ \\
Relative phase & $0$ & $\pi/2$ & $\pm \pi/2$ \\ 
Waveform Eqn & Eqn~(\ref{eqn:TTVwaveform}) & Eqn~(\ref{eqn:TDVwaveform}) & Eqn~(\ref{eqn:TIPwaveform}) \\
RMS Amplitude Eqn & Eqn~(\ref{eqn:TTVrms}) & Eqn~(\ref{eqn:TDVrms}) & Eqn~(\ref{eqn:TIPrms}) \\ [1ex]
\hline\hline 
\end{tabular}
\label{tab:ttvsummary} 
\end{table}

\begin{table*}
\caption[Transit timing amplitudes for a selection of known transiting planets,
for a hypothetical 1\,$M_{\oplus}$ exomoon]
{\emph{Predicted TTV and TDV (both V- \& TIP- components) RMS amplitudes 
due to a 1\,$M_{\bigoplus}$ exomoon at 1/3 the Hill radius, for a selection of 
candidate transiting planets.  System parameters are taken from various 
references, which are shown.}} 
\centering 
\begin{tabular}{c c c c c} 
\hline\hline 
Planet & $\delta_{TTV}$/s & $\delta_{TDV-V}$/s & $\delta_{TDV-TIP}$/s & Reference \\ [0.5ex] 
\hline 
HAT-P-11b & 19.19 & 22.54 & 0.40 & \citet{bakos2010} \\
GJ 436b & 14.12 & 13.68 & 1.30 & \citet{torres2008} \\
CoRoT-4b & 7.58 & 9.15 & 0.00 & \citet{aigrain2008} \\
OGLE-TR-111b & 4.63 & 7.32 & 0.11 & \citet{diaz2008} \\
HAT-P-1b & 4.58 & 6.82 & 0.47 & \citet{johnson2008} \\
HD 149026b & 3.61 & 9.76 & 0.00 & \citet{winn2008} \\
Lupus-TR-3b & 3.28 & 5.19 & 0.07 & \citet{weldrake2008} \\
WASP-7b & 3.26 & 5.88 & 0.00 & \citet{hellier2009a} \\
HD 17156b & 3.07 & 1.06 & 0.43 & \citet{barbieri2007} \\
TrES-1b & 3.04 & 5.95 & 0.05 & \citet{winn2007a} \\
HD2 09458b & 2.97 & 5.95 & 0.07 & \citet{kipping2008} \\
XO-5b & 2.65 & 4.69 & 0.17 & \citet{burke2008} \\
HAT-P-4b & 2.54 & 8.34 & 0.00 & \citet{kovacs2007} \\
HD 189733b & 1.52 & 2.96 & 0.16 & \citet{beaulieu2008} \\ 
XO-3b & 0.41 & 0.87 & 0.07 & \citet{winn2008b} \\ [1ex]
\hline\hline 
\end{tabular}
\label{tab:ttvexamples} 
\end{table*}

\subsection{Prograde vs Retrograde}

It was shown earlier in \S\ref{sec:phasedifference} how a retrograde orbit
does not alter the phase shift between TTV and TDV-V. Retrograde orbits are 
defined by $\pi<i_S<2\pi$, 
or alternatively as when the sense of the moon's orbital motion around the planet
is counter to that of the planet around the star. A retrograde orbit would
indicate a capture origin for the exomoon and thus its measurement would be
a major milestone in the road to understanding satellite formation outside
of the Solar System. The TDV-TIP effect offers a way to make this determination.

Consider the planet-moon barycentre moving in the $+\hat{X}$-direction in 
Figure~\ref{fig:tipexplan}. For a prograde orbit, the velocity of the planet 
around the planet-moon barycentre must be in the $+\hat{X}$-direction when it is 
at position (1).  At position (1), the transit impact parameter has increased 
and thus the transit duration has shortened.  At the same time, the planet's 
reflex velocity is additive to the planet-moon barycentre velocity around the 
host star, and so the transit duration is further shortened.  Thus for prograde 
orbits, it can be seen that the TIP- and V-components are additive. The 
reverse logic is true for retrograde orbits.

Whilst this intuitive explanation is useful, the validation may be provided
by considering the $\Lambda$ terms for each timing effect.
For the conditions used earlier in \S\ref{sec:phasedifference}, a prograde
moon has:

\begin{align}
\lim_{e_P,e_S\to0} \lim_{\Omega_S\to0} \lim_{i_S\to\pi/2} \Lambda_{\mathrm{TTV}} &= \cos f_S \nonumber \\
\lim_{e_P,e_S\to0} \lim_{\Omega_S\to0} \lim_{i_S\to\pi/2} \Lambda_{\mathrm{TDV-V}} &= \sin f_S \nonumber \\
\lim_{e_P,e_S\to0} \lim_{\Omega_S\to0} \lim_{i_S\to\pi/2} \Lambda_{\mathrm{TDV-TIP}} &= \sin f_S \cos i_P
\end{align}

But for a retrograde moon, this becomes:

\begin{align}
\lim_{e_P,e_S\to0} \lim_{\Omega_S\to0} \lim_{i_S\to3\pi/2} \Lambda_{\mathrm{TTV}} &= \cos f_S \nonumber \\
\lim_{e_P,e_S\to0} \lim_{\Omega_S\to0} \lim_{i_S\to3\pi/2} \Lambda_{\mathrm{TDV-V}} &= \sin f_S \nonumber \\
\lim_{e_P,e_S\to0} \lim_{\Omega_S\to0} \lim_{i_S\to3\pi/2} \Lambda_{\mathrm{TDV-TIP}} &= -\sin f_S \cos i_P
\end{align}

So the TDV-TIP effect can be seen to flip its phase shift. For prograde orbits,
the TDV-V and TDV-TIP effects are in phase and thus constructively interfere.
For a retrograde orbit, destructive interference occurs. If one assumes
that in general the TDV-V component is larger than the TDV-TIP component,
which can be justified by the typical examples given in 
Table~\ref{tab:ttvexamples}, then for a retrograde orbit, the constant term in 
$\eta$ becomes negative rather than positive:

\begin{align}
\lim_{e_S \to 0} \lim_{i_S \to \pi/2} \eta &= \frac{\tilde{T}_B}{P_S} \pm \frac{\tilde{T}_B}{P_B} \Bigg(\frac{b_{P,T}^2}{1-b_{P,T}^2}\Bigg)
\label{eqn:etaretro}
\end{align}

Since the constant term is controlled only by the planetary properties, this
raises the possibility of measuring the sense of orbital motion. For any
set of measurements, two versions of $\eta$ can be constructed, one for
prograde moons and one for retrograde. The orbital period is derived using
$\eta$ and then this orbital period could be checked against the periodogram
for both signals. The version of $\eta$ which provides the closest agreement
to a periodogram peak would be accepted as the real one, and thus the sense
of orbital motion could be determined (or at least some odds ratio of prograde
to retrograde).

\section{Conclusions}

In this chapter, I have provided a self-consistent model for the transit
timing effects due to an extrasolar moon. The approximations and assumptions
have been carefully considered and outlined, yet the model maintains a full
three-dimensional framework including the effects of orbital eccentricity. The
predicted effects are transit timing variations (TTV), velocity induced transit
duration variations (TDV-V) and transit impact parameter induced transit 
duration variations (TDV-TIP).

TTV and TDV-TIP are due to the changes in position of the planet whereas
TDV-V is a velocity effect. TTV by itself provides $M_S a_S$ only due to
sampling constraints, but combining TTV with TDV-V breaks the mass degeneracy
and allows one to measure $M_S$ and $a_S$ separately. Further, the two effects
exhibit a $\pi/2$ shift allowing one to unambiguously identify the signals as
being due to an exomoon, solving another long-standing problem with TTV alone.

I have also discussed non-coplanar effects, notably TDV-TIP, which is due
to kicks out of the plane. TDV-TIP is generally an order-of-magnitude (or more)
smaller than the TDV-V component (e.g. Table~\ref{tab:ttvexamples}) and so is 
usually a second-order effect. However, even for moons which are aligned to the 
planet's orbital plane, the effect persists but fortunately it does not 
invalidate the mass and period determination for the moon, it merely complicates
the analysis somewhat. In fact, combining the TDV to TTV amplitude ratio
with some harmonic information from the periodogram allows one to determine
the orbital sense of motion of the moon. Naturally, this would require a large
signal to noise in the timing effects though.

The final expressions for the RMS amplitudes of the TTV, TDV-V and TDV-TIP
effects are given in Equations~(\ref{eqn:TTVrms}), (\ref{eqn:TDVrms}) \& 
(\ref{eqn:TIPrms}) respectively and the combined TDV RMS amplitude in
Equation~(\ref{eqn:tottdv2}). Also see Table~\ref{tab:ttvsummary} for a concise
summary.

This chapter completes the theoretical aspect of TTV and TDV, but one question
which remains is how feasible a search for exomoons using these techniques
would actually be. In Chapter~\ref{ch:Chapt7}, I will present a feasibility
study using the recently launched \emph{Kepler Mission}.


%% file: Chapt7.tex
\chapter{Detectability of Habitable Exomoons with \emph{Kepler}-Class Photometry}
\label{ch:Chapt7}

\vspace{1mm}
\leftskip=4cm

{\it ``
Imagination will often carry us to worlds that never were. \\
But without it we go nowhere.
''} 

\vspace{1mm}

\hfill {\bf --- Carl Sagan, \emph{Cosmos}, 1980} 

\leftskip=0cm


\section{Introduction}
\label{sec:intro7}

In Chapter~\ref{ch:Chapt6}, I discussed how extrasolar moons may be detected
around transiting planets by searching for timing deviations in both
transit minima and duration (TTV and TDV respectively). Whilst the theory
and modelling of these effects has been covered in detail in the previous
chapter, there has been little discussion of how feasible this enterprise
actually is.

The goal of this chapter will be to provide quantitative estimates of the minimum
exomoon mass which can be detected with current technology and facilities. 
To measure these small timing
deviations, one requires a highly precise photometric instrument with
uninterrupted temporal coverage on the years time-scale. The best instrument
up to this challenge is \emph{Kepler}, which was launched in $7^{\mathrm{th}}$ 
March 2009. \emph{Kepler} is a mission designed to detect the transit 
of an Earth across the Sun with its highly sensitive photometric camera; more 
details can be found in \citet{basri2005} and \citet{koch2007}, as well as on 
the mission website (http://www.kepler.nasa.gov/sci).

In this chapter, I will evaluate the range of exomoons that the 
\emph{Kepler Mission} or \emph{Kepler}-class photometry (KCP) could detect 
through transit timing effects, with particular attention to habitable-zone 
exomoons. This chapter is based upon the paper \citet{kippingetal2009}, which
was co-authored with colleagues S. Fossey and G. Campanella.

I emphasise the use of \emph{Kepler}-class photometry (KCP) throughout this
work, due to the increasingly impressive results being obtained from the ground 
which are matching space-based photometry, for example \citet{johnson2009}. 
Furthermore, ground-based observations are often more ideally 
suited for transit-timing studies due to the fewer constraints placed on the 
system, such as telemetry-limited data-download speeds.

\section{Modelling the Detectability of Exomoons}
\subsection{Confidence of Detection}

In order to explore a large range of parameter space, it is more convenient and 
efficient to employ analytic expressions rather than repeated individual 
simulations for thousands of different scenarios.  Therefore, one needs 
analytic expressions for the following:

\begin{enumerate}
\item[{\tiny$\blacksquare$}] TTV \& TDV signal amplitudes
\item[{\tiny$\blacksquare$}] Times of transit minimum and transit duration errors
\item[{\tiny$\blacksquare$}] Confidence of detection, based upon signal-to-noise
\end{enumerate}

The TTV and TDV root mean square (RMS) amplitudes have been derived in 
Chapter~\ref{ch:Chapt7}, notably in 
Equations~(\ref{eqn:TTVrms})~\&~(\ref{eqn:tottdv2}).  However, one also requires 
expressions for the timing errors, which are critical in evaluating the 
signal-to-noise.

To address this, I will use the analytic expressions for the uncertainty on the 
time of transit minimum ($\tau_T$) and duration ($\tilde{T}$) as derived by 
\citet{carter2008} using a Fisher-analysis of a trapezoid-approximated 
circular-orbit light curve.

For the purposes of TDV measurements, the primary requirement is to use a 
measure of transit duration which has the lowest possible uncertainty.  As 
discussed earlier in \S\ref{sec:Tduration}, by calculating the covariances of 
the light curve, \citet{carter2008} were able to show that $\tilde{T}$ can be 
calculated more precisely than either $T_{1,4}$ or $T_{2,3}$ and so I 
select $\tilde{T}$ as a robust duration parameter to explore the TDV effect. The 
uncertainties on transit depth, $\delta$, transit duration, $\tilde{T}$, and 
the time of transit minimum, $\tau_T$, were derived by \citet{carter2008} to be:

\begin{align}
\sigma_{\delta} &= W^{-1} \delta \\ 
\sigma_{\tilde{T}} &= W^{-1} \sqrt{2 \tilde{T} T_{1,2}} \\
\sigma_{\tau} &= W^{-1} \sqrt{\tilde{T} T_{1,2}/2} \\
W &= \delta \sqrt{\Gamma_{\mathrm{ph}} \tilde{T}}
\end{align}

Where $T_{1,2}$ is the ingress/egress duration, $\Gamma_{\mathrm{ph}}$ is the 
photon collection rate, and $\delta$ is the transit depth.
These expressions do not hold for a poorly sampled ingress or egress and 
therefore I assume a cadence of 1\,minute, corresponding to \emph{Kepler's} 
short-cadence mode.  

The equations of \citet{carter2008} require the ingress 
duration\footnote{Defined as the duration between contact points 1 \& 2, which 
is generally equivalent to the egress duration}, $T_{1,2}$, and the transit 
duration, $\tilde{T}$, as inputs, which were provided in 
Equations~(\ref{eqn:ingressdurcirc})~\&~(\ref{eqn:tildeTdurcirc}) for circular
orbits.  Note, that since circular orbits will be assumed throughout this 
chapter, then the $T_{x,y}^{\mathrm{one}}$ expressions are equivalent to the
$T_{x,y}^{\mathrm{SMO}}$ expressions. For the \emph{Kepler Mission} or KCP, I 
employ the same estimate for $\Gamma_{\mathrm{ph}}$ as that of 
\citet{borucki2005} (B05) and \citet{yee2008}:

\begin{equation}
\Gamma_{\mathrm{ph}} = 6.3 \times 10^{8} \, \mathrm{hr}^{-1} \, 10^{-0.4 (m-12)}
\end{equation} 

Where $m$ is the apparent magnitude.
For a normal transit depth observed $n$ times, the confidence, $C$, to which the 
transit is detected, in terms of the number of standard deviations, is defined 
by B05 as:

\begin{equation}
C(\mathrm{photometric}) = \frac{d}{\sigma_d} \sqrt{n}
\end{equation}

The transit timing signals due to an exomoon are periodic in nature and so 
require a different detection method.  Typically, this problem is tackled by 
searching for significant peaks in a periodogram, as often employed for radial 
velocity searches (e.g. \citet{butler2002}).  However, this approach is less 
useful for transit timing effects due to an exomoon, since the frequency one is 
trying to detect will always be much higher than the sampling frequency, as 
pointed out in \S\ref{sec:ttvproblems}.  The only available method is to 
therefore search for statistically significant excess variance and then use the 
$\chi^2$-distribution to calculate the confidence of signal detection. 
In order for this method to be applicable, one 
requires a) that the uncertainty estimates are robust and accurate; and b) that 
the period of the signal may be derived from amplitude information alone (which 
may then be compared to a periodogram to further refine the frequency).

The first of these requirements can be seen to be valid as several 
investigations have verified.  \citet{holman2006} derived the uncertainties of 
the times of transit minimum for four transits of XO-1b using three different 
methods: i) $\Delta \chi^2 = 1$ perturbation of the best-fit; ii) Monte Carlo 
bootstrapping; iii) Markov-Chain Monte Carlo (MCMC).  The authors found that all 
three methods produced very similar uncertainties, which implies the uncertainty 
estimates are highly robust.  Another example is that of 
\citet{carter2008}, who showed that the uncertainties derived using an 
MCMC-analysis were very similar to those predicted using analytic arguments.

The second requirement was validated in 
\S\ref{sec:moonmassdetermination}, where it was 
shown that the ratio of the TTV and TDV signal amplitudes may be used to obtain 
the period of the exomoon.  This period may then be compared to the set of 
possible harmonic frequencies derived from a periodogram in order to obtain a 
highly reliable estimate.  I therefore conclude that a search for excess 
variance is the most appropriate strategy for searching for exomoons through 
transit timing effects.  The confidence of detection may be found by integrating 
the probability density function (PDF) of the $\chi^2$-distribution.

\begin{equation}
C(\mathrm{timing}) = \sqrt{2} \mathrm{erf}^{-1}\Big[1-\int_{\alpha^2}^{\infty} \frac{x^{(n/2)-1} \exp^{-x/2}}{2^{n/2} \Gamma(n/2)}  \mathrm{d}x\Big]
\end{equation}

Where $n$ is the number of transits observed and $\alpha^2$ is the observed 
value of $\chi^2$, given by:

\begin{equation}
\alpha^2 = n\Big(1 + \frac{\delta_x^2}{\sigma_{\tau}^2}\Big)
\end{equation} 

Where $\delta_x$ is the RMS amplitude of the transit timing signal and 
$\sigma_{\tau}$ is the uncertainty on the time of transit minimum/transit 
duration. Integrating and making the above substitution, the confidence, $C$,
in detecting a timing signal in units of standard deviations is:

\begin{equation}
C(\mathrm{timing}) = \sqrt{2} \mathrm{erf}^{-1}\Big[1-\mathrm{Q}\Big\{\frac{n}{2},\frac{n}{2} \Big(1+\frac{\delta_x^2}{\sigma_{\tau}^2}\Big)\Big\}\Big]
\end{equation} 

Where $\mathrm{Q}\{a,b\}$ is the incomplete upper regularized 
Gamma function.  I summarise the assumptions below:

\begin{itemize}
\item[{\tiny$\blacksquare$}] Only one exomoon exists around the gas giant 
exoplanet of interest.
\item[{\tiny$\blacksquare$}] The moon and planet are both on circular orbits and 
the moon's orbit is prograde.
\item[{\tiny$\blacksquare$}] The moon's orbital plane is coaligned to that of 
the planet-star plane which is itself perpendicular to the line-of-sight of the 
observer, i.e. $i_S = i_P = \pi/2$
\item[{\tiny$\blacksquare$}] If a planet is within the habitable-zone, then any 
moon around that planet may also be considered to be ``habitable''.
\item[{\tiny$\blacksquare$}] A transiting planet must be detected to 8-$\sigma$ 
confidence to be accepted as genuine.
\item[{\tiny$\blacksquare$}] An exomoon must be detected through either a) TTV 
to 8-$\sigma$ and TDV to 3-$\sigma$ confidence or b) TTV to 3-$\sigma$ and TDV 
to 8-$\sigma$ confidence, in order to be accepted as genuine.
\item[{\tiny$\blacksquare$}] The \emph{Kepler Mission} or KCP will be used in 
short cadence mode for the transit timing of a target of interest for $\simeq 4$ 
years.
\item[{\tiny$\blacksquare$}] $n = M/P_P$ where $M$ is the mission duration and 
$P_P$ is the period of transiting planet.  
\item[{\tiny$\blacksquare$}] At least three transits are needed to detect both a 
planet and a moon.
\end{itemize}

The assumption of co-alignment nullifies any TDV-TIP effect and thus the TDV
signal is not enhanced by this extra contribution. In this sense, the 
calculations presented in this work can be considered conservative.

In most of the cases I will consider, many more than three transits will be 
detected and three can be seen to be the limiting case for G0V stars, where the 
habitable zone is sufficiently distant to only permit three transits in a 4-year 
timespan.  Although statistically speaking three transits is sufficient, there 
is a risk of an outlier producing a false positive.  I therefore consider 
detections of habitable exomoons in early G-type star systems to be described as 
``tentative'', whereas once four transits are detected, for stars of spectral 
type G5V and later, this risk can be considered to be reduced.

The nominal mission length of \emph{Kepler} is 3.5\,years and it may be extended 
to up to 6\,years, which justifies the choice of 4\,years of transit timing 
observations.  A ground-based search achieving KCP may easily be operational 
for 4\,years or more.  I choose 8-$\sigma$ as the signal detection threshold 
since this is the same as that used by \emph{Kepler}.  The second signal may be 
detected to lower significance since it is only used to confirm the phase 
difference between the two and also derive the exomoon period.

\subsection{The Total Noise}

The expressions of \citet{carter2008} only consider shot noise through the 
$\Gamma_{\mathrm{ph}}$ parameter.  However, if one assumes that the impact on 
$\sigma_{\tilde{T}}$, $\sigma_{\tau}$ and $\sigma_{\delta}$ are approximately 
equivalent for additional uncorrelated noise and for correlated noise, then one 
may simply modify $W$ to absorb the effects of red noise.  For uncorrelated 
noise, I add the additional sources of noise in quadrature.  Note, that this is 
the same treatment utilised in the design technical documents for 
\emph{Kepler}, for example see B05.

In general, there are expected to be three major sources of noise present in the 
\emph{Kepler} data in the form of shot noise, instrument noise and stellar 
variability.  Instrument noise is due to a variety of effects and has been 
modelled in depth by B05 (and \citet{koch2004}) to quantify its effect as a 
function of magnitude.  With all three noise sources, I modify $W$ to $W'$, 
given by:

\begin{equation}
\frac{1}{W'} = \frac{1}{\delta} \sqrt{\frac{1}{\Gamma_{\mathrm{ph}} \tilde{T}} + I^2 + S^2}
\end{equation} 

\begin{figure}
\begin{center}
\includegraphics[width=10.0 cm]{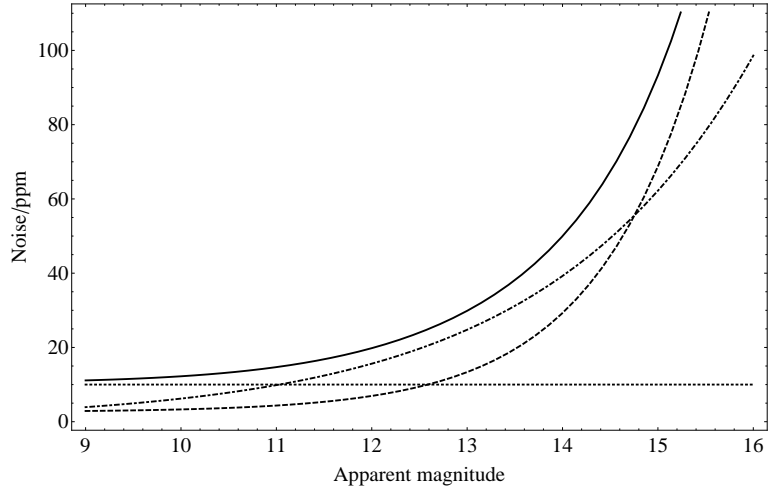}
\caption[Noise sources affecting \emph{Kepler} photometry]
{\emph{Noise sources predicted to affect \emph{Kepler} photometry as a 
function of magnitude.  Instrument noise is dashed, photon noise is dot-dashed, 
stellar variability is dotted and the total is solid.  Values obtained from 
Bill Borucki in personal communication.}} 
\label{fig:kep1}
\end{center}
\end{figure}

Where $I$ is the instrument noise and $S$ is the stellar variability.
$I$ is a function of magnitude which may be calculated using the model of B05 
and I show all three noise sources plotted as a function of magnitude in 
Figure~\ref{fig:kep1}.  I assume a constant value for stellar variability of 
10\,ppm across all spectral types, a reasonable assumption, given that 
65--70\% of F7-K9 main-sequence stars in the \emph{Kepler} field are likely to 
have similar or lower intrinsic variability than the Sun (\citet{batalha2002}; 
B05) on timescales important to transit detections.  I also note that this 
equation is equivalent to the formulation used in the original technical design 
papers for \emph{Kepler}, for example see Equation~(1) of B05.

\subsection{The Habitable-Zone}

I choose to consider a moon-hosting gas-giant exoplanet around a variety of 
main-sequence stars as shown in Table~\ref{tab:kep1}.  For each star, I 
calculate the habitable-zone orbital distance, $a_{\mathrm{hab}}$, to be defined 
as the distance where a planet would receive the same insolation as the Earth.
This straightforwardly permits a reasonable estimate of the habitable-zone for 
each star type. For a more in-depth consideration of habitability of exomoons 
see \citet{williams1997}.

\begin{equation}
a_{\mathrm{hab}} = \sqrt{L_*/L_{\odot}} \, \mathrm{AU}
\end{equation}

For each planet-moon system considered, the period of the transiting planet is 
calculated using Kepler's Third Law:

\begin{equation}
P_{\mathrm{hab}} = 2 \pi \sqrt{\frac{a_{\mathrm{hab}}^3}{G (M_*+M_P + M_S)}}
\end{equation}

I choose to work in the time domain, rather than the orbital-distance 
formulation, since a major limiting factor in this study is the 
\emph{Kepler Mission} duration.

\subsection{Properties of Host Star}

In this analysis, I will only consider single main-sequence stars which offer 
the best potential for hosting habitable environments.  I will consider 
spectral types from M5V to F0V and assume for each an approximate mass, radius, 
and effective temperature as given by \citet{cox2000}, and a luminosity derived 
from data therein.  I use the \emph{Kepler} bandpass to calculate the absolute 
magnitude of these stars\footnote{Guidelines available from 
http://keplergo.arc.nasa.gov/proposal.html}.

For each stellar type, I assume the stars are not young and may be considered 
to be slow rotators.  Since stellar variability is correlated to rotational 
period \citep{dorren1994}, I therefore limit the study to quiet stars. 
This is the same assumption used for \emph{Kepler}'s ability to detect 
Earth-like planets since very active stars will be too variable for the 
detection of such bodies.  In Table~\ref{tab:kep1}, I list the different star 
properties.

\begin{table*}
\caption[Properties of main-sequence stars]
{\emph{Properties of stars used in these calculations.  Values taken from 
\citet{cox2000}.  Absolute magnitudes in the \emph{Kepler} bandpass calculated 
using guidelines on the mission website.}} 
\centering 
\begin{tabular}{c c c c c c} 
\hline\hline 
Star type & $M_*$/$M_{\odot}$ & $R_*$/$R_{\odot}$ & $L_*$/$L_{\odot}$ & $T_{\mathrm{eff}}$/K & $M_{\mathrm{Kep}}$ \\ [0.5ex] 
\hline 
M5V & 0.21 & 0.27 & 0.0066 & 3170 & 11.84 \\
M2V & 0.40 & 0.50 & 0.0345 & 3520 & 9.49 \\
M0V & 0.51 & 0.60 & 0.0703 & 3840 & 8.42 \\
K5V & 0.67 & 0.72 & 0.1760 & 4410 & 7.06 \\
K0V & 0.79 & 0.85 & 0.4563 & 5150 & 5.78 \\
G5V & 0.92 & 0.92 & 0.7262 & 5560 & 5.02 \\
G2V & 1.00 & 1.00 & 1.0000 & 5790 & 4.63 \\
G0V & 1.05 & 1.10 & 1.3525 & 5940 & 4.34  \\
F5V & 1.4 & 1.3 & 2.9674 & 6650 & 3.47 \\
F0V & 1.6 & 1.5 & 5.7369 & 7300 & 2.71 \\ [1ex]
\hline\hline 
\end{tabular}
\label{tab:kep1} 
\end{table*}

\subsection{Properties of Exomoons}

Although no exomoons have yet been discovered, it is possible to calculate the 
range of exomoons which are dynamically stable around each planet. 
\citet{barnes2002} addressed this problem and developed a set of analytic 
expressions, which can be shown to provide excellent agreement to numerical 
simulations, for the stability of exomoons around exoplanets.  Assuming an 
exomoon has to be stable for at least 5 Gyr, one is able to calculate the 
maximum allowed exomoon mass in each case (see Equations~(7) to (9) of 
\citet{barnes2002}).  I assume Jupiter-like values for the tidal dissipation 
factor, $Q_P = 10^5$, and for the tidal Love number, $k_{2p} = 0.51$, as used 
by \citet{barnes2002}.

One is also able to estimate the range of allowed values for the planet-moon 
separation, in units of Hill radii, $\mathfrak{D}$.  \citet{domingos2006} 
presented the relevant expressions, which again can be shown to provide 
excellent agreement to numerical simulations.  Using their Equation~(5), one is
able to estimate the maximum distance at which an exomoon is stable for prograde 
orbits.

For the minimum distance, I calculate the Roche limit of the planet in all 
cases.  If this value is greater than $2 R_P$ then I use the Roche limit as the 
minimum distance, otherwise $2 R_P$ is adopted.  I assume that there is no 
reason for a moon to exist at any particular value of $\mathfrak{D}$ and thus 
the prior distribution is flat.

\section{Light Curve Simulations}
\subsection{Light Curve Generation}

A critical assumption in this paper is the use of the equations of 
\citet{carter2008} for the uncertainties on the transit duration and the times
of transit minimum.  The authors tested their expressions using synthetic light 
curves and an MCMC fitting procedure as well as a numerical Fisher-analysis. 
They report excellent agreement between their expressions and the derived errors 
but find the greatest departure for poorly sampled ingresses and near-grazing 
transits. The choice of assumptions made in this work avoids both of these 
issues.

Note that these tests were run for shot noise only and in this work I 
consider the effects of both instrument and stellar noise.  Having modified $W$ 
to $W'$ to account for these additional noise sources, I proceed to test this 
modified formulation through generation of synthetic light curves.

I use the \citet{mandel2002} code to generate light curves accounting for limb 
darkening based upon a non-linear law from \citet{claret2000}.  Light curves are
generated for a Neptune, Saturn and Jupiter-like planet in the habitable-zone of 
a $m_{\mathrm{Kep}}=12$, G2V Sun-like star with $i_P = \pi/2$ and 
$e_P = 0$. Each light curve is generated to have 1000 data points evenly spaced 
with a 1-minute cadence centred on the transit minimum.

The first noise source added is shot noise generated by taking a random real 
number from a normal distribution of mean zero and standard deviation given by:

\begin{equation}
\sigma_{shot} = \frac{1}{\sqrt{t_{exp} \Gamma_{\mathrm{ph}}}}
\end{equation} 

Where $t_{exp}$ is the time between each consecutive measurement, 
which is assumed to be one minute in these calculations.

For the stellar noise, 1000 sinusoidal waveforms are generated with varying 
periods randomly selected between one minute up to 24 hours.  These waveforms 
are then coadded to give a single synthetic stellar signal designed to mimic the 
non-stochastic nature of real stellar variability.  The RMS amplitude of the 
combined signal is then scaled to $10$\,ppm, which matches the amplitude 
prediction of B05 for a G2V star.

I consider that the instrument noise is composed of hundreds of different noise 
sources periodic in nature varying on timescales from one minute to one day.  
There is no benefit of including timescales longer than this since transit 
events will not last longer than $\sim 1$\,day.  The combined RMS amplitude 
of the instrument noise is set to be given by the values calculated by Bill 
Borucki (personal communication) and is a function of visual magnitude.  For  
example, a $m_{\mathrm{Kep}}=12$ target has an RMS instrument noise of 
7.47\,ppm.

\subsection{Light Curve Fitting}

The noisy light curves are generated 10,000 times with the correlated noises and 
photon noise being randomly generated in all cases.  The light curves are then 
passed onto a light curve fitting code used to obtain best-fit values for 
$\tilde{T}$ and $\tau_T$.  In all cases, I fit for $a_P/R_*$, $i_P$, $R_P/R_*$ 
and $\tau_P$ and therefore assume that the out-of-transit baseline is well known 
and its errors are essentially negligible.  This is a reasonable assumption for 
high quality photometry with large amounts of out-of-transit data.

The fitting code finds the best-fit to the light curve by utilising a genetic 
algorithm\footnote{Available from http://whitedwarf.org/parallel/}, {\tt PIKAIA}
\citep{metcalfe2003}, to get close to a minimum in $\chi^2$.  This 
approximate solution is then used as a starting point for a 
$\chi^2$-minimisation performed with the AMOEBA routine \citep{press1992}. 
The AMOEBA solution is tested by randomly perturbing it and refitting in 20 
trials.

In each subsequent fitting run of the 10,000 light curves, the AMOEBA routine 
starts from the original best-fit.  The resulting values of $\tau_T$ and 
$\tilde{T}$ are binned and plotted as a histogram (Figure~\ref{fig:kep2}) and 
then compared to the distribution expected from the modified 
\citet{carter2008} expressions.

\subsection{Comparison to the Analytic Expressions}

\begin{figure}
\begin{center}
\includegraphics[width=10.0 cm]{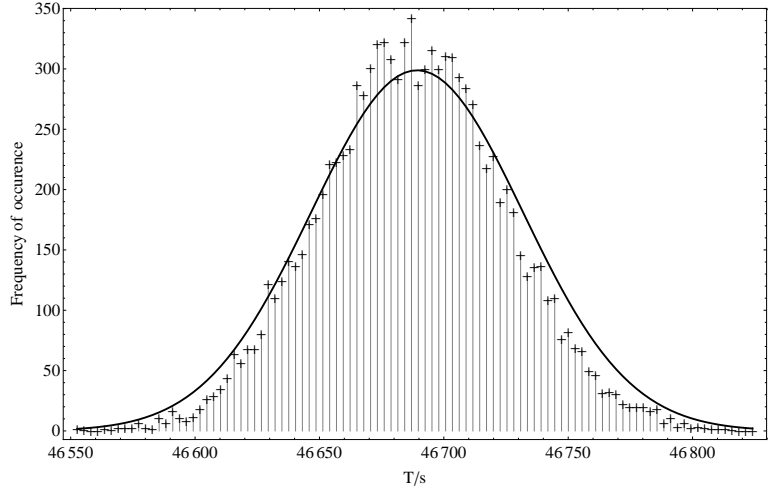}
\caption[Comparison of analytic expression for uncertainty in $\tilde{T}$ versus
that found through light curve fits]
{\emph{Comparison of the distribution in $\tilde{T}$ found using Monte 
Carlo simulations (scattered points) of a transit of a Saturn-like planet versus 
that from theoretical prediction (smooth line).  The theoretical prediction for 
the timing error (42.6\,s) is slightly larger than that found in numerical 
simulations (37.3\,s).}}
\label{fig:kep2}
\end{center}
\end{figure}

For the three cases of a Neptune, a Saturn and a Jupiter, I obtained 10,000 
estimates of $\tilde{T}$ and $\tau_T$ in each case.  One may compare the 
distribution of $\tau_T$ and $\tilde{T}$ to the predicted distribution from the 
expressions of \citet{carter2008}. In all cases, I find excellent agreement 
between the predicted uncertainties and the theoretical values.  In 
Figure~\ref{fig:kep2}, I show a histogram of the results for the Saturn-case 
$\tilde{T}$ values, and overlay the predicted value of $\sigma_{\tilde{T}}$ for 
comparison where the quality of the agreement is evident from the plot.  Note 
that this overlaid Gaussian is not a fit but a theoretical prediction.

I find excellent agreement for both $\sigma_{\tilde{T}}$ and $\sigma_{\tau}$, 
and if anything the theoretical expressions slightly overestimate the 
uncertainties. Thus any results from this study can in fact be considered to be 
slightly conservative.  It is preferred to adopt a conservative approach in the 
analysis since there may be unexpected sources of noise which increase the 
timing uncertainties.

\section{Implementation of the Expressions}
\subsection{Earth-Transit Magnitude Limit}

I first consider the magnitude limit in detecting an Earth-like transit which 
will allow one to compare the efficiency of detecting Earth-like planets and 
moons.  \emph{Kepler} is designed to look at $\sim 10^5$ stars between 
$6^{\mathrm{th}}$ to $16^{\mathrm{th}}$ 
magnitude\footnote{See http://kepler.nasa.gov/sci}.  For each spectral type in 
Table~\ref{tab:kep1}, I compute the faintest visual magnitude for which a 
habitable-zone Earth-like transit can be detected. \emph{Kepler} was conceived 
with the goal of detecting an Earth-Sun transit of 6.5\,hours in duration, which 
represents about half the duration if the Earth has an impact parameter of zero.  
However, the maximum magnitude to which an Earth-like transit could be detected 
will be for the cases of equatorial transits as these maximise the transit 
duration and so too the net flux decrement.  The following criteria for a 
reliable transit detection are defined:

\begin{itemize}
\item[{\tiny$\blacksquare$}] Each transit must be detected to $\geq$ 1-$\sigma$ 
confidence.
\item[{\tiny$\blacksquare$}] The folded light curve must have a significance of 
$\geq$ 8-$\sigma$.
\item[{\tiny$\blacksquare$}] The time between contact points 2 \& 3 $\geq$ 
1\,hour, in order to be detected with \emph{Kepler's} long-cadence mode of 
30\,minutes.
\item[{\tiny$\blacksquare$}] At least 3 transits must be detected over the 
mission duration.
\end{itemize}

With these criteria, I compute the magnitudes of the faintest stars hosting a 
habitable, transiting Earth which can be detected with KCP, for both full 
transit durations (i.e. equatorial transits) and half-durations (i.e. impact 
parameter chosen to be such that the transit duration is half the full transit). 
The results are given in Table~\ref{tab:kep2}.  As expected, smaller host stars 
can be fainter due to their smaller radius and hence deeper transit depths. One 
also picks up more transits towards smaller stars since the orbital period of 
the habitable-zone decreases.

\begin{table*}
\caption[Faintest magnitude for which \emph{Kepler} could detect a habitable
Earth]
{\emph{Faintest stars for which a habitable transiting Earth could be 
detected for different star types.  Final column gives maximum distance of such 
a star with a full transit duration ($T_{1,4}^{\mathrm{full}}$), based on 
absolute magnitude in the \emph{Kepler} bandpass.  A blank indicates that no 
magnitude can satisfy the detection criteria.}} 
\centering 
\begin{tabular}{c c c c c} 
\hline\hline 
Star type & $T_{1,4}^{\mathrm{half}}$ $m_{\mathrm{Kep}}^{\mathrm{max}}$ & $T_{1,4}^{\mathrm{full}}$ $m_{\mathrm{Kep}}^{\mathrm{max}}$ & $d_{\mathrm{min}}$/pc & $d_{\mathrm{max}}$/pc \\ [0.5ex] 
\hline 
M5V & - & 18.111 & 0.68 & 179.56 \\
M2V & 15.992 & 16.190 & 2.00 & 218.78 \\
M0V & 15.206 & 15.465 & 3.28 & 256.45 \\
K5V & 14.278 & 14.618 & 6.14 & 324.79 \\
K0V & 13.286 & 13.712 & 11.07 & 385.83 \\
G5V & 12.762 & 13.240 & 15.70 & 440.56 \\
G2V & 12.236 & 12.763 & 18.79 & 423.25 \\
G0V & 11.520 & 12.106 & 21.48 & 357.44 \\
F5V & - & - & - & - \\
F0V & - & - & - & - \\ [1ex]
\hline\hline 
\end{tabular}
\label{tab:kep2} 
\end{table*}

Even if \emph{Kepler} was extended to a mission length of 6 years, an F5V star 
would have a habitable zone so far out that detecting three transits within this 
region would be impossible.  Hence, I do not consider F type stars in this
analysis.  I also consider $6^{\mathrm{th}}$-magnitude stars to be the bright 
limit, giving us $d_{\mathrm{min}}$, the minimum distance for each spectral type 
(see Table~\ref{tab:kep2}), since the \emph{Kepler} field is specifically chosen 
to avoid such bright targets.

\subsection{Jupiters vs Saturns vs Neptunes}

One may use the approximate analytic expressions to get a handle on the general 
trends in exomoon detection.  The first question one may pose is- what is the 
optimum planet to search for moons around, out of the three classes of Neptunes, 
Saturns and Jupiters? For each of these planets I simulate the 
detectability of the TTV signal as a function of planetary orbital period, 
$P_P$, for an $0.2 M_{\oplus}$ exomoon with $\mathfrak{D} = 0.4895$.  The star
is fixed to a G2V spectral type with $m_{\mathrm{Kep}}=12$ and I work with 
$i_P = \pi/2$ for simplicity.

\begin{figure}
\begin{center}
\includegraphics[width=10.0 cm]{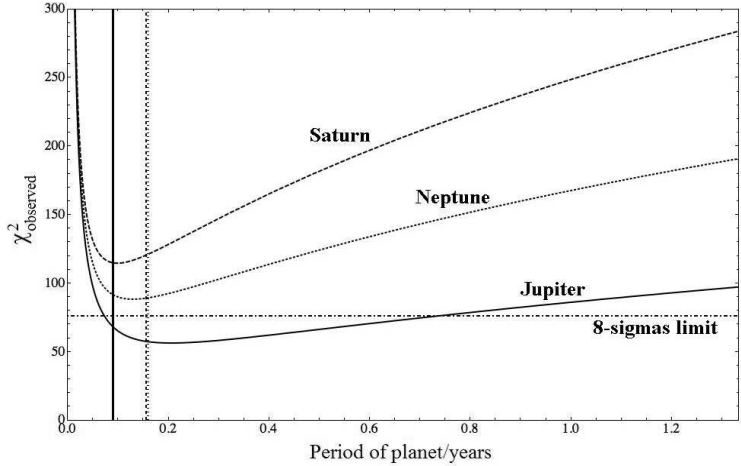}
\caption[Detectability of an Earth-mass moon around a habitable-zone Jupiter,
Saturn and Neptune]
{\emph{Example plot showing the TTV detectability of a 0.2\,$M_{\oplus}$ 
($\mathfrak{D}=0.4895$) exomoon around a Jupiter, Saturn and Neptune-like 
exoplanet for a G2V, $m_{\mathrm{Kep}}=12$ star.  Values of $\chi^2 \gg 76$ are 
detectable at $\geq$ 8-$\sigma$ confidence.  Saturns present the strongest 
signal due to their low density.  The vertical lines represent the stability 
limit of such a moon, calculated using the model of \citet{barnes2002}.}} 
\label{fig:kep3}
\end{center}
\end{figure}

I calculate the signal amplitude and time of transit minimum uncertainty in all 
cases and hence find the $\chi^2$ value for a range of orbital periods.  In 
Figure~\ref{fig:kep3}, $\chi^2$ is plotted as a function of host planet's 
orbital period.  The figure reveals that Jupiters are the hardest to search for 
exomoons around whilst Saturns are the easiest.  This is due to Saturn's low 
density meaning a large transit depth but a low enough mass such that an exomoon 
still perturbs it significantly.  I find the same sequence of detectability 
consistently in many different orbital configurations and for TDV as well. Note
that Saturn would not be considered a low-density planet from an exoplanet
perspective since a significant population of much lower density exoplanets
have now been found (see http://www.exoplanet.eu).

\subsection{$\mathfrak{D}$-$m_{\mathrm{Kep}}$ Parameter Space}

I now consider an exomoon in the case of a Saturn hosting a 
single satellite in the habitable-zone of the host star.  The detectability of 
the timing signals depends on the planet-moon separation, 
$a_S = \mathfrak{D} R_{\mathrm{Hill}}$, the apparent magnitude of the host star 
in the \emph{Kepler} bandpass, $m_{\mathrm{Kep}}$, mass of the exomoon, $M_S$, 
and finally the host star's mass and radius (i.e. spectral type).

TDV increases with lower values of $\mathfrak{D}$ and TTV increases with higher 
values of $\mathfrak{D}$, but one must maintain a balance such that at least one 
of the effects is detected to 8-$\sigma$ and the other to 3-$\sigma$ confidence 
(dictated by the previous detection criteria).  To find the limits of interest, 
I assume a zero-impact-parameter transit as the default planet-moon arrangement. 
The orbital period of the host planet is always given by $P_{\mathrm{hab}}$. 

\begin{figure}
\begin{center}
\includegraphics[width=10.0 cm]{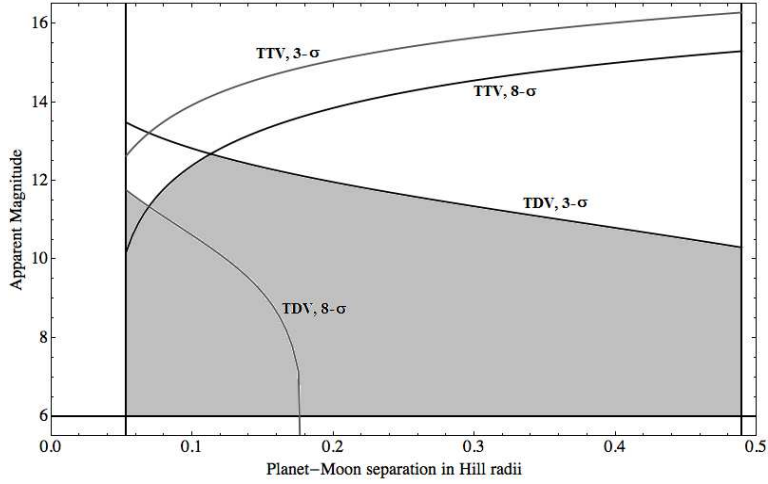}
\caption[Detectable range of a $\frac{1}{3} M_{\oplus}$ habitable exomoon around 
an M0V star]
{\emph{Detectable range of a habitable $\frac{1}{3} M_{\oplus}$ exomoon 
for an M0V star with respect to apparent magnitude and $\mathfrak{D}$ parameter 
space, with \emph{Kepler}-class photometry.  The grey area represents the area 
satisfying the detection criteria of detecting TTV or TDV to 8-$\sigma$ and the 
other timing effect to 3-$\sigma$, shown by the four curves. Additional 
constraints are the lower Roche limit, the upper dynamical stability limit and 
the bright-star magnitude limit. Note how the 8-$\sigma$ TDV line drops very
sharply after $m\sim9$ since \emph{Kepler's} photometry does not improve
appreciably past this magnitude due to instrument noise.}} 
\label{fig:kep4}
\end{center}
\end{figure}

For a given star type and exomoon mass, I plot the contours of 
$m_{\mathrm{Kep}}$ and $\mathfrak{D}$, which provide TTVs of 3- and 8-$\sigma$ 
confidence and then repeat for TDVs of the same confidences.  There are two 
possible acceptance criteria: i) TTV confidence is $\geq 3$-$\sigma$ and TDV 
confidence is $\geq 8$-$\sigma$; ii) TTV confidence is $\geq 8$-$\sigma$ and TDV 
confidence is $\geq 3$-$\sigma$.  The loci of points below these lines 
represents the potential detection parameter space.  For each star type and 
exomoon mass, these loci will be different.  In Figure~\ref{fig:kep4}, a typical 
example is shown for an M0V-type star and $M_S = \frac{1}{3} M_{\oplus}$.

There are certain cases which constrain the stars of interest.  The 
\citet{barnes2002} limit may be calculated for a Saturn harbouring an exomoon in 
a system of 5\,Gyr age.  This suggests that the maximum moon mass that a 
habitable-zone Saturn could hold around a M5V star would be $\sim 0.3$\,Ganymede 
masses, assuming the maximum prograde orbital distance is 0.4895\,Hill radii, as 
calculated by \citet{domingos2006}.  In contrast, an M2V star allows for a 
habitable-zone Saturn to hold onto a $0.4$\,Earth-mass moon for over 5\,Gyr.

A second lower limit exists from tidal forces and the Roche limit.  An upper 
limit is given by the fact one requires three transits in a 4\,year observation 
duration and thus the most distant habitable-zone assumed here corresponds to a 
period of 1.33\,years, which excludes the F0V and F5V stellar types.

\subsection{Magnitude Quartiles}

For Figure~\ref{fig:kep4}, one may convert the plot into three key numbers by 
assuming that the probability distribution of exomoons with respect to 
$\mathfrak{D}$ is approximately uniform. This simple a-priori approximation 
allows one to get a handle on some typical magnitude limits.  This assumption 
effectively converts the two-dimensional array of 
$\mathfrak{D}_i$-$m_{\mathrm{Kep},i}$ into a one-dimensional list of just 
$m_{\mathrm{Kep},i}$.  One then simply takes the quartiles of this 
list to obtain estimates for the 25\%, 50\% and 75\% ``catch-rate'' values of 
$m_{\mathrm{Kep}}$.  Physically speaking, we are saying that, for example, at 
the 50\% catch-rate value of $m_{\mathrm{Kep}}$, there is a 50\% probability of 
detecting an exomoon if an exomoon exists with a uniform prior distribution 
between $\mathfrak{D}_{\mathrm{min}}$ and $\mathfrak{D}_{\mathrm{max}}$.

Accordingly, one may calculate the magnitude limits to detect 25\%, 50\% or 75\% 
of the exomoons in the given sample, which I label as the upper, middle and 
lower quartiles respectively.  For the case shown in Figure~\ref{fig:kep4}, the 
quartile values are $m_{\mathrm{Kep}}=12.0$, $11.5$ and $10.9$ respectively.

\begin{figure}
\begin{center}
\includegraphics[width=10.0 cm]{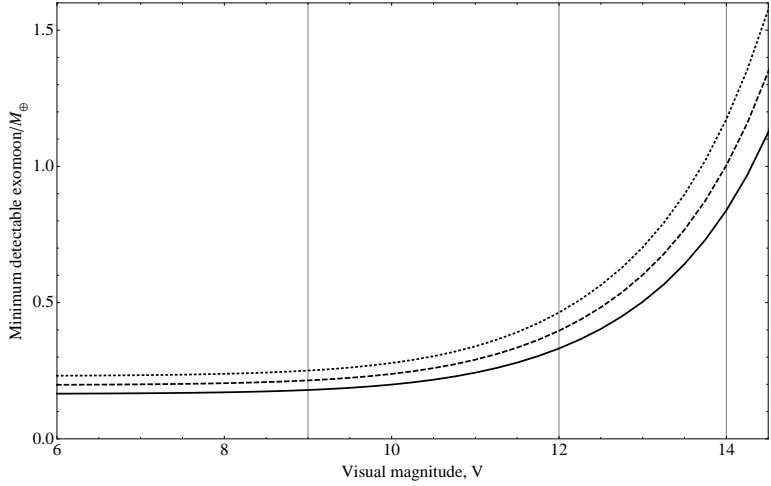}
\caption[Minimum detectable exomoon mass around a habitable-zone Saturn for an
M0V star]
{\emph{Minimum detectable exomoon mass around a Saturn-like planet host 
in the habitable-zone of an M0V star.  The top line is the 75\% catch-rate, the 
middle is 50\% and the lowest is for a 25\% exomoon catch-rate.  I also mark 
three key magnitudes of interest; $9^{\mathrm{th}}$, $12^{\mathrm{th}}$ and 
$14^{\mathrm{th}}$ magnitudes.}} 
\label{fig:kep5}
\end{center}
\end{figure}

One may then calculate these quartile values for different exomoon masses, for a 
given star type.  Effectively, one may determine the minimum detectable 
exomoon mass as a function of magnitude for three different detection yields: 
i) 25\%, ii) 50\%, and iii) 75\%.  In Figure~\ref{fig:kep5}, I show an example 
of one of these plots for an M2V star with a habitable-zone Saturn host planet.
The contours have been calculated by a spline-interpolation of 20 evenly
spaced exomoon masses. Magnitudes brighter than $6^{\mathrm{th}}$ are not 
considered, since \emph{Kepler's} field-of-view has been selected to omit such 
stars. 

\subsection{Minimum Detectable Habitable Exomoon Masses}

\begin{figure}
\begin{center}
\includegraphics[width=10.0 cm]{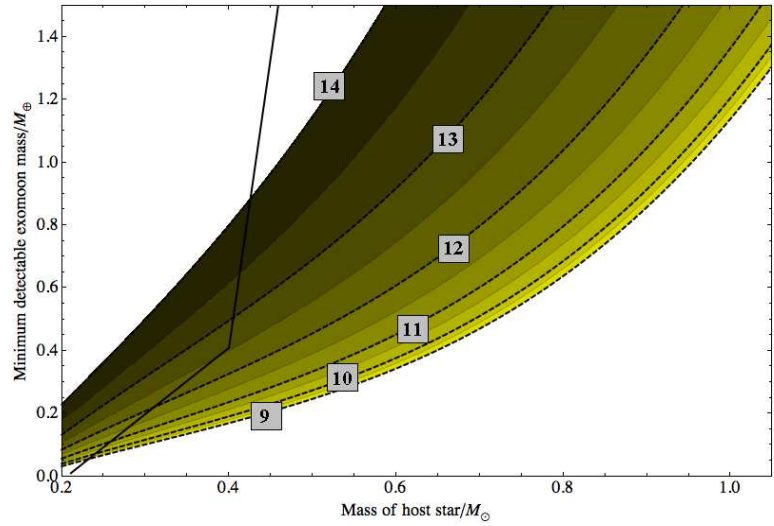}
\caption[Minimum detectable exomoon mass versus stellar mass for \emph{Kepler}]
{\emph{Minimum detectable exomoon mass around a Saturn-like planet host 
in the habitable-zone of various star types.  Contours show the different 
magnitude limits for a 25\% exomoon-catch-rate and overlaid is the mass 
stability limit (black, solid), found using \citet{barnes2002}.}} 
\label{fig:kep6}
\end{center}
\end{figure}

To complete the picture, one wishes to see how these values differ for various 
star types.  I repeat the analysis performed above for M5V, M2V, M0V, K5V, K0V, 
G5V, G2V and G0V-type stars.  In Figure~\ref{fig:kep6}, I plot the minimum 
detectable exomoon mass, for several contours of visual magnitude, as a function 
of stellar mass, in the 25\% detection-yield case.  I also show the upper limit 
on moon mass calculated from \citet{barnes2002} but this is only an 
approximation since I have assumed values of $Q_P$ and $k_{2p}$.

The $m_{\mathrm{Kep}}=12.5$ limit almost intersects the \citet{barnes2002} limit 
at 0.4\,$M_{\odot}$ and $\sim 0.4\,M_{\oplus}$ for the exomoon corresponding to 
an M2V star $\sim 40$\,pc away. For comparison, within $10,000\,\mathrm{pc}^3$ 
(a spherical shell of radius 13.4 pc) there are $\sim 800$ main-sequence stars, 
$\sim 630$ of which are M-dwarfs \citep{ledrew2001}.

If one sets a lower limit of 10\,pc for the target star, then the lowest-mass 
habitable-zone exomoon which could be detected with KCP around a M5V star would 
be 0.09\,$M_{\oplus}$, which is above the maximum allowed stable mass limit of 
0.008\,$M_{\oplus}$.  However, moving up to an M2V star I find a minimum 
detectable habitable exomoon mass of 0.18\,$M_{\oplus}$ which is below the 
stability limit of 0.41\,$M_{\oplus}$.  I therefore conclude that the minimum 
detectable habitable-zone exomoon mass with KCP is $\sim 0.2\,M_{\oplus}$.

\subsection{1\,$M_{\oplus}$ Limit}

\begin{figure}
\begin{center}
\includegraphics[width=10.0 cm]{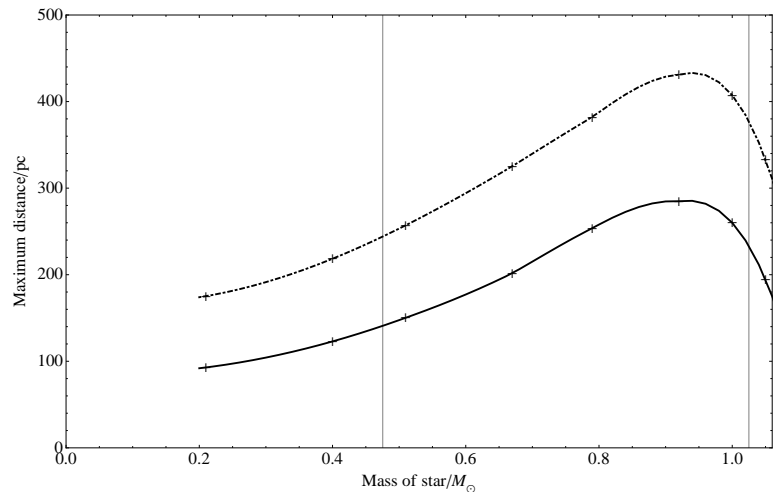}
\caption[Distance limit for detecting an 1\,$M_{\oplus}$ habitable exomoon versus
a 1\,$R_{\oplus}$ habitable exoplanet with \emph{Kepler}]
{\emph{Distance limit for detecting a 1\,$M_{\oplus}$ habitable exomoon 
(solid) and a 1\,$R_{\oplus}$ exoplanet (dot-dashed) as a function of stellar 
mass, for main sequence stars.  I assume a Saturn-like planet as the host for 
the exomoon case.  I also mark the lower stellar mass stability limit of 
$\sim 0.475$\,$M_{\odot}$ and the upper mass limit of $\sim 1$\,$M_{\odot}$ 
imposed by the requirement to observe $\geq 3$ transits.}} 
\label{fig:kep7}
\end{center}
\end{figure}

For the 25\% detection yield case, one may convert the magnitude limit for 
detecting a 1\,$M_{\oplus}$ habitable exomoon into a distance limit by making 
use of the absolute magnitudes for each star type.  The distance limit for 
detecting a 1\,$M_{\oplus}$ habitable exomoon may then be compared to the limit 
found for a transiting 1\,$R_{\oplus}$ habitable exoplanet.  In 
Figure~\ref{fig:kep7}, I compare the two distance limits, which reveals that the 
distance limit for moons takes the same shape as the transit limit but is 
$\sim 1.6$ times lower. This translates to a volume space diminished by a factor 
of $\sim 4$.

It is not the intention of this work to estimate accurately how many observable 
stars might be contained within this volume, but one may deduce an 
order-of-magnitude estimate. Assuming that there are about $10^5$ useable stars 
in \emph{Kepler}'s field-of-view, I estimate roughly that 25,000 stars would be 
within range for habitable-zone exomoon detection. Extrapolating \emph{Kepler}'s 
field-of-view to a band centred on the Galactic plane between $\pm15^\circ$, and 
simplistically assuming the same number of target stars per square degree, one 
obtains about $2\times 10^6$ stars that are accessible to KCP for a search for 
habitable-zone exomoons, if each star had a habitable-zone gas giant similar to 
Saturn.

\subsection{Detectability of an Earth-Moon Analogue}

Finally, I consider the detectability of an Earth-Moon system.  Using the 
orbital parameters of the Sun-Earth-Moon system and assuming the most favourable 
configuration of $i_P=\pi/2$, I find that the Earth would exhibit a TTV RMS 
amplitude of 112.4\,seconds and a TDV of 13.7\,seconds.  The TDV is particularly 
low due to the large value of $\mathfrak{D}$ for the Moon.

In comparison, KCP of a $V=6$ star would yield timing errors of 
$\sigma_{t_c} = 389.6$ seconds and $\sigma_{T} = 779.2$ seconds.  Even for the 
most favourable case then, the TTV has a signal-to-noise of 0.3, which would be 
quite undetectable.  I therefore conclude that KCP cannot detect a 
Sun-Earth-Moon analogue through transit timing effects.

\section{Conclusions}

In this chapter, I have presented a feasibility study which finds that habitable 
exomoons are detectable up to $\sim 100$-$200$\,pc away around early-M, K and 
later-G dwarf stars with the \emph{Kepler Mission} or photometry of equal 
quality (KCP).  This photometric quality should be sensitive down to 
0.2\,$M_{\oplus}$ habitable exomoons in the idealised cases and could survey 
$\sim 10^6$ stars for $1\,M_{\oplus}$ habitable exomoons, with a full Galactic 
Plane survey.  This number is around 25,000 stars for \emph{Kepler}'s 
field-of-view.

Saturn-like planets are found to be the ideal host candidates for detection, due 
to their large radius to mass ratio.  Additionally, I find lower-mass exomoons 
may be found around M-dwarfs due to the closer habitable-zone permitting a 
larger number of transits in a 4-year window.

The exciting prospect of discovering a habitable exomoon is well within the 
grasp of KCP and whether such worlds exist or whether they will be classed as 
truly habitable worlds are questions we can hope to answer in the coming years. 
It is interesting to note that $\sim 0.2\,M_{\oplus}$ is the minimum habitable 
mass for a planet or moon used by several other sources in the literature, 
including \citet{raymond2007} ($0.3\,M_{\oplus}$) and \citet{williams1997} 
($0.12\,M_{\oplus}$).  It has also been proposed by \citet{scharf2006} that the 
habitable-zone could be extended for exomoons due to tidal heating and so these
calculations may in fact be an underestimate.

I find that \emph{Kepler} will be incapable of finding Ganymede-mass moons 
around Saturn-like planets.  \citet{canup2006} have suggested that the maximum 
moon mass which could form from a planetary debris disk is $2 \times 10^{-4}$ of 
that of the primary's and therefore, if this rule holds true, it unlikely KCP 
would detect any moons which formed around a planet in such a way.  However, 
moons above this mass limit are still dynamically stable (according to 
\citet{barnes2002}) and could have been captured by a planet or formed through 
an impact, for example like the formation of Triton and the Moon respectively
(see Chapter~\ref{ch:Chapt2} for a more detailed discussion). Whether or not 
such objects are common is unknown but KCP could make the first in-depth search.

These results suggest it is easier to detect an Earth-like exoplanet than an 
Earth-like exomoon around a gas giant.  However, there are no statistics to draw 
upon to estimate which of these scenarios is more common.  If a roughly equal 
number of both are discovered, it would indicate that the latter is more common 
due to the detection bias.

These results highlight the promising opportunity of making the first exomoon 
detection using the \emph{Kepler} telescope, or photometry of equivalent 
quality, especially the feasibility of detecting habitable-zone exomoons.  
All-sky surveys focussing on bright M-dwarf stars would be ideally placed to 
search for habitable exomoons in greater depth and thus a telescope like 
\emph{TESS} or \emph{PLATO} could continue the search after the 
\emph{Kepler Mission} ends.



%% file: Chapt8.tex
\chapter{Conclusions \& Future Work}
\label{ch:Chapt8}

\vspace{1mm}
\leftskip=4cm

{\it ``
For tomorrow belongs to the people who prepare for it today
''} 

\vspace{1mm}

\hfill {\bf --- African Proverb} 

\leftskip=0cm


\section{Summary}

Twenty years ago, humanity knew of just a handful of planets, most of which
had been known since antiquity. Since this time, a new branch of astronomy
has blossomed and bore many fruits. At the time of writing, over 500 exoplanets
are known to exist and the breadth and depth of understanding of those
worlds has matured from measuring just minimum masses to characterizing their
atmospheres, dynamics and temperatures. The rich history of planet and exoplanet
detection is overviewed in Chapter~\ref{ch:Chapt1}.

The ultimate goal of many exoplanetary scientists is to detect an Earth-like
world; a habitable, temperate abode where beings such as ourselves may just
reside and ask the same questions humanity has asked since the dawn of
civilization. Only now can such worlds be detected and early hints suggest
Earths may be frequent \citep{howard2010} and a detection is round the corner 
\citep{arbesman2010}.

The search for an Earth-like world is centered almost exclusively upon the
search for an Earth-like planet, but this may be unnecessarily restrictive - 
Earth-like moons may also be common throughout the Universe. A moon in the
habitable-zone with a mass $\geq 0.3$\,$M_{\oplus}$ could retain an atmosphere
and satisfy all of the criteria usually associated with habitable planets 
\citep{williams1997}. In Chapter~\ref{ch:Chapt2}, I reviewed the likely 
properties of extrasolar moons based upon physical constraints. The conclusion
was that for a moon to be $\geq 0.3$\,$M_{\oplus}$, a capture or impact origin
is probably necessary. With only two known examples of large satellites 
originating in such a way, the frequency of habitable moons is unknown and
a Copernican view does not favour such a hypothesis. However, it must also
be acknowledged that a Copernican model of planetary systems is an extremely
poor one, with most exoplanetary systems bearing little resemblance to our own.

With every reason to conduct a search for such bodies, the missing piece in
the puzzle is a method to actually detect moons. Whilst several methods have
been proposed for non-transiting systems, which are reviewed in 
Chapter~\ref{ch:Chapt2}, it is the transiting system which has always been
the most generous donor of information for distant worlds. This forms the 
principal goal of the thesis presented here - a method to detect exomoons in 
transiting systems.

Before one can search for the tiny perturbations due to a moon buried within the
planet's transit signal, one must first describe a detailed model of transiting
planets by themselves. In Chapter~\ref{ch:Chapt3}, I presented the basics of
transiting planets, reviewing the current literature on the subject. This
chapter introduces the various assumptions made in the model too, and discusses
the consequences of breaking each one.

It was shown later that exomoons can be detected through changes in the transit
times and durations. Therefore, a necessary step is a detailed understanding of
the transit duration too. Chapter~\ref{ch:Chapt4} is dedicated to the times
of transit minima and the transit duration. It is shown that in both cases an
exact, simple, closed-form expression is elusive and one is forced to deal
with solutions from a bi-quartic equation. Due to the root selection issues
introduced by the bi-quartic, it is preferable to adopt alternative approaches.

For the transit minima, I have presented a series expansion solution up to
$6^{\mathrm{th}}$-order in $\cos^2i_P$ which rapidly converges upon the true
minima solution (Equations~(\ref{eqn:etasT}) \& (\ref{eqn:etasO}) for the 
transit and occultation minima respectively). This series was previously known 
to the second-order, but the additional terms allow one to achieve much higher 
precisions which are frequently required in modern exoplanet timing.

For the transit duration, a physically motivated new approximate formula is
derived which is shown to yield the highest accuracy out of all previously
proposed formulas (Equation~{\ref{eqn:T1}). The equation is easily 
differentiated allowing one to evaluate the effects of secular parameter change,
such as apsidal precession, nodal precession, etc (Equations~(\ref{eqn:apsidal}) 
to (\ref{eqn:eccvar})).

In Chapter~\ref{ch:Chapt3}, three assumptions in the transit model were 
identified as being frequently invalid and to have significant consequences in 
such a scenario. These are A0) a uniform source star; 
A4/A5) the planet emits no flux/there are no background luminous objects; 
A10) the integration times are small. A0 can be compensated for by using the 
\citet{mandel2002} code to account for limb darkening and generally avoid
observing spotty stars. In Chapter~\ref{ch:Chapt5}, I addressed the consequences
of methods of compensation for assumptions A4/5 and A10.

The first of these is the case of a blended source of light in addition to the
host star. The blend source could be a background object or even the planet
itself for very hot planets observed at infrared wavelengths. In all cases,
the blend leads to a dilution of the transit depth. In \S\ref{sec:nightside},
I presented a model for compensating for blend sources, with particular focus on
the self-blend scenario of a hot-planet, since this was previously unconsidered.

For the other assumption, A10, it is shown in \S\ref{sec:binning} that 
long-integration times lead to a smearing of the transit light curve signal.
This smearing severely compromises the parameter retrieval, unless accounted 
for. In the example of the 30-minute integrations of the \emph{Kepler Mission},
the effects are so severe that many planets would be rejected as unphysical
unless the integration time is included in the model. \S\ref{sec:binning}
describes the appropriate model to accomplish this and how to quickly
evaluate the required numerical integration needed.

With the transit model now well-understood, one may return to the original
problem of this thesis - how to detect exomoons in transiting systems.
In Chapter~\ref{ch:Chapt6}, I introduced the transit timing variations (TTV)
method first proposed by \citet{sartoretti1999}. The premise is that the moon
causes the planet to exhibit reflex motion on top of the orbital motion around
the star leading to variations in the position of the planet. These positional
changes manifest as offsets to the times of transit minimum. It is shown that the 
TTV signal is always heavily undersampled due to the orbital constraints of the 
moon, which means one cannot determine a period for the exomoon. Without this
information, TTV only provides mass of the moon multiplied by the distance to
the planet, $a_S M_S$. This was identified as one of two critical problems with
TTV in \S\ref{sec:ttvproblems}. The other problem was the plethora of other
effects which also produce TTV, meaning that one requires a method to 
discriminate between an exomoon TTV and the other sources.

Before tackling these problems, \S\ref{sec:kipttv} provides an updated model
for the TTV accounting for many effects not given in the \citet{sartoretti1999}
model, such as orbital eccentricities and non-coplanarity. I also provide a very
careful and detailed outline of the various assumptions required in the model.
Next, the prediction of velocity induced transit duration variations (TDV-V)
is presented in \S\ref{sec:TDVV}. The planet's reflex motion results in changes
in not only position but also velocity and this causes a broadening and 
shrinking of the transit duration.

It is shown that TDV-V solves both of the problems with TTV alone. Firstly,
TDV-V scales as $M_S/P_S$ meaning the ratio of the TDV-V to TTV RMS amplitudes
(which is determined through excess variance analysis) provides $P_S$ and $M_S$ 
separately. Second, TDV-V exhibits a $\pi/2$ phase shift from TTV which gives 
the observer a unique exomoon signature to hunt for.

In \S\ref{sec:tip}, a second-order effect is explored which I denote as
transit impact parameter induced transit duration variations (TDV-TIP). Here,
the moon causes the planet to exhibit a component of its reflex motion
orthogonal to the planet's trajectory across the stellar face. These orthogonal
kicks cause the transit impact parameter to vary, which the duration is very
sensitive to. The effect is maximized for highly inclined moons, but such
systems are expected to be rare due to the contracted region of Hill stability
\citep{donnison2010}. For co-aligned moons, the effect is still present but
generally an order-of-magnitude less than the TDV-V effect. Still, high
signal-to-noise measurements may be able to exploit the effect to determine the
sense of orbital motion. This is because for prograde orbits TDV-V and TDV-TIP
are in phase and thus constructively interfere, but for retrograde orbits the
phase difference is $\pi$ causing destructive interference. I direct the reader
to Table~\ref{tab:ttvsummary} for a concise summary of the properties of the
various timing effects.

The TTV and TDV effects therefore offer a way of detecting extrasolar moons
in transiting systems, satisfying the original goal of this thesis. However,
until this point, it remains unclear how feasible such an enterprise actually
is. Therefore, in Chapter~\ref{ch:Chapt7}, I present a feasibility study for
detecting exomoons with the \emph{Kepler Mission}, or photometry of equal
quality. The study focusses exclusively on habitable-zone exomoons, since
this not only limits the parameter space of investigation, but also is the case
of greatest interest (Figure~\ref{fig:habmoon}).

The results find that habitable-zone exomoons down to 0.2\,$M_{\oplus}$ may be 
detectable with \emph{Kepler}. The ideal host planets are low-density planets
due to the large transit signal but high reflex motion e.g. Saturn. The ideal
host stars are small and cool to give a large transit depth and short period
habitable-zone e.g. an M-dwarf. The analysis finds that $\sim$25,000 stars may
be surveyed within \emph{Kepler's} field-of-view for 1\,$M_{\oplus}$
habitable-zone exomoons. This exciting prospect means that \emph{Kepler} will
be able to obtain meaningful constraints on the frequency of habitable-zone 
exomoons in the coming years. It also validates TTV and TDV as detection 
techniques, showing that these methods are not just theoretical oddities but 
plausible, viable detection techniques.

\begin{figure}
\begin{center}
\includegraphics[width=16.0 cm]{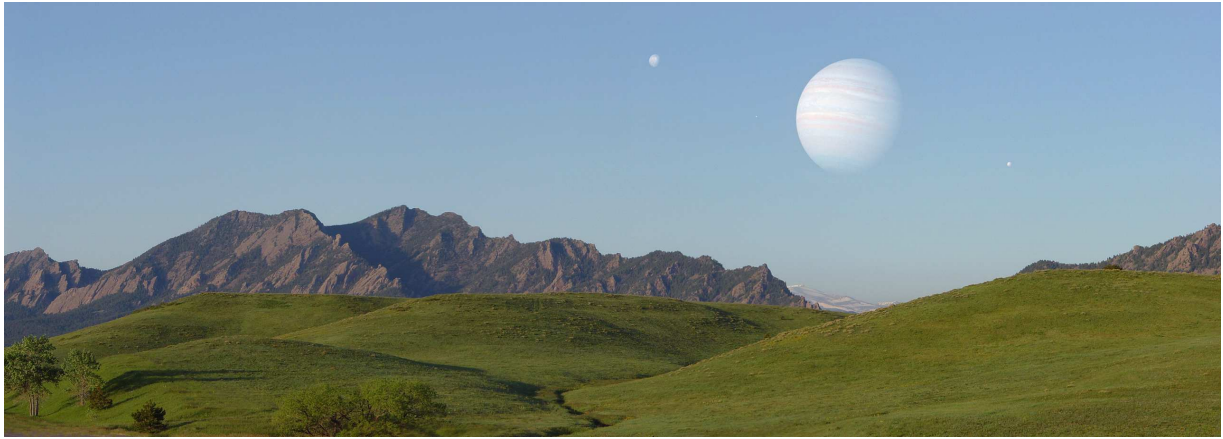}
\caption[Artist's impression of a habitable exomoon. Credit: Dan Durda]
{\emph{Artist's impression of a habitable exomoon. \emph{Kepler} may well
detect such an object in the near future. Credit: Dan Durda}} 
\label{fig:habmoon}
\end{center}
\end{figure}

One of the limitations of \emph{Kepler} is that the telescope is a deep-field
survey meaning most of the target stars are faint by the standards of radial
velocity surveys. Since the transit method cannot provide the mass of a planet,
only its radius, then if and when \emph{Kepler} detects an Earth-sized planet
there will be no way of knowing what the mass of the planet is. As a result,
the composition will be completely unknown which is clearly key to understanding
the likely habitability. In fact, without radial velocity confirmation the
planetary nature of the signal is also questionable with numerous astrophysical
false positives being equally viable. It is in this respect that exomoons
will have a potent advantage.

Extrasolar moons can have their masses determined through the TTV and TDV
effects. The only requirement is a prior knowledge of the planetary mass
through radial velocity measurements, but given that the planetary host may
well be a Jupiter-sized object, RV confirmation will be trivial relative to
the Earth-mass case. Indeed, recently \citet{kipping2011} has gone further
by showing that the exomoon systems could even be used to weigh the mass of the
star directly which removes any stellar evolution dependencies which are usually
invoked. The TTV/TDV signals may be used to compile an exomoon ephemeris and
predict the moments when the moon transits. This would allow for the radius
of the moon to be measured as well and thus the mass and radius would be known.
Finally then, one can see that the composition and structure of the moon could
be inferred by combining the moon's mass and radius with structural models
(e.g. \citet{valencia2006}). In this sense then, not only would the exomoon
detection be more secure, it would also provide a great deal more information
about the nature of the object.

\section{Future Work}

For exolunar science, there is a great deal of work to be done in both the
theoretical and observational aspects. The observational work is fairly
obvious since no exomoons have been detected yet; a dedicated program
to search for their existence is required. I will here though consider the 
theory side, which is most relevant to the theme of this thesis. The model for 
TTV and TDV presented in this thesis can be considered as really the first step 
in the long chain of future developments which can be envisaged and ultimately 
required. I list here some aspects which I identify as requiring further work 
for the TTV and TDV theory specifically:

\begin{itemize}
\item[{\tiny$\blacksquare$}] Generalization of TTV and TDV to multiple moons
\item[{\tiny$\blacksquare$}] Generalization to multiple planet systems, which 
can also induce TTV/TDV
\item[{\tiny$\blacksquare$}] Extension of TTV and TDV to more general three-body 
cases, including long-term variations e.g. nodal precession
\item[{\tiny$\blacksquare$}] Solutions for the exact integrals of the TTV and 
TDV RMS amplitudes using sky-projected distances rather than just the dominant 
Cartesian component
\end{itemize}

One major problem which remains unsolved is the modelling of the transit
light curve of a planet with a moon in a completely analytic manner. This model
needs to include mutual events, eccentricity, the full three-dimensional orbital 
elements, stellar limb darkening and inherently account for TTV and TDV. This
demanding list of requirements is challenging enough, but the entire model
must be completely analytic for a light curve fitting code to have any chance
of expediently fitting real observations. Whilst numerical techniques such as
pixelating the stellar disc are much easier to write down, the analytic solution
is the real prize offering the potential for rapid exploration of the large
numbers of model parameters.

Although not a part of this thesis, I mention here a work in preparation which
has indeed solved this problem. The framework and code for a completely analytic
planet-moon transit modelling routine, called \luna, has been developed by
myself in recent months. \luna\ is a major step forward as it will not only 
model the TTVs and TDVs, but also previously unaccounted for effects such as 
velocity induced ingress/egress asymmetry. Further, \luna\ models the exomoon's 
light curve dip and thus the moon's own reflex motion (including the moon's own 
version of TTV, TDV-V, TDV-TIP, etc) is automatically accounted for. Finally, 
\luna\ also allows one to measure the radius of the moon through the depth of 
the transit features it imparts. Put simply, when implemented in a fitting 
routine, \luna\ will extract every ounce of information imparted onto the light 
curve by the moon. It will be a highly potent weapon in exomoon searches
(for example see Figures~\ref{fig:closemoon} \& \ref{fig:farmoon}).

\begin{figure}
\begin{center}
\includegraphics[width=15.0 cm]{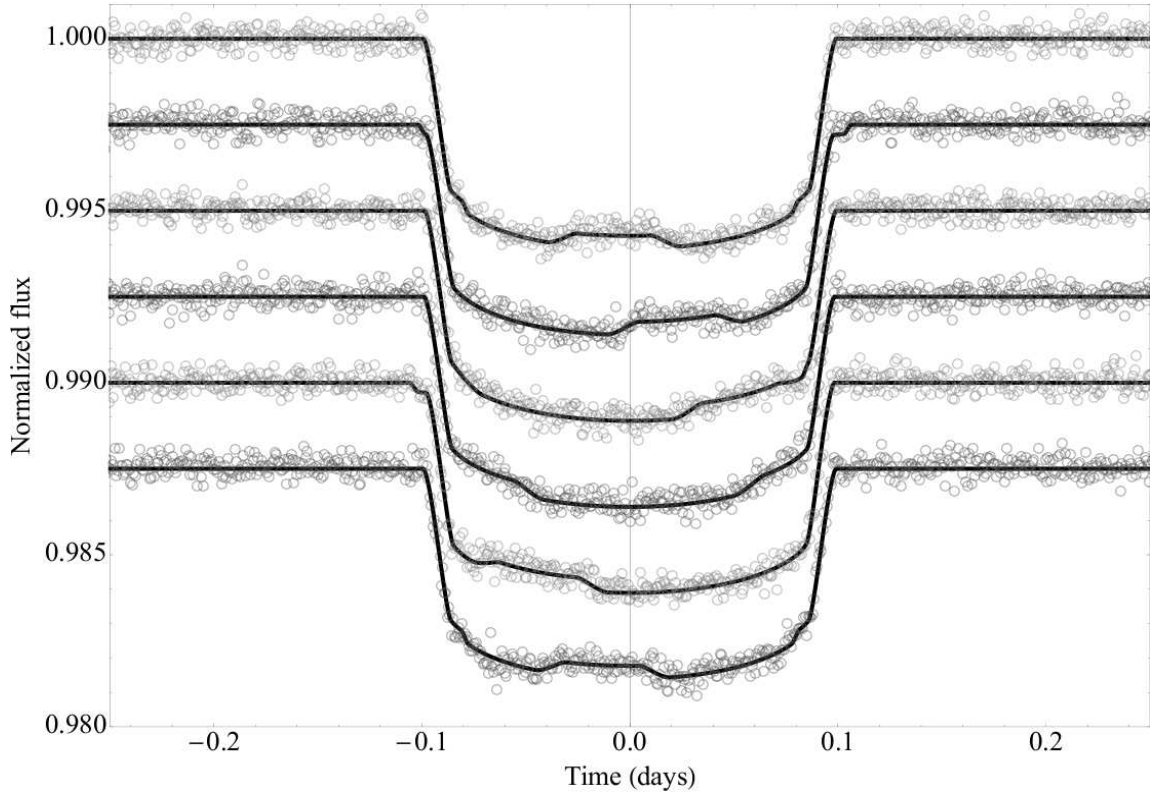}
\caption[Example light curve from the \luna\ code for a close-in moon]
{\emph{Example light curve from the \luna\ code for a moon at 5\% of the Hill
radius around a habitable-zone Neptune around an M2V star. Normal TTV/TDV
methods are invalid here due to the moon's very short period but the
moon can be easily detected through the crossing events as it eclipses the
planet during the transits. This is a work in preparation.}} 
\label{fig:closemoon}
\end{center}
\end{figure}

\begin{figure}
\begin{center}
\includegraphics[width=15.0 cm]{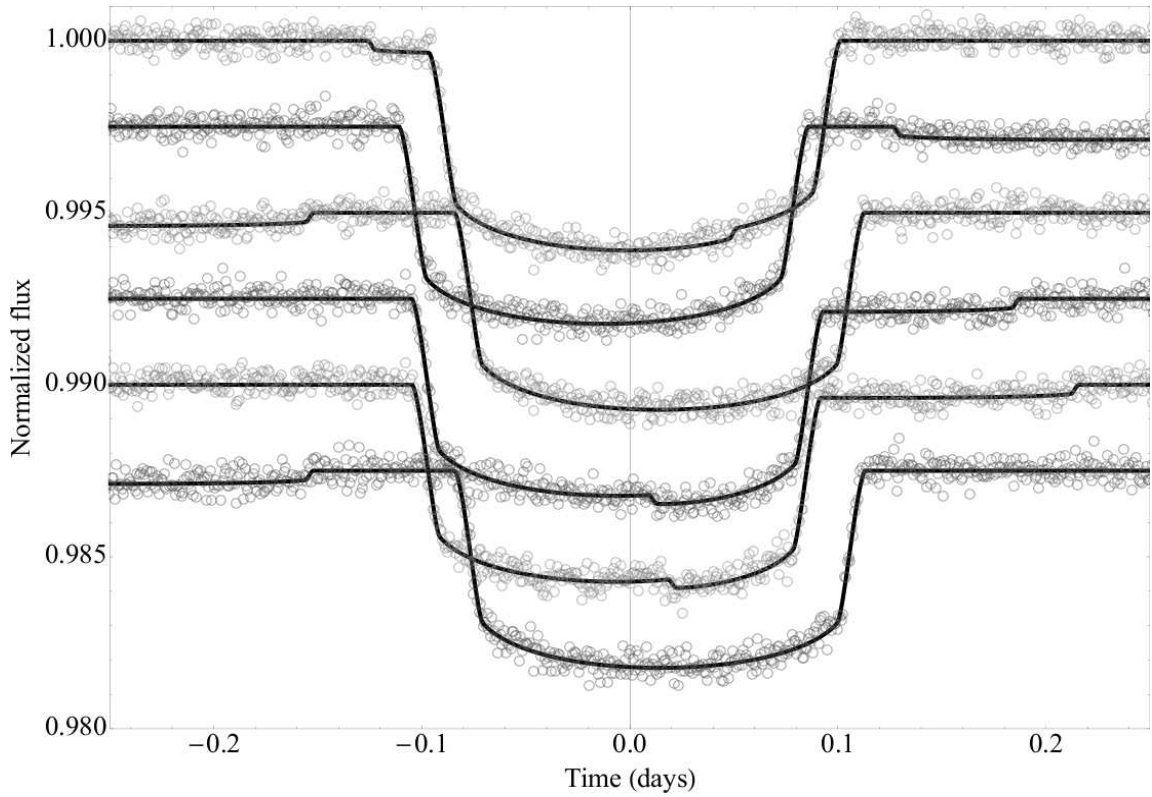}
\caption[Example light curve from the \luna\ code for a far-out moon]
{\emph{Example light curve from the \luna\ code for a moon at the Hill
radius around a habitable-zone Neptune around an M2V star. The TTV effect
is visible by eye but crossing events are less probable due to the moon's
much longer period. This is a work in preparation.}} 
\label{fig:farmoon}
\end{center}
\end{figure}

Despite the obvious power of \luna, TTV and TDV will remain as very useful
methods for finding exomoons. This is because \luna\ is a moon-specific 
algorithm whereas TTV and TDV can be obtained by just modelling a planetary 
transit. It is now becoming a standard aspect of new observations of transiting 
systems to provide the transit timings allowing one to use archived timings to 
search for exomoons, perturbing planets or whatever else an observer is 
interested in. In contrast, \luna\ requires the full photometric time series and 
it is clearly a more specialized and dedicated algorithm.

\luna\ is currently just in the beta-phase and models the transits for a planet
with a single moon, so another obvious extension is to include multiple 
moons. For the time being, a more paramount goal is probably to implement
\luna\ on \emph{Kepler} photometry, which I will be doing in the future. 

This then is a time to be hopeful for finding habitable worlds. An Earth-like
exoplanet could be announced any day now, and the methods for finding Earth-like
moons are now becoming mature enough for effective implementation in a survey
style. Then it seems great insight will be soon forthcoming regarding the 
question which has haunted great thinkers for millennia - \emph{are we alone?}

%% file: AppenA.tex
\chapter{Notations \& Acronyms}
\label{app:notation}

%
%
%
%


\section{Discussion of Notation Used}
\subsection{Introduction}

During the writing of this thesis, a great effort was spent in finding a
self-consistent notation spanning the full breadth of transiting planets and
moons. Within the field of transiting exoplanets, a wide berth of notation
and symbolic representation is employed. Some of the choices are historic, some
convenient, and others enigmatic. The goal of the scheme I used, was to be able
to convey parameters unambiguously, even if that comes at the cost of a somewhat
more elaborate notation scheme.

\subsection{Remaining Ambiguities}

Despite my efforts, some ambiguities do still exist. For example, early on in
the thesis, whilst discussing a planet orbiting a star only, the planetary
properties, such as the period and semi-major axis, are given the subscript 
``P'' to denote planet. However, once a planet-moon system is introduced, the
period of the planet around the star becomes an ambiguous definition and really
one now has two periods: $P_B$ the period of the planet-moon barycentre around 
the star and $P_S$, the period of the satellite around the planet-moon 
barycentre (which must equal the period of the planet around the planet-moon 
barycentre). Whilst $P_B$ is a more encompassing definition, it would be 
confusing to employ $P_B$ rather than $P_P$ early on in the thesis as the 
barycentre would be in reference to the barycentre of the planet itself, which 
is implicit with the standard definition of $P_P$ anyway. Therefore, to remain
consistent, $P_P$ is used for the orbital period of the planet-moon barycentre
around the star later on too.

\subsection{Subscript Scheme}

Subscripts to various parameters come in three flavours. The first is the simple
case of a single subscript letter e.g. $P_P$. In this case, the subscript letter
refers to a type of object e.g. ``P'' for planet, ``S'' for satellite, ``*''
for star, ``b'' for planet b, ``c'' for planet c, etc.

The second type of subscript used is two letters without any separation e.g.
$S_{P*}$, $X_{S*}$, etc. Both letters again refer to objects but here the term
in general means object A to object B. For example, $S$ is the sky-projected
separation between two objects, so for $S_{P*}$ one is referring to the
sky-projected planet-star separation.

The third type of subscript used is with a comma separating the two letters
e.g. $b_{P,T}$, $\varrho_{P,O}$, etc. This refers to the parameter in reference
to the object as the first letter but evaluated at or in reference to some 
special instance denoted by the second letter. For example, $f_{P,T}$ refers to
the true anomaly of the planet, ``P'', evaluated at the time of transit minimum,
which is always given by the subscript ``T''. The other instance commonly used
is ``O'' for occultation.

\subsection{Superscripts}

In some cases, superscripts are used to further the definition of a 
parameter. This is generally done because otherwise the subscripts would become
very messy whereas the superscripts remain an empty slot. In this thesis,
superscripts generally denote some additional property of a parameter and there
is a wide range of its usage. For example, $O_{xy}^{\mathrm{one}}$ denote that
$O_{xy}$ is calculated using the ``one''-term model for the transit duration.
Other options include the ``two''-term model and references to different
authors. Another example would be $\rho_*^{\mathrm{circ}}$, which refers to the
density of the star evaluated for circular orbits (``circ'').

\section{Nomenclature}

There are also various naming schemes employed throughout this thesis which
I believe require some explanation.

\subsection{``Satellite'' vs ``Moon''}

Throughout this thesis, the companions of extrasolar planets are referred to
as ``exomoons'' or ``moons'' rather than ``exosatellites'' or ``satellites''.
Whilst a satellite can denote anything in orbit of a planet, a moon refers
specifically to a natural satellite and therefore is the more apt description
for what I am searching for.

Despite this, the subscript ``S'' is always used in reference to satellite. The
reason for this choice is that ``S'' is the same terminology used by numerous
other authors in regard to exomoons, starting with \citet{sartoretti1999} and
consistently used since. This is an example of a historically consistent 
nomenclature.

\subsection{``Occultation'' vs ``Secondary Eclipse''}

In the exoplanet literature, it is generally more common to denote the 
anti-transit as the secondary eclipse rather than the occultation. However, in 
this thesis, occultation is always used. If one uses secondary eclipse, then the 
transit is dubbed the primary transit. Therefore, a natural choice of subscripts 
would be ``P'' for primary and ``S'' for secondary. However, both ``P'' and 
``S'' subscripts are already used in reference to the planet and satellite. 
Therefore, this is a not a viable choice for our purposes.

Instead, ``transit'' and ``occultation'' give the subscripts ``T'' and ``O''
which are unambiguous against other subscript terms.

\subsection{``Transit Minimum'' vs ``Mid-Transit Time''}

In the exoplanet literature, the timing of the transit is usually given by
what is called the ``mid-transit time''. Throughout this thesis, the 
nomenclature of ``mid-transit time'' is carefully avoided and ``transit
minimum'' or time of the transit minimum is used throughout. This is because the
mid-transit time is highly ambiguous. For an eccentric orbit, the halfway point
(i.e. the mid-point) between the first and last contacts is not the same as the
instant of inferior conjunction. Further, the instant of inferior conjunction is
not the same as the instant when the sky-projected planet-star separation is
minimized. For a limb-darkened star, the minimum of the transit curvature occurs
when $S_{P*}$ is minimized and this represents a completely unambiguous 
definition.

For this reason, ``transit minimum'' is always used. This invalidates several
of the standard terms used to reference this point in time e.g. $t_C$ 
referring to central time and $t_{\mathrm{mid}}$ referring to mid-time. In light
of the ambiguity of mid-times, these terms can be seen to be unsatisfactory
descriptors of the instant of the transit minimum. For that reason, a new
notation is devised, $\tau$, which alleviates these problems.

\section{List of Acronyms}

Table~\ref{tab:acronyms} provides a list of the various acronyms used
throughout this thesis.

\begin{table*}
\caption[List of important acronyms used throughout this thesis]
{\emph{List of important acronyms used throughout this thesis.}} 
\centering 
\begin{tabular}{l l} 
\hline\hline 
Acronym & Definition \\ [0.5ex] 
\hline 
TTV & Transit timing variations \\
TDV & Transit duration variations \\
TDV-V & Velocity induced transit duration variations \\
TDV-TIP & Transit impact parameter induced transit duration variations \\
MCMC & Markov chain monte carlo \\
RMS & Root-mean-square \\
SNR & Signal-to-noise \\ 
FAP & False alarm probability \\
RV & Radial velocity \\ [1ex]
\hline\hline 
\end{tabular}
\label{tab:acronyms} 
\end{table*}

\section{List of Parameters}

Table~\ref{tab:parameters} provides a list of the various key parameters used
throughout this thesis.

\begin{center}
\begin{longtable}{ll}
\caption[List of important parameters used in this paper]
{\emph{List of important parameters used in this paper.}} \\ 
\hline\hline 
\textbf{Parameter} & \textbf{Definition} \\ [0.5ex] 
\hline 
$S'$ & Sky-projected separation between two objects \\
$S$ & Sky-projected separation between two objects, in units of $R_*$ \\
$X'$ & Sky-projected $X$-component position of an object \\
$X$ & Sky-projected $X$-component position of an object, in units of $R_*$ \\
$Y'$ & Sky-projected $Y$-component position of an object \\
$Y$ & Sky-projected $Y$-component position of an object, in units of $R_*$ \\
$Z'$ & Sky-projected $Z$-component position of an object \\
$Z$ & Sky-projected $Z$-component position of an object, in units of $R_*$ \\
$S_{P*}'$ & Sky-projected planet-star separation \\
$S_{P*}$ & Sky-projected planet-star separation, in units of $R_*$ \\
$X_{P*}'$ & Sky-projected $X$-component of planet-star separation \\
$X_{P*}$ & Sky-projected $X$-component of planet-star separation, in units of $R_*$ \\
$Y_{P*}'$ & Sky-projected $Y$-component of planet-star separation \\
$Y_{P*}$ & Sky-projected $Y$-component of planet-star separation, in units of $R_*$ \\
$Z_{P*}'$ & Sky-projected $Z$-component of planet-star separation \\
$Z_{P*}$ & Sky-projected $Z$-component of planet-star separation, in units of $R_*$ \\
$S_{S*}'$ & Sky-projected satellite-star separation \\
$S_{S*}$ & Sky-projected satellite-star separation, in units of $R_*$ \\
$X_{S*}'$ & Sky-projected $X$-component of satellite-star separation \\
$X_{S*}$ & Sky-projected $X$-component of satellite-star separation, in units of $R_*$ \\
$Y_{S*}'$ & Sky-projected $Y$-component of satellite-star separation \\
$Y_{S*}$ & Sky-projected $Y$-component of satellite-star separation, in units of $R_*$ \\
$Z_{S*}'$ & Sky-projected $Z$-component of satellite-star separation \\
$Z_{S*}$ & Sky-projected $Z$-component of satellite-star separation, in units of $R_*$ \\
$S_{S*}'$ & Sky-projected satellite-star separation \\
$S_{S*}$ & Sky-projected satellite-star separation, in units of $R_*$ \\
$X_{S*}'$ & Sky-projected $X$-component of satellite-star separation \\
$X_{S*}$ & Sky-projected $X$-component of satellite-star separation, in units of $R_*$ \\
$Y_{S*}'$ & Sky-projected $Y$-component of satellite-star separation \\
$Y_{S*}$ & Sky-projected $Y$-component of satellite-star separation, in units of $R_*$ \\
$Z_{S*}'$ & Sky-projected $Z$-component of satellite-star separation \\
$Z_{S*}$ & Sky-projected $Z$-component of satellite-star separation, in units of $R_*$ \\
$S_{PS}'$ & Sky-projected planet-satellite separation \\
$S_{PS}$ & Sky-projected planet-satellite separation, in units of $R_*$ \\
$X_{PS}'$ & Sky-projected $X$-component of planet-satellite separation \\
$X_{PS}$ & Sky-projected $X$-component of planet-satellite separation, in units of $R_*$ \\
$Y_{PS}'$ & Sky-projected $Y$-component of planet-satellite separation \\
$Y_{PS}$ & Sky-projected $Y$-component of planet-satellite separation, in units of $R_*$ \\
$Z_{PS}'$ & Sky-projected $Z$-component of planet-satellite separation \\
$Z_{PS}$ & Sky-projected $Z$-component of planet-satellite separation, in units of $R_*$ \\
$f$ & True anomaly of an object around its primary \\
$E$ & Eccentric anomaly of an object around its primary \\
$M$ & Mean anomaly of an object around its primary \\
$f_P$ & True anomaly of the barycentre of the planet (+ any satellites) around the host star \\
$E_P$ & Eccentric anomaly of the barycentre of the planet (+ any satellites) around the host star \\
$M_P$ & Mean anomaly of the barycentre of the planet (+ any satellites) around the host star \\
$f_S$ & True anomaly of the satellite around the host planet \\
$E_S$ & Eccentric anomaly of the satellite around the host planet\\
$M_S$ & Mean anomaly of the satellite around the host planet \\
$e$ & Orbital eccentricity of an object around its primary \\
$e_P$ & Orbital eccentricity of the barycentre of the planet (+ any satellites) around the host star \\
$e_S$ & Orbital eccentricity of the satellite around the host planet \\
$\omega$ & Argument of periapsis of an object around its primary \\
$\omega_P$ & Argument of periapsis of the barycentre of the planet (+ any satellites) around the host star \\
$\omega_S$ & Argument of periapsis of the satellite around the host planet \\
$k_P$ & Lagrangian eccentricity parameter $k$ for the planet; $k_P = e_P \cos \omega_P$ \\
$h_P$ & Lagrangian eccentricity parameter $h$ for the planet; $h_P = e_P \sin \omega_P$ \\
$k_P$ & Lagrangian eccentricity parameter $k$ for the satellite; $k_S = e_S \cos \omega_S$ \\
$k_P$ & Lagrangian eccentricity parameter $h$ for the satellite; $h_S = e_S \sin \omega_S$ \\
$i$ & Orbital inclination of an object around its primary \\
$i_P$ & Orbital inclination of the barycentre of the planet (+ any satellites) around the host star \\
$i_S$ & Orbital inclination of the satellite around the planet-moon barycentre \\
$\Omega$ & Longitude of the ascending node of an object around its primary \\
$\Omega_P$ & Longitude of the ascending node of a planet, relative to the sky-plane \\
$\Omega_S$ & Longitude of the ascending node of a satellite, relative to the orbital plane of the planet \\
$\varpi_S$ & Longitude of the periapsis of a satellite, defined as $\varpi_S = \omega_P + \Omega_S$ \\
$R$ & Radius of an object \\
$R_*$ & Radius of the host star \\
$R_P$ & Radius of the planet \\ 
$R_S$ & Radius of the satellite \\ 
$a$ & Semi-major axis of an object around its primary \\
$a_{P*} = a_P$ & Semi-major axis of the planet around the host star \\
$a_{B*}$ & Semi-major axis of the planet-moon barycentre around the host star \\
$a_{SP} = a_S$ & Semi-major axis of the satellite around the host planet \\
$a_{SB}$ & Semi-major axis of the satellite around the planet-moon barycentre \\
$a_{PB}$ & Semi-major axis of the planet around the planet-moon barycentre \\
$P$ & Orbital period of an object around its primary \\
$P_P$ & Orbital period of the barycentre of the planet (+ any satellites) around the host star \\
$P_S$ & Orbital period of the satellite around the host planet \\
$M$ & Mass of an object \\
$M_*$ & Mass of the star \\
$M_P$ & Mass of the planet \\
$M_S$ & Mass of the satellite \\
$\rho$ & Mean density of an object \\
$\rho_*$ & Mean density of the star \\
$\rho_P$ & Mean density of the planet \\
$\rho_S$ & Mean density of the satellite \\
$r$ & Separation of an object from its primary \\
$r_{P*} = r_P$ & Separation of the planet from the host star \\
$r_{B*}$ & Separation of the planet-moon barycentre from the host star \\
$r_{SP} = r_S$ & Separation of the satellite from the host planet \\
$r_{SB}$ & Separation of the satellite from the planet-moon barycentre \\
$r_{PB}$ & Separation of the satellite from the planet-moon barycentre \\
$\varrho$ & Separation of an object from its primary, in units of $a$ \\
$\varrho_{P*} = r_P$ & Separation of the planet from the host star, in units of $a_{P*}=a_P$ \\
$\varrho_{B*}$ & Separation of the planet-moon barycentre from the host star, in units of $a_{P*}=a_P$ \\
$\varrho_{SP} = r_S$ & Separation of the satellite from the host planet, in units of $a_{SP}=a_S$ \\
$\varrho_{SB}$ & Separation of the satellite from the planet-moon barycentre, in units of $a_{SB}$ \\
$\varrho_{PB}$ & Separation of the satellite from the planet-moon barycentre, in units of $a_{PB}$ \\
$p$ & Ratio of the planet's radius to the stellar radius ($R_P/R_*$) \\
$s$ & Ratio of the satellite's radius to the stellar radius ($R_S/R_*$) \\
$\delta$ & Defined as $p^2$ \\
$\tau_{T}$ & Instant when $\mathrm{d}S_{P*}/\mathrm{d}t = 0$ near inferior conjunction \\
$\tau_{O}$ & Instant when $\mathrm{d}S_{P*}/\mathrm{d}t = 0$ near superior conjunction \\
$b_{P}$ & Impact parameter of the planet, defined as $r_{B*}\cos i_P/R_*$ \\
$b_{S}$ & Impact parameter of the satellite, defined as $r_{SB}\cos i_S/R_P$ \\
$t_{I}$ & Instant when $S_{B*} = 1+p$ and $\dot{S_{B*}} < 0$ \\
$t_{II}$ & Instant when $S_{B*} = 1-p$ and $\dot{S_{B*}} < 0$ \\
$t_{III}$ & Instant when $S_{B*} = 1-p$ and $\dot{S_{B*}} > 0$ \\
$t_{IV}$ & Instant when $S_{B*} = 1+p$ and $\dot{S_{B*}} > 0$ \\
$T_{xy}$ & Time for a planet to move between contact points x and y \\
$\tilde{T}$ & Time between the planet's centre crossing the stellar limb to exiting under the same condition \\
$\delta_{\mathrm{TTV}}$ & RMS amplitude of TTV signal \\
$\delta_{\mathrm{TDV-V}}$ & RMS amplitude of TDV-V signal \\
$\delta_{\mathrm{TDV-TIP}}$ & RMS amplitude of TDV-TIP signal \\
$\delta_{\mathrm{TDV}}$ & RMS amplitude of total TDV signal \\
$\Lambda_{\mathrm{TTV}}$ & Waveform of TTV signal \\
$\Lambda_{\mathrm{TDV-V}}$ & Waveform of TDV-V signal \\
$\Lambda_{\mathrm{TDV-TIP}}$ & Waveform of TDV-TIP signal \\
$\Lambda_{\mathrm{TDV}}$ & Waveform of total TDV signal \\
$\Phi_{\mathrm{TTV}}$ & Enhancement factor of TTV signal \\
$\Phi_{\mathrm{TDV-V}}$ & Enhancement factor of TDV-V signal \\
$\Phi_{\mathrm{TDV-TIP}}$ & Enhancement factor of TDV-TIP signal \\
$\Phi_{\mathrm{TDV}}$ & Enhancement factor of total TDV signal \\
$\eta$ & Equal to $\delta_{\mathrm{TDV}}/\delta_{\mathrm{TTV}}$ \\ [1ex]
\hline\hline 
\label{tab:parameters} 
\end{longtable}
\end{center}

%% file: thesis.bbl
\begin{thebibliography}{208}
\bibitem[{Pearson}(1900)]{pearson1900}
K.~{Pearson}.
\newblock {On the criterion that a given system of deviations from the probable
  in the case of a correlated system of variables is such that it can be
  reasonably supposed to have arisen from random sampling}.
\newblock \emph{Phil. Mag.}, 50:\penalty0 157--175, 1900.
\end{thebibliography}
